# Chemically Induced Dynamic Nuclear Polarization of $^{19}$F Nuclei

A thesis submitted for the degree of
Doctor of Philosophy
by

**Ilya Kuprov**

Trinity term 2005

Corpus Christi College
Oxford University

# Chemically Induced Dynamic Nuclear Polarization of $^{19}$F Nuclei

A thesis submitted for the degree of
Doctor of Philosophy
by

**Ilya Kuprov**
Corpus Christi College
Trinity term 2005

This study explores, both theoretically and experimentally, the photochemically induced dynamic nuclear polarization (photo-CIDNP) of $^{19}$F nuclei, the associated spin relaxation, cross-relaxation and cross-correlation effects, as well as potential applications of $^{19}$F CIDNP to protein structure and folding problems.

It was demonstrated that in the flavin mononucleotide / 3-fluorotyrosine system and in the whole class of structurally related photochemical systems the $^{19}$F spin polarization exceeds the Boltzmann level by nearly two orders of magnitude and may be sustained in this elevated state indefinitely by continuous laser irradiation of the sample.

A theoretical model describing photo-CIDNP magnetization evolution in a continuously photochemically pumped fluorine-proton nuclear system was assembled and tested. It was found that an accurate account of DD-CSA cross-correlation is essential for the correct description of the temporal behaviour and relaxation of the chemically pumped $^{13}$C, $^{15}$N and $^{19}$F nuclear magnetic systems. The sign and the amplitude of the observed $^{19}$F CIDNP effect were found to depend strongly on the effective rotational diffusion correlation time, giving an *in situ* probe of this parameter for these rapidly evolving systems. The origin of this effect has been traced to electron-nuclear cross-relaxation effects in the intermediate radicals.

To counteract light intensity decay in optically dense samples, an efficient and inexpensive method for *in situ* laser illumination of NMR samples inside the narrow superconducting magnet bore has been developed. It utilises a stepwise tapered optical fibre to deliver light uniformly along the axis of a 5 mm NMR tube.

Application of the theoretical and experimental framework described above to the uniformly fluorotyrosinated Green Fluorescent Protein and Trp-cage protein allowed the characterization of the protein surface structure in the immediate vicinity of the fluorotyrosine residues, and provided information on the motional regimes of all the labelled residues.

The results demonstrate the power and versatility of CIDNP based structure investigation methods, particularly for mapping out the solvent-accessible amino acid residues and their surroundings and giving an order of magnitude sensitivity increase in $^{19}$F nuclear magnetic resonance experiments.

# Acknowledgements

I would like to thank Professor Peter Hore for helping me to escape from the ruins of Soviet Union, for the very encouraging and wise supervision, and for continuing support and lots of good advice. I am grateful to Professor Martin Goez, who shared with me his vast practical experience with flash-photolysis and NMR hardware. Many thanks to Prof. Valeri Kopytov, Prof. Margarita Kopytova, Prof. Viktor Bagryansky, Prof. Brian Howard, Prof. Peter Atkins, Prof. Nick Trefethen from Oxford and Novosibirsk, whose excellent lectures have formed the core of what I presently know and a point of reference for my own teaching endeavours. I am indebted to Prof. Ludmila Krylova, Dr. Anatoly Golovin and Dr. Viktor Mamatyuk, for the wise supervision and providing the much needed spectrometer time in the early days.

I thank the Oxford Supercomputing Centre for providing about 2 years' worth of CPU time, Nick Soffe for help with the spectrometer hardware, Pete Biggs and David Temple for keeping the IT infrastructure running, Tim Craggs, Farid Khan and Sophie Jackson for preparing the fluorinated GFP, Prof. Niels Andersen for cooking the fluorinated Trp-cage protein.

Loud cheers go to the jolly and diverse PJH and CRT group crowd. To Nicola Wagner and Chris Rodgers for always being there to share a laugh. To Iain Day for showing me around and helping to get started. To Ken Hun Mok for very helpful discussions, for volunteering to verify my $^{15}N$ and $^{19}F$ relaxation theory calculations, and jointly with Alexandra Yurkovskaya, for showing that science is not all roses. Many thanks to Corpus Christi College for being my home for the last three years.

The financial support from Scatcherd Europeran Foundation and Hill Foundation is gratefully acknowledged with special thanks to Anthony Smith, Alastair Tulloch and Susan Richards of Hill Foundation for always being there to help.

Special thanks to Olga for making the world a much better place to be.

## Publications arising directly from work described in this thesis

1. Chemically amplified $^{19}F$-$^{1}H$ nuclear Overhauser effects
   I. Kuprov, P.J. Hore, *J. Magn. Reson.* **168**, 1-7 (2004).

2. Uniform illumination of optically dense NMR samples
   I. Kuprov, P.J. Hore, *J. Magn. Reson.* **171**, 171-175 (2004).

3. Design and performance of a microsecond time-resolved photo-CIDNP add-on for a high-field NMR spectrometer
   I. Kuprov, M. Goez, P.A. Abbott, P.J. Hore, *Rev. Sci. Instrum.* **77**, 084103 (2005)

4. Increasing the sensitivity of time-resolved photo-CIDNP experiments by multiple laser flashes and temporary storage in the rotating frame
   M. Goez, I. Kuprov, P.J. Hore, *J. Magn. Reson.* **177**, 146-152 (2005)

## Publications related to the work described in this thesis

1. Bidirectional electron transfer in Photosystem I: determination of two distances between $P_{700}^{+}$ and $A_1^{-}$ in spin-correlated radical pairs
   S. Santabarbara, I. Kuprov, W.V. Fairclough, P.J. Hore, S. Purton, P. Heathcote, M.C.W. Evans, *Biochemistry*, **44**, 119-2128 (2005).

# Contents

# Chapter 1
# *Theoretical framework*

This Chapter provides the theoretical and phenomenological background for the rest of this Thesis. It is assumed that the reader is closely familiar with the density operator description of quantum mechanical ensembles and general magnetic resonance.

## 1.1 Brief review of spin relaxation theory in non-viscous liquids

### *1.1.1. General formalism*

The IUPAC Gold Book [1] defines relaxation in the following way: *"if a system is disturbed from its state of equilibrium, it relaxes to that state, and the process is referred to as relaxation. The branch of kinetics concerned with such processes is known as relaxation kinetics"*. Relaxation and its formal description known as relaxation theory are ubiquitous in magnetic resonance.

Beyond providing us with the NMR signal as such, relaxation yields valuable structural information: inter-particle distances (from Overhauser effects [2], residual dipolar couplings [3], paramagnetic shifts and relaxation rates [4]), relative orientations and angles (from cross-correlations between anisotropic interactions [5]), degrees of motion restriction as well as local and global molecular motion correlation times (from magnetic field dependence of relaxation rates [6]). Most of the spin relaxation analysis techniques rely on a remarkably compact and elegant general treatment known as Bloch-Redfield-Wangsness theory [7]. BRW theory is essentially a second-order time-dependent perturbation theory, and although it works extremely well in most cases of practical importance, it is notorious for its bulky intermediate expressions. In the treatment below, however, the tedious mathematics has been relegated to *Mathematica* [8], so we can concentrate on ideas and answers rather than on overcoming algebraic hurdles.

The starting point in the original formulation of relaxation theory [9-11] with a semi-classical treatment of the lattice is the Liouville-von Neumann equation governing the evolution of the density operator $\hat{\rho}$ of an ensemble of systems evolving under a Hamiltonian $\hat{H}$

$$\frac{\partial\hat{\rho}}{\partial t}=-\frac{\mathrm{i}}{\hbar}[\hat{H},\hat{\rho}] \tag{1.1}$$

Without loss of generality, the spin system Hamiltonian can be separated into a static part $\hat{H}_0$, and dynamic part $\hat{H}_1(t)$ so that the ensemble average of the dynamic part is zero

$$\hat{H}(t)=\hat{H}_0+\hat{H}_1(t), \quad \overline{\hat{H}_1(t)}=\hat{0} \tag{1.2}$$

A move to the interaction representation (sometimes called the 'Dirac representation', hence the 'D' superscript) then yields

$$\hat{\rho}^{\mathrm{D}}(t)=e^{\frac{\mathrm{i}}{\hbar}\hat{H}_0 t}\hat{\rho}(t)e^{-\frac{\mathrm{i}}{\hbar}\hat{H}_0 t} \quad \hat{H}_1^{\mathrm{D}}(t)=e^{\frac{\mathrm{i}}{\hbar}\hat{H}_0 t}\hat{H}_1(t)e^{-\frac{\mathrm{i}}{\hbar}\hat{H}_0 t} \quad \frac{\partial\hat{\rho}^{\mathrm{D}}}{\partial t}=-\frac{\mathrm{i}}{\hbar}[\hat{H}_1^{\mathrm{D}},\hat{\rho}^{\mathrm{D}}] \tag{1.3}$$

A formal integration of the last equation in (1.3) yields

$$\hat{\rho}^{\mathrm{D}}(t)=\hat{\rho}^{\mathrm{D}}(0)-\frac{\mathrm{i}}{\hbar}\int_0^t[\hat{H}_1^{\mathrm{D}}(t'),\hat{\rho}_i^{\mathrm{D}}(t')]dt' \tag{1.4}$$

Re-inserting (1.4) into (1.3) yields the following integro-differential equation, which is still exact and fully equivalent to (1.1)

$$\frac{\partial\hat{\rho}^{\mathrm{D}}(t)}{\partial t}=-\frac{\mathrm{i}}{\hbar}[\hat{H}_1^{\mathrm{D}}(t),\hat{\rho}^{\mathrm{D}}(0)]-\frac{1}{\hbar^2}\int_0^t[\hat{H}_1^{\mathrm{D}}(t),[\hat{H}_1^{\mathrm{D}}(t'),\hat{\rho}^{\mathrm{D}}(t')]]dt' \tag{1.5}$$

Taking an ensemble average of this equation will cause the first term on the right hand side to vanish because of condition (1.2) on the ensemble average of $\hat{H}_1(t)$. We then have

$$\frac{\partial\overline{\hat{\rho}^{\mathrm{D}}(t)}}{\partial t}=-\frac{1}{\hbar^2}\int_0^t\overline{[\hat{H}_1^{\mathrm{D}}(t),[\hat{H}_1^{\mathrm{D}}(t'),\hat{\rho}^{\mathrm{D}}(t')]]}dt' \tag{1.6}$$

We continue by expanding the dynamic part of the Hamiltonian in terms of a complete orthonormal set of operators, writing

$$\hat{H}_1(t)=\sum_j f_j(t)\hat{K}_j=\sum_j f_j^*(t)\hat{K}_j^\dagger \tag{1.7}$$

In this expansion the basis operators are static and the coefficients are stochastic functions of time. In solution magnetic resonance, where the relaxation is often caused by rotational modulation of anisotropic interactions, the most convenient choice is irreducible spherical tensor operators [12] for $\hat{K}_j$ and Wigner matrix elements [13] for $f_j(t)$. After making this substitution, the equation (1.6) becomes

$$\frac{\partial \overline{\hat{\rho}^{\mathrm{D}}(t)}}{\partial t}=-\frac{1}{\hbar^2}\sum_{i,j}\int_0^t \overline{f_i(t)f_j^*(t')[\hat{K}_i^{\mathrm{D}}(t),[\hat{K}_j^{\mathrm{D}\dagger}(t'),\hat{\rho}^{\mathrm{D}}(t')]]}dt' \tag{1.8}$$

In the general case this equation is still just as difficult to solve as (1.1). In non-viscous liquids however, we can make further progress by assuming that the ensemble averages may be performed independently on the rotational function part and the spin operator part of (1.8), this approximation was experimentally found to be valid [7]. After noting that the expansion operators $\hat{K}_j^{\mathrm{D}}(t)$ are the same throughout the ensemble and dropping the overbar on the density matrix for convenience, we can write

$$\frac{\partial \hat{\rho}^{\mathrm{D}}(t)}{\partial t}=-\frac{1}{\hbar^2}\sum_{i,j}\int_0^t \overline{f_i(t)f_j^*(t')}[\hat{K}_i^{\mathrm{D}}(t),[\hat{K}_j^{\mathrm{D}\dagger}(t'),\hat{\rho}^{\mathrm{D}}(t')]]dt' \tag{1.9}$$

If the rotational functions in (1.7) are stationary random functions (which is normally the case in an equilibrated fluid), their correlation functions $\overline{f_i(t)f_j^*(t')}$ only depend on the time offset $\tau=t'-t$, so that we can write

$$\overline{f_i(t)f_j^*(t')}=\overline{f_i(t)f_j^*(t+\tau)}=g_{ij}(\tau) \tag{1.10}$$

For molecules with an effective dynamic radius smaller than 50 Å in common solvents the correlation functions (1.10) decay with characteristic times (called *correlation times*) of the order of pico- to nanoseconds. To eliminate the integral in (1.9) and facilitate the actual calculation, the so-called *coarse-graining* of time is introduced, in which it is assumed that the entire integral that wraps $\hat{\rho}^{\mathrm{D}}(t')$ can be approximated by a constant superoperator acting on $\hat{\rho}^{\mathrm{D}}(t)$. This seemingly gross approximation is actually quite accurate so long as the resulting relaxation time is much longer than the correlation time of the stochastic functions (1.10). The latter assumption also allows us to extend the upper limit of the integral to infinity. After changing the variables so that $\tau=t'-t$, we obtain

$$\frac{\partial \hat{\rho}^{\mathrm{D}}(t)}{\partial t}=-\frac{1}{\hbar^2}\sum_{i,j}\int_0^\infty g_{ij}(\tau)[\hat{K}_i^{\mathrm{D}}(t),[\hat{K}_j^{\mathrm{D}\dagger}(t+\tau),\hat{\rho}^{\mathrm{D}}(t)]]d\tau \tag{1.11}$$

After moving back to Schrödinger representation the equation becomes

$$\frac{\partial \hat{\rho}(t)}{\partial t}=-\frac{\mathrm{i}}{\hbar}[\hat{H}_0,\hat{\rho}(t)]-\frac{1}{\hbar^2}\sum_{i,j}\int_0^\infty g_{ij}(\tau)[\hat{K}_i,[\left(e^{\frac{\mathrm{i}}{\hbar}\hat{H}_0\tau}\hat{K}_j e^{-\frac{\mathrm{i}}{\hbar}\hat{H}_0\tau}\right)^\dagger,\hat{\rho}(t)]]d\tau \tag{1.12}$$

The integral in the expression (1.12) is a superoperator-valued Fourier transform over positive times of the correlation functions $g_{ij}(\tau)$. While it may be technically difficult to compute, all the fundamental difficulties have now been eliminated. To account for a finite lattice temperature, and therefore a non-zero equilibrium state, we must phenomenologically replace $\hat{\rho}(t')$ by $\hat{\rho}(t')-\hat{\rho}_{\text{eq}}$. This last step can only be justified in the full quantum mechanical treatment of the lattice [7], which is outside the scope of this thesis. Since the equilibrium density matrix commutes with the static Hamiltonian, we also insert it into the first commutator on the right hand side to make the expression uniform

$$\frac{\partial\hat{\rho}(t)}{\partial t}=-\frac{\mathrm{i}}{\hbar}[\hat{H}_0,\hat{\rho}(t)-\hat{\rho}_{\text{eq}}]-\frac{1}{\hbar^2}\sum_{i,j}\int_0^\infty g_{ij}(\tau)[\hat{K}_i,[e^{\frac{\mathrm{i}}{\hbar}\hat{H}_0\tau}\hat{K}_j^\dagger e^{-\frac{\mathrm{i}}{\hbar}\hat{H}_0\tau},\hat{\rho}(t)-\hat{\rho}_{\text{eq}}]]d\tau \quad (1.13)$$

This is now a first order ordinary differential equation in time. It can be solved directly, both analytically and numerically, if the expressions for the correlation functions $g_{ij}(\tau)$ are known. The exact form of $g_{ij}(\tau)$ depends on the choice of functions in the expansion (1.7) and on the motional dynamics model used [14]; among the common choices are the isotropic rotational diffusion approximation [15] and the Lipari-Szabo restricted motion model [16-18]. After the integral in (1.13) is evaluated, the final density matrix element evolution equations have the following form

$$\left\{\frac{\partial\rho_{ij}}{\partial t}=\sum_{k,l}R_{ijkl}(\rho_{lk}-\rho_{lk}^{\text{eq}})\right. \quad (1.14)$$

The optional last step in the derivation is to translate this system of equations into the equations for the physical observables using any convenient operator basis set. The choice of the basis is a question of taste, but there are a few which are particularly convenient [12]. Taking the trace with the chosen operator basis gives a final equation for the evolution of the corresponding full system of observables $A_i$

$$\left\{\frac{d}{dt}\langle A_i\rangle=\sum_j k_{ij}\left(\langle A_j\rangle-\langle A_{0,j}\rangle\right)\right. \quad (1.15)$$

in which the analytical expressions for the coefficients are usually rather complicated, especially for systems with multiple cross-correlations [14]. The reader may notice that by allowing for an arbitrary basis set in the operator expansion of the density matrix we implicitly allow complex observables. Philosophers of science might be annoyed, but complex-valued observables (e.g. transverse magnetization corresponding to non-Hermitian $\hat{I}_\pm$ operators) make perfect sense in NMR.

### *1.1.2 Relaxation and cross-relaxation caused by magnetic dipole interaction*

The magnetic dipole-dipole interaction[1] is the most ubiquitous relaxation-inducing mechanism in magnetic resonance of spin-½ systems, both diamagnetic and paramagnetic. In the diamagnetic case, nuclear relaxation is caused by translational and rotational modulation of the inter-nuclear dipolar interaction. In paramagnetic molecules, nuclear and electron relaxation occur largely due to electron-nucleus and inter-electron dipolar interactions.

[1] The magnetic dipolar interaction is henceforth referred to as just the *dipolar interaction*, as we will not be considering electric dipolar interaction anywhere in this Thesis.

The phenomenon of inter-nuclear dipolar cross-relaxation is usually referred to as the *Nuclear Overhauser Effect (NOE)* even though the original work by Albert Overhauser considered an electron-nucleus system [19]. The genuine *Overhauser Effect*, which is electron-nuclear cross-relaxation, is called *Dynamic Nuclear Polarization (DNP)*. DNP is caused by the modulation not only of the dipolar interaction, but also of the electron-nucleus contact interaction. Both NOE and DNP are exploited in modern NMR spectroscopy: NOE is one of the cornerstones of NMR structure elucidation [20] and the almost forgotten DNP has recently been dusted off and used for NMR sample pre-polarization [21].

We start by considering nuclear paramagnetic relaxation arising from electron-nucleus interactions. There are two principal magnetic interactions between an unpaired electron and a nucleus: *contact* and *dipolar* [22]. Both have a common origin – electron-nuclear dipolar interaction: the isotropic contact term reflects the interaction at close distances where the finite size of the nucleus is important, and the anisotropic traceless dipolar term arises from the (effectively point dipole) electron-nucleus interaction at large distances.

To be a cause of spin relaxation, an interaction has to be modulated with a non-zero fluctuation spectral power density at the Zeeman frequency. For rigid molecules the contribution to relaxation from the contact mechanism is negligibly small, because the modulation frequency lies in the infra-red. The dipolar mechanism, on the contrary, is nearly always active, because the dipolar interaction tensor is aligned with molecular structure and the interaction is modulated at the rate of molecular rotational diffusion. The latter has a timescale of nanoseconds and usually has some spectral density at NMR and EPR transition frequencies. The anisotropic electron-nuclear dipole interaction is therefore one of the dominant interactions determining spin relaxation rates in conformationally rigid radicals [4].

The energy of interaction between two classical point magnetic dipoles $\vec{\mu}_1$, $\vec{\mu}_2$ separated by a distance $\vec{r}$ is given by a well-known relation [22]

$$E = -\frac{\mu_0}{4\pi}\left(\frac{(\vec{\mu}_1 \cdot \vec{\mu}_2)}{r^3} - \frac{3(\vec{\mu}_1 \cdot \vec{r})(\vec{\mu}_2 \cdot \vec{r})}{r^5}\right) = \vec{\mu}_1 \cdot \mathbf{A} \cdot \vec{\mu}_2 , \tag{1.16}$$

in which $\mathbf{A}$ is a traceless second rank tensor, called the *dipolar interaction tensor*. For the interaction of two point dipoles this tensor is axial. For our purposes the nucleus may be considered, to a very high accuracy, to be a point magnetic dipole. The electron density however, is in most cases delocalized, so the expression for the spin interaction energy should involve an integral over the electron density distribution

$$E = \vec{\mu}_{\mathrm{n}} \cdot \left( \int_V \mathbf{A}(\vec{r}) \left| \psi_{\mathrm{e}}(\vec{r}) \right|^2 dV \right) \cdot \vec{\mu}_{\mathrm{e}} \tag{1.17}$$

in which $\vec{r}$ is the electron-nucleus distance vector. We may take $\vec{\mu}_{\mathrm{e}}$ out of the integral here because the electron magnetic moment is related to the spin and does not depend on coordinate and spatial distribution. Consequently, only the magnitude of $\vec{\mu}_{\mathrm{e}}$ is modulated in space, but not direction. With each particular point of the electron distribution the electron-nucleus dipolar interaction is axial, but adding differently oriented axialities may yield a tensor with non-zero rhombicity.

In a quantum-mechanical formulation, for an electron-nucleus two-spin system in a strong magnetic field the spin Hamiltonian has the following form

$$\hat{H} = \omega_{\mathrm{e}} \hat{I}_Z + \omega_{\mathrm{n}} \hat{S}_Z + A\left(\hat{\vec{I}} \cdot \hat{\vec{S}}\right) + \hat{\vec{I}} \cdot \mathbf{A} \cdot \hat{\vec{S}} \tag{1.18}$$

in which the first two terms describe electron and nuclear Zeeman interaction with the magnetic induction applied, as usual, along *Z*-axis of the laboratory frame, and the dipolar interaction explicitly split into the isotropic and anisotropic parts. The anisotropic part of the electron-nucleus dipolar interaction is a traceless second rank tensor that acts as an operator in a spin space and at the same time transforms according to the $l = 2$ irreducible representation of the SU(2) group[2] under molecular rotation in a Cartesian space [23]. The evaluation of the nuclear magnetic relaxation rates becomes particularly simple if we choose to exploit the properties of SU(2).

Any tensor may be decomposed into a linear combination of the special type of tensors that have ordered rotation properties – the irreducible spherical tensors, so called because they form irreducible representations of SU(2) [12]. The irreducible spherical tensor operators for a single spin ½ are:

$$\hat{T}_{1,-1} = \begin{pmatrix} 0 & 0 \\ \frac{1}{\sqrt{2}} & 0 \end{pmatrix}; \quad \hat{T}_{1,0} = \begin{pmatrix} \frac{1}{2} & 0 \\ 0 & -\frac{1}{2} \end{pmatrix}; \quad \hat{T}_{1,1} = \begin{pmatrix} 0 & -\frac{1}{\sqrt{2}} \\ 0 & 0 \end{pmatrix}; \tag{1.19}$$

and higher rank operators may be obtained from (1.19) by constructing linear combinations of their direct products. This procedure is completely identical to the highest weight construction procedure of adding angular momenta, and the coefficients are the familiar 3*j*-symbols [23]:

$$\hat{T}_{k,q} = \sum_{q_1,q_2} (-1)^{-k_1+k_2-q} \sqrt{2k+1} \begin{pmatrix} k_1 & k_2 & k \\ q_1 & q_2 & -q \end{pmatrix} \hat{T}_{k_1,q_1} \hat{T}_{k_2,q_2}$$
$$q = q_1 + q_2, \quad k = |k_1 - k_2|, \ |k_1 - k_2| + 1, \ \ldots, k_1 + k_2 \tag{1.20}$$

[2]Special unitary group SU(2) (also called unitary unimodular group) is a Lie group of 2×2 unitary matrices with determinant +1. In the corresponding Lie algebra, called su(2), the convenient generators were suggested by Wolfgang Pauli and are known as Pauli matrices. SU(2) is homomorphous with SO(3), which is a Lie group or all rotations of a three-dimensional Cartesian space.

In the case of magnetic dipolar interaction between two spins ½, these are specific combinations of two-spin operators

$$\begin{aligned}
\hat{T}_{2,2} &= +\frac{1}{2}\hat{S}_+\hat{I}_+ \\
\hat{T}_{2,1} &= -\frac{1}{2}\left(\hat{S}_Z\hat{I}_+ + \hat{S}_+\hat{I}_Z\right) \\
\hat{T}_{2,0} &= +\sqrt{\frac{2}{3}}\left(\hat{S}_Z\hat{I}_Z - \frac{1}{4}\left(\hat{S}_+\hat{I}_- + \hat{S}_-\hat{I}_+\right)\right) \\
\hat{T}_{2,-1} &= +\frac{1}{2}\left(\hat{S}_Z\hat{I}_- + \hat{S}_-\hat{I}_Z\right) \\
\hat{T}_{2,-2} &= +\frac{1}{2}\hat{S}_-\hat{I}_-
\end{aligned} \tag{1.21}$$

The following relation allows straightforward translation of two-spin product operator notation into the irreducible spherical tensor notation

$$a\hat{S}_X\hat{I}_X + b\hat{S}_Y\hat{I}_Y + c\hat{S}_Z\hat{I}_Z = \frac{a-b}{2}\hat{T}_{2,-2} + \frac{a-b}{2}\hat{T}_{2,2} + \frac{2c-(a+b)}{\sqrt{6}}\hat{T}_{2,0} \tag{1.22}$$

The three coefficients on the right hand side are related to axiality and rhombicity of the hyperfine interaction tensor.

The irreducible spherical tensors of a given rank transform only into each other under any rotation

$$\hat{\hat{R}}(\alpha,\beta,\gamma)\hat{T}_{l,m} = \sum_{m'} \hat{T}_{l,m'} \mathfrak{D}^{(l)}_{m',m}(\alpha,\beta,\gamma) \tag{1.23}$$

In this expression $\hat{\hat{R}}(\alpha,\beta,\gamma)$ denotes a molecular rotation with Euler angles $\alpha,\beta,\gamma$ and the coefficients $\mathfrak{D}^{(l)}_{m',m}(\alpha,\beta,\gamma)$ are known as *Wigner matrix elements*. In the isotropic rotational diffusion approximation the correlation functions of Wigner matrix elements may be obtained analytically [13], making it particularly simple to evaluate the correlation functions in (1.10)

$$\left\langle \mathfrak{D}^{(l)*}_{m,m'}\left(\alpha(0),\beta(0),\gamma(0)\right)\mathfrak{D}^{(k)}_{n,n'}\left(\alpha(t),\beta(t),\gamma(t)\right)\right\rangle = \frac{\delta_{m,n}\delta_{m',n'}\delta_{l,k}}{2k+1} e^{-\frac{t}{\tau_c}} \tag{1.24}$$

The characteristic decay time appearing in the autocorrelation function is called the *correlation time*. Other rotational diffusion models result in slightly more complicated formulae, but in most cases they are still very straightforward to deal with. Although the Wigner matrix elements $\mathfrak{D}^{(l)}_{m',m}\left(\alpha,\beta,\gamma\right)$ are well known and tabulated, in the relaxation theory treatment we do not need to know their explicit functional form, because only the correlation functions like those in (1.24) appear in the result.

In the relaxation theory treatment described above the Hamiltonian (1.18) must be split up into the static part $\hat{H}_0$ and the time-dependent interaction part $\hat{H}_1\left(t\right)$, resulting in the following expressions

$$\begin{aligned}
&\hat{H}_0 = \omega_{\mathrm{e}}\hat{I}_Z + \omega_{\mathrm{n}}\hat{S}_Z + A(\hat{\vec{I}}\cdot\hat{\vec{S}})\\
&\hat{H}_1\left(t\right) = \frac{Rh}{2}\sum_{m'=-2}^{2}\hat{T}_{2,m'}\mathfrak{D}^{(2)}_{m',-2}\left(t\right) + \frac{Rh}{2}\sum_{m'=-2}^{2}\hat{T}_{2,m'}\mathfrak{D}^{(2)}_{m',2}\left(t\right) + \frac{Ax}{\sqrt{6}}\sum_{m'=-2}^{2}\hat{T}_{2,m'}\mathfrak{D}^{(2)}_{m',0}\left(t\right)\\
&A = \frac{A_{XX}+A_{YY}+A_{ZZ}}{3} \qquad Ax = 2A_{ZZ} - \left(A_{XX}+A_{YY}\right) \qquad Rh = A_{YY} - A_{XX}
\end{aligned} \tag{1.25}$$

The Wigner coefficients in (1.25) inherit their time dependence from the Euler angles specifying molecular orientation. Solving equation (1.13) with this Hamiltonian in the isotropic rotational diffusion approximation for the longitudinal nuclear magnetization, yields the following relaxation rate (its reciprocal quantity is known as the $T_1$ *time*):

$$\begin{aligned}
\frac{1}{T_1} = &\frac{\left(Ax^2+3Rh^2\right)\tau_{\mathrm{c}}}{360}\times\\
&\left(\frac{1}{1+\left(\omega_{\mathrm{n}}-\omega_{\mathrm{e}}\right)^2\tau_{\mathrm{c}}^2} + \frac{6}{1+\left(\omega_{\mathrm{n}}+\omega_{\mathrm{e}}\right)^2\tau_{\mathrm{c}}^2} + \frac{1}{2}\left(\frac{3}{1+\left(\omega_{\mathrm{n}}-A/4\right)^2\tau_{\mathrm{c}}^2} + \frac{3}{1+\left(\omega_{\mathrm{n}}+A/4\right)^2\tau_{\mathrm{c}}^2}\right)\right)
\end{aligned} \tag{1.26}$$

Although the isotropic part of the hyperfine coupling does enter the frequency-dependent terms, we may neglect it because even in the extreme case of fluorine in the 4-fluorophenoxyl radical the $A/4$ term is just 3% of the nuclear Larmor frequency $\omega_{\mathrm{n}}$. Furthermore, the nuclear Larmor frequency may be neglected compared to the electron one. These two approximations result in a substantial simplification to equation (1.26):

$$\frac{1}{T_1} = \frac{\left(Ax^2+3Rh^2\right)\tau_{\mathrm{c}}}{360}\left(\frac{7}{1+\omega_{\mathrm{e}}^2\tau_{\mathrm{c}}^2} + \frac{3}{1+\omega_{\mathrm{n}}^2\tau_{\mathrm{c}}^2}\right) \tag{1.27}$$

This equation will be used in Chapter 8 to compare the experimental paramagnetic relaxation rates to those calculated using *ab initio* hyperfine tensors. Chapter 8 also contains an extensive analysis and complete tables of relaxation pathways and rates in an electron-nucleus two-spin system.

The derivation outlined above considers a general (non-zero rhombicity) case of dipolar interaction tensor. In the inter-nuclear case, the dipolar interaction may be considered to originate from point-dipoles to a very high accuracy. The rhombicity of such an interaction is zero. The role of scalar interaction part is played by *J*-coupling[3], which is always very small compared to the nuclear Zeeman frequency. In the inter-nuclear case equation (1.26) transforms into:

$$\frac{1}{T_1}=\frac{Ax^2\tau_c}{360}\left(\frac{3}{1+\omega_1^2\tau_c^2}+\frac{6}{1+\left(\omega_1+\omega_2\right)^2\tau_c^2}+\frac{1}{1+\left(\omega_1-\omega_2\right)^2\tau_c^2}\right) \quad (1.28)$$

in which indices enumerate the nuclei and the expression for the dipolar tensor axiality may be obtained from the definition of the magnetic dipolar coupling (1.16):

$$Ax=6\frac{\gamma_1\gamma_2\hbar}{r^3}\frac{\mu_0}{4\pi} \quad (1.29)$$

so that well-known formulae are obtained for the longitudinal relaxation rate:

$$\frac{1}{T_1}=\frac{\gamma_1^2\gamma_2^2\hbar^2}{10}\left(\frac{\mu_0}{4\pi}\right)^2\frac{\tau_c}{r^6}\left(\frac{3}{1+\omega_1^2\tau_c^2}+\frac{6}{1+\left(\omega_1+\omega_2\right)^2\tau_c^2}+\frac{1}{1+\left(\omega_1-\omega_2\right)^2\tau_c^2}\right) \quad (1.30)$$

and the inter-nuclear dipolar cross-relaxation rate:

$$\sigma_{12}=\frac{\gamma_1^2\gamma_2^2\hbar^2}{10}\left(\frac{\mu_0}{4\pi}\right)^2\frac{\tau_c}{r^6}\left(\frac{6}{1+\left(\omega_1+\omega_2\right)^2\tau_c^2}-\frac{1}{1+\left(\omega_1-\omega_2\right)^2\tau_c^2}\right) \quad (1.31)$$

These expressions for the dipolar auto- and cross-relaxation rates are quoted without derivation in quite a few books [24, 25]. The enclosed DVD contains a *Mathematica* program that performs the rather bulky analytical derivation.

The third and a much less common type of dipolar relaxation is caused by the rotational and translational modulation of the zero-field splitting. It is specific to the systems containing more than one unpaired electron and originates from the inter-electron dipolar interaction. In the case of a radical pair, not only angles, but also the inter-electron distance is a stochastic function of time. Therefore the expressions for the correlation functions are usually more complicated than either simple exponentials or Lipari-Szabo biexponentials [16, 26]. The ZFS mechanism is only operational in long-lived triplet and higher multiplicity states, biradicals and confined radical systems. Such systems will not be encountered in the present thesis, and the interested reader is therefore referred to the original papers on the matter [27-29].

[3] *J*-coupling, very strictly speaking, is also tensorial due to spin-orbit and spin-dipole contributions. However, the commonly encountered *J*-couplings are dominated by the Fermi contact terms, and therefore may be considered isotropic to a very high accuracy.

### *1.1.3 Relaxation caused by rotational modulation of the Zeeman interaction*

The nuclear and electron Zeeman interaction anisotropy, that is, the orientation dependence of the spin energy level splitting in an external magnetic field, is another common cause of spin relaxation in both radicals and neutral molecules. The electron *g*-tensor acquires its orientation dependence predominantly from the spin-orbit coupling, because the orbital contribution to the electron magnetic moment is anisotropic [30], and the nuclear magnetic shielding essentially inherits the symmetry of the surrounding electron shell [31].

Before embarking on the relaxation analysis, a few more words must be said about the chemical shift tensor. The theoretical explanation of the phenomenon of chemical shift as a partial magnetic screening of a nucleus by its surrounding electrons was given as early as 1950 by Ramsey [31-33], who considered the chemical shift of the nuclear precession frequency to be an annoying small correction term arising in the measurements of nuclear magnetic moments. He developed the theoretical framework necessary to compute the chemical shifts, but the actual calculation proved challenging for multi-atomic systems, and it was not until around 1990 that robust and computationally feasible methods started to emerge [34]. Adequate computing power was reached by common desktop computers in around the year 2000. In this work we use *ab initio* chemical shift tensor computation to complement experimental measurements. As of present, solid state NMR results and calculations based on X-ray diffraction data are still much more accurate [35].

In the vast majority of cases the chemical shift tensor is strongly anisotropic, even for protons. The typical anisotropy of the proton chemical shift is a few ppm, and in the case of $^{15}N$ or $^{19}F$ nuclei it can reach hundreds of ppm and starts to play a major role in determining the relaxation behaviour of the spin system. A particular case of the influence of chemical shielding anisotropy (CSA) on nuclear relaxation which is relevant to the present work is briefly outlined below.

The Hamiltonian for a single spin-½ with an anisotropic Zeeman interaction with the external magnetic field contains the constant term $H_0$, corresponding to either the *g*-factor or the chemical shift, and the time-dependent term $H_1(t)$ resulting from the rotational modulation of the anisotropy:

$$\hat{H}_0 = \omega_{\mathrm{n}} \hat{I}_Z$$
$$\hat{H}_1(t) = \frac{Rh}{2} \sum_{m'=-2}^{2} \hat{T}_{2,m'} \mathfrak{D}^{(2)}_{m',-2}(t) + \frac{Rh}{2} \sum_{m'=-2}^{2} \hat{T}_{2,m'} \mathfrak{D}^{(2)}_{m',2}(t) + \frac{Ax}{\sqrt{6}} \sum_{m'=-2}^{2} \hat{T}_{2,m'} \mathfrak{D}^{(2)}_{m',0}(t) \tag{1.32}$$

The axiality and rhombicity parameters are defined in the same way as in (1.25), but now refer to the Zeeman interaction tensor. Because the magnetic field is assumed to be directed along the *Z*-axis of the laboratory frame, most of the terms in the irreducible spherical tensors will vanish:

$$\begin{aligned}
&\hat{T}_{2,2} = 0 \\
&\hat{T}_{2,1} = -\frac{1}{2}\hat{I}_{+}B_Z \\
&T_{2,0} = +\sqrt{\frac{2}{3}}\hat{I}_Z B_Z \\
&\hat{T}_{2,-1} = +\frac{1}{2}\hat{I}_{-}B_Z \\
&\hat{T}_{2,-2} = 0
\end{aligned} \tag{1.33}$$

Evaluating the right hand side of equation (1.13) with the Hamiltonian (1.32) results in the following relaxation rates for the longitudinal and transverse magnetization:

$$\begin{aligned}
\frac{1}{T_1} &= \frac{Ax^2 + 3Rh^2}{30}\frac{\tau_c}{1+\tau_c^2\omega_Z^2} \\
\frac{1}{T_2} &= \frac{\left(Ax^2 + 3Rh^2\right)\tau_c}{180}\left(4 + \frac{3}{1+\tau_c^2\omega_Z^2}\right)
\end{aligned} \tag{1.34}$$

It is worth noting that the frequency-independent term in the transverse relaxation rate is the chief cause of the sensitivity and line width problems that NMR spectroscopy encounters for large molecules. The chemical shielding anisotropy of $^{13}C$ and $^{15}N$ nuclei is of the order of 100 ppm, and for the molecules with rotational correlation times greater than about 50 ns the transverse relaxation becomes extremely fast, leading to intractably broad lines and huge magnetization losses during coherence transfer stages.

### *1.1.4 Cross-correlated relaxation*

One of the less apparent features of NMR relaxation theory as applied to rigid or conformationally restricted molecules is the fact that molecular rotation does not drive *all* correlation functions to zero. In fact, in a tumbling rigid molecule all the interaction tensors keep constant orientation with respect to one another, which means that the orientational functions for the corresponding interactions may not be averaged independently.

Although the proper theoretical groundwork has been laid a long time ago [36], cross-correlated relaxation is a current fashion in biomolecular NMR [37]. The reason is that progressively more accurate NMR structures are required, and spectroscopists gradually run out of experimental constraints on molecular geometry, which were traditionally obtained from scalar couplings and nuclear Overhauser effects. Cross-correlated relaxation data provide a wealth of additional orientational constraints, and pulse sequences harvesting them are now in active development [38-40].

Let us consider a nuclear two-spin system with a magnetic dipolar interaction between the spins and an axially anisotropic magnetic shielding of one of the nuclei[4]. The Hamiltonian, comprising nuclear Zeeman interactions with the anisotropy included for the *S* spin and point dipole interaction between the spins, can be written as follows

$$\hat{H} = \omega_1 \hat{I}_Z + \omega_2 \hat{S}_Z + \hat{\vec{I}} \cdot \mathbf{D} \cdot \hat{\vec{S}} + \hat{\vec{S}} \cdot \mathbf{C} \cdot \vec{B} \tag{1.35}$$

Using equation (1.22) to move to the irreducible spherical tensor notation, and assuming as usual that the magnetic induction is applied along the $Z$ axis of the laboratory frame of reference, we obtain the following expression for the spin-Hamiltonian:

$$\hat{H} = \omega_1 \hat{I}_Z + \omega_2 \hat{S}_Z - \frac{\Delta D}{\sqrt{6}} \sum_{k=-2}^{2} \hat{T}_{2,k}^{DD} \mathfrak{D}_{k,0}^{(2)}(t) - \frac{\Delta C}{\sqrt{6}} \sum_{k=-2}^{2} \sum_{n=-2}^{2} \hat{T}_{2,k}^{CSA} \mathfrak{D}_{k,n}^{(2)}(t) \mathfrak{D}'^{(2)}_{n,0}(\theta) \tag{1.36}$$

where $\theta$ is the angle between the dipolar and the shielding tensor axis, $\Delta D = \gamma_I \gamma_S \hbar \mu_0 / 4\pi r^3$ is the dipolar interaction anisotropy, $\Delta C = \gamma_S B_0 \Delta\sigma / 3$ is the nuclear Zeeman interaction anisotropy, $\Delta\sigma = \sigma_{\parallel} - \sigma_{\perp}$ is the chemical shift anisotropy and the spherical tensor operators for Zeeman and dipolar interactions defined in (1.33) and (1.21). The Hamiltonian (1.36) contains two sets of Wigner coefficients the $\mathfrak{D}'^{(2)}_{n,k}(\theta)$ set describing the rotation of the CSA tensor axis relative to the dipole tensor axis and $\mathfrak{D}^{(2)}_{n,k}(t)$ describing the overall molecular reorientation.

Substituting the Hamiltonian (1.36) into Equation (1.13), taking into account property (1.24) and letting *Mathematica* do the pencilwork results in the following equations for relaxation and cross-relaxation of longitudinal magnetization:

$$\frac{d}{dt}\begin{bmatrix} 1 \\ I_z \\ S_z \\ 2I_z S_z \end{bmatrix} = -\begin{bmatrix} 0 & 0 & 0 & 0 \\ 0 & \rho_{II} & \sigma_{IS} & 0 \\ 0 & \sigma_{IS} & \rho_{SS} & \delta_{S,IS} \\ 0 & 0 & \delta_{S,IS} & \rho_{ISIS} \end{bmatrix} \begin{bmatrix} 1 \\ \Delta I_z \\ \Delta S_z \\ 2I_z S_z \end{bmatrix} \tag{1.37}$$

with the following expressions for the coefficients

[4]The description of the cross-correlation between the nuclear dipole interaction and the chemical shift anisotropy is mathematically identical to the description of the cross-correlation between the anisotropy of the hyperfine coupling and the *g*-tensor anisotropy in an electron-nucleus system. In the latter case the second spin would be that of an electron. The electron-nucleus cross-correlated relaxation theory is exploited in Chapter 8.

$$
\begin{aligned}
\rho_{II} &= \frac{\gamma_I^2\gamma_S^2\hbar^2}{10}\left(\frac{\mu_0}{4\pi}\right)^2\frac{\tau_c}{r_{IS}^6}\left(\frac{3}{1+\omega_I^2\tau_c^2}+\frac{6}{1+(\omega_I+\omega_S)^2\tau_c^2}+\frac{1}{1+(\omega_I-\omega_S)^2\tau_c^2}\right)\\
\rho_{SS} &= \frac{\gamma_I^2\gamma_S^2\hbar^2}{10}\left(\frac{\mu_0}{4\pi}\right)^2\frac{\tau_c}{r_{IS}^6}\left(\frac{3}{1+\omega_S^2\tau_c^2}+\frac{6}{1+(\omega_I+\omega_S)^2\tau_c^2}+\frac{1}{1+(\omega_I-\omega_S)^2\tau_c^2}\right)+\frac{2}{15}\frac{\gamma_S^2\tau_c(\Delta\sigma)^2B_0^2}{1+\omega_S^2\tau_c^2}\\
\rho_{ISIS} &= \frac{3}{10}\left(\frac{\mu_0}{4\pi}\right)^2\frac{\gamma_I^2\gamma_S^2\hbar^2\tau_c}{r_{IS}^6}\left(\frac{1}{1+\omega_I^2\tau_c^2}+\frac{1}{1+\omega_S^2\tau_c^2}\right)+\frac{2}{15}\frac{\gamma_S^2\tau_c(\Delta\sigma)^2B_0^2}{1+\omega_S^2\tau_c^2}\\
\sigma_{IS} &= \frac{\gamma_I^2\gamma_S^2\hbar^2}{10}\left(\frac{\mu_0}{4\pi}\right)^2\frac{\tau_c}{r_{IS}^6}\left(\frac{6}{1+(\omega_I+\omega_S)^2\tau_c^2}-\frac{1}{1+(\omega_I-\omega_S)^2\tau_c^2}\right)\\
\delta_{S,IS} &= -\frac{2}{5}\left(\frac{\mu_0}{4\pi}\right)\frac{\gamma_I\gamma_S^2\hbar B_0}{r_{IS}^6}\frac{\tau_c}{1+\omega_S^2\tau_c^2}\Delta\sigma_g
\end{aligned}
\qquad (1.38)
$$

where $\Delta\sigma_g = \Delta\sigma\left(3\cos^2\theta - 1\right)/2$ is the so-called geometrically weighted shielding anisotropy parameter, which in a (somewhat bulkier) case of a shielding tensor with three different principal values becomes [36]

$$
\Delta\sigma_g = \frac{1}{2}\sigma_{XX}\left(3\cos^2\theta_{X,IS}-1\right)+\frac{1}{2}\sigma_{YY}\left(3\cos^2\theta_{Y,IS}-1\right)+\frac{1}{2}\sigma_{ZZ}\left(3\cos^2\theta_{Z,IS}-1\right) \qquad (1.39)
$$

where the three angles are the angles between the inter-nuclear vector and the three eigenvectors of the CSA tensor.

These SI units expressions for the relaxation rates are remarkable for their absence in the classical literature. This is partly due to the ever ongoing transition from the CGS units beloved by theorists to SI units in which current university courses are taught. This creates a fair deal of confusion, and after some unsuccessful attempts to develop a unit conversion system, the author concluded that it is simpler to just re-derive the expressions from scratch.

We will not consider a general multi-spin system here. Although the corresponding expressions are straightforward to derive, they are far too cumbersome to be reported here. Readers requiring relaxation rate expressions for systems with multiple dipolar interactions and multiple chemical shift anisotropies will probably find the enclosed *Mathematica* program useful as a starting point.

## 1.2 Origins of the photo-CIDNP effect

CIDNP (Chemically Induced Dynamic Nuclear Polarization) is defined by IUPAC [1] as a *Non-Boltzmann nuclear spin state distribution produced in thermal or photochemical reactions, usually from colligation and diffusion, or disproportionation of radical pairs, and detected by NMR spectroscopy by enhanced absorption or emission signals.* Detected as enhanced absorptive or emissive signals in the NMR spectra of the reaction products, CIDNP has been exploited for the last 30 years to characterise transient free radicals and their reaction mechanisms [41, 42]. The theory of CIDNP is well developed and allows extraction of radical reaction rate constants, hyperfine couplings, *g*-values and nuclear paramagnetic relaxation rates [43, 44]. In certain cases, CIDNP also offers the possibility of large improvements in NMR sensitivity. The principal application of this photo-CIDNP technique, as devised by Kaptein in 1978 [45], has been to proteins in which the aromatic amino acid residues histidine, tryptophan and tyrosine can be polarized using flavins or other aza-aromatics as photosensitizers [46, 47]. The key feature of the method is that only solvent-accessible histidine, tryptophan and tyrosine residues can undergo the radical pair reactions that result in nuclear polarization. Photo-CIDNP has thus been used to probe the surface structure of proteins, both in native and partially folded states, and their interactions with molecules that modify the accessibility of the reactive side chains [46-50]. Photo-CIDNP experiments on proteins using an argon ion laser as the light source and flavin mononucleotide (FMN) as the photosensitizer routinely exhibit $^{1}H$ NMR signal enhancements of up to 10-fold. Heavier spin-½ nuclei ($^{13}C$, $^{15}N$, $^{19}F$) often show larger polarizations [48, 51]. A 10-100 fold $^{19}F$ hyperpolarization can be created and sustained for several seconds in photo-CIDNP experiments on fluorinated aromatic systems [52].

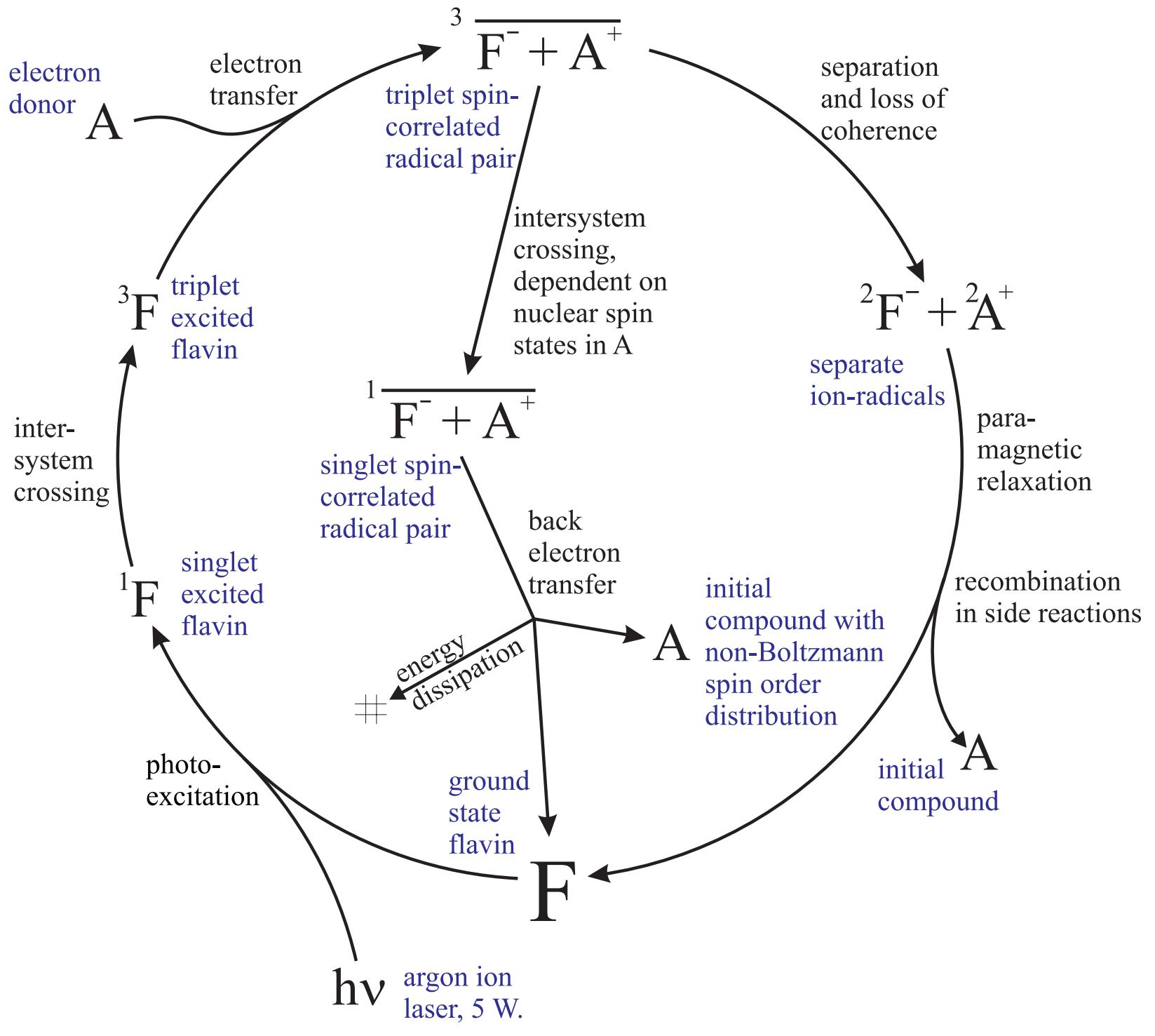

**Figure 1.1.** The scheme of sensitized photolysis of the donor molecule **A** with the photosensitizer **F**. The photosensitizer molecule is excited by a photon of the incident light and undergoes intersystem crossing into the triplet state, which is a very strong electron acceptor. On encounter with the target molecule **A** the electron transfer occurs, forming a triplet spin-correlated radical pair. The intersystem crossing in this radical pair is nuclear spin dependent through the mechanism outlined in the text, resulting in the nuclear spin sorting during recombination/separation processes. The upper left indices indicate spin multiplicity.

We shall limit our consideration of the CIDNP effect to the simplest case of a "strong" external magnetic field, i.e. when the electron Zeeman interaction in the intermediate radicals is much stronger than the hyperfine interaction. The reason for this is twofold: the fact that the strong field case is the only case encountered in this work, and the present state of the low-field CIDNP theory, which in its current formulation is hardly adequate, and is computationally prohibitive for real-life systems [53, 54].

The generation of CIDNP in the systems considered in this work is a cyclic photochemical process shown schematically in Figure 1.1. The chain of reactions is initiated by a blue light photon, which excites the flavin mononucleotide photosensitizer to the $\pi \to \pi^*$ singlet excited state. The fluorescence quantum yield of this state is rather low [55], and approximately half of the molecules undergo intersystem crossing into the long-lived triplet state. Triplet flavin has a remarkable electron affinity: 364 kJ/mol, or 3.77 eV [56]. If a molecule with a low ionization potential (e.g. phenols, polyaromatics) is present in the system, the diffusion-limited electron transfer reaction forms a spin-correlated triplet electron transfer state – a radical pair. The actual kinetics are rather complicated and may involve multiple (de)protonations and hence exhibit pH dependence [57, 58].

The radical pair may either cross over to a singlet electron state and then recombine, or separate and perish in side reactions. The relative probability of these two pathways for a given radical pair depends on the nuclear spin state via the mechanism outlined below. Nuclear magnetization generation in the strong field case may conveniently be illustrated using the Schrödinger equation rather than the more powerful density matrix formalism[5]. We start from the Schrödinger equation for the spin part of the wave function

$$\mathrm{i}\hbar \frac{\partial}{\partial t}\psi = \hat{H}\psi \tag{1.40}$$

and assume that the two molecules are already sufficiently far apart that the inter-electron exchange interaction is negligibly small. The spin part of the system Hamiltonian is then decoupled from the spatial part and may be written as

$$\hat{H} = \omega_{\mathrm{e}1}\hat{S}_{1Z} + \omega_{\mathrm{e}2}\hat{S}_{2Z} + \hat{S}_{1Z}\sum_i a_{1i}\hat{I}_{1Z,i} + \hat{S}_{2Z}\sum_j a_{2j}\hat{I}_{2Z,j} \tag{1.41}$$

in angular frequency units with the usual notation for all the variables and operators [12]. Isotropic hyperfine couplings and *g*-tensors are assumed in the Hamiltonian (1.41), and nuclear Zeeman interaction is neglected. If we choose the longitudinal electron spin subspace basis set to be $\{|\alpha\alpha\rangle, |\alpha\beta\rangle, |\beta\alpha\rangle, |\beta\beta\rangle\}$, the electron spin subspace that interests us would be spanned by the $\{|\alpha\beta\rangle, |\beta\alpha\rangle\}$ subspace (or, equivalently, $S - T_0$ subspace in the coupled notation), which is closed with respect to the strong-field Hamiltonian operator. The electron singlet and zero-projection triplet states are defined as:

$$\begin{aligned} |S\rangle &= \frac{1}{\sqrt{2}}(|\alpha\beta\rangle - |\beta\alpha\rangle) \\ |T_0\rangle &= \frac{1}{\sqrt{2}}(|\alpha\beta\rangle + |\beta\alpha\rangle) \end{aligned} \tag{1.42}$$

[5]Density matrix formulation is obligatory in realistic models of radical spin dynamics, especially when the coupling between the spatial and the spin degrees of freedom is accounted for.

The radical pair is initially created in the triplet state. Assuming the nuclear spin configuration to be stationary, we can calculate the time dependence of the probability to find the system in an electronic singlet state. The wave function is

$$\psi(t) = e^{-\frac{\mathrm{i}}{\hbar}\hat{H}t} |T_0\vec{n}\rangle \tag{1.43}$$

where $\vec{n}$ denotes a nuclear spin part of the wave function, which may vary from system to system in an ensemble. In the chosen electron state basis the Hamiltonian is diagonal

$$\hbar^{-1}\hat{H} = diag(\omega_{\alpha\alpha}, \omega_{\alpha\beta}, \omega_{\beta\alpha}, \omega_{\beta\beta}) \tag{1.44}$$

with the following expressions for the eigenvalues:

$$\begin{aligned} \omega_{\alpha\alpha} &= -\omega_{\beta\beta} = \frac{\omega_{\mathrm{e}1} + \omega_{\mathrm{e}2}}{2} + \frac{1}{2}\sum_i a_{1i} m_{1i} + \frac{1}{2}\sum_j a_{2j} m_{2j} \\ \omega_{\alpha\beta} &= -\omega_{\beta\alpha} = \frac{\omega_{\mathrm{e}1} - \omega_{\mathrm{e}2}}{2} + \frac{1}{2}\sum_i a_{1i} m_{1i} - \frac{1}{2}\sum_j a_{2j} m_{2j} \end{aligned} \tag{1.45}$$

Starting from the triplet state and computing the time dependence of singlet and triplet probabilities

$$\begin{aligned} \psi(0) &= |\mathrm{T}_0\rangle = \frac{1}{\sqrt{2}}\left(|\alpha\beta\rangle + |\beta\alpha\rangle\right) \\ \psi(t) &= \frac{1}{\sqrt{2}}\left(e^{-i\omega_{\alpha\beta}t}|\alpha\beta\rangle + e^{-i\omega_{\beta\alpha}t}|\beta\alpha\rangle\right) \end{aligned} \tag{1.46}$$

$$\begin{aligned} \langle\psi(t)|T_0\rangle &= \frac{1}{2}\left(e^{-i\omega_{\alpha\beta}t} + e^{-i\omega_{\beta\alpha}t}\right) \qquad p_{T_0}(t) = |\langle\psi(t)|T_0\rangle|^2 = \cos^2\left(\frac{\omega_{\alpha\beta} - \omega_{\beta\alpha}}{2}t\right) \\ \langle\psi(t)|S\rangle &= \frac{1}{2}\left(e^{-i\omega_{\alpha\beta}t} - e^{-i\omega_{\beta\alpha}t}\right) \qquad p_S(t) = |\langle\psi(t)|S\rangle|^2 = \sin^2\left(\frac{\omega_{\alpha\beta} - \omega_{\beta\alpha}}{2}t\right) \end{aligned} \tag{1.47}$$

Thus the frequency of oscillation between a triplet and a singlet state is

$$\begin{aligned} &\omega_{ST_0} = \omega_{\alpha\beta} - \omega_{\beta\alpha} = \Delta\omega_e + \sum_i a_{1i} m_{1i} - \sum_j a_{2j} m_{2j} \\ &\Delta\omega_e = \omega_{e2} - \omega_{e1} \end{aligned} \tag{1.48}$$

which is dependent on the nuclear spin configuration. Radical pair recombination occurs from the singlet state, so we must expect a certain amount of nuclear spin selection in the reaction products. Molecules with a favourable nuclear spin state configuration have a greater chance of recombination and yielding neutral molecules, especially when the difference in *g*-factors is small. The radical pairs that diffuse apart have the polarisation of opposite sign, but that one partially vanishes due to fast paramagnetic relaxation. Thus, after all the radicals in the system have disappeared (a few milliseconds after the excitation flash), only neutral molecules with nuclear spin states perturbed from the equilibrium remain. Including anisotropies in the discussion above will introduce an orientation dependence of the reaction yield. This orientation dependence is now believed to be a possible mechanism responsible for the magnetic field direction sense in some biological species [59].

Based on equation (1.48) for the speed of the singlet-triplet interconversion, a set of rules may be formulated for prediction of the sign of the net CIDNP effect on a given nucleus (the Kaptein rules [60]). For the case of the dye-sensitized photo-CIDNP experiment

$$\begin{aligned} &\Gamma_{\text{net}}(i) = \mu \cdot \varepsilon \cdot \text{sign}(\Delta g) \cdot \text{sign}(A_i) = \begin{Bmatrix} + \text{ absorptive} \\ - \text{ emissive} \end{Bmatrix}; \\ &\mu = \begin{Bmatrix} + \text{ triplet precursor} \\ - \text{ singlet precursor} \end{Bmatrix}; \quad \varepsilon = \begin{Bmatrix} + \text{ recombination products} \\ - \text{ escape products} \end{Bmatrix} \end{aligned} \tag{1.49}$$

where $\Delta g$ is the difference between *g*-factors of donor and photosensitizer radicals (radical – photosensitizer), and $A_i$ is the hyperfine coupling constant of the nucleus in question. These rules (and similar rules for multiplet CIDNP) should be treated with caution, as they are only obeyed by a rather narrow range of CHNO molecules, and are fairly often violated for systems containing heavier atoms, as will be demonstrated in Chapters 5-8.

### 1.3 Secondary radial kinetics in the photo-CIDNP generation

After the geminate spin-sorting process is complete, those radicals that have escaped the primary cage recombination engage in secondary reactions leading to further change in the overall nuclear magnetization. The particular case of dye-sensitized pulsed photo-CIDNP experiment has been treated by Vollenweider and Fischer [43, 44] with the following system of equations suggested for description of the kinetics of escaped radicals and the associated magnetization dynamics

$$\begin{cases} R(t) = \dfrac{R_0}{1 + k_t R_0 t} \\ \dfrac{dP}{dt} = -k_t PR - k_t \beta R^2 - \dfrac{P}{T_1} - k_{ex} CP \\ \dfrac{dQ}{dt} = k_t PR + k_t \beta R^2 + k_{ex} CP \end{cases} \tag{1.50}$$

These equations resemble kinetic equations for the radical concentrations, but have been modified to describe dynamics of *magnetization* rather *concentration* [44]. The first equation in (1.50) describes the escaped radical recombination process, which is a second order reaction

$$A^+ + F^- \xrightarrow{k_t} A + F$$

with equal initial concentrations $R_0$ of donor cations and photosensitizer anions. This step is typically followed by fast vibrational relaxation which may involve solvent molecules. The second and third equations describe the magnetization of a given nucleus in the pool of radicals ($P$) and neutral molecules ($Q$) with the first term describing recombination in singlet encounters and the second describing F-pair encounters that generate additional magnetization, hence a factor of $\beta = \gamma P^G / R_0$. The initial conditions $P(0) = -P^G$, $Q(0) = P^G$ reflect the spin-sorting nature of the radical pair mechanism. The last terms of the second and third equation in (1.50) describe the transfer of magnetization between radicals and neutral molecules (present in concentration *C*) by the degenerate electron exchange. See Figure 1.1 for the complete photochemical reaction chain schematic.

Of the two polarized ensembles described by equations (1.50) only the neutral molecule magnetization is detected in a time-resolved CIDNP experiment (which is described in detail in Chapter 7), therefore only $Q(t)$ is available experimentally. The task of parameter estimation consists therefore in solving (1.50) for $Q(t)$ and fitting the solution to the experimental data. Let us consider some limiting solutions to the system (1.50).

In the case of slow relaxation ($T_1 \to \infty$) and in the absence of degenerate exchange ($k_{ex} = 0$) the system (1.50) is trivially solved

$$\begin{cases} P(t) = -P^G \dfrac{1 + \gamma \ln(1 + k_t R_0 t)}{1 + k_t R_0 t} \\ Q(t) = -P(t) \end{cases} \tag{1.51}$$

but not applicable to any real problem, the reason being that the nuclear paramagnetic relaxation time in common radicals is almost always comparable or faster than the characteristic time of the chemical kinetics.

In the case of fast relaxation ($T_1 \to 0$) and lack of degenerate exchange ($k_{\mathrm{ex}} = 0$) the asymptotic solution has the following form

$$\begin{cases} P(t) = -P^{\mathrm{G}} e^{-\frac{t}{T_1}} \\ Q(t) = P^{\mathrm{G}} \left( 1 + \gamma \dfrac{k_{\mathrm{t}} R_0 t}{1 + k_{\mathrm{t}} R_0 t} \right) \end{cases} \tag{1.52}$$

This solution to equations (1.50) may be useful in the case of $^{13}$C, $^{15}$N, $^{19}$F and heavier nuclei, for which the relaxation is often much faster than the chemical kinetics. Fast relaxation however implies that there may be a relaxation contribution to the geminate dynamics, therefore the initial conditions have to be unlocked from each other, i.e. $P(0) \neq -Q(0)$, resulting in the following solution

$$\begin{cases} P(t) = -P^{\mathrm{G}} e^{-\frac{t}{T_1}} \\ Q(t) = Q^{\mathrm{G}} + \gamma P^{\mathrm{G}} \dfrac{k_{\mathrm{t}} R_0 t}{1 + k_{\mathrm{t}} R_0 t} \end{cases} \tag{1.53}$$

These expressions, along with the full numerical solution of the system (1.50) will be used in Chapter 8 to analyze the time-resolved magnetization kinetics in fluorinated photo-CIDNP systems.

In the third limiting case, when we assume that the relaxation is slow, $T_1 \to \infty$, the exchange term leads to some non-elementary functions, but the solution is still quite neat

$$\begin{cases} P(t) = -P^{\mathrm{G}} \dfrac{e^{-\frac{k_{\mathrm{ex}} C (1 + k_{\mathrm{t}} R_0 t)}{k_{\mathrm{t}} R_0}} \left( e^{\frac{k_{\mathrm{ex}} C}{k_{\mathrm{t}} R_0}} - \gamma \mathrm{Ei}\left( \dfrac{k_{\mathrm{ex}} C}{k_{\mathrm{t}} R_0} \right) + \gamma \mathrm{Ei}\left( \dfrac{k_{\mathrm{ex}} C (1 + k_{\mathrm{t}} R_0 t)}{k_{\mathrm{t}} R_0} \right) \right)}{1 + k_{\mathrm{t}} R_0 t} \\ Q(t) = -P(t) \end{cases} \tag{1.54}$$

however, just as in the case of solution (1.51) this expression is hardly applicable to practically important cases because the assumption $T_1 \to \infty$ rarely holds.

The fourth case of finite relaxation time and lack of degenerate exchange is something that is useful in practice and applicable to systems where the electron transfer stage is followed by rapid deprotonation (e.g. phenols and tyrosine). With some effort the solution may be obtained in a closed algebraic form

$$Q(k_{\mathrm{t}},R_0,\gamma,T_1,P^{\mathrm{G}},t)=\frac{P^{\mathrm{G}}}{k_{\mathrm{t}}R_0T_1(1+k_{\mathrm{t}}R_0t)}\times \\ \times\left(\begin{array}{l} k_{\mathrm{t}}R_0T_1\left(1-e^{-\frac{1+k_{\mathrm{t}}R_0t}{k_{\mathrm{t}}R_0T_1}}\left(\gamma\mathrm{Ei}\left[\frac{1}{k_{\mathrm{t}}R_0T_1}\right]-\gamma\mathrm{Ei}\left[\frac{1+k_{\mathrm{t}}R_0t}{k_{\mathrm{t}}R_0T_1}\right]\right)\right)- \\ (1+k_{\mathrm{t}}R_0t)\left(\begin{array}{l} e^{\frac{1}{k_{\mathrm{t}}R_0T_1}}\left(\mathrm{Ei}\left[-\frac{1}{k_{\mathrm{t}}R_0T_1}\right]-\mathrm{Ei}\left[-\frac{1+k_{\mathrm{t}}R_0t}{k_{\mathrm{t}}R_0T_1}\right]\right)+ \\ \gamma\left(\begin{array}{l} -\mathrm{G}_{2\,3}^{3\,1}\left[\frac{1}{k_{\mathrm{t}}R_0T_1}\middle|_{0\,0\,0}^{0\,1}\right]+\mathrm{G}_{2\,3}^{3\,1}\left[\frac{1+k_{\mathrm{t}}R_0t}{k_{\mathrm{t}}R_0T_1}\middle|_{0\,0\,0}^{0\,1}\right]+ \\ \left(\mathrm{Ei}\left[\frac{1}{k_{\mathrm{t}}R_0T_1}\right]-\mathrm{Ei}\left[\frac{1+k_{\mathrm{t}}R_0t}{k_{\mathrm{t}}R_0T_1}\right]\right)\left(\mathrm{Chi}\left[\frac{1+k_{\mathrm{t}}R_0t}{k_{\mathrm{t}}R_0T_1}\right]-\mathrm{Shi}\left[\frac{1+k_{\mathrm{t}}R_0t}{k_{\mathrm{t}}R_0T_1}\right]\right)\end{array}\right)\end{array}\right)\end{array}\right) \tag{1.55}$$

with the following notation for the non-elementary functions:

$$\mathrm{Ei}(z)=-\int_{-z}^{\infty}\frac{e^{-t}}{t}dt;\quad \mathrm{Chi}(z)=\gamma_{\mathrm{E}}+\ln(z)+\int_0^z\frac{\cosh(t)-1}{t}dt;\quad \mathrm{Shi}(z)=\int_0^z\frac{\sinh(t)}{t}dt$$

$$\mathrm{G}_{p,q}^{m,n}\left(x\middle|_{b_1,\dots,b_q}^{a_1,\dots,a_p}\right)=\frac{1}{2\pi\mathrm{i}}\int_{\gamma_{\mathrm{L}}}\frac{\prod_{j=1}^{m}\Gamma(b_j-s)\prod_{j=1}^{n}\Gamma(1-a_j+s)}{\prod_{j=n+1}^{p}\Gamma(a_j-s)\prod_{j=m+1}^{q}\Gamma(1-b_j+s)}x^s ds;\quad \Gamma(z)=\int_0^{\infty}t^{z-1}e^{-t}dt$$

in which $\gamma_{\mathrm{E}}$ is Euler's constant and $\gamma_{\mathrm{L}}$ contour lies between the poles of $\Gamma(1-a_i-s)$ and the poles of $\Gamma(b_i+s)$ [61]. This expression, although it looks formidable, is in fact straightforward to evaluate, although the abundance of non-elementary functions makes the evaluation rather slow. The brute force numerical solution (adaptive fourth order Runge-Kutta) to the differential equations (1.50) is also quite slow, especially for least squares inverse problems where a minimizer normally requires a few thousand error function evaluations to converge to a desired accuracy. Some approximations are therefore required, if not to obtain an exact match, then to approximately locate a least squares minimum to save time on full minimization. One simple approach would be to attempt solving (1.50) in power series around $t=0$:

$$Q(t)=\sum_{n=1}^{\infty}\frac{1}{n!}\left.\frac{d^nQ}{dt^n}\right|_{t=0}t^n \tag{1.56}$$

The coefficients of this expansion are trivial to find:

$$\begin{cases} \left.\dfrac{d^{n+1}P}{dt^{n+1}}\right|_{t=0} = -\dfrac{1}{T_1}\left.\dfrac{d^nP}{dt^n}\right|_{t=0} - \displaystyle\sum_{k=1}^{n}\dfrac{(-1)^{n-k}\, n!\,k_t^{\,n-k+1}R_0^{n-k+1}}{k!}\left.\dfrac{d^kP}{dt^k}\right|_{t=0} + (-1)^n (n+1)!\,k_t^{\,n+1}\gamma P^{\mathrm{G}} R_0^{n+1} \\ \left.\dfrac{d^{n+1}Q}{dt^{n+1}}\right|_{t=0} = -\dfrac{1}{T_1}\left.\dfrac{d^nP}{dt^n}\right|_{t=0} - \left.\dfrac{d^{n+1}P}{dt^{n+1}}\right|_{t=0} \\ \left.\dfrac{dQ}{dt}\right|_{t=0} = k_t P^{\mathrm{G}} R_0(\gamma-1) \\ \left.\dfrac{dP}{dt}\right|_{t=0} = k_t P^{\mathrm{G}} R_0(1-\gamma) + \dfrac{P^{\mathrm{G}}}{T_1} \\ P(0) = -P^{\mathrm{G}} \\ Q(0) = P^{\mathrm{G}} \end{cases} \tag{1.57}$$

Quite disappointingly, however, this series diverges for a wide variety of parameters, even for small values of $t$.

Another, and a more successful approach is to use a Padé approximant around $t = 0$. The (1,2) approximant generally looks similar to the target function, but has an incorrect $t^{-1}$ asymptotic behaviour; the (2,3) approximant is unsatisfactory for small times and has the same erroneous asymptote; the (3,2) approximant expectedly diverges at $t \to \infty$; the (2,2) approximant reproduces all the essential features of the function, being off by ~10%, mostly in the asymptotic region; the (3,3) approximant is rather good, with a relative error of less than 5%. More accurate higher order approximants such as (4,4) are notationally bulky, but are still a lot faster than both analytical and numerical solution of (1.50).

The (4,4) Padé approximant, which was used for initial least squares minimum location, provides very satisfactory results with the approximation inaccuracy of less than 3% within the operational interval and five orders of magnitude speed-up compared to numerical ODE solution. Its full expression is too cumbersome to be reported here, the reader will find it inside the `ik_fit.m` program on the enclosed DVD.

Finally, the case when both relaxation and degenerate exchange are present, yields a solution of the following form:

$$\frac{P^{\mathrm{G}}}{a_1a_2} e^{-\frac{a_1a_2}{a_3}} \left( \begin{array}{l} -a_1 e^{\frac{(1+a_1)a_2}{a_3}} \mathrm{Ei}\left[-\dfrac{a_2}{a_3}\right] + \left(a_3 + a_1 e^{\frac{a_1a_2}{a_3}} \mathrm{Ei}\left[-\dfrac{a_1a_2}{a_3}\right]\right)\left(e^{\frac{a_2}{a_3}} - \gamma \mathrm{Ei}\left[\dfrac{a_2}{a_3}\right] + \gamma \mathrm{Ei}\left[\dfrac{a_1a_2}{a_3}\right]\right) + \\ +a_1\gamma e^{\frac{a_1a_2}{a_3}} \left( \mathrm{G}_{2\,3}^{3\,1}\left[\dfrac{a_2}{a_3}\middle|\begin{matrix}0\;1\\0\;0\;0\end{matrix}\right] - \mathrm{G}_{2\,3}^{3\,1}\left[\dfrac{a_1a_2}{a_3}\middle|\begin{matrix}0\;1\\0\;0\;0\end{matrix}\right]\right) \end{array} \right) \tag{1.58}$$

with the following shorthands: $a_1 = 1 + k_t R_0 t$; $a_2 = 1 + k_{ex} C T_1$; $a_3 = k_t R_0 T_1$ and others as in (1.55). For this function obtaining a Padé approximation around $t = 0$ is considerably more difficult, because Meijer's $G$ functions no longer cancel each other at zero time and $\mathrm{G}_{2\,3}^{3\,1}\left[ z \middle|_{0\,0\,0}^{0\,1} \right]$ cannot be expressed through functions much simpler than the definition [61]. The fastest way to solve (1.50) when degenerate exchange is present appears to be numerical; some time can still be saved, however, by using (1.55) and its Padé approximation to obtain initial estimates.

# Chapter 2

# $^{19}$*F nucleus: exploratory calculations and photo-CIDNP experiments*

## 2.1 Introduction

The purpose of the experiments and calculations described in this Chapter is to get a general idea of the magnitude, properties and potential applications of $^{19}$F photo-CIDNP and secondary effects, specifically of CIDNP-pumped $^{19}$F-$^{1}$H dipolar cross-relaxation. Specific questions answered herein are:

> *What are the essential conditions that the reaction system must meet to be informative and useful in* $^{19}$*F photo-CIDNP experiments?*
>
> *What is the best available reaction system for application of* $^{19}$*F CIDNP to proteins, considering cost, effect magnitude and photostability?*
>
> *How many of the essential photo-CIDNP system parameters may be computed ab initio and what is the accuracy of such computations?*
>
> *Can a simple extension of Redfield's relaxation theory [7, 9] be constructed for describing CW photo-CIDNP pumped nuclear spin systems?*

The prior work on the $^{19}$F CIDNP effect is very sparse and can be condensed to two facts. One is that $^{19}$F CIDNP is extremely strong, typically about five times larger than the $^{1}$H CIDNP effect in the same position of the same molecule [51, 62, 63], primarily due to larger hyperfine couplings in the intermediate radicals. The second is that the $^{19}$F CIDNP effect often disobeys Kaptein's rules [60] and has been shown to have a strong contribution from electron-nuclear cross-relaxation and/or other relaxation-related processes [64-69].

The work described in this Chapter uses *ab initio* quantum chemical calculations essentially as a method of investigation equal in most respects to the experimental practice. Provided that due care is taken, the *"in silico experiment"* constitutes a faithful representation of reality and can be thought of as a pocket Universe. The systematic searches performed in recent years have uncovered robust and reliable recipes for calculation of just about any molecular property, provided the molecule is sufficiently small to make the calculation affordable [70, 71]. As one of my colleagues put it: *"Quantum chemistry has made some truly spectacular advances. You can actually trust it now."*

The simplest fluorinated aromatic compounds that have an ionisation potential low enough to engage in an electron transfer with a photosensitizer triplet are fluorophenols. These compounds are cheap, water-soluble, and do not have titration points around neutral pH. Parafluorophenol also has four important tensors aligned to the same eigenframe: diffusion, $^{19}$F CSA, $^{19}$F HFC and *g*-tensor. The small size and the flat geometry of the intermediate radicals ($C_S$ symmetry for 2- and 3-fluorophenoxyl, $C_{2V}$ symmetry for 4-fluorophenoxyl) also makes them a convenient target for *ab initio* calculation of molecular properties. The three isomeric fluorophenols (Figure 2.1) were therefore chosen as test compounds for initial evaluation and reconnaissance experiments.

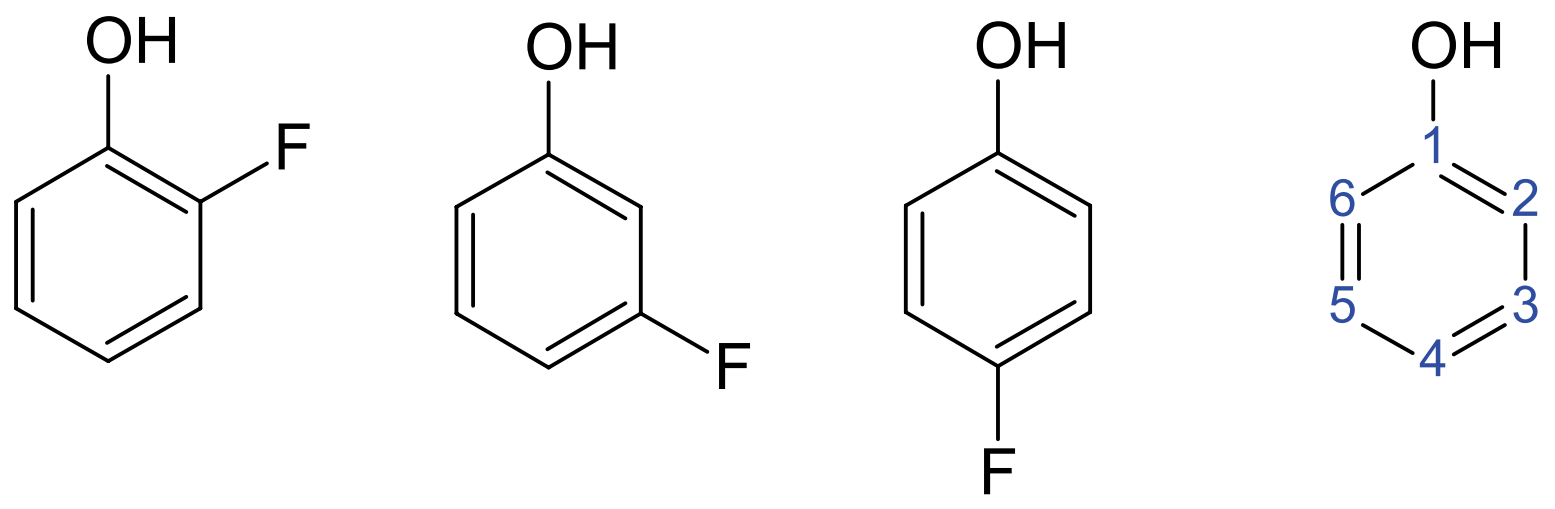

**Figure 2.1.** Three fluorophenols and the numbering scheme in the aromatic ring.

The reader will probably remember that at the dawn of the XXI century every research work in natural sciences, from mathematical all the way down to philosophical, had either to claim a biological application, or face no chance of ever being funded. Most fortunately, as outlined in Section 1.3, the photo-CIDNP effect has such an application, and is used with some success for investigation of protein surface dynamics during folding or state change processes [49, 72]. Introducing the fluorine nucleus in biomolecules has also proved to be remarkably informative, mostly due to the large chemical shift dispersion of the $^{19}$F nucleus and the sensitivity of the fluorine chemical shift to the changes in the molecular environment [51, 73]. The combination of the two techniques, that is, $^{19}$F photo-CIDNP of fluorine-labelled proteins is evaluated in this Thesis. Because fluorinated proteins are not available commercially and rather difficult to make, it seems reasonable in reconnaissance experiments to try out individual amino acids first.

Of the three natural amino acids that are known to exhibit detectable photo-CIDNP polarization (Tyr, Trp, His), histidine is unlikely to be a good target for fluorination because of the relatively slow electron transfer to photosensitizer and consequently weak polarization [57], as well as unfavourable competition effects from tyrosine and tryptophan [57, 74]. Only tyrosine and tryptophan therefore need to be considered. Excluding optical isomers, there are seven possible positions of single fluorination in the tryptophan amino acid and four in the tyrosine. In the resulting 11 molecules the observed CIDNP amplitude and sensitivity will depend primarily on the $^{19}$F hyperfine coupling constant in the intermediate radical (neutral phenoxyl radical in the case of tyrosine and molecular cation radical in the case of tryptophan).

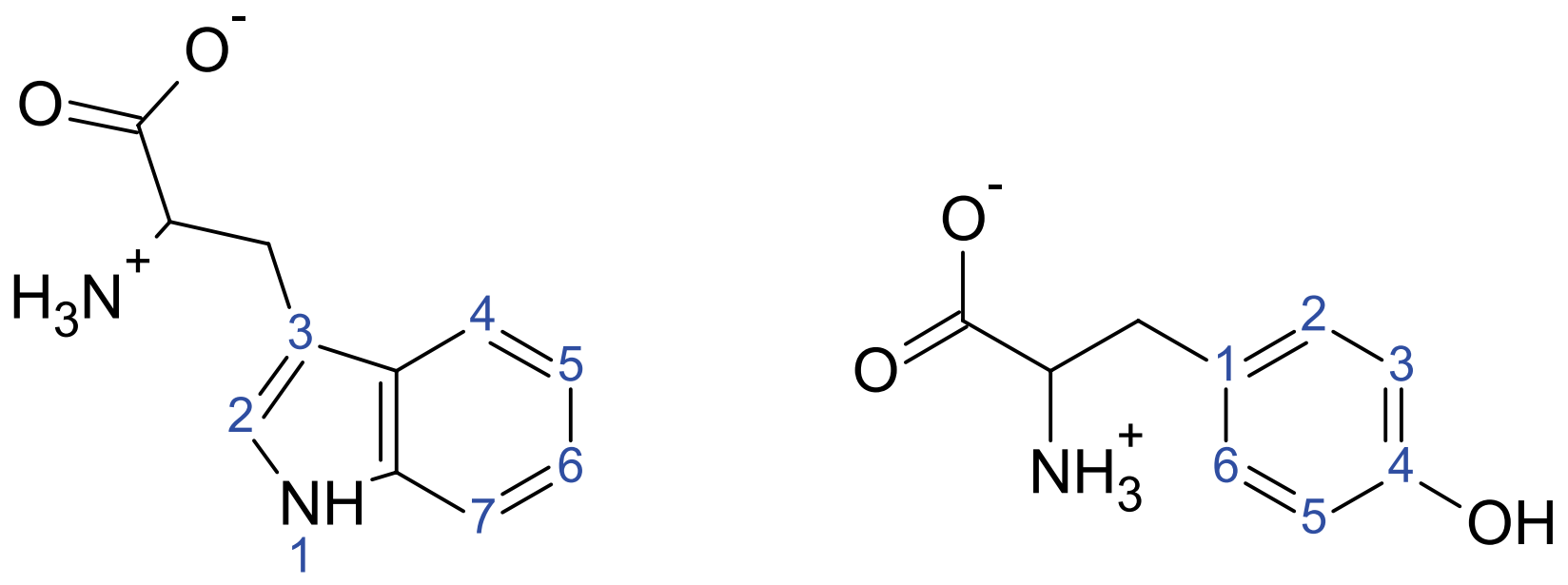

**Figure 2.2.** Numbering scheme in tyrosine and tryptophan aromatic rings.

All isomers of fluorotyrosine and fluorotryptophans are sufficiently small to enable a high quality *ab initio* calculation of molecular properties. Should the *ab initio* results appear promising, all of these compounds are also either commercially available or straightforward to synthesize.

## 2.2 Experimental details

The reagents were used as received from Apollo Scientific ($D_2O$), Lancaster (fluorophenols, 3-fluorotyrosine) and Sigma-Aldrich (fluorotryptophans, FMN). $^{19}$F and $^{1}$H NMR spectra were recorded either on a home-built 600 MHz NMR spectrometer, or on a Varian Inova 600 NMR system. In all experiments the NMR sample contained 4.0 mM target compound (fluorophenol, fluorotyrosine or fluorotryptophan) and 0.20 mM FMN photosensitizer in $D_2O$ at pH 5.0 (uncorrected for deuterium isotope effect) at 25.0 °C. Unless otherwise stated, all NMR and photo-CIDNP spectra were recorded in one scan with sufficient sweep width to incorporate all signals without reflections. The acquisition time in all cases was set to longer than five times the reciprocal width of the narrowest signal. The hardware oversampling rate was in all cases set to the instrumental maximum. All FIDs were weighted by an exponential or a Gaussian pseudoecho window function matched to the narrowest peak and zero filled to twice the original length prior to the Fourier transformation.

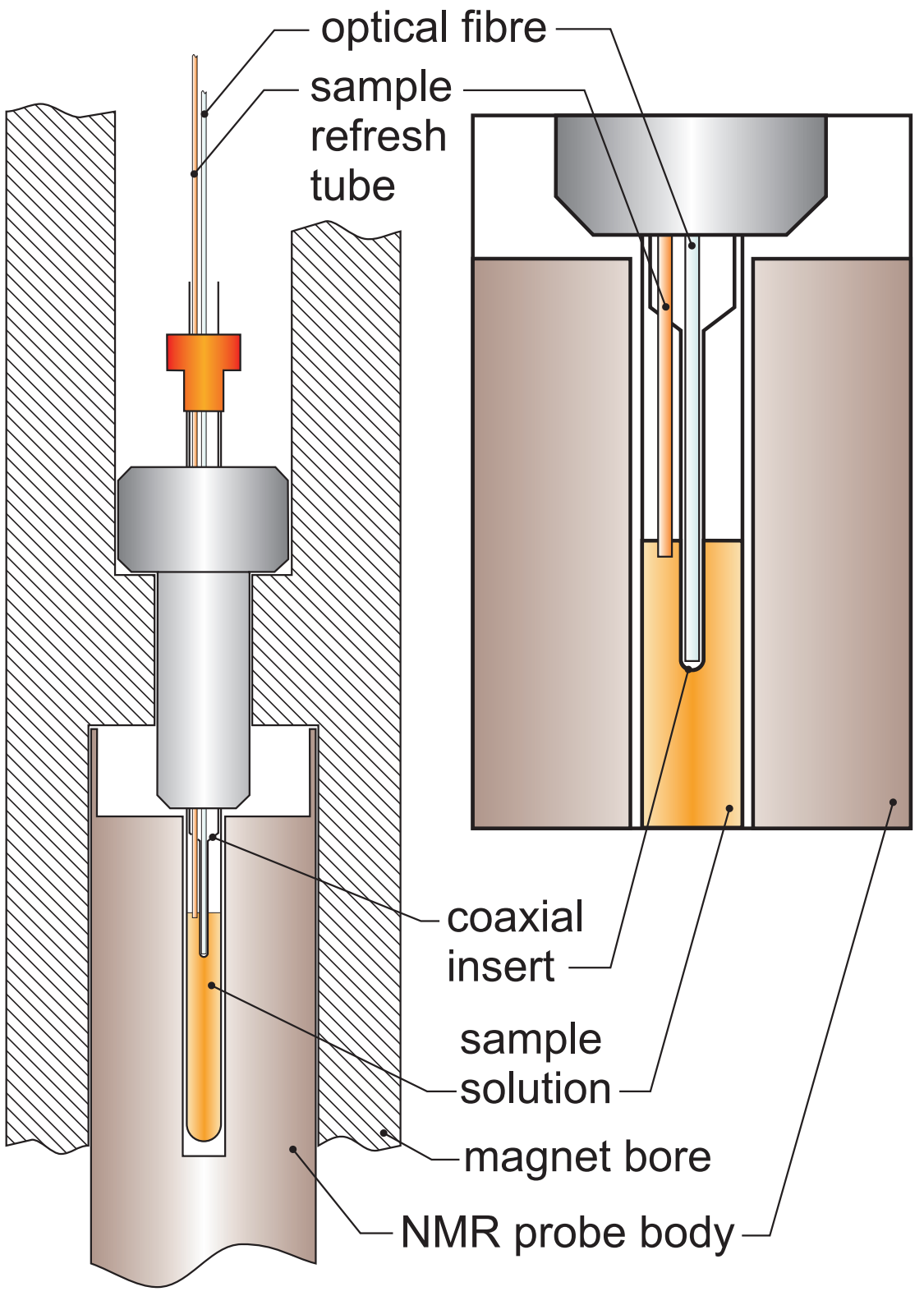

**Figure 2.3.** A schematic drawing of the CW photo-CIDNP sample assembly located inside a commercial 600 MHz NMR magnet.

Our continuous-wave photo-CIDNP setup (assembled jointly by the author and Dr. Iain Day) utilizes a Spectra-Physics BeamLok 2080 argon ion laser, operating in either a single-line (488 nm) or a multi-line (principally at 488 and 514 nm) mode, with 25 W maximum output power. A mechanical shutter controlled by the spectrometer is used to produce light pulses of 10-$10^4$ ms duration. The Newport M-5X objective lens is used to focus the light into a 6 m length of Newport F-MBE optical fibre ( ≥ 70% transmission efficiency), which transmits the light to the SMA connector at the platform mounted to the ceiling beside the spectrometer magnet. The optical fibre attached to the other end of the SMA connector then routes the light inside the magnet to the NMR sample assembly, which is shown schematically in Figure 2.3. The optical fibre is housed inside a Wilmad WGS-5BL coaxial insert. This sample illumination system, first suggested by Scheffler et al. [75], is very robust and convenient to use, provided that the sample optical density is below ~2 (dense samples are dealt with in Chapter 4).

In each CIDNP experiment the sample solution was purged with argon for 30 minutes prior to loading into the air-tight re-injection refreshing system, comprising a syringe linked to the sample refresh tube (Figure 2.3). A spectrometer-controlled stepper motor moves the syringe plunger so as to mix the solution contained in the coil region with the layers above and below it, as well as with about 3 ml of the sample contained in the syringe. The sample solution was thus remixed after each scan, and the samples were replaced as necessary to keep photosensitizer degradation below 10%.

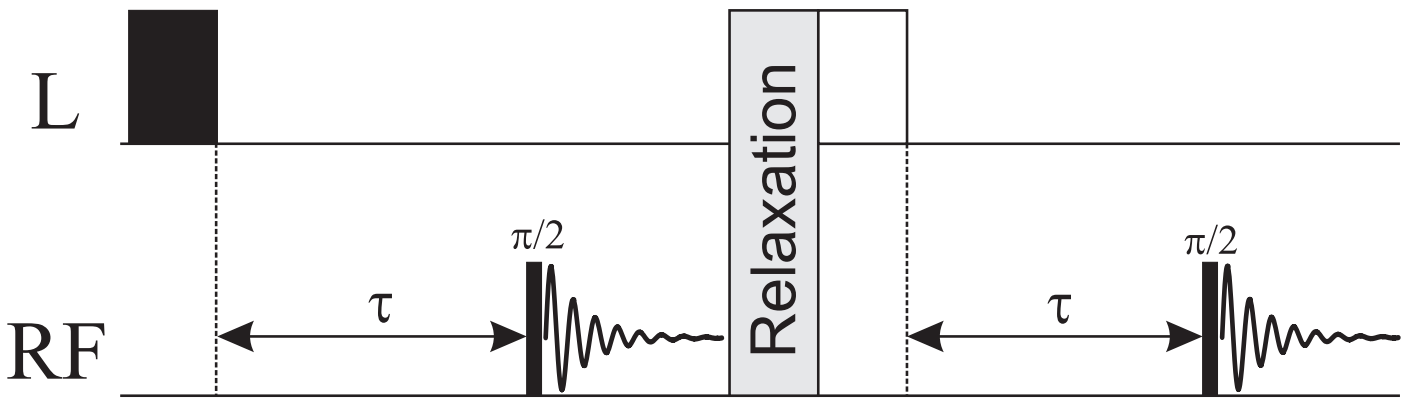

**Figure 2.4.** A scheme of a typical continuous-wave photo-CIDNP experiment. **L** stands for light output from the laser and **RF** is the spectrometer radiofrequency input/output.

A typical continuous-wave photo-CIDNP experiment is schematically shown in Figure 2.4. The laser flash of 10 ms to 10 s duration is followed by an optional mixing time, after which a 90-degrees pulse makes the magnetization observable. After a relaxation delay (with optional sample refreshment) a “dark” spectrum is recorded to serve as a signal intensity and position reference. In the case of very crowded spectra, such as those of proteins, the “dark” spectrum can be subtracted from the illuminated spectrum so as to leave just the signals whose intensity is perturbed by the photo-CIDNP effect.

## 2.3 Computational details

The *ab initio* calculations were performed using Gaussian03 [76], GAMESS [77] or DALTON [78] running either on local workstations (Intel Pentium 4 Northwood 3.0 GHz / 2 GB DDR400 2.5-3-3-6 / Linux) or on the Sun Fire 6800 shared memory system installed at Oxford Supercomputing Centre. When running on the latter, 6-12 CPUs were normally used together with 15-20 GB of RAM. Gaussian03 was run in either single-CPU (workstation) or SHMEM parallel (supercomputer) mode, GAMESS and DALTON were either run on a single CPU or in parallel using either built-in TCP/IP DDI interface (GAMESS) or MPI parallel environment (DALTON). A local Gigabit network was assembled to run independently of the departmental network and provide interlink for GAMESS and DALTON parallel runs on local workstations.

The first approximation for the geometry optimization runs was obtained with unrestricted PM3 runs to save CPU time on subsequent DFT optimization. The accurate geometry optimization was normally performed with restricted (if necessary, open-shell) DFT method using B3LYP exchange-correlation functional in a 6-311++G(2d,2p) basis set either in vacuum or in a PCM solvent. Unless otherwise stated, the property calculations were performed in vacuum or PCM solvent using: unrestricted DFT B3LYP in an EPR-III basis set for hyperfine coupling tensors, unrestricted GIAO DFT B3LYP using EPR-III basis set for *g*-tensors, unrestricted GIAO DFT or HF methods in a 6-311++G(2d,2p) basis for chemical shielding tensors. The tolerance for SCF convergence for single-point property runs and at every geometry optimization iteration was set to: $10^{-9}$ (energy), $10^{-6}$ (max density matrix), $10^{-8}$ (RMS density matrix). The geometry convergence tolerance was set to: $4.5\times10^{-4}$ (max force), $3.0\times10^{-4}$ (RMS force), $1.8\times10^{-3}$ (max displacement), $1.2\times10^{-3}$ (RMS displacement) for PCM optimizations and to $1.5\times10^{-5}$ (max force), $1.0\times10^{-5}$ (RMS force), $6.0\times10^{-5}$ (max displacement), $4.0\times10^{-5}$ (RMS displacement) for vacuum optimizations.

A program was written in Matlab 7.0 to parse Gaussian 98/03 single-point output and draw visual representations of hyperfine tensors using 3D ellipsoid plots. The version current at the time of writing can be found on the enclosed DVD, the most recent one is likely to be available from Matlab Central. Every ellipsoid is drawn in the following way:

1. A unit sphere in a Cartesian space is scaled by $|A_{11}|$ in the $X$ direction, $|A_{22}|$ in the $Y$ direction and $|A_{33}|$ in the $Z$ direction, where $A_{ii}$ are the eigenvalues of the dipolar tensor in units of milliTesla. The sphere is tinted in shades of red and made translucent.

2. A set of axes is drawn inside the sphere with red axis for positive eigenvalues and blue for negative ones.

3. The sphere is translated to the point of corresponding atom and rotated into the molecular frame.

There certainly is a more mathematically consistent way of representing the anisotropy of a tensor, that is, when for negative eigenvalues the ellipsoid is actually pulled inside out through zero and the resulting plot looks like (and may be shown to be related to, see Chapter 1) a superposition of $Y_{2,m}(\theta,\varphi)$ spherical harmonics. It was found however, that with this faithful representation the picture gets cluttered and far less easy to comprehend. Since the hyperfine tensor itself operates in a direct product of two four-dimensional spin spaces, which fundamentally have no classical analogue, the details of its representation in Cartesian space are a matter of convenience.

## 2.4 *Ab initio* hyperfine coupling constants

### *2.4.1 General notes*

The electron-nucleus dipolar interaction that gives rise to hyperfine coupling naturally falls into two categories: the *anisotropic* part corresponding to pure dipolar interaction between electron and nucleus at large separations and the *isotropic* part resulting from a short-range dipolar interaction and stemming from the fact that because of finite size of the nucleus, the spherical average of point-to-point dipolar interaction is not zero. Because this second part results from short-range interactions, it is often referred to as *contact interaction* [22].

The expression for the anisotropic part of hyperfine coupling involves an integral over the spatial distribution of the unpaired electron, which is relatively easy to compute accurately even at a relatively low level of theory [79]. The contact term, however, includes a delta-function that chips out the wave function amplitude at the nucleus point. The latter is quite difficult to compute both because standard Gaussian basis sets do not reproduce the wave function cusp at the nucleus point and because additional flexibility has to be introduced into the core part of the basis to account for the now essential core-valence interaction [80].

The basis sets and methods adequate for computing contact hyperfine couplings are relatively recent; the best general-purpose methods (DFT B3LYP with EPR-II or EPR-III basis set) were introduced in 1995 [81] and 1996 [82]. EPR-II is a double-ζ and EPR-III a triple-ζ basis set augmented by polarization and diffuse functions with core s-type Gaussians uncontracted and a number of tight s-type functions added. Furthermore, contraction coefficients were specifically optimized to reproduce experimental isotropic hyperfine couplings in a standard set of molecules. Notably the basis sets have been optimized for computing hyperfine couplings specifically with B3LYP exchange-correlation functional, and an attempt to use a different functional, even a fairly good one like B3PW91 or PBE1PBE, usually leads to poorly predicted contact interactions. The anisotropic components however are always satisfactory.

The important aspect of *ab initio* hyperfine coupling calculations is the variation of this parameter during molecular vibrations. In the paper that introduces the EPR-II and EPR-III basis sets, Barone and co-workers show that the equilibrium geometry calculation of hyperfine couplings underestimates the large couplings, which are rectified after taking a vibrational average [82]. The practical difficulties of performing this averaging even for medium-sized molecules are however formidable, with a need to compute a Hessian to perform even the simplest vibrational average. Even though the code performing this procedure was available to the author, the necessary calculations were well out of reach of even the SunFire 6800 supercomputer, with the result that the larger of the computed hyperfine couplings in the tables reported below are noticeably smaller than the experimental values. Annoying as these small deviations might be, the estimates obtained at the equilibrium geometry are still very reliable [83].

### *2.4.2 The role of the solvent interactions*

While it is quite clear from a large volume of literature on the subject that the DFT U-B3LYP / EPR-III method will provide reliable hyperfine couplings [83], the account of the solvent interactions is a more ambiguous question. A small additional inquiry has been made below into the solvent electrostatic effect on the computed hyperfine couplings. This benchmarking was performed on 4-fluorophenoxyl because of its high symmetry and therefore relatively short calculation times.

Two DFT B3LYP 6-311++G(2d,2p)/EPR-III calculations, one in vacuum and the other in PCM water were performed with the purpose of assessing the importance of solvent electrostatic effects in HFC calculations of fluorinated aromatic radicals. The results, along with experimental values for isotropic HFCs, are presented in Table 2.1. For both protons and fluorine the PCM water HFC values are closer to the experiment than the vacuum values. Including PCM water in the calculation does not increase the computation time, so it was decided to include PCM solvent in all HFC calculations.

The use of PCM water also shifts the hyperfine tensor axiality and rhombicity of those atoms which bear strong partial charges, a result of charge compensation by the polarized solvent. No experimental data is available on anisotropy parameters of HFC tensors in these systems, but the literature data on other molecules and the fact that the values seem to have got saturated (see EPR-III vs. aug-cc-pVQZ comparison below), suggest that these parameters are accurate.

**Table 2.1.** Experimental and calculated hyperfine coupling tensor parameters (in units of Gauss) in 4-fluorophenoxyl radical[†].

| | Exp[a] | Exp[b] | DFT B3LYP 6-311++G(2d,2p)/EPR-III in PCM water | | | DFT B3LYP 6-311++G(2d,2p)/EPR-III in vacuum | | | DFT B3LYP 6-311++G(2d,2p)/ aug-cc-pVQZ in PCM water | | |
|---|---|---|---|---|---|---|---|---|---|---|---|
| Nucleus | Iso | Iso | Iso | Ax | Rh | Iso | Ax | Rh | Iso | Ax | Rh |
| $^{17}$O(1) | – | – | −9.5 | 123.0 | 0.0 | −9.7 | 135.3 | 0.0 | −4.6 | 123.0 | 0.0 |
| **$^{19}$F(4)** | **27.5** | **26.7** | **22.4** | **242.2** | **5.2** | **22.1** | **236.7** | **5.4** | **15.1** | **241.8** | **5.0** |
| $^{1}$H(2,6) | −6.5 | −6.9 | −6.5 | 11.7 | 2.4 | −6.9 | 12.0 | 2.6 | −6.2 | 11.8 | 2.5 |
| $^{1}$H(3,5) | 1.5 | 2.0 | 2.1 | 3.5 | 1.0 | 2.4 | 3.8 | 1.0 | 2.0 | 3.5 | 1.0 |
| $^{13}$C(1) | – | – | −9.7 | 3.2 | 0.1 | −11.8 | 8.4 | 3.9 | −9.7 | 3.1 | 0.6 |
| $^{13}$C(2,6) | – | – | 6.1 | 43.7 | 0.5 | 7.0 | 46.7 | 0.4 | 4.1 | 43.6 | 0.5 |
| $^{13}$C(3,5) | – | – | −7.9 | 8.3 | 7.5 | −8.3 | 9.4 | 8.4 | −7.1 | 8.3 | 7.5 |
| $^{13}$C(4) | – | – | 12.1 | 68.1 | 0.1 | 12.2 | 66.9 | 0.1 | 9.1 | 68.0 | 0.1 |

[†]See (1.25) for HFC tensor axiality and rhombicity definitions.
[a]Reference [84], EPR spectra recorded in aqueous sulphuric acid.
[b]Reference [85], EPR spectra recorded in a solid adamantane matrix.

### *2.4.3 Tight basis functions*

The inadequacy of general-purpose basis sets for computing isotropic hyperfine couplings is illustrated on the example of DFT B3LYP 6-311++G(2d,2p)/aug-cc-pVQZ calculation. A very large and chemically accurate aug-cc-pVQZ basis set ideally reproduces the overall unpaired electron distribution and hence the anisotropy parameters (which are almost identical to EPR-III data), but fails to reproduce experimental value for $^{19}$F contact coupling. Because the wave function cusp at the nucleus is smoothed out by aug-cc-pVQZ, the resulting isotropic HFCs, especially the larger ones, come out wrong by a large margin.

The experimental [84] *versus* theoretical plot for the hyperfine couplings in fluorophenoxyl radicals that are experimentally available is shown on Figure 2.5. The agreement is quite good, however, as discussed above, the large hyperfine couplings are underestimated by the *ab initio* calculation due to the lack of vibrational averaging.

As a side note, while it is extremely good at computing hyperfine couplings, the GIAO DFT B3LYP 6-311++G(2d,2p)/EPR-III method appears to be inadequate for calculation the radical g-factors, in particular it is certainly not accurate enough to distinguish the three isomeric fluorophenoxyl radicals:

| Radical | Experiment [84] | GIAO DFT B3LYP 6-311++G(2d,2p)/EPR-III calculation in PCM water |
|---|---|---|
| 2-fluorophenoxyl | 2.00475 | 2.00559 |
| 3-fluorophenoxyl | 2.00456 | 2.00555 |
| 4-fluorophenoxyl | 2.00503 | 2.00594 |

The error looks systematic rather than random, and is small enough to permit calculation of expected CIDNP intensities if $\Delta g$ value is larger than about 0.002.

### *2.4.4 Sharp hyperfine couplings*

Before we proceed to take a look at the computed hyperfine tensors for fluorophenoxyl radical, one other phenomenon should be discussed, specifically, the very peculiar shape of many $^{13}C$ hyperfine coupling tensors in aromatic radicals. In all fluorophenoxyl radicals (and also in a large number of other aromatic radicals that were computed at different stages of the work reported herein) the carbons that bear substantial *p*-type HOMO density all have very large hyperfine tensor components along the eigenaxis perpendicular to the aromatic ring plane and almost zero eigenvalues along the axes lying in the ring plane. In other words, these HFC tensors are very stretched in one direction and almost needle-like. This peculiar shape is easy to rationalise if we consider a dipolar interaction tensor between a nucleus and a cloud of electron density located above and below, but not around it (Figure 2.4). Evaluating the dipolar integral results in a tensor, which does indeed have a very elongated shape in the ellipsoid plot.

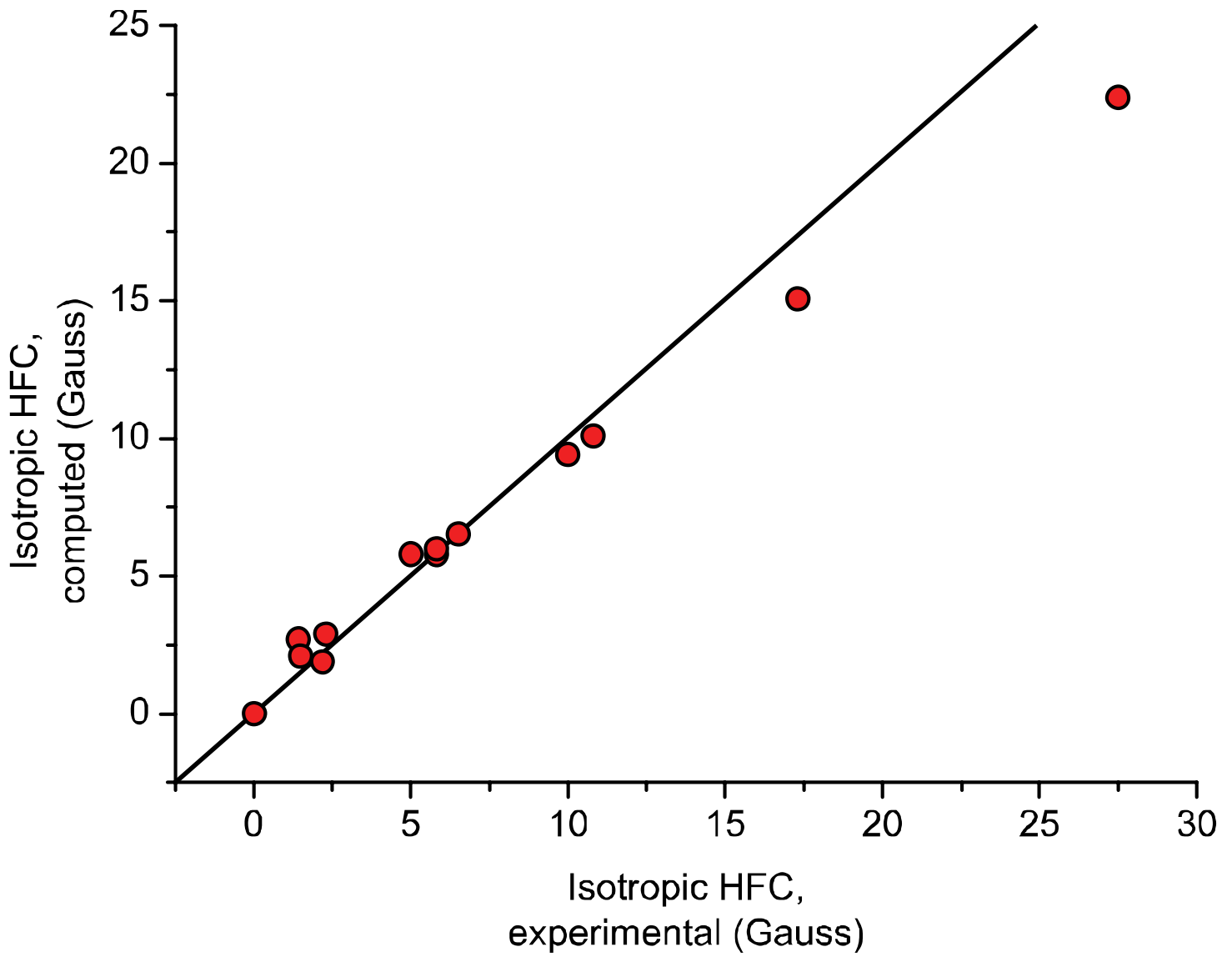

**Figure 2.5.** Experimental isotropic hyperfine couplings in three fluorophenoxyl radicals, obtained from spectra recorded in aqueous $H_2SO_4$ *versus* DFT B3LYP 6-311++G(2d,2p)/EPR-III calculation in PCM water.

Beyond just being a peculiar phenomenon, these extremely anisotropic HFC tensors (depicted in quantities on Figures 2.5-2.7 and throughout this Thesis) are likely to cause strong orientation dependence of recombination probability in the spatially ordered radical pairs [59]. This anisotropy becomes even more pronounced, if we notice that all such HFC tensors have parallel axes. In a suitably chosen molecule this could well result in overall HFC anisotropy of the order of kiloGauss for certain nuclear entanglements.

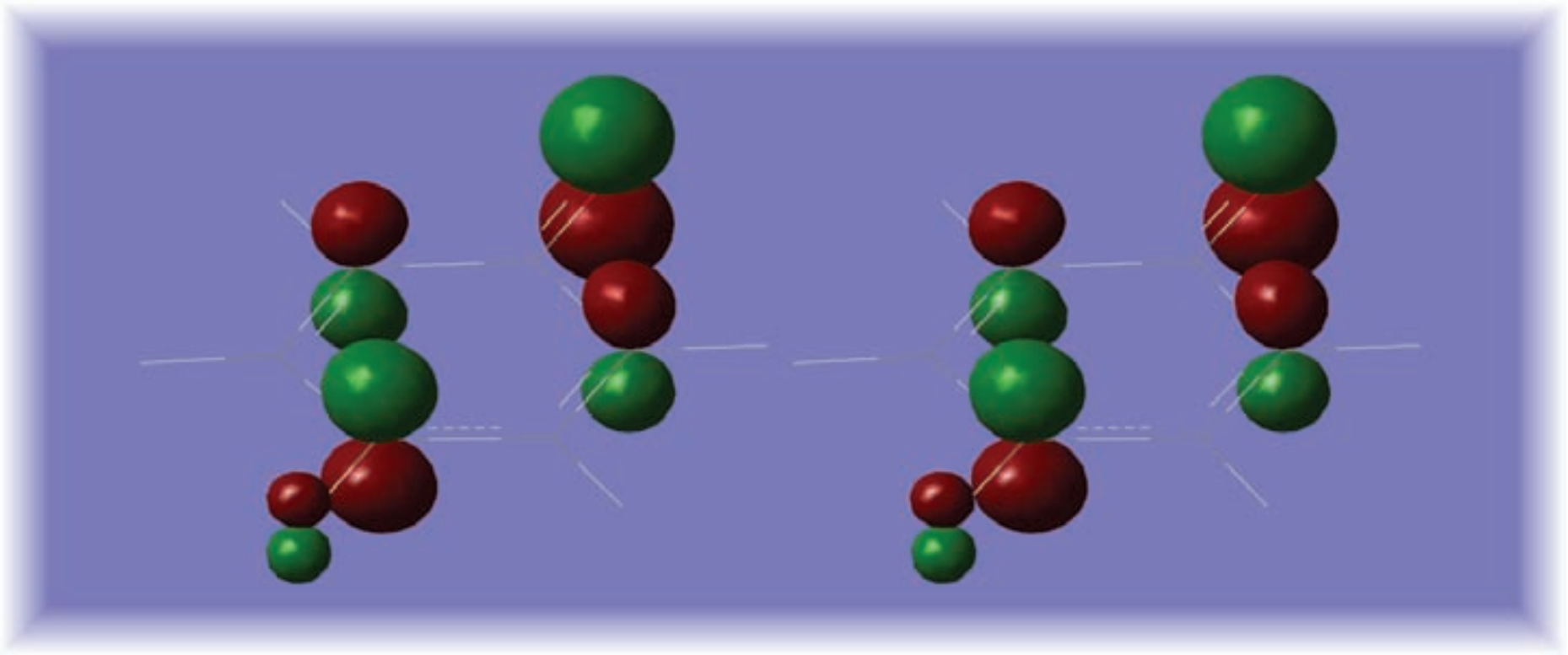

**Figure 2.4.** Stereo view of HOMO $\psi = 0.10$ isosurface in 4-fluorophenoxyl radical (DFT B3LYP 6-311G++(2d,2p)/EPR-III in PCM water). Those carbon atoms that show sharp hyperfine tensors (e.g. Figure 2.6) have large clouds of spin density above and below them.

### *2.4.5 Fluorophenoxyl radicals*

The computed hyperfine coupling tensors for the three fluorophenoxyl radicals are shown as ellipsoid plots on Figures 2.5-2.8 and reported in Tables 2.2 and 2.3. There are several things to note. First of all, to a very high accuracy the $^{19}$F HFC tensors in 2-fluorophenoxyl and 4-fluorophenoxyl radicals are axial. This is a pleasant finding, because it results in a substantial simplification to the theoretical treatment further on, removing a number of irreducible spherical tensors from the calculation and simplifying Wigner rotation matrices in the relaxation theory setup. Second, the isotropic $^{19}$F HFC of 2- and 4-fluorophenoxyls is extremely large, explaining earlier findings that $^{19}$F photo-CIDNP effect is very strong. A single dominant hyperfine coupling also means that a single-nucleus radical is likely to be a good approximation for describing photo-CIDNP effect in these systems (the *ab initio* computation was performed for the $^{17}$O isotopomer, but the natural abundance of magnetic isotopes of oxygen is 0.24% [86]) .

Another notable finding is the remarkably high anisotropy of $^{19}$F hyperfine tensors, which is as high as 242 Gauss in 4-fluorophenoxyl radical. Rotational modulation of hyperfine interaction is the primary nuclear relaxation mechanism in radicals, furthermore, the relaxation rate is quadratic in HFC anisotropy. We should therefore expect to find an extremely fast nuclear paramagnetic relaxation, at least for some correlation times, and expect relaxation processes to play a major role in $^{19}$F CIDNP formation. The complicated relaxation channels of the photo-CIDNP effect are explored in detail in Chapter 8.

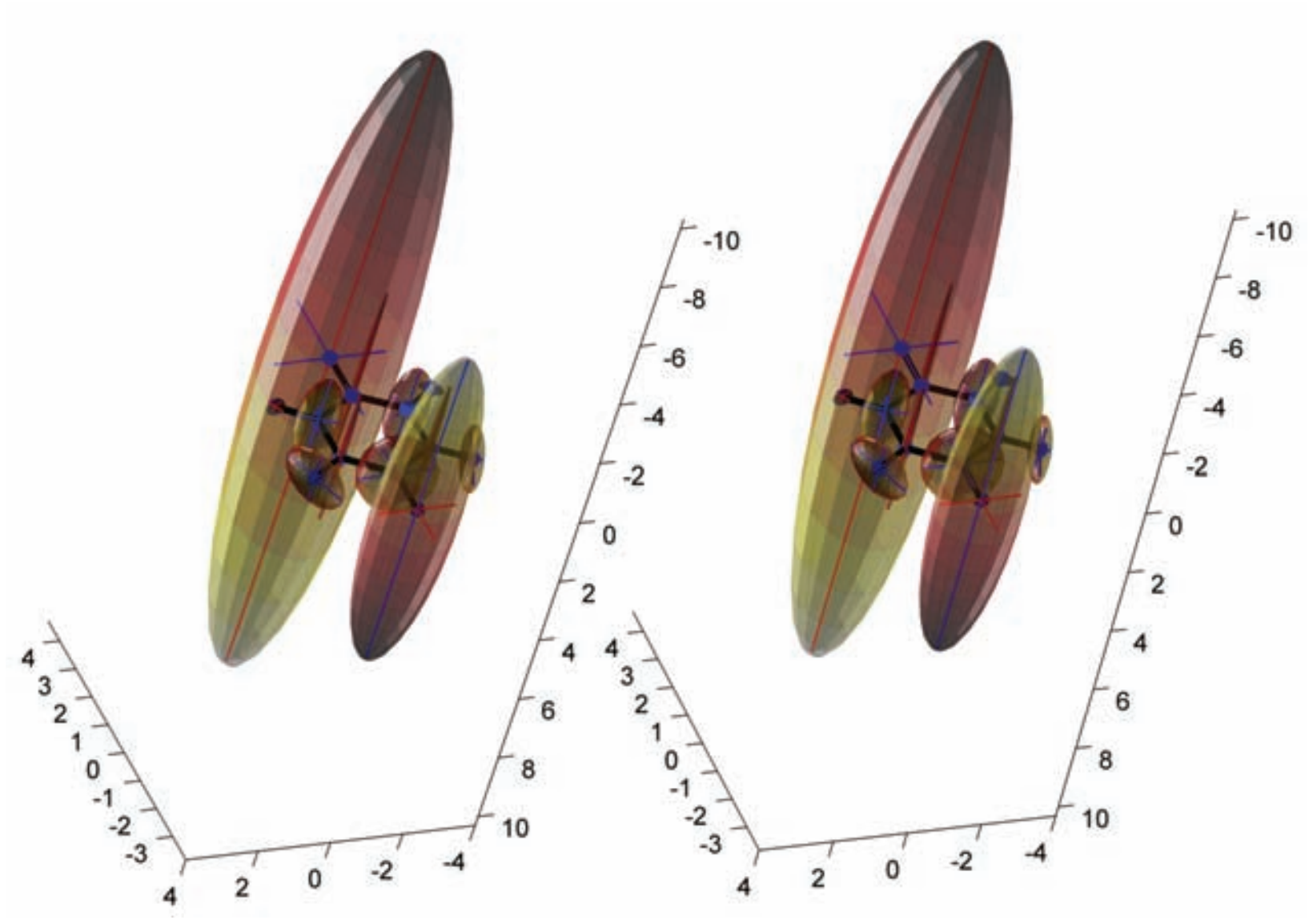

**Figure 2.5.** Stereo view of an ellipsoid plot of hyperfine tensors in 4-fluorophenoxyl radical with geometry obtained form a DFT B3LYP/6-311++G(2d,2p) calculation in PCM water and hyperfine tensors from a DFT B3LYP/EPR-III in PCM water. The largest ellipsoid belongs to fluorine.

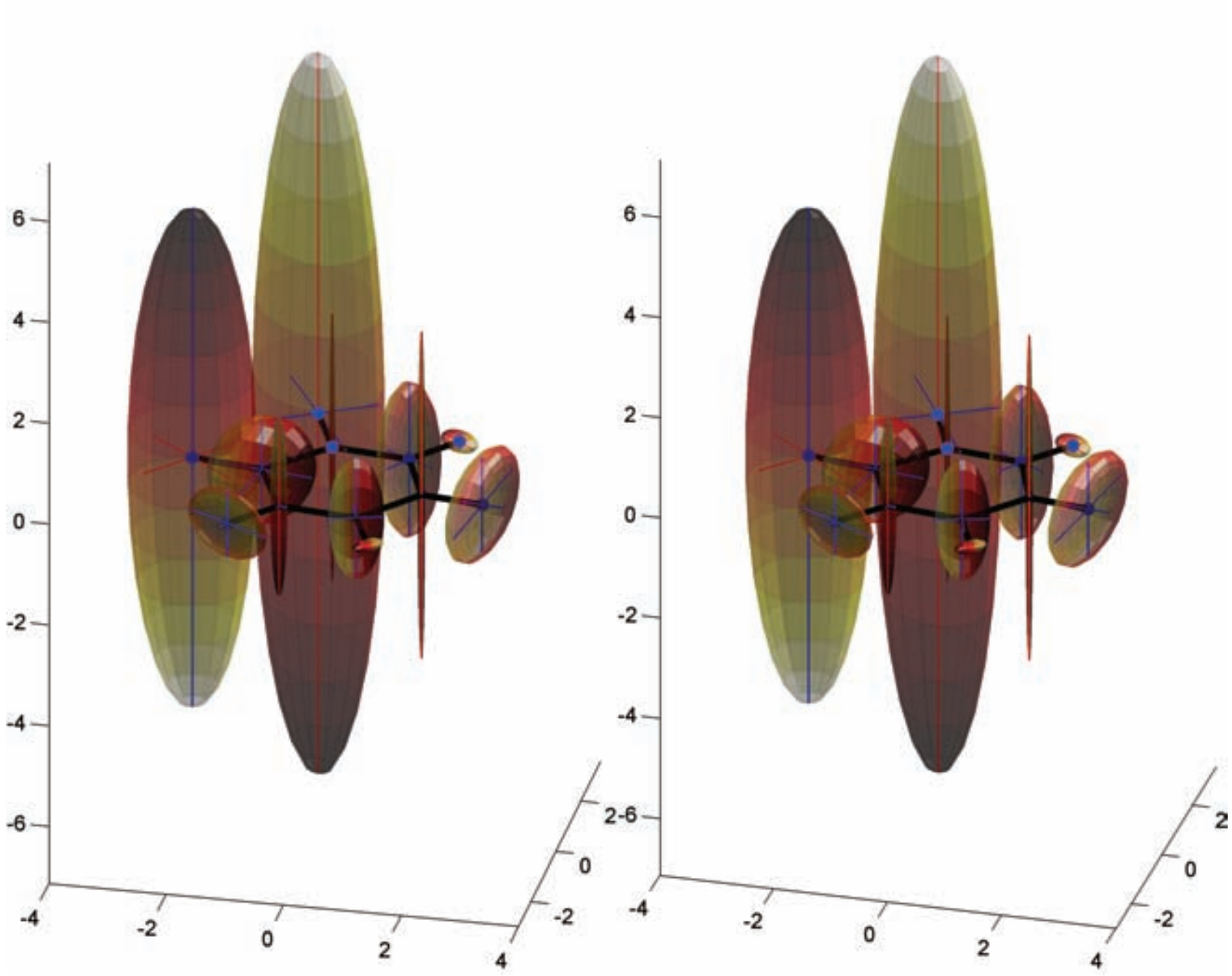

**Figure 2.6.** Stereo view of an ellipsoid plot of hyperfine tensors in 2-fluorophenoxyl radical with geometry obtained form a DFT B3LYP/6-311++G(2d,2p) calculation in PCM water and hyperfine tensors from a DFT B3LYP/EPR-III in PCM water. The largest ellipsoid belongs to fluorine.

From the *ab initio* data it looks like 4-fluorophenol with its high symmetry, large hyperfine coupling and huge HFC anisotropy is a compound of choice for exploring the mechanistic and theoretical side of the $^{19}$F photo-CIDNP phenomenon.

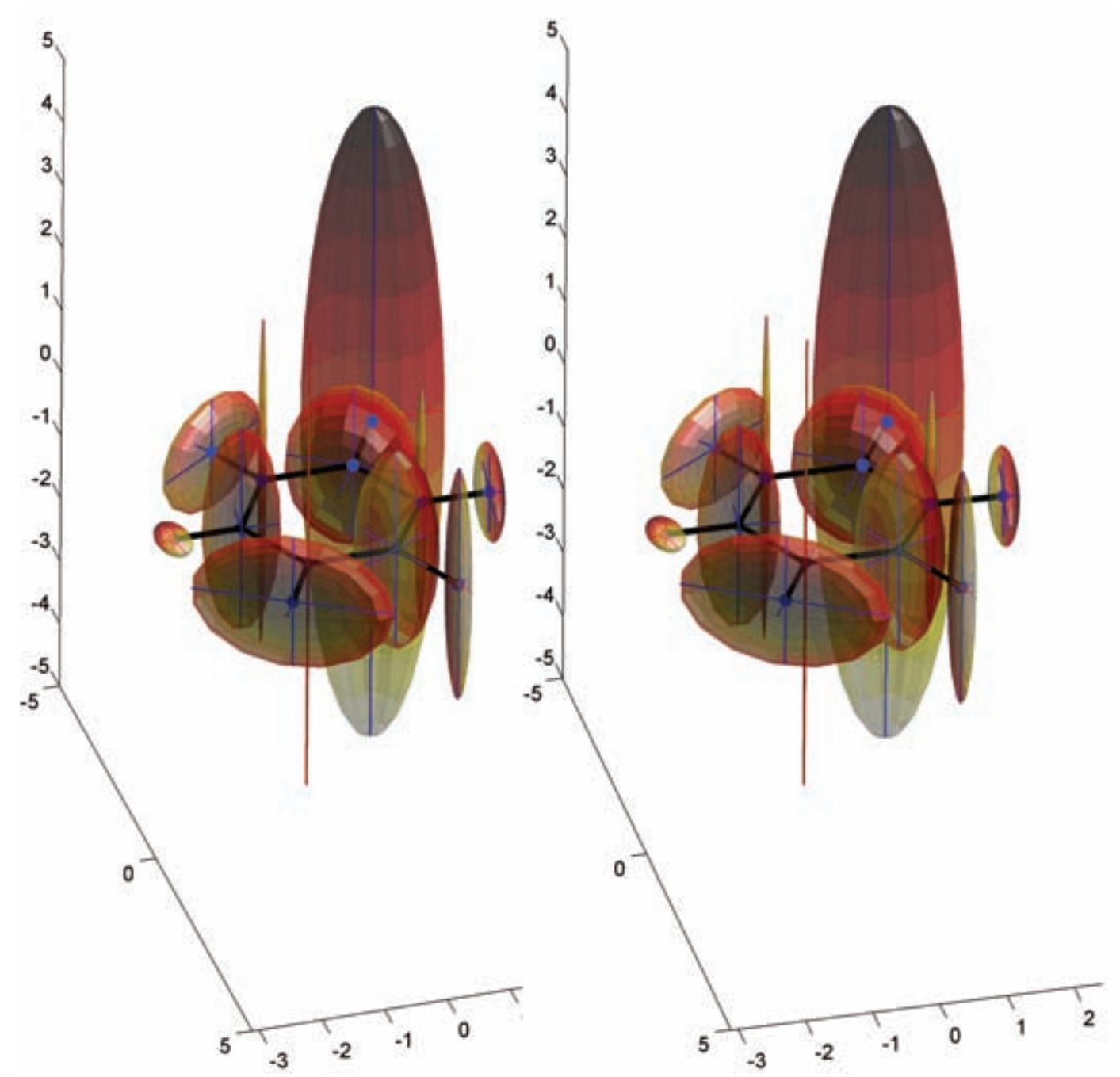

**Figure 2.7.** Stereo view of an ellipsoid plot of hyperfine tensors in 3-fluorophenoxyl radical with geometry obtained form a DFT B3LYP/6-311++G(2d,2p) calculation in PCM water and hyperfine tensors from a DFT B3LYP/EPR-III in PCM water. The largest ellipsoid belongs to oxygen.

**Table 2.2** Experimental and calculated hyperfine coupling tensor parameters (in units of Gauss) in 2-fluorophenoxyl and 3-fluorophenoxyl radicals[†].

| 2-fluorophenoxyl | | | | | 3-fluorophenoxyl | | | | |
|---|---|---|---|---|---|---|---|---|---|
| Nucleus | Exp. | DFT B3LYP 6-311++G(2d,2p)/EPR-III in PCM water | | | Nucleus | Exp. | DFT B3LYP 6-311++G(2d,2p)/EPR-III in PCM water | | |
| | Iso | Iso | Ax | Rh | | Iso | Iso | Ax | Rh |
| $^{17}$O(1) | – | −9.2 | 120.4 | 0.1 | $^{17}$O(1) | – | −9.3 | 122.4 | 0.0 |
| **$^{19}$F(2)** | **17.3** | **15.1** | **168.8** | **4.7** | $^{1}$H(2) | −5.0 | −5.8 | 10.7 | 2.1 |
| $^{1}$H(3) | 1.4 | 2.7 | 3.9 | 0.9 | **$^{19}$F(3)** | **−5.8** | **−6.0** | **25.0** | **16.7** |
| $^{1}$H(4) | −10.0 | −9.4 | 16.4 | 5.3 | $^{1}$H(4) | −10.8 | −10.1 | 17.8 | 5.7 |
| $^{1}$H(5) | 2.2 | 1.9 | 3.5 | 0.9 | $^{1}$H(5) | 2.3 | 2.9 | 4.3 | 0.9 |
| $^{1}$H(6) | −5.8 | −5.8 | 10.8 | 2.1 | $^{1}$H(6) | −7.9 | −7.7 | 13.6 | 3.1 |
| $^{13}$C(1) | – | −9.7 | 2.7 | 1.3 | $^{13}$C(1) | – | −10.4 | 5.1 | 1.4 |
| $^{13}$C(2) | – | 8.5 | 53.8 | 0.4 | $^{13}$C(2) | – | 5.5 | 39.6 | 0.4 |
| $^{13}$C(3) | – | −8.7 | 10.8 | 9.7 | $^{13}$C(3) | – | −9.6 | 9.5 | 8.7 |
| $^{13}$C(4) | – | 11.0 | 64.1 | 0.7 | $^{13}$C(4) | – | 11.9 | 70.5 | 0.6 |
| $^{13}$C(5) | – | −7.2 | 7.7 | 6.8 | $^{13}$C(5) | – | −9.2 | 11.5 | 10.0 |
| $^{13}$C(6) | – | 5.1 | 38.4 | 0.5 | $^{13}$C(6) | – | 8.0 | 53.0 | 0.4 |

[†]See (1.25) for HFC tensor axiality and rhombicity definitions.

[a]Reference [84], EPR spectra recorded in aqueous sulphuric acid.

### *2.4.6 Fluorotyrosyl radicals*

The ring-fluorinated tyrosines have been used for a long time in protein $^{19}$F spectroscopy, providing information on enzymatic reaction mechanisms [87-89], biological ligand binding [90-93], protein structure and dynamics [94-97]. Fluorotyrosines are relatively straightforward to incorporate into a protein, either biosynthetically, or using solid phase peptide synthesis. If useful data can be extracted from $^{19}$F photo-CIDNP experiments in fluorotyrosines, the established fluorinated proteins community should certainly benefit from it.

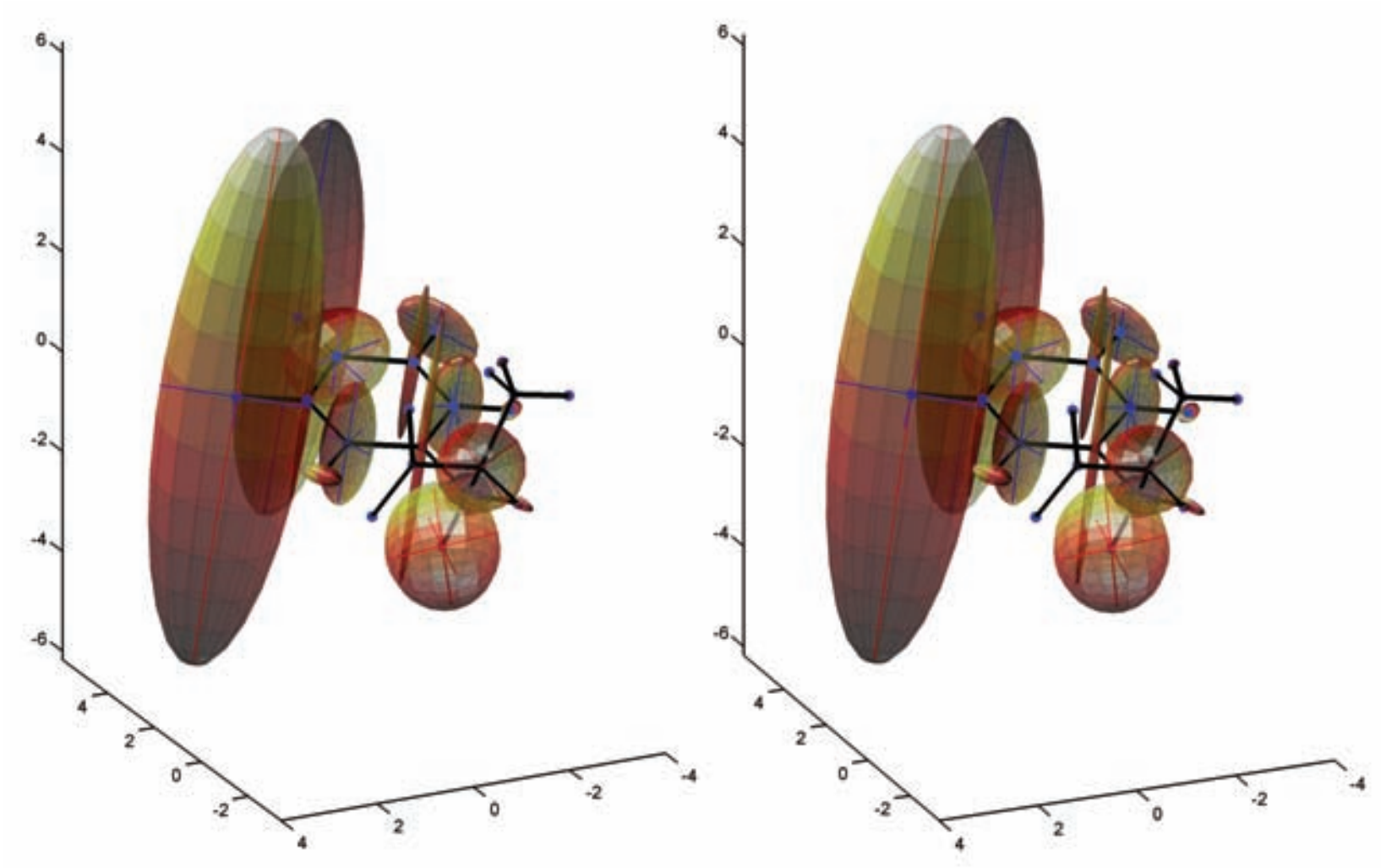

**Figure 2.8.** Stereo view of an ellipsoid plot of hyperfine tensors in 3-fluorotyrosyl radical with geometry obtained form a DFT B3LYP/6-311++G(2d,2p) calculation in PCM water and hyperfine tensors from a DFT B3LYP/EPR-III in PCM water. The largest ellipsoid belongs to fluorine.

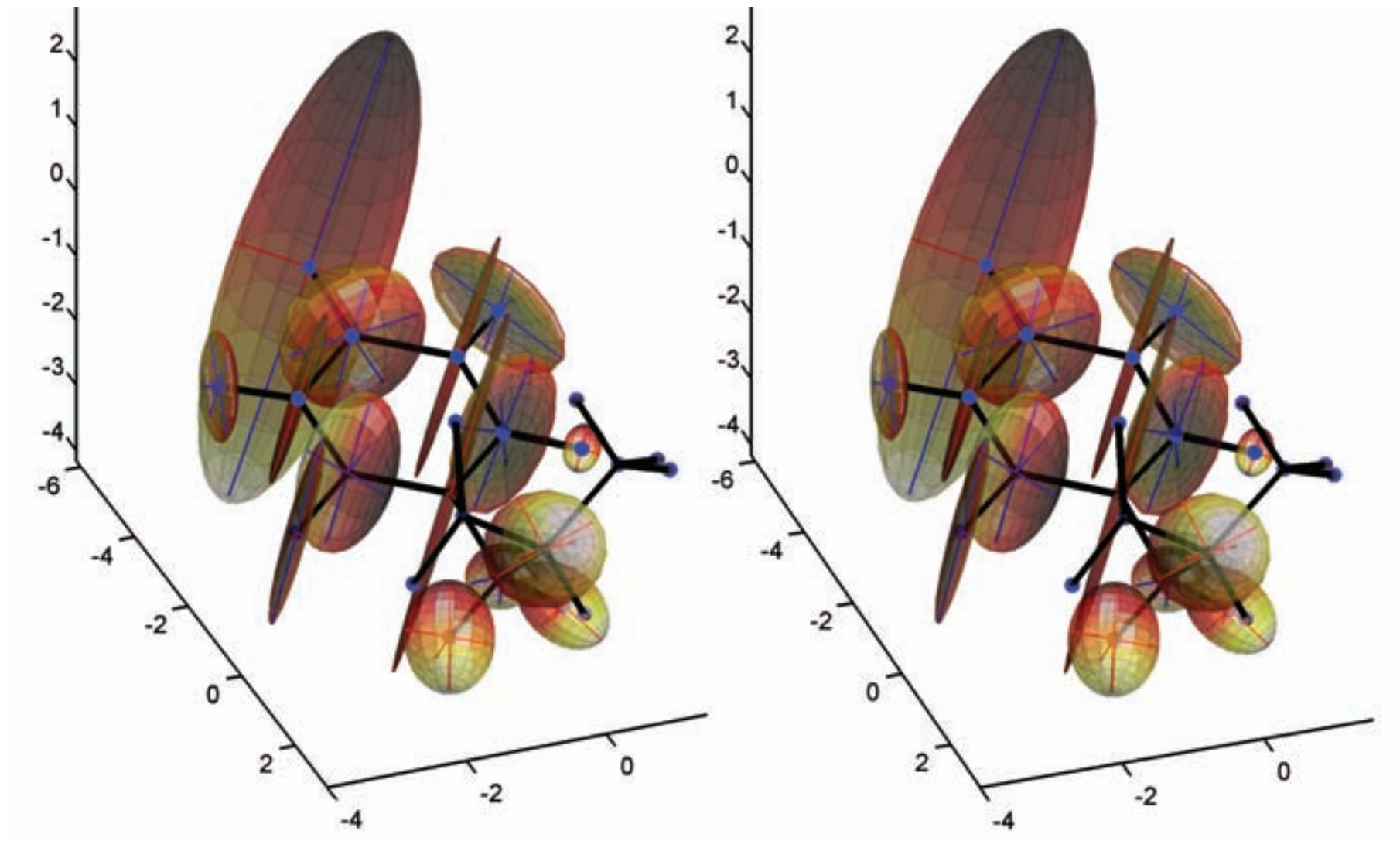

**Figure 2.9.** Stereo view of an ellipsoid plot of hyperfine tensors in 2-fluorotyrosyl radical with geometry obtained form a DFT B3LYP/6-311++G(2d,2p) calculation in PCM water and hyperfine tensors from a DFT B3LYP/EPR-III in PCM water.

It is clear from the data presented in Table 2.3 that 2-fluorotyrosine is unlikely to be a good target for $^{19}$F photo-CIDNP experiments, owing to a small hyperfine coupling. Since 2-fluorotyrosine is also considerably more difficult to synthesize than its 3-fluorinated counterpart, there's no reason whatsoever to even try 2-fluorotyrosine, and the commercially available (and therefore widely used for protein incorporation) 3-fluorotyrosine will be chosen for the experiments described later in this Thesis.

**Table 2.3** Calculated hyperfine coupling tensor parameters (in units of Gauss) in 2-fluorotyrosyl and 3-fluorotyrosyl radicals†.

| 2-fluorotyrosyl | | | | 3-fluorotyrosyl | | | |
|---|---|---|---|---|---|---|---|
| Nucleus | DFT B3LYP 6-311++G(2d,2p)/EPR-III in PCM water | | | Nucleus | DFT B3LYP 6-311++G(2d,2p)/EPR-III in PCM water | | |
| | Iso | Ax | Rh | | Iso | Ax | Rh |
| $^{17}$O(4) | −9.0 | 117.8 | 0.0 | $^{17}$O(4) | −9.2 | 118.0 | 0.1 |
| $^{13}$C(1) | 12.9 | 71.4 | 0.5 | $^{13}$C(1) | 11.7 | 64.1 | 0.5 |
| $^{13}$C(2) | −9.4 | 8.9 | 8.2 | $^{13}$C(2) | −8.4 | 10.4 | 9.3 |
| $^{13}$C(3) | 5.2 | 37.7 | 0.4 | $^{13}$C(3) | 7.8 | 49.5 | 0.4 |
| $^{13}$C(4) | −10.0 | 4.9 | 1.2 | $^{13}$C(4) | −9.6 | 2.8 | 1.3 |
| $^{13}$C(5) | 7.8 | 51.9 | 0.4 | $^{13}$C(5) | 5.4 | 40.5 | 0.5 |
| $^{13}$C(6) | −9.1 | 11.1 | 9.5 | $^{13}$C(6) | −7.4 | 7.6 | 6.6 |
| $^{13}$C($CH_2$) | −4.7 | 0.9 | 0.5 | $^{13}$C($CH_2$) | −4.2 | 0.9 | 0.5 |
| $^{13}$C(CH) | 8.9 | 6.2 | 0.1 | $^{13}$C(CH) | 9.2 | 5.0 | 0.1 |
| **$^{19}$F(2)** | **−5.8** | **25.2** | **17.4** | $^{1}$H(2) | 2.6 | 3.9 | 0.9 |
| $^{1}$H(3) | −5.5 | 10.3 | 2.0 | **$^{19}$F(3)** | **13.9** | **157.0** | **4.4** |
| $^{1}$H(5) | −7.5 | 13.3 | 3.1 | $^{1}$H(5) | −6.1 | 11.2 | 2.2 |
| $^{1}$H(6) | 2.7 | 4.2 | 1.0 | $^{1}$H(6) | 1.8 | 3.3 | 1.0 |
| $^{1}$H($CH_2$) | 5.7 | 4.2 | 0.2 | $^{1}$H($CH_2$) | 1.6 | 3.9 | 0.2 |
| $^{1}$H($CH_2$) | 6.8 | 4.2 | 0.2 | $^{1}$H($CH_2$) | 11.4 | 4.0 | 0.3 |
| $^{1}$H(CH) | 0.2 | 2.7 | 0.4 | $^{1}$H(CH) | 0.1 | 2.4 | 0.3 |

†See (1.25) for HFC tensor axiality and rhombicity definitions.

### *2.4.7 Fluorotryptophan radicals*

The sheer number of possible mono- and polyfluorinated tryptophans, combined with the typical cost of £1000 per gram of enantiomerically pure compound, clearly means that *ab initio* evaluation has to be performed before trying out the fluorotryptophans experimentally or preparing proteins labelled with them. The high-quality *ab initio* calculations of molecular geometry and hyperfine couplings of fluorotryptophan are on the brink of the capability of the processing hardware that was available to the author. For each of the molecules tabulated below, the full geometry optimization and HFC calculation (see Section 2.3 for details) has taken about 30 days of CPU time on the SunFire 6800 supercomputer. With 12 processors working in parallel and 20GB of RAM allocated for the two-electron integral storage, the calculation took about 60 wall clock hours per molecule, which is acceptable.

**Table 2.4.** Calculated (DFT B3LYP 6-311++G(2d,2p)/EPR-III in PCM water) hyperfine coupling tensor parameters (in units of Gauss) for five possible ring-fluorinated tryptophan cation radicals.

| 2−fluorotryptophan cation | | | | 4−fluorotryptophan cation | | | | 5−fluorotryptophan cation | | | | 6−fluorotryptophan cation | | | | 7−fluorotryptophan cation | | | |
|---|---|---|---|---|---|---|---|---|---|---|---|---|---|---|---|---|---|---|---|
| Nuc | Iso | Ax | Rh | Nuc | Iso | Ax | Rh | Nuc | Iso | Ax | Rh | Nuc | Iso | Ax | Rh | Nuc | Iso | Ax | Rh |
| N(1) | 0.35 | 8.14 | 0.16 | N(1) | 2.10 | 16.68 | 0.13 | N(1) | 2.50 | 19.33 | 0.16 | N(1) | 1.22 | 13.20 | 0.15 | N(1) | 2.47 | 18.41 | 0.13 |
| C(2) | −0.79 | 36.10 | 0.86 | C(2) | −2.02 | 23.62 | 0.57 | C(2) | −3.28 | 20.86 | 0.64 | C(2) | 0.44 | 36.77 | 0.59 | C(2) | −2.98 | 19.12 | 0.59 |
| C(3) | 8.95 | 67.54 | 0.91 | C(3) | 9.11 | 64.23 | 0.53 | C(3) | 11.35 | 76.59 | 0.51 | C(3) | 8.04 | 61.51 | 0.61 | C(3) | 9.78 | 66.63 | 0.48 |
| C(3′) | −7.78 | 6.18 | 3.66 | C(3′) | −7.22 | 7.12 | 4.45 | C(3′) | −8.10 | 8.50 | 5.63 | C(3′) | −6.82 | 6.15 | 3.84 | C(3′) | −7.61 | 8.27 | 5.50 |
| C(4) | 6.57 | 39.75 | 0.23 | C(4) | 6.80 | 42.17 | 0.23 | C(4) | 6.31 | 40.22 | 0.05 | C(4) | 5.57 | 33.29 | 0.22 | C(4) | 7.16 | 45.28 | 0.03 |
| C(5) | −6.14 | 7.78 | 6.86 | C(5) | −4.73 | 3.46 | 2.87 | C(5) | −5.36 | 4.50 | 3.77 | C(5) | −5.44 | 6.76 | 6.17 | C(5) | −4.86 | 4.04 | 3.11 |
| C(6) | 6.41 | 40.44 | 0.36 | C(6) | 3.14 | 27.10 | 0.22 | C(6) | 3.07 | 23.47 | 0.23 | C(6) | 5.48 | 37.31 | 0.28 | C(6) | 3.50 | 28.77 | 0.26 |
| C(7) | −2.97 | 1.11 | 0.43 | C(7) | 0.65 | 17.2 | 0.33 | C(7) | −0.47 | 8.40 | 0.35 | C(7) | −1.59 | 8.9 | 0.38 | C(7) | 0.18 | 15.93 | 0.46 |
| C(7′) | 2.37 | 15.50 | 0.47 | C(7′) | −1.34 | 1.95 | 0.80 | C(7′) | −0.52 | 4.10 | 0.72 | C(7′) | 0.83 | 12.25 | 0.73 | C(7′) | −1.74 | 1.51 | 0.09 |
| C(β) | −4.19 | 1.36 | 0.55 | C(β) | −4.12 | 1.71 | 0.53 | C(β) | −4.87 | 1.95 | 0.60 | C(β) | −4.01 | 1.34 | 0.57 | C(β) | −4.26 | 1.55 | 0.52 |
| C(α) | 5.24 | 7.91 | 0.07 | C(α) | 5.06 | 7.55 | 0.07 | C(α) | 5.70 | 9.65 | 0.08 | C(α) | 4.74 | 7.55 | 0.06 | C(α) | 4.81 | 8.11 | 0.07 |
| H(1) | −2.42 | 8.31 | 0.91 | H(1) | −4.69 | 13.36 | 2.23 | H(1) | −5.43 | 15.33 | 2.75 | H(1) | −3.79 | 11.66 | 1.66 | H(1) | −5.15 | 14.26 | 2.54 |
| **F(2)** | **11.87** | **146.49** | **0.97** | H(2) | −3.61 | 9.61 | 1.47 | H(2) | −3.24 | 9.61 | 1.21 | H(2) | −5.57 | 12.41 | 2.50 | H(2) | −2.91 | 8.73 | 1.08 |
| H(4) | −5.76 | 9.91 | 2.39 | **F(4)** | **14.25** | **159.62** | **3.56** | H(4) | −5.67 | 9.62 | 2.24 | H(4) | −4.91 | 8.39 | 1.93 | H(4) | −6.40 | 11.01 | 2.74 |
| H(5) | 1.89 | 2.86 | 0.65 | H(5) | 0.70 | 2.70 | 0.59 | **F(5)** | **−2.70** | **9.68** | **5.19** | H(5) | 1.68 | 2.49 | 0.63 | H(5) | 0.80 | 2.9 | 0.54 |
| H(6) | −5.87 | 10.88 | 3.20 | H(6) | −4.05 | 8.34 | 2.07 | H(6) | −3.41 | 7.07 | 1.81 | **F(6)** | **13.29** | **156.29** | **2.39** | H(6) | −4.21 | 8.60 | 2.11 |
| H(7) | −0.06 | 2.90 | 0.52 | H(7) | −2.30 | 5.67 | 0.59 | H(7) | −1.11 | 3.34 | 0.18 | H(7) | −1.15 | 4.1 | 0.16 | **F(7)** | **5.47** | **67.55** | **0.60** |
| H(β) | 3.68 | 5.03 | 0.39 | H(β) | 0.73 | 4.71 | 0.22 | H(β) | 2.28 | 5.24 | 0.31 | H(β) | 2.07 | 4.92 | 0.29 | H(β) | 1.97 | 4.72 | 0.27 |
| H(β) | 7.88 | 4.28 | 0.59 | H(β) | 14.81 | 4.34 | 0.61 | H(β) | 13.39 | 4.86 | 0.65 | H(β) | 10.28 | 4.24 | 0.55 | H(β) | 11.56 | 4.17 | 0.66 |
| H(α) | −0.75 | 3.36 | 0.91 | H(α) | −0.71 | 2.87 | 0.98 | H(α) | −0.91 | 3.51 | 1.14 | H(α) | −0.74 | 3.00 | 0.93 | H(α) | −0.79 | 2.88 | 1.08 |

†See (1.25) for HFC tensor axiality and rhombicity definitions.

**Table 2.5.** Calculated (DFT B3LYP 6-311++G(2d,2p)/EPR-III in PCM water) hyperfine coupling tensor parameters (in units of Gauss) for five possible ring-fluorinated tryptophan neutral radicals.

| 2−fluorotryptophan | | | | 4−fluorotryptophan | | | | 5−fluorotryptophan | | | | 6−fluorotryptophan | | | | 7−fluorotryptophan | | | |
|---|---|---|---|---|---|---|---|---|---|---|---|---|---|---|---|---|---|---|---|
| Nuc | Iso | Ax | Rh | Nuc | Iso | Ax | Rh | Nuc | Iso | Ax | Rh | Nuc | Iso | Ax | Rh | Nuc | Iso | Ax | Rh |
| N(1) | 1.66 | 14.16 | 0.08 | N(1) | 3.28 | 24.72 | 0.04 | N(1) | 3.90 | 29.09 | 0.03 | N(1) | 2.91 | 22.65 | 0.06 | N(1) | 3.74 | 27.56 | 0.04 |
| C(2) | -7.19 | 7.68 | 0.77 | C(2) | -7.90 | 3.84 | 2.46 | C(2) | -9.35 | 6.73 | 5.16 | C(2) | -6.89 | 1.02 | 0.46 | C(2) | -8.72 | 5.80 | 4.30 |
| C(3) | 15.59 | 95.52 | 0.32 | C(3) | 15.25 | 89.79 | 0.21 | C(3) | 16.38 | 96.11 | 0.11 | C(3) | 15.14 | 90.74 | 0.26 | C(3) | 15.30 | 90.10 | 0.13 |
| C(3´) | -10.09 | 9.89 | 7.04 | C(3´) | -8.83 | 8.79 | 6.05 | C(3´) | -8.46 | 7.44 | 4.69 | C(3´) | -8.95 | 8.93 | 6.38 | C(3´) | -8.27 | 8.00 | 5.33 |
| C(4) | 6.53 | 37.02 | 0.10 | C(4) | 6.14 | 34.93 | 0.34 | C(4) | 4.89 | 31.80 | 0.18 | C(4) | 5.40 | 30.60 | 0.08 | C(4) | 5.39 | 33.89 | 0.14 |
| C(5) | -5.70 | 6.87 | 6.00 | C(5) | -3.74 | 2.27 | 1.76 | C(5) | -3.99 | 1.93 | 1.33 | C(5) | -4.75 | 5.25 | 4.71 | C(5) | -3.58 | 1.71 | 1.02 |
| C(6) | 5.85 | 35.51 | 0.38 | C(6) | 3.18 | 25.52 | 0.23 | C(6) | 2.31 | 19.07 | 0.21 | C(6) | 4.83 | 31.31 | 0.21 | C(6) | 2.87 | 23.85 | 0.22 |
| C(7) | -3.72 | 2.77 | 1.77 | C(7) | -0.86 | 8.41 | 0.43 | C(7) | -0.86 | 5.38 | 0.42 | C(7) | -2.21 | 3.81 | 0.45 | C(7) | -0.66 | 9.07 | 0.47 |
| C(7´) | 3.35 | 19.7 | 0.41 | C(7´) | -0.53 | 7.55 | 0.75 | C(7´) | -1.21 | 5.34 | 0.72 | C(7´) | 0.98 | 14.60 | 0.70 | C(7´) | -1.63 | 3.23 | 0.79 |
| C(β) | -6.04 | 1.08 | 0.73 | C(β) | -5.76 | 1.04 | 0.57 | C(β) | -6.20 | 1.02 | 0.64 | C(β) | -5.92 | 1.01 | 0.55 | C(β) | -5.83 | 0.94 | 0.58 |
| C(α) | 12.65 | 8.40 | 0.09 | C(α) | 11.14 | 25.52 | 0.09 | C(α) | 11.41 | 7.78 | 0.13 | C(α) | 10.91 | 7.41 | 0.13 | C(α) | 10.73 | 7.36 | 0.12 |
| **F(2)** | **0.39** | **50.68** | **0.94** | H(2) | -0.05 | 5.05 | 0.68 | H(2) | 0.67 | 4.57 | 1.12 | H(2) | -0.97 | 6.16 | 0.23 | H(2) | 0.49 | 4.46 | 1.01 |
| H(4) | -5.39 | 8.79 | 2.09 | **F(4)** | **10.35** | **107.90** | **3.50** | H(4) | -4.56 | 7.60 | 1.57 | H(4) | -4.55 | 7.33 | 1.61 | H(4) | -4.87 | 8.10 | 1.78 |
| H(5) | 1.70 | 2.73 | 0.61 | H(5) | 0.42 | 2.49 | 0.51 | **F(5)** | **-1.09** | **1.89** | **1.30** | H(5) | 1.31 | 2.08 | 0.65 | H(5) | 0.24 | 2.73 | 0.35 |
| H(6) | -5.20 | 9.49 | 3.01 | H(6) | -3.82 | 7.61 | 2.06 | H(6) | -2.80 | 5.94 | 1.53 | **F(6)** | **9.84** | **109.23** | **2.23** | H(6) | -3.50 | 7.20 | 1.86 |
| H(7) | 0.62 | 1.61 | 1.03 | H(7) | -1.12 | 3.30 | 0.33 | H(7) | -0.76 | 2.29 | 0.63 | H(7) | -0.48 | 2.43 | 0.76 | **F(7)** | **2.45** | **31.30** | **0.78** |
| H(β) | 7.79 | 6.08 | 0.32 | H(β) | 1.37 | 5.64 | 0.18 | H(β) | 2.06 | 5.84 | 0.20 | H(β) | 2.32 | 5.80 | 0.19 | H(β) | 2.36 | 5.55 | 0.19 |
| H(β) | 7.67 | 5.32 | 0.47 | H(β) | 17.25 | 5.31 | 0.47 | H(β) | 16.67 | 5.67 | 0.48 | H(β) | 14.84 | 5.42 | 0.43 | H(β) | 14.59 | 5.26 | 0.48 |
| H(α) | -0.96 | 4.38 | 0.89 | H(α) | -0.94 | 3.48 | 0.84 | H(α) | -1.06 | 3.75 | 0.96 | H(α) | -1.02 | 3.57 | 0.91 | H(α) | -1.01 | 3.47 | 0.95 |

†See (1.25) for HFC tensor axiality and rhombicity definitions.

Arguably, the molecular geometry for tryptophans could have been computed in a smaller basis set (such as 6-31G**) without introducing too much of an error, but given the past controversy about the values of the hyperfine couplings in tryptophan radicals and the fact that HFCs are quite sensitive to distortions in molecular geometry, the author thought he should take no chances.

In neutral and basic tryptophan solutions the CIDNP generation mechanism involves an electron transfer from the tryptophan to the photosensitizer, followed by the deprotonation of the cation to form a neutral radical [98]. Depending on the acidity of the medium the deprotonation may occur with different speeds or not at all, meaning that there are two kinds of radical that can contribute, solely or in part, to the photo-CIDNP generation – neutral radical and cation radical. A test of involvement of cation radical in the photo-CIDNP generation on tryptophans may be suggested based on the computed hyperfine couplings from Tables 2.4 and 2.5: the hyperfine coupling on F(2) in 2-flurotryptophan is close to zero in the neutral radical, whereas the cation radical has a healthy 12 Gauss of isotropic HFC. If the photo-CIDNP effect is observed experimentally on either $^{19}$F in 2-fluorotryptophan, or H(2) in common tryptophan, this is an indication that the spin evolution occurs, to a considerable extent, in a tryptophan cation radical. The experimental data and its implications are analyzed later in this Chapter.

The computed hyperfine couplings in the intermediate cation radicals lead to several conclusions. The first is that out of five possible ring-fluorinated tryptophans, only 2-, 4- and 6-fluorinated isomers are feasible for use in $^{19}$F photo-CIDNP studies. Out of those three, the $^{19}$F CIDNP polarization in 2-fluorotryptophan is likely to have a strong pH-dependence due to different hyperfine couplings in the neutral and cation radical. Sadly, the synthetically most accessible (and therefore the cheapest) 5-fluorotryptophan appears useless. The second conclusion is that, just in the case of fluorophenols, the distinctive feature of fluorine is an extremely large HFC anisotropy, likely leading to very fast nuclear paramagnetic relaxation in these radicals. Lastly, although the $^{15}$N HFC on the aromatic nitrogen is quite small, the relative $^{15}$N photo-CIDNP enhancement will be rather large because of low equilibrium polarization of $^{15}$N nuclei. This finding is consistent with experimental observation made by Lyon [63] that nitrogen CIDNP polarization in tryptophan/FMN system is approximately two orders of magnitude larger than the equilibrium polarization.

### *2.4.8 Rotational dependence of the hyperfine couplings*

One of the obvious fluorination positions in all CIDNP-active amino acids is a β-$CH_2$ group, featuring large proton hyperfine couplings, at least for some conformations. A $CH_2$ proton positioned orthogonally to the aromatic ring plane is known to possess hyperfine coupling as large as 28 Gauss in a tryptophan cation [59]. It is expected that $^{19}$F hyperfine coupling in the same position would be at least twice as big, potentially making the $CH_2$ linker the best fluorination position from the photo-CIDNP effect magnitude point of view. The dihedral angle dependence of $^{19}$F hyperfine coupling in structurally similar anion radicals (fluorine-substituted nitrobenzene derivatives) has been addressed experimentally and using early quantum chemical methods in 1972-1993 by Polenov [99-101] and in 1990 by Beregovaya [102], who found the expected sinusiodal dependence of the $^{19}$F HFC on the dihedral angle and very large hyperfine couplings for certain values of that angle. Anion radical results, although instructive, are of little use for the present work, in which the target amino acid is present in either neutral or cation radical form. A separate calculation was therefore performed to clear up the question.

The correct analysis of $^{19}$F hyperfine couplings in β-fluorinated tyrosine or tryptophan radical requires scanning the dihedral angle corresponding to the $CH_2$ group rotation with constrained reoptimization of geometry at every step. The computation time required for accurate calculation of this kind on a whole amino acid is too large, so a smaller system was analysed, for which the calculation is affordable.

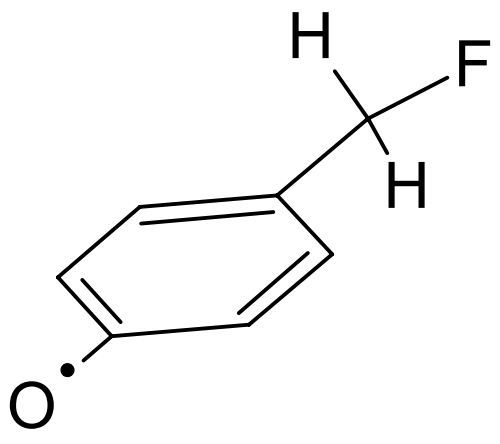

The simplest model system for β-fluorinated tyrosine is parafluoromethylphenoxyl radical (see the inset for structure), which is small enough for both fast and accurate *ab initio* calculation and possesses convenient symmetry. Ten constrained DFT B3LYP 6-311G(2d,2p) optimization runs were performed in PCM water with CCCF dihedral angle restricted to values from 0° to 90°. Hyperfine couplings were then computed for each optimized geometry with the basis set changed to EPR-III. Energy differences due to basis set superposition error [103, 104] have been neglected in the geometry optimization.

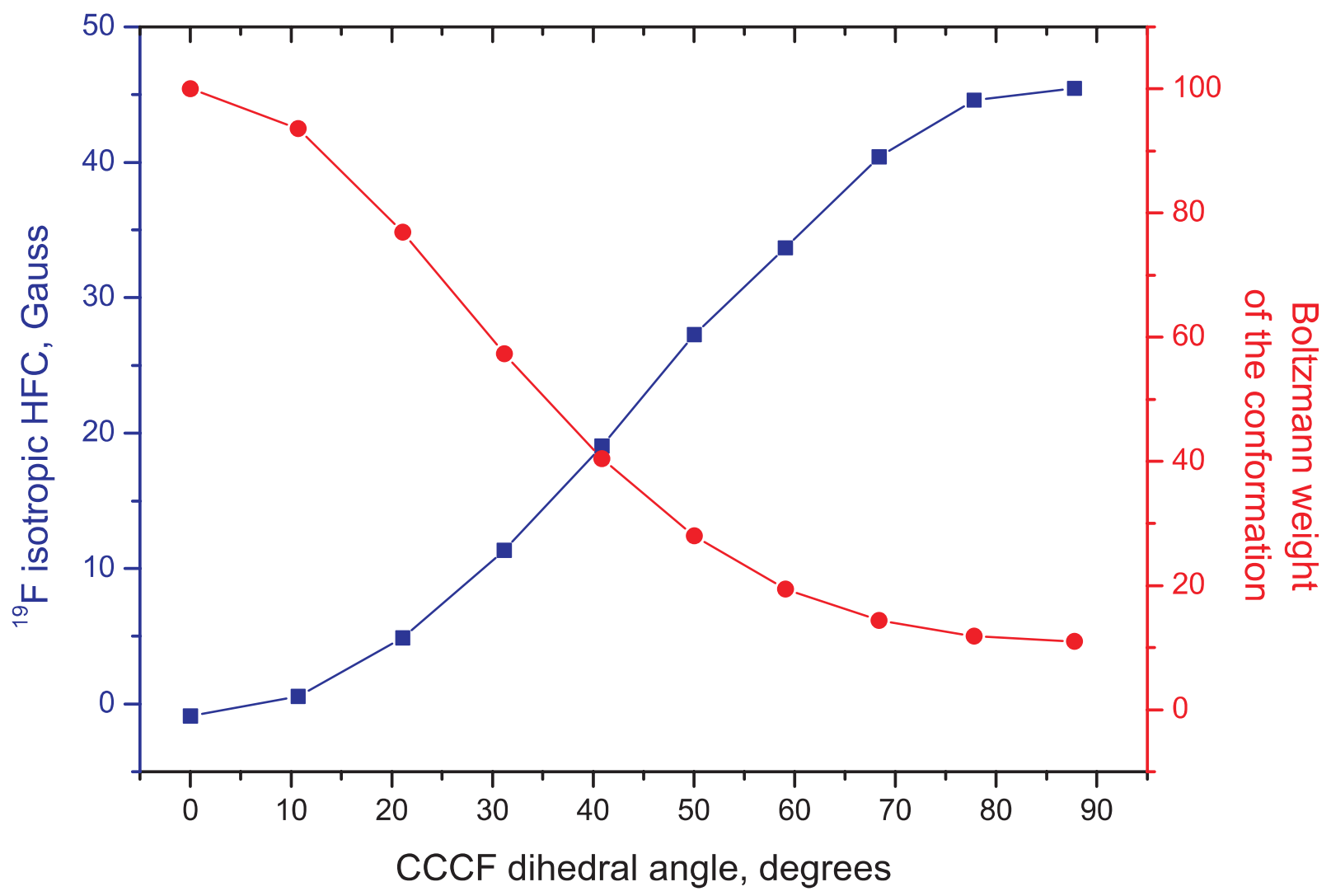

**Figure 2.10.** CCCF dihedral angle dependence of isotropic $^{19}$F hyperfine coupling in a parafluoromethylphenoxyl radical and the Boltzmann weights of corresponding conformations at 298 K.

The computation results are presented on Figures 2.10-2.12. Indeed, as can be seen from Figure 2.10, for certain conformations the $^{19}$F hyperfine coupling is in excess of 40 Gauss, which is nearly twice as much as the largest HFC found in the ring-fluorinated radicals. However, the consideration of the conformation energies shows that the high-HFC conformations are the least favourable, with Boltzmann weights of about 10%, and the energetically favourable confirmations have low HFCs. The Boltzmann average for the $^{19}$F isotropic hyperfine coupling at room temperature works out to 10.6 Gauss, which is smaller than the hyperfine couplings found in 3-fluorotyrosine and 4- and 6-fluorotryptophans.

Given the synthetic difficulty of preparing β-fluorinated aromatic amino acids, and the above mentioned fact that the average hyperfine coupling does not offer any benefits compared to ring-fluorinated amino acids, it was decided not to try β-fluorinated amino acids experimentally.

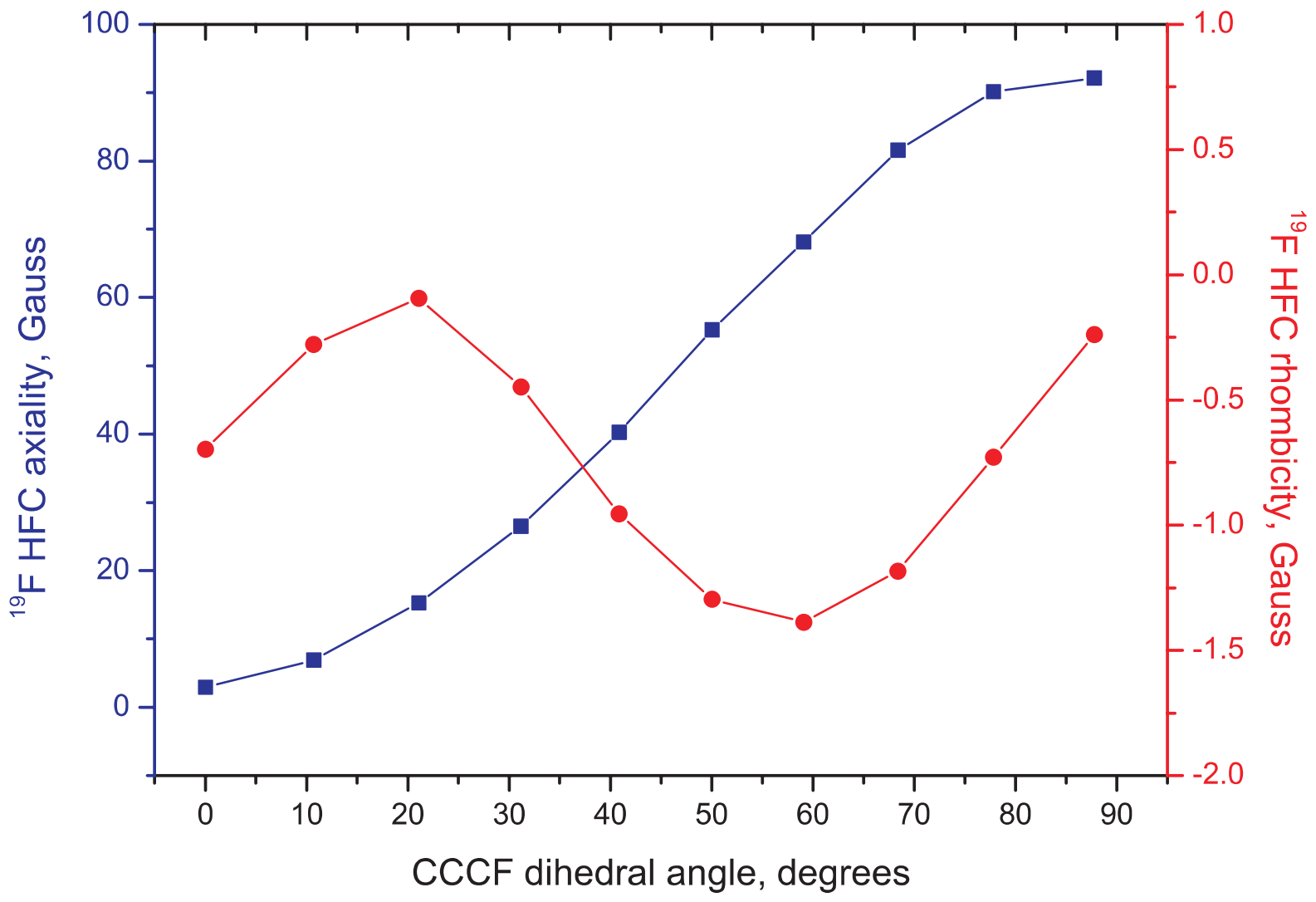

**Figure 2.11.** CCCF dihedral angle dependence of axiality and rhombicity of $^{19}$F hyperfine coupling in a parafluoromethylphenoxyl radical.

Not doing any experiments would not prevent us from harvesting some interesting conclusions from the calculations though. A potential difficulty (or a potential source of information if handled properly) with fluorine in a β-position of an amino acid is its relaxation behaviour. For ring nuclei of aromatic systems the isotropic hyperfine coupling is either constant (in a naive rigid approximation), or modulated with frequencies far removed from NMR frequency range (if we account for molecular vibrations). The primary relaxation mechanism therefore is purely rotational modulation of otherwise static anisotropies. The β-position, however, has both the isotropic HFC and HFC anisotropy *modulated in amplitude*, as a function of the dihedral angle. Because all three modulations (isotropic magnitude, anisotropic magnitude, anisotropic orientation) are functions of the same angles and at least two of them (isotropic and anisotropic magnitude) are strongly correlated, this leads to a very complex cross-correlated behaviour, which, although it could be interesting in its own right, it is not author's wish to untangle[6].

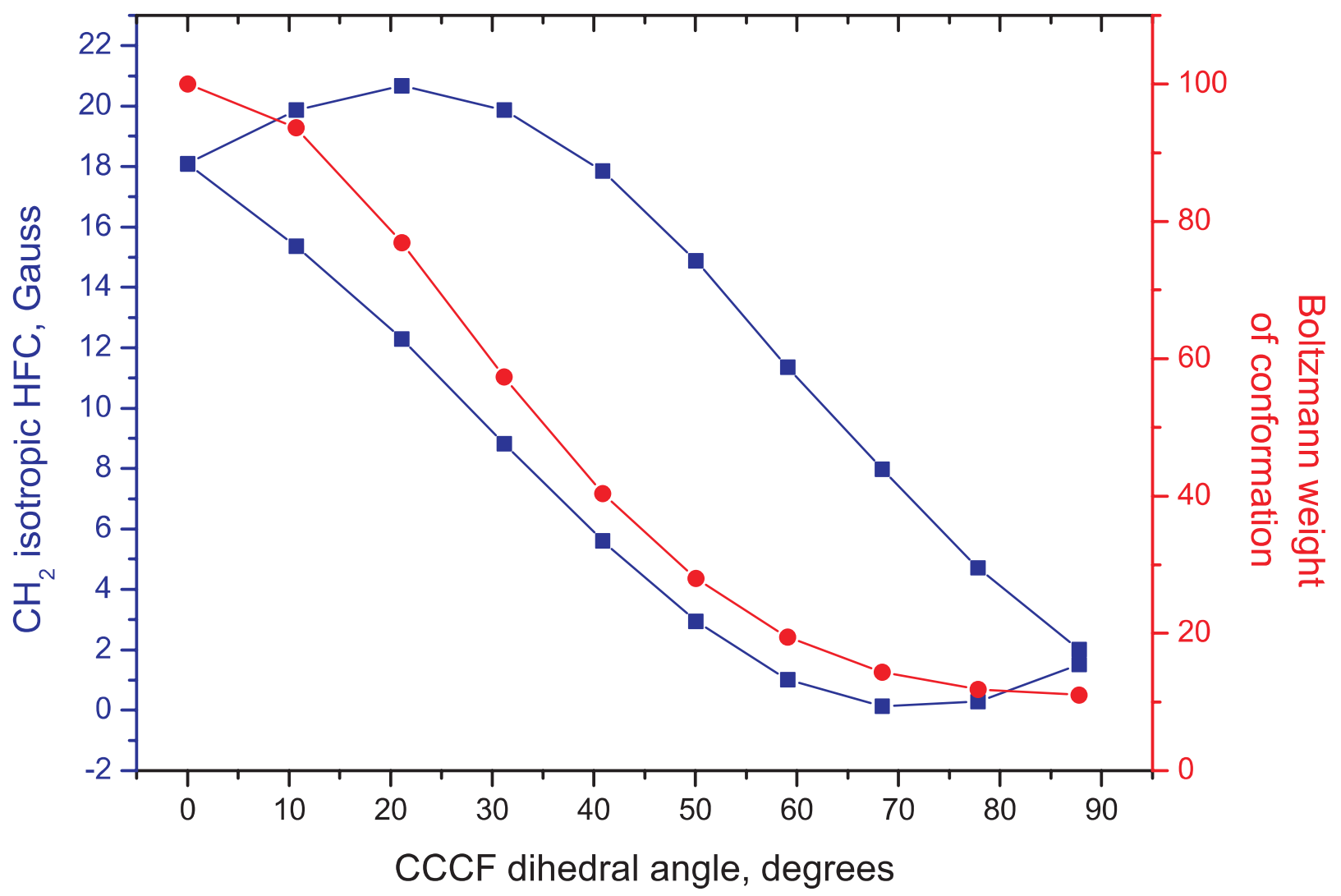

**Figure 2.12.** CCCF dihedral angle dependence of isotropic $^{1}$H hyperfine coupling for two protons of the fluoromethyl group parafluoromethylphenoxyl radical and the Boltzmann weights of corresponding conformations at 298 K.

While the complicated paramagnetic relaxation behaviour may be of no consequence for qualitative CW photo-CIDNP studies, and certainly is of no consequence if no CIDNP experiments are planned, the rotational dependence of magnetic shielding tensor in neutral fluoromethylbenzene suggests that the diamagnetic relaxation description is likely to be just as complicated [105].

## 2.5 Experimental reconnaissance

Armed with the hyperfine couplings and *g*-factors in the intermediate radicals, we can proceed to explore the $^{19}$F photo-CIDNP effect experimentally. There are several experimentally determined conditions that the compound should satisfy to be useful in high-accuracy photo-CIDNP experiments. It should be photochemically stable, i.e. should survive as many photocycles (Figure 1.1) as possible without degrading or bleaching the photosensitizer. The chemical shift of the fluorine nucleus should be insensitive to temperature variation within about 10°C, so that the signal multiplicity is still possible to analyze after prolonged laser irradiation. The compound should have a simple spin system, be cheap and chemically well-behaved (meaning toxicity, acid-base equilibria, stability, etc.). This section will enumerate the compounds tested.

[6]The situation may be somewhat simpler if the amino acid is rigidly immobilized in the protein core, but all the interesting CIDNP-active amino acids are likely to be exposed and have their $CH_2$ groups tumbling to some extent with very complicated rotational correlation functions.

### *2.5.1 2-fluorophenol*

The spin system of 2-fluorophenol is reasonably simple, with all signals resolved at 600 MHz and mostly weak *J*-coupling between the nuclei. Least squares fitting of the theoretical ABCDX spectrum to the observed $^1$H and $^{19}$F spectra pulls out an unremarkable set of *J*-couplings and chemical shifts (Figure 2.13).

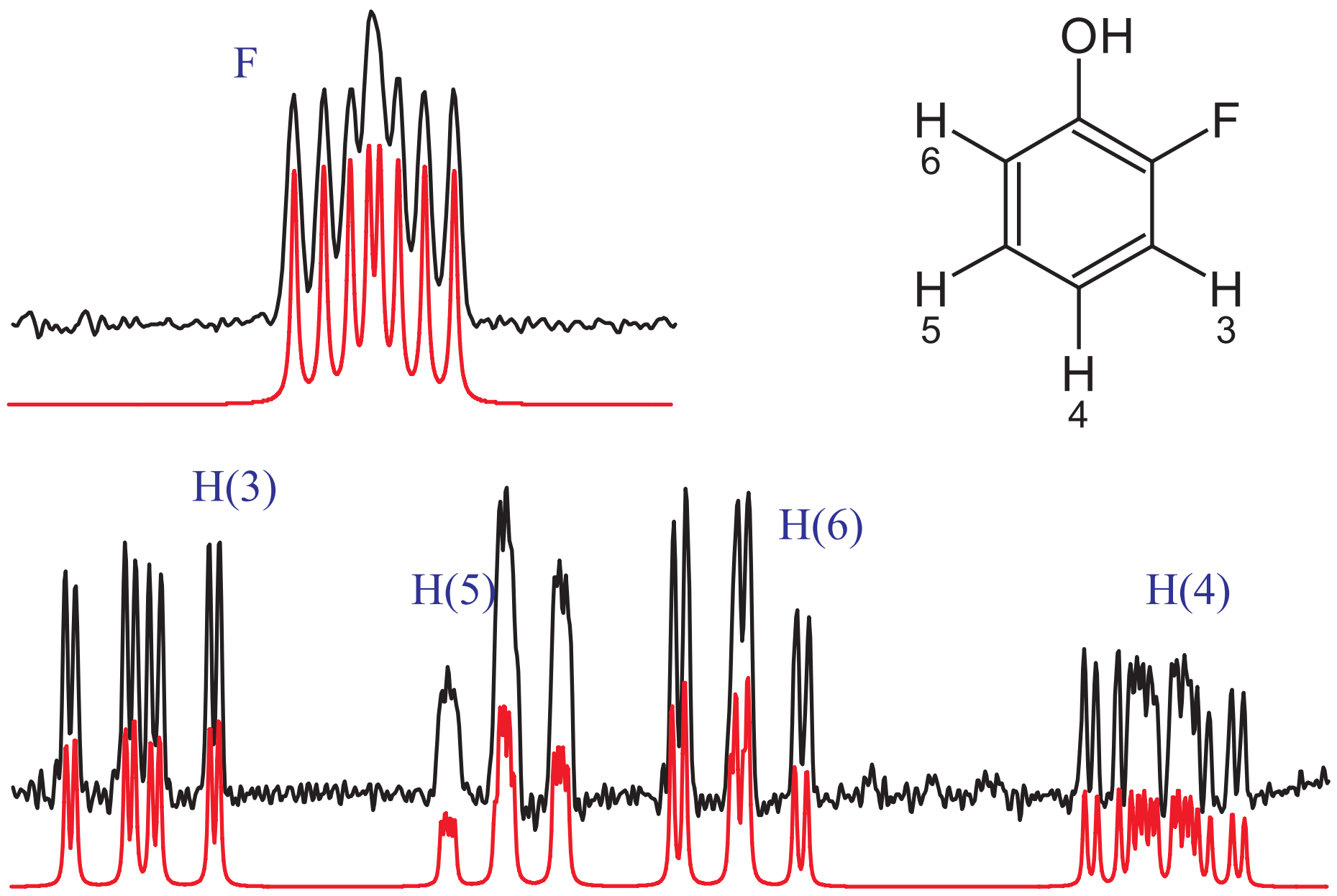

**Figure 2.13.** 600 MHz $^1$H and 564 MHz $^{19}$F NMR spectra of 2-fluorophenol. Spectra shown in red (with lines made narrower for clarity) are simulated with the following scalar coupling constants: $J_{H(3)\text{-}F}$=11.6 Hz, $J_{H(4)\text{-}F}$=4.7 Hz, $J_{H(6)\text{-}F}$=8.9 Hz, $J_{H(3)\text{-}H(4)}$=8.2 Hz, $J_{H(3)\text{-}H(5)}$=1.2 Hz, $J_{H(4)\text{-}H(5)}$=7.5 Hz, $J_{H(4)\text{-}H(6)}$=1.6 Hz, $J_{H(5)\text{-}H(6)}$=8.1 Hz. The coupling constants were obtained by least squares fitting of the model spectrum to the experimental data. The spectral assignment and the positive sign of all *J*-coupling constants were confirmed by a GIAO DFT B3LYP aug-cc-pVTZ-J *ab initio* calculation. The fitted line width (a variable parameter during the fitting procedure) is 0.44 Hz ($^1$H) and 1.37 Hz ($^{19}$F).

Based on the known hyperfine couplings and *g*-factors for both 2-fluorophenoxyl and flavosemiquinone radicals, the simple calculation described in Section 1.3 predicts a strong absorptive enhancement for F(2), moderate emissive enhancements for H(4) and H(6), and small absorptive enhancements for H(3) and H(5) in 2-fluorophenol. The CIDNP enhancement of the H(3,5) protons should be very small in a steady-state photo-CIDNP experiment because of excessive bulk recombination cancellation resulting from slow paramagnetic relaxation, which in turn results from small anisotropy of hyperfine coupling on H(3) and H(5) nuclei (Table 2.2). The behaviour of H(3) and H(5) magnetization on a CW CIDNP experiment is therefore likely to be governed almost entirely by dipolar cross-relaxation and DD-CSA cross-correlation with strongly polarised nearby nuclei. This is indeed the case, as shown on Figures 2.14 and 2.15. The only seeming anomaly – the emissive enhancement and multiplet distortion on H(5), which is positioned far from the fluorine, gets explained away once the strong *J*-coupling between H(5) and H(6) is taken into account, and it is realised that it is actually the strong *J*-coupling that makes H(5) signal follow the nearby H(6) and get inverted with a distorted multiplet structure. This behaviour is demonstrated more clearly in theoretical spectra shown in Figure 2.16. It is caused by transverse bilinear terms that are present in the strong *J*-coupling Hamiltonian, and unless due care is taken, these can be easily confused with NOEs or DD-CSA cross-correlation effects.

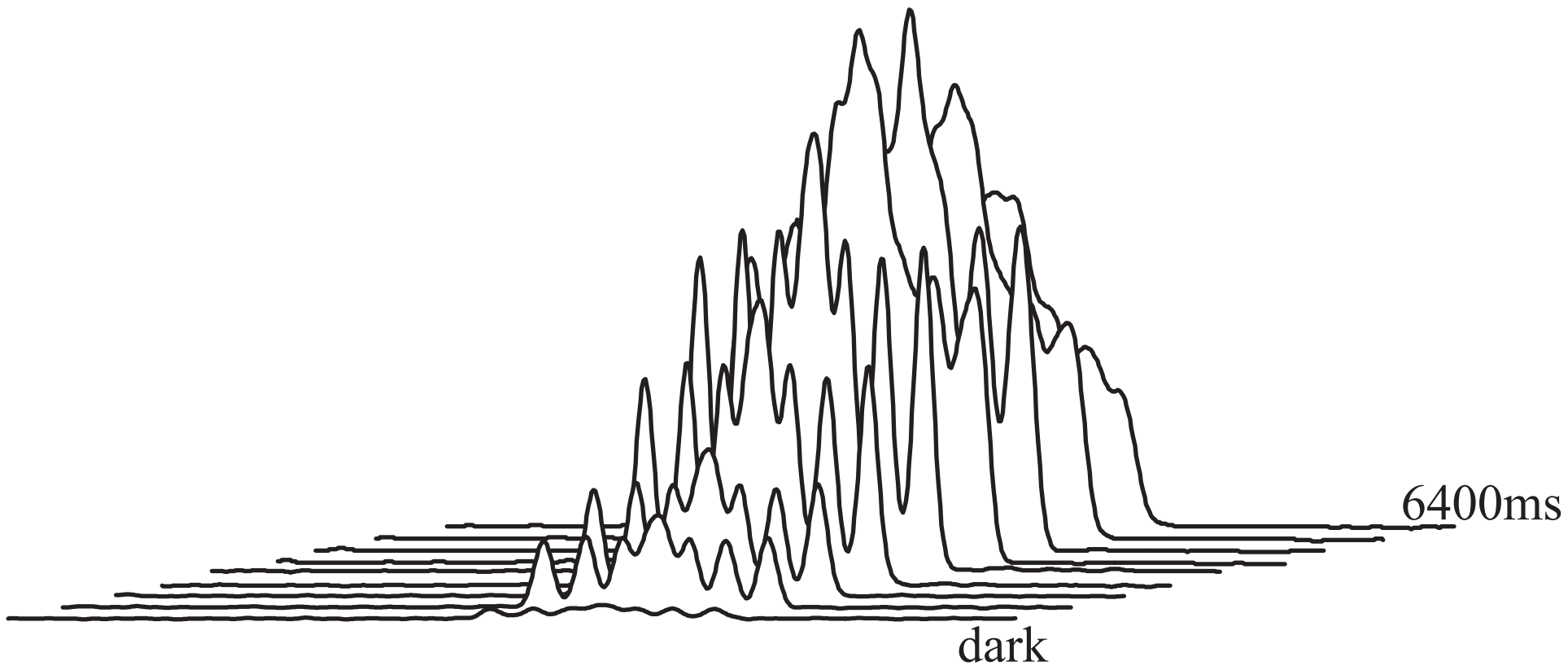

**Figure 2.14.** 564 MHz $^{19}$F photo-CIDNP spectrum of 2-fluorophenol as a function of irradiation time. The spectrum labelled "dark" corresponds to the normal NMR spectrum. The irradiation times are 50, 100, 200, 400, 800, 1600, 3200, 6400 ms. The resolution deterioration at large irradiation times is caused by non-uniform sample heating by the intense laser light.

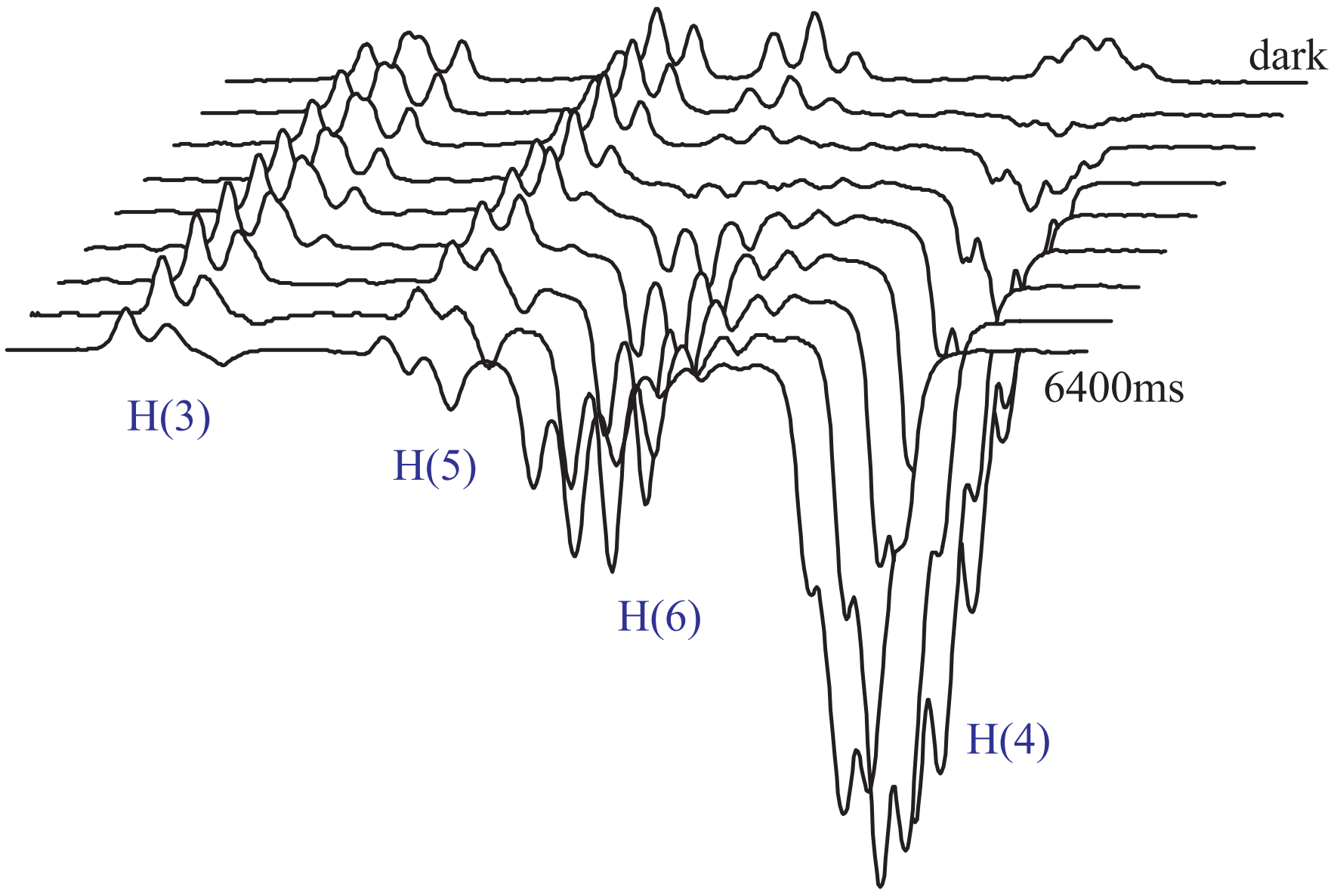

**Figure 2.15.** 600 MHz proton spectrum of 2-fluorotyrosine as a function of irradiation time. The spectrum labelled "dark" corresponds to the normal NMR spectrum. The irradiation times are 50, 100, 200, 400, 800, 1600, 3200, 6400 ms.

Fitting the magnetization dynamics (Figure 2.17) to a linear longitudinal model similar to (1.37), with constant pumping terms and cross-relaxation between the adjacent nuclei, results in the following CIDNP pumping rates: –20.8 $s^{-1}$ for H(4), –6.4 $s^{-1}$ for H(6), 117 $s^{-1}$ for F(2). Although H(3) and H(5) dynamics is fitted reasonably well (Figure 2.17), the resulting values are unlikely to be correct because the model fitted does not include the above mentioned strong coupling effects.

There are two more things to note about Figures 2.14-2.17. One is that long irradiation times lead to a severe resolution deterioration in the $^{19}$F spectrum, likely resulting from non-uniform sample heating by the laser light. The other is that the simple relaxation model with constant pumping terms appears to be sufficient to describe (or at least to fit) the observed magnetization dynamics.

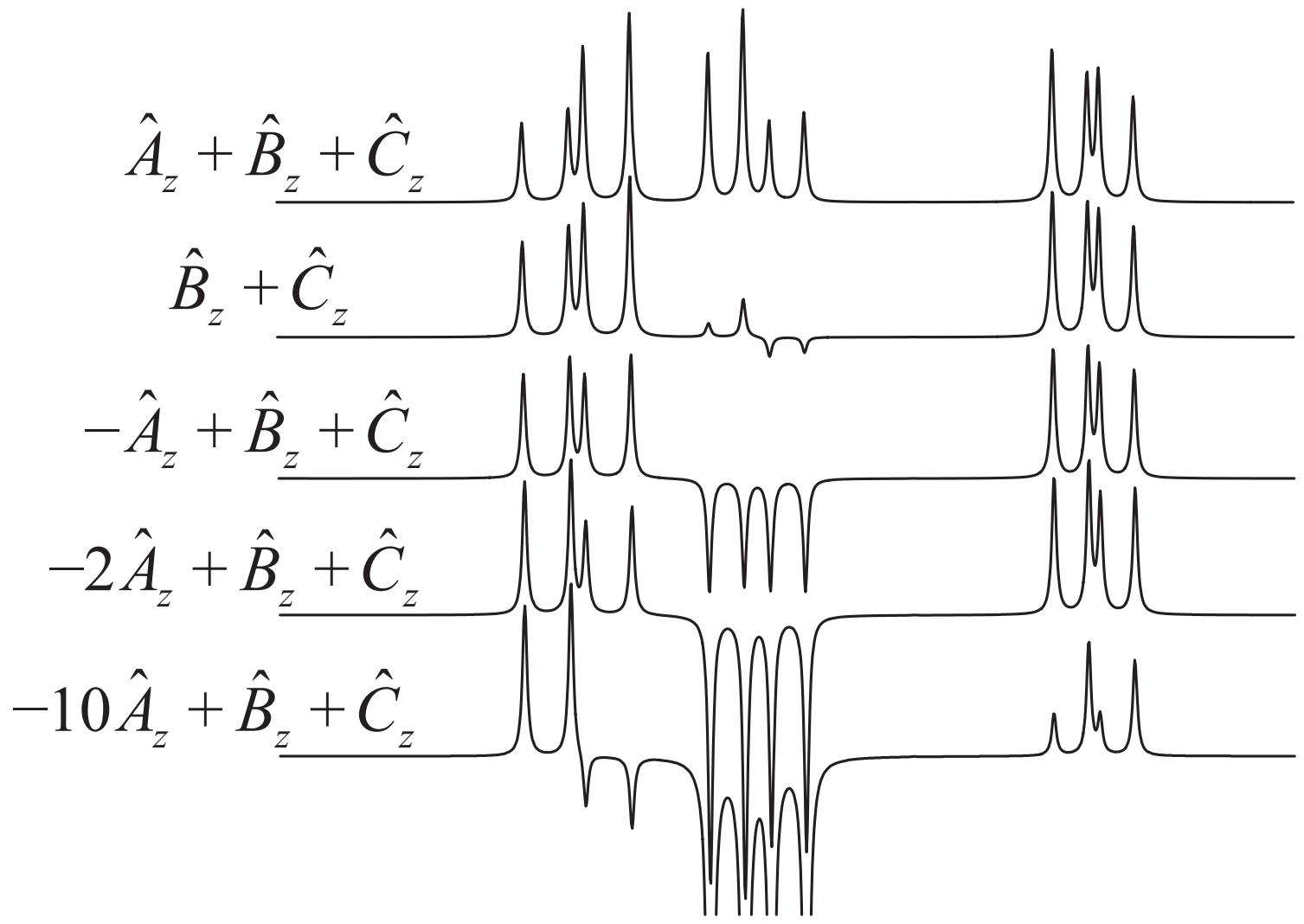

**Figure 2.16.** Theoretical NMR spectra (90°-detection) of a strongly coupled three-spin system with the initial populations of one spin perturbed from the equilibrium. The initial system density matrices are printed on the left.

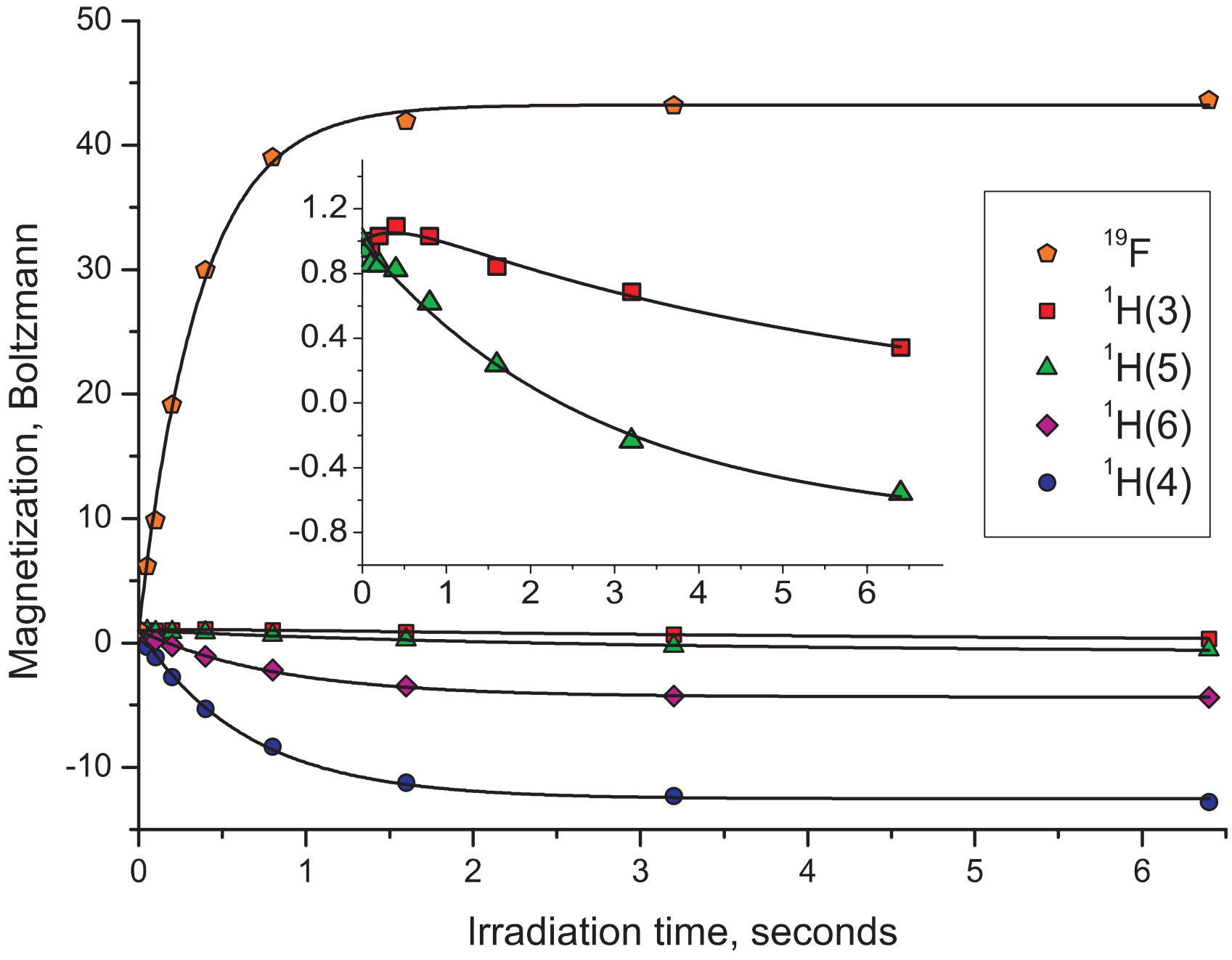

**Figure 2.17.** Dependence of longitudinal nuclear magnetisation in 2-fluorophenol on the laser irradiation time. Solid lines: least squares fitting to a linear longitudinal model.

Lastly, the photostability of 2-fluorophenol/FMN system is very high, with the photodegradation curves essentially repeating those shown in Chapter 3 (Figure 3.9) for the 4-fluorophenol/FMN system, meaning that hundreds of flashes may be performed on this system without significant photodegradation.

### 2.5.2 3-fluorophenol

$^{19}$F CIDNP effect magnitude observed in 3-fluorophenol is relatively small compared to the other two fluorophenols because of smaller spin density on the fluorine nucleus in the intermediate radical as well as smaller HFC anisotropy, which leads to greater extent of recombination cancellation in the secondary dynamics (this effect is analyzed in detail in chapter 8). The photo-CIDNP pumping rate measured for the $^{19}$F nucleus is $-11.5\ s^{-1}$, which is ten and thirteen times less than the pumping rates in 2- and 4-fluorophenols respectively. Thus, although the total fluorine polarisation in 3-fluorophenol can be pumped up to not insignificant 7 times the equilibrium value, the potential for generating long-range $^{19}$F-$^{1}$H NOEs is relatively small for this compound. The changes that fluorine and proton NMR signals of 3-fluorophenol undergo as the irradiation time gets longer are shown in Figures 2.19, 2.20 and analyzed in Figure 2.21.

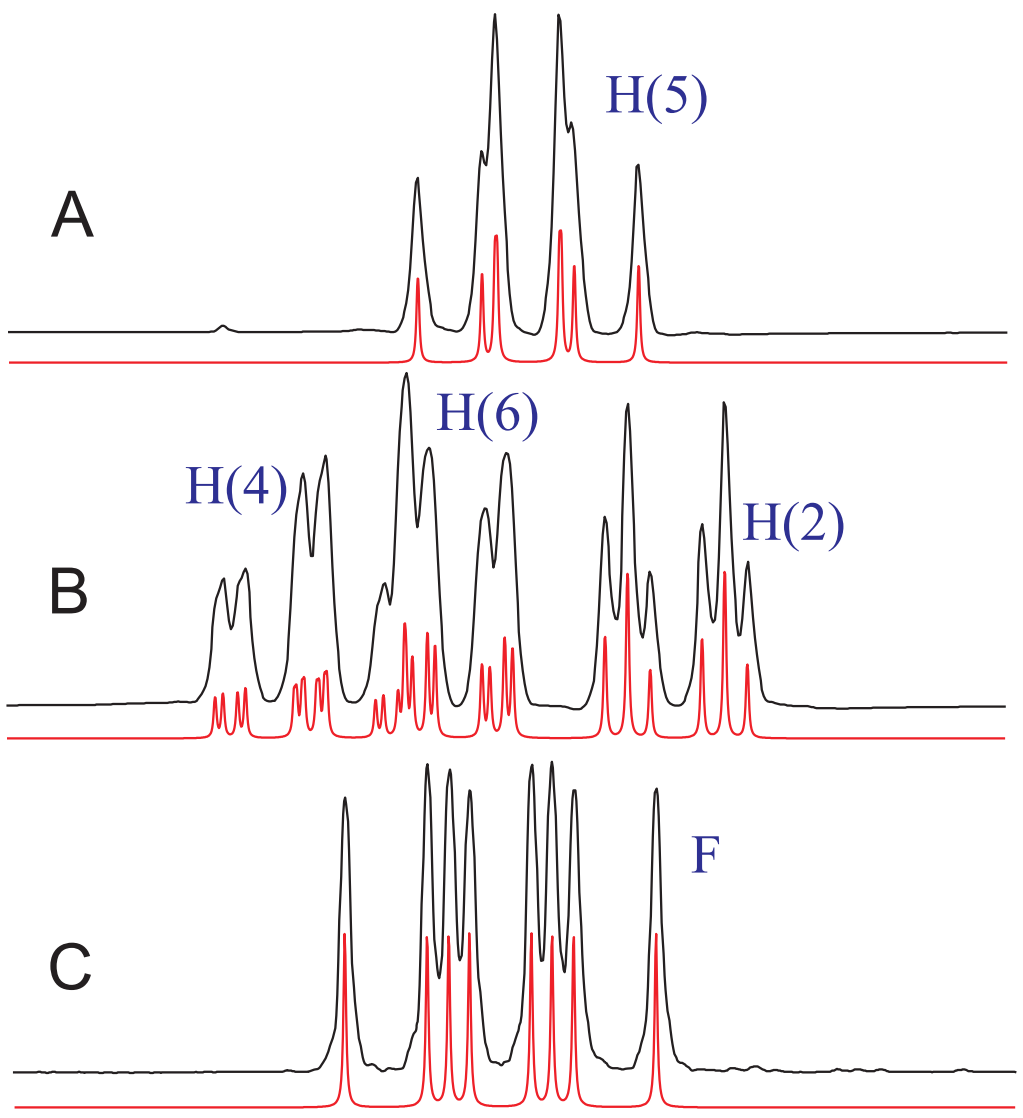

**Figure 2.18.** Regions of the high-resolution 600 MHz $^{1}$H **(A, B)** spectrum and 564 MHz $^{19}$F **(C)** spectrum of 3-fluorophenol. Least squares fits of the theoretical spectra are drawn in red with lines made narrower for clarity. The fitted line width (a variable parameter during the fitting procedure) is 1.02 Hz ($^{1}$H) and 1.14 Hz ($^{19}$F). The spectral assignment and the positive sign of all *J*-coupling constants were confirmed by a GIAO DFT B3LYP aug-cc-pVTZ-J *ab initio* calculation. $J_{H(2)-F}$=10.3Hz, $J_{H(2)-H(4)}$=2.4Hz, $J_{H(2)-H(6)}$=2.4Hz, $J_{F-H(4)}$=8.6Hz, $J_{H(4)-H(6)}$=0.8Hz, $J_{H(4)-H(5)}$=8.3Hz, $J_{F-H(5)}$=6.8Hz, $J_{H(5)-H(6)}$=8.2Hz.

The CIDNP magnetization pumping rate for H(5) proton in 3-fluorophenol is close to zero, but the signal does gain some amplitude at long irradiation times (Figures 2.19, 2.21) due to nuclear Overhauser effect from nearby H$^{6}$ and H$^{4}$ which have strong emissive CIDNP polarisation. The pumping rate is $-7.4\ s^{-1}$ for H(2) and approximately –16 $s^{-1}$ for the overlapping signals of H(4) and H(6). The accumulation of longitudinal multi-spin orders is clearly manifested on the fluorine spectra, a result of either direct pumping (CIDNP multiplet effect) or $^{19}$F-$^{1}$H DD-CSA cross-correlation. This effect is analyzed in detail in Chapter 4.

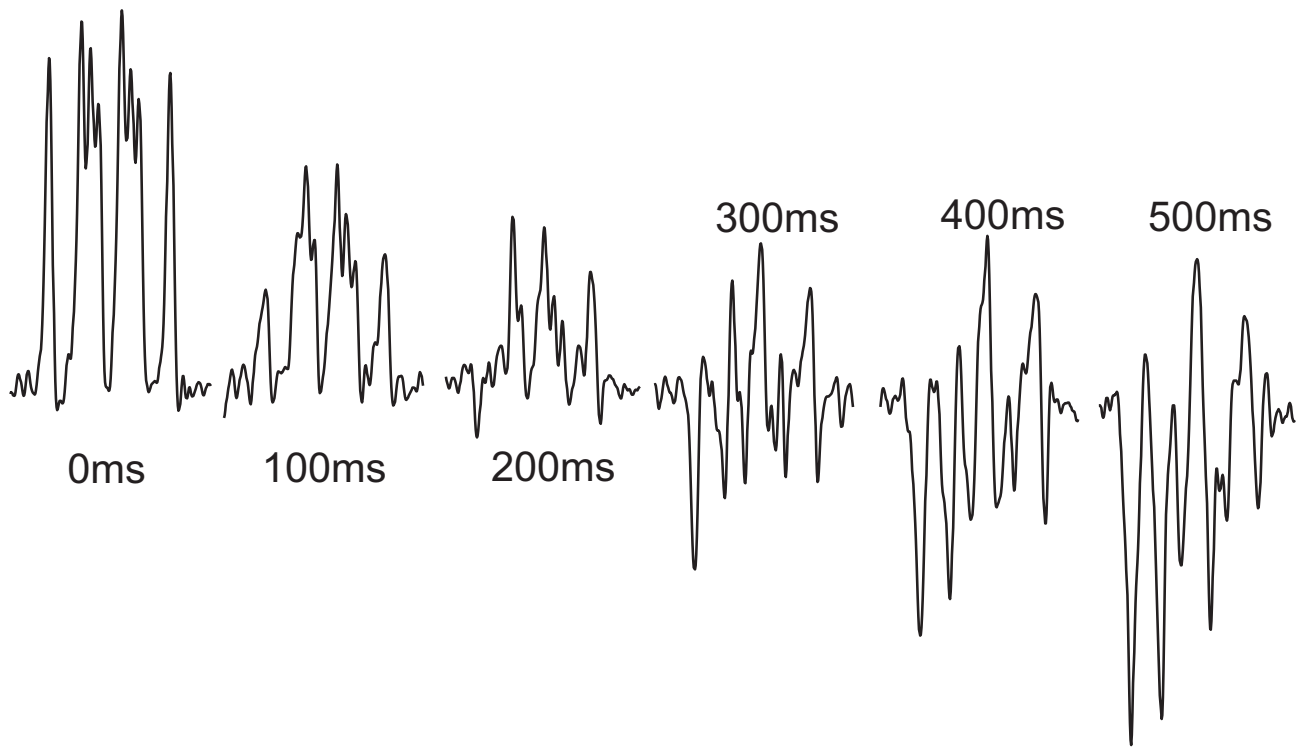

**Figure 2.20.** Fluorine NMR signal of 3-fluorophenol as a function of laser irradiation time. The end of the optical fibre is placed 4 mm above the edge of the receiver coil. The relatively high optical fibre elevation allows to obtain CIDNP spectra with maximum resolution, but the effect magnitude is somewhat weaker.

Just as in the case of 2-fluorophenol, the photostability of 3-fluorophenol/FMN system is very high, with hundreds of flashes necessary to get a noticeable photobleaching. Again, $^{19}$F spectral resolution becomes a problem for long laser flashes, indicating that the chemical shift of 3-fluorophenol also has substantial temperature dependence.

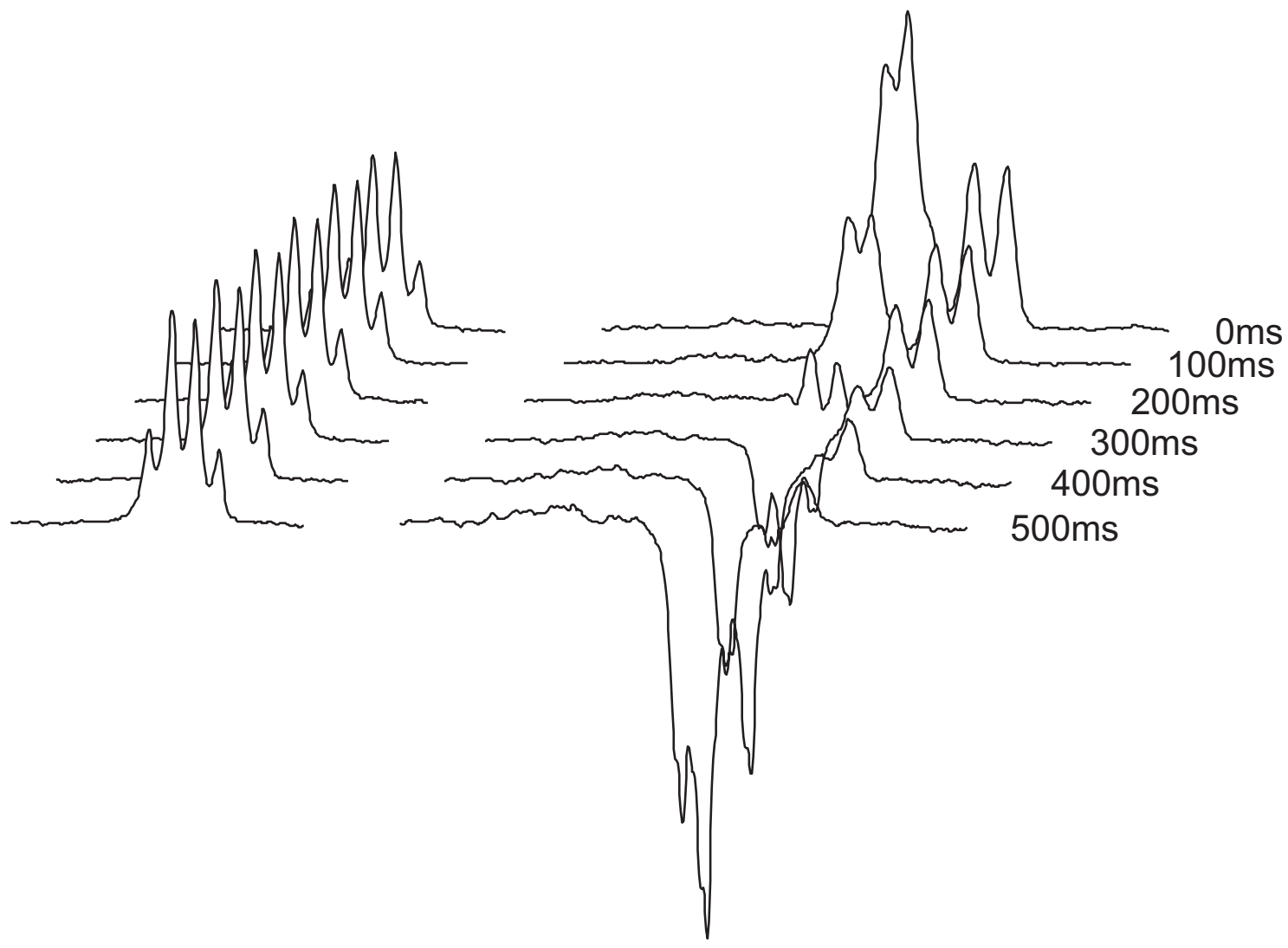

**Figure 2.19.** Proton NMR spectrum of 3-fluorophenol as a function of laser irradiation time.

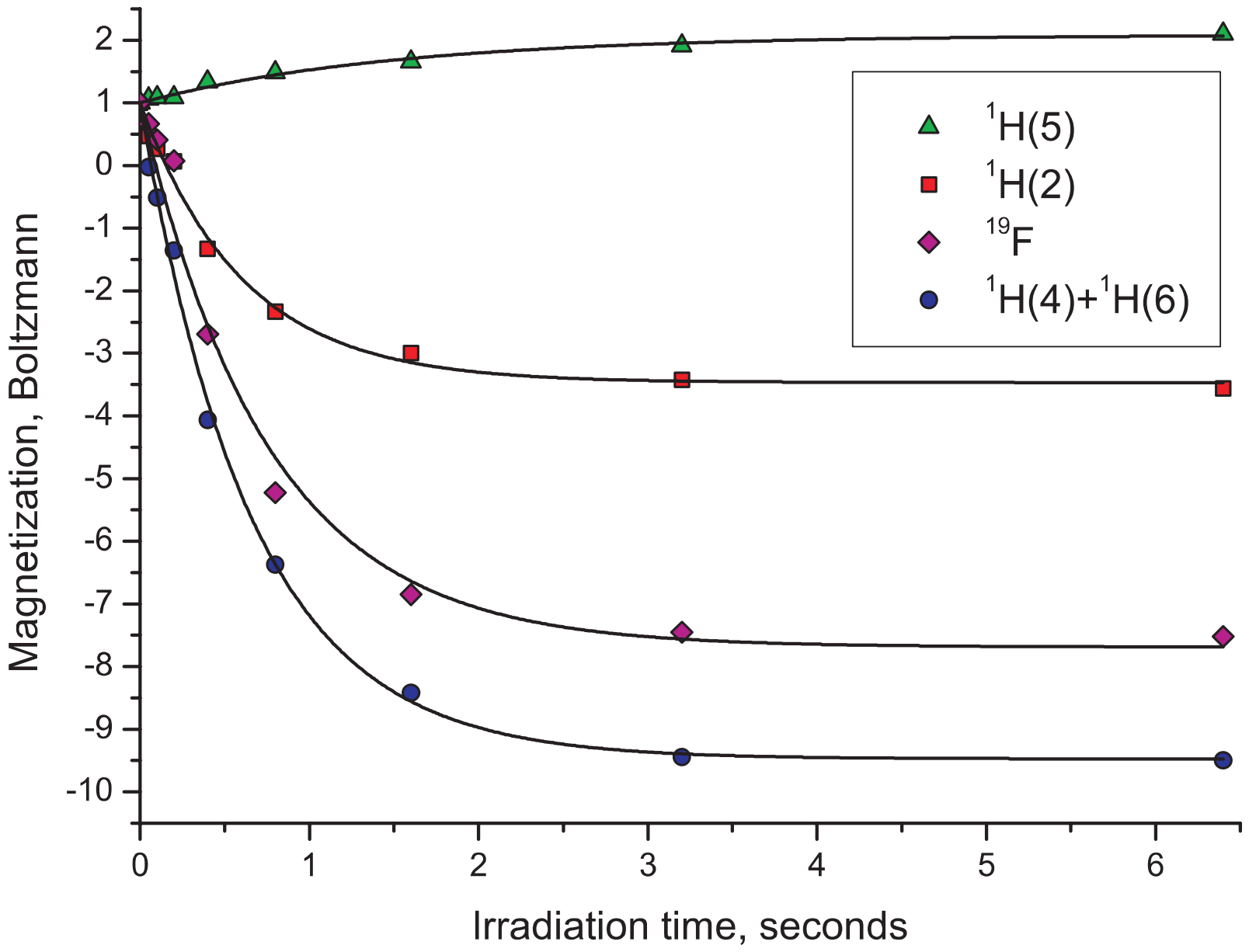

**Figure 2.21**. Dependence of longitudinal nuclear magnetisation in 3-fluorophenol on the laser irradiation time. Solid lines: least squares fitting to a linear longitudinal model.

In the absence of confusing effects resulting from strong *J*-couplings, it is clearly seen that the nuclear cross-relaxation in chemically pumped systems (e.g. H(4,6)→H(5) in the present case of 3-fluorophenol, Figure 2.21) appears to be unaffected by the exotic origin of the nuclear magnetization and proceeds just as the usual cross-relaxation and cross-correlation in chemically static systems, likely because the concentration of radical intermediates at any point in time is vanishingly small compared to the overall sample concentration. Figure 2.21 shows that the initial slope of the H(5) magnetization is zero, indicating the lack of direct pumping, but there is some later magnetization pickup from cross-relaxation, which proceeds with the rate constant of about –0.016 $s^{-1}$, which is a common number for an NOE between the adjacent protons in small aromatic molecules at 600 MHz.

The observation that the dipolar relaxation does not seem to inherit the complexity of the CIDNP generation processes has far reaching implications. Specifically, this means that the standard relaxation theoretical approaches may be applicable to the CIDNP-pumped systems, the only change being that the constant "magnetization source" terms need to be introduced into the density matrix evolution equation for the duration of the laser flash. This conclusion is investigated in further detail in Chapter 3.

### *2.5.3 4-fluorophenol*

Consistent with the very large hyperfine coupling in the intermediate radical, 4-fluorophenol exhibits a record $^{19}$F CIDNP amplitude with the magnetization pumping rate of 148 $s^{-1}$. The polarisation attained after about 2 seconds of irradiation is 55-fold (Figures 2.22, 2.23). H(2) and H(6) have got –8.3 $s^{-1}$ CIDNP pumping rate. The magnetization dynamics of H(3,5) is a superposition of very weak direct photo-CIDNP, cross-relaxation and *J*-coupling effects, therefore, although it is fitted reasonably well by the linear model, we should not expect the resulting parameters to be of any physical significance.

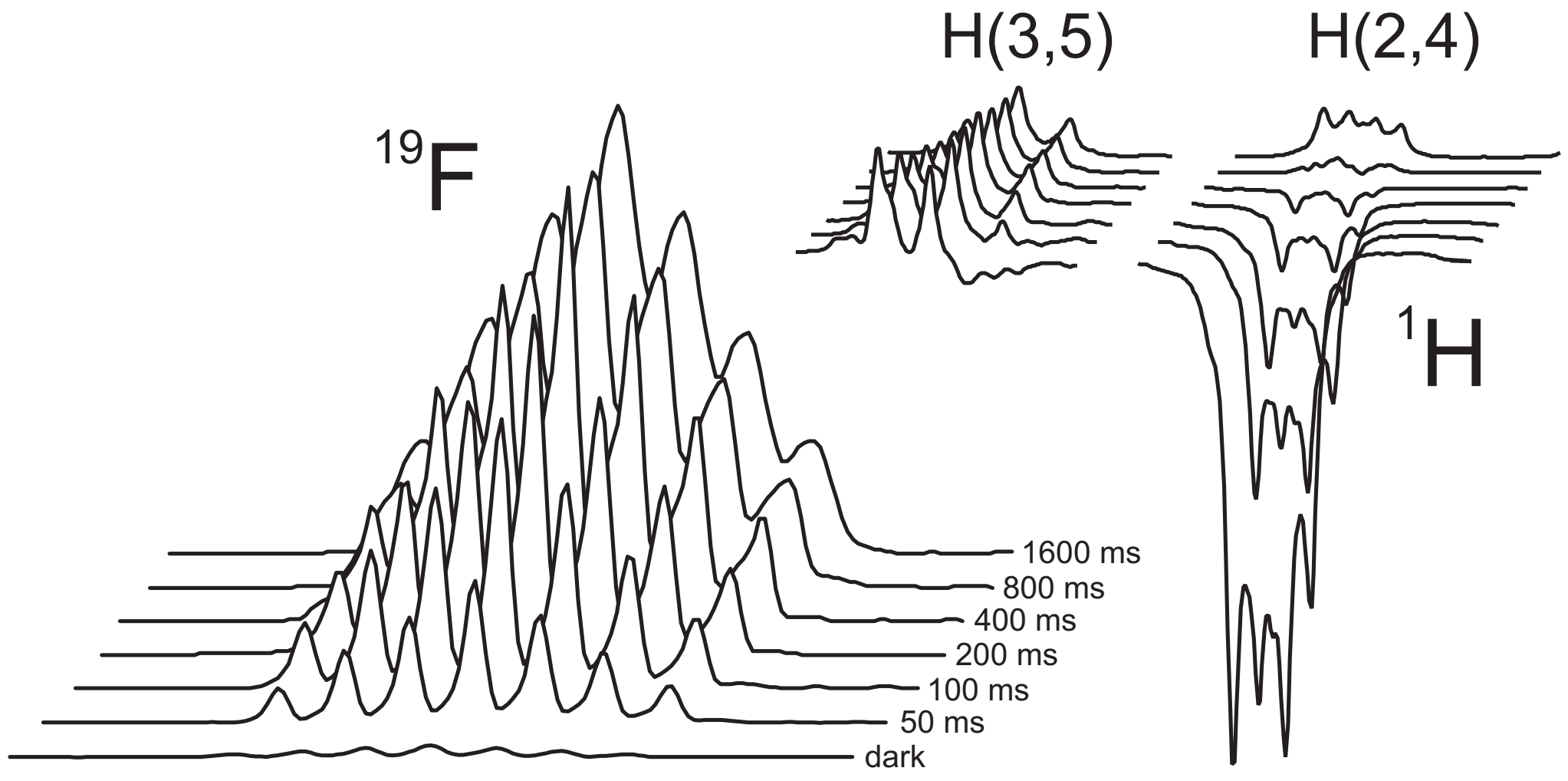

**Figure 2.22.** 564 MHz $^{19}$F and 600 MHz $^{1}$H photo-CIDNP spectra of 4-fluorophenol as a function of irradiation time. The spectrum labelled "dark" corresponds to the conventional NMR spectrum.

The photodegradation curves for the 4-fluorophenol/FMN system are presented in Chapter 3 (Figures 3.8-3.10) in conjunction with the uniform sample illumination setup. Just as the other fluorophenols, 4-fluorophenol appears to be highly resistant to photodegradation with a few hundred flashes needed to achieve noticeable photobleaching.

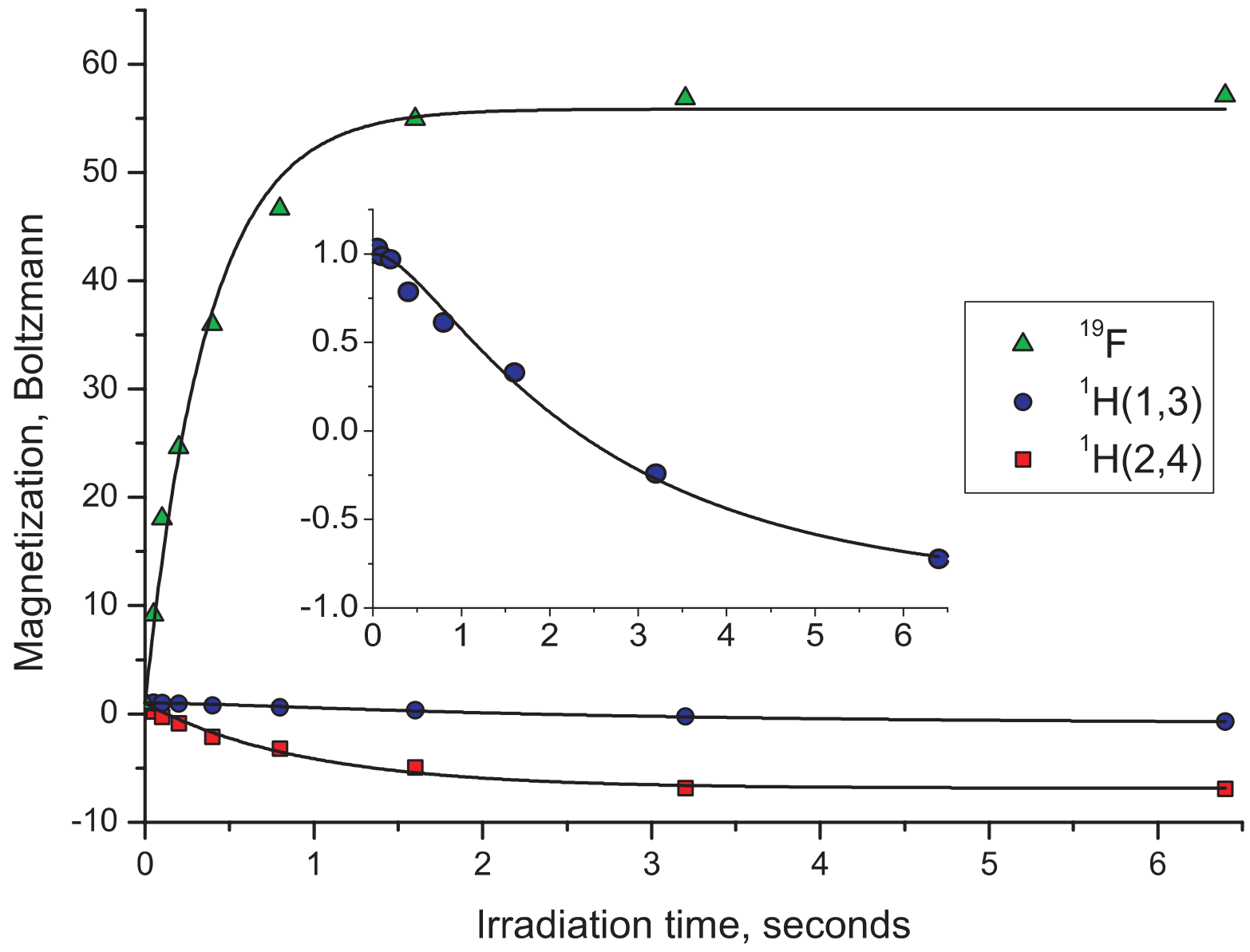

**Figure 2.23.** Dependence of longitudinal nuclear magnetisation in 4-fluorophenol on the laser irradiation time. Note the spectacular CIDNP magnitude observed on fluorine. The curves are theoretical fits to a linear longitudinal model.

As a side note, 4-fluorophenol appears to associate in DMSO, leading to additional splitting in $^{19}$F spectrum, presumably due to formation of a hydrogen bond.

### *2.5.4 Fluorotryptophans*

As we have concluded in Section 2.4, out of a large number of possible fluorinated tryptophans, 4- and 6-fluorinated isomers deserve a closer attention. As a test of a negative prediction, the relatively cheap 5-fluoro-DL-tryptophan, which was theoretically predicted to have a zero $^{19}$F photo-CIDNP, was also included. Experiments and processing identical to those described in Sections 2.5.1-2.5.3 for fluorophenols have been performed on three fluorotryptohans. An example of the observed magnetization dynamics is shown in Figure 2.24 and the resulting photo-CIDNP magnetization pumping rates are reported in Table 2.6.

As predicted above (see Tables 2.4 and 2.5), the $^{19}$F photo-CIDNP in 5-fluorotryptophan is close to zero, while the other two isomers demonstrate the expected high magnetization pumping rates. A notable feature of the magnetization accumulation curves is the fact that although the equilibrium CIDNP of 6-fluorophenol is greater than that of 4-fluorophenol, the pumping rate is actually lower. This is due to the fact that the fluorine in 4-position of the indole ring is positioned very close to $CH_2$ protons which induce rapid relaxation via DD and DD-CSA mechanisms. On the other hand, the fluorine atom of 6-fluorotryptophans has fewer protons around, so it relaxes slower and consequently has a greater equilibrium CIDNP.

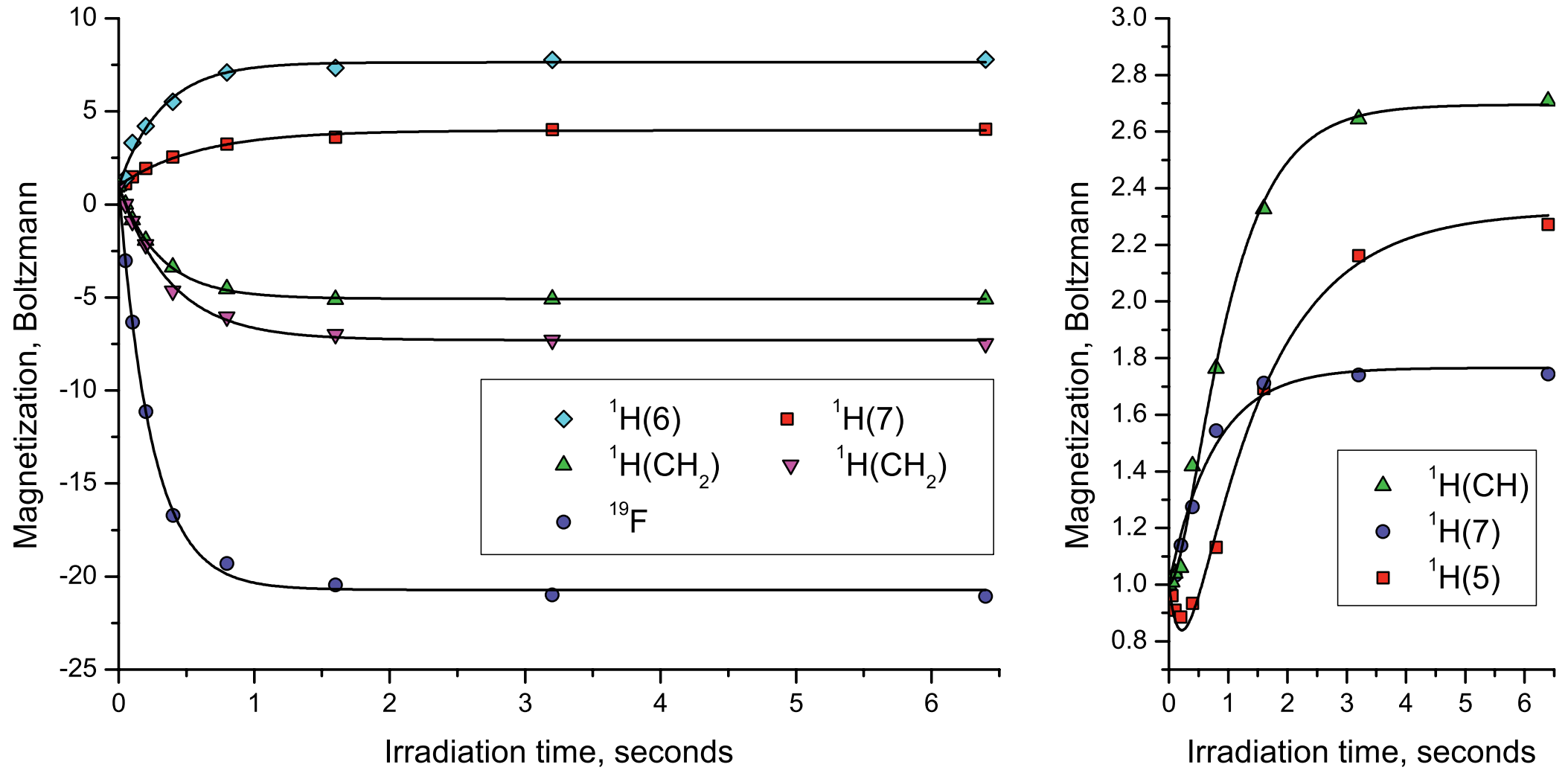

**Figure 2.24.** Dependence of longitudinal nuclear magnetisation in 4-fluorotryptophan on the laser irradiation time. Fresh sample was used for each point to avoid excessive photobleaching. The curves are theoretical fits to a linear longitudinal model

In all experimentally studied fluorotryptophans the $^{19}$F chemical shift appears to be strongly dependent on temperature. Because temperature gradients are inevitable in continuous-wave photo-CIDNP experiments, the resolution degradation is sometimes very pronounced. While this does not cause any problems with overall signal intensity quantification, the multiplet structure is severely obscured. This makes fluorinated tryptophans unsuitable for chemically pumped cross-relaxation studies, because an accurate account of DD-CSA cross-correlation requires multiplet structure to be resolved at all times.

**Table 2.6** Experimental photo-CIDNP magnetization pumping rates (s$^{-1}$) for hydrogen and fluorine nuclei in three isomeric fluorotryptophans.

| 4-fluorotryptophan | | 5-fluorotryptophan | | 6-fluorotryptophan | |
|---|---|---|---|---|---|
| H(2) | 4.3 | H(2) | 2.3 | H(2) | 6.5 |
| **F(4)** | **–97.0** | H(4) | 8.0 | H(4) | 9.9 |
| H(5) | –0.7 | **F(5)** | **2.0** | H(5) | –0.7 |
| H(6) | 13.5 | H(6) | 6.6 | **F(6)** | **–81.0** |
| H(7) | 0.7 | H(7) | 0.2 | H(7) | 0.3 |
| H(β) | –9.4 | H(β) | –10.0 | H(β) | –12.4 |
| H(β) | –12.3 | H(β) | –12.7 | H(β) | –15.0 |
| H(α) | 0.0 | H(α) | 0.0 | H(α) | 0.0 |

The photostability of fluorotryptophan/FMN systems is very low, with only 16 laser flashes (see Section 2.2 for experimental details) necessary to achieve a near-complete bleaching of the photosensitizer, likely due to irreversible photooxidation of the tryptophan indole ring.

### 2.5.5 Fluorotyrosines

Of the two ring-fluorinated tyrosines, the 2-fluorinated isomer has been ruled out based on the computed hyperfine couplings (Table 2.3). The 3-fluorinated isomer exhibits a $^{19}$F photo-CIDNP magnetization pumping rate of 34 $s^{-1}$ with an attainable polarization of over 20 times the Boltzmann level.

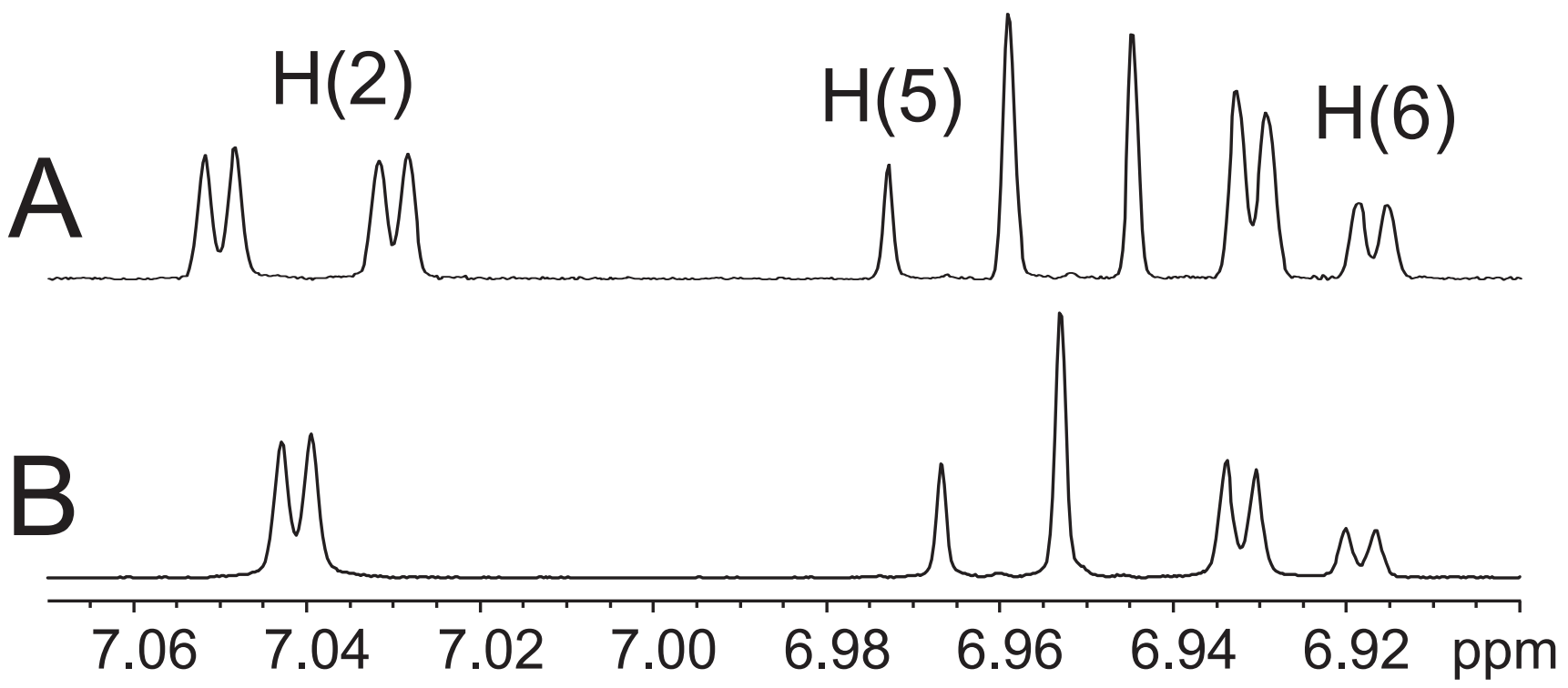

**Figure 2.25. (A) A**romatic region of 3-fluorotyrosine $^{1}$H NMR spectrum. **(B)** The same spectrum recorded with $^{19}$F decoupling. See Figure 3.1 for $^{19}$F NMR spectrum and Figure 2.2 for atom numbering.

The $^{19}$F chemical shift of 3-fluorotyrosine has a very weak temperature dependence with a result that $^{19}$F spectra retain high resolution even after very prolonged laser irradiation periods. The 3-fluorotyrosine/FMN system also possesses remarkable photostability, enduring at least 50 laser flashes without noticeable photodegradation. These observations, and also the fact that 3-fluorotyrosine has a convenient spin system with a well-resolved *J*-coupling (Figures 2.25 and 3.1), mean that, of all the compounds studied so far, 3-fluorotyrosine is probably best suited for exploring further behaviour of chemically pumped nuclear magnetization.

A simple way to get an idea of the magnitude of the photo-CIDNP driven $^{19}$F-$^{1}$H cross-relaxation effect is to monitor the proton relaxation pattern in a saturation-recovery type experiment. If protons are decoupled during the irradiation period (Figure 2.26), this prevents the accumulation of $^{1}$H CIDNP magnetization. At the end of the irradiation period the decoupler is switched off allowing the proton nuclear subensemble to relax in the presence of strongly polarized $^{19}$F nuclei. Any powerful manifestation of $^{19}$F-$^{1}$H cross-relaxation should distort the proton relaxation picture, which is made observable by the final 90-degree pulse.

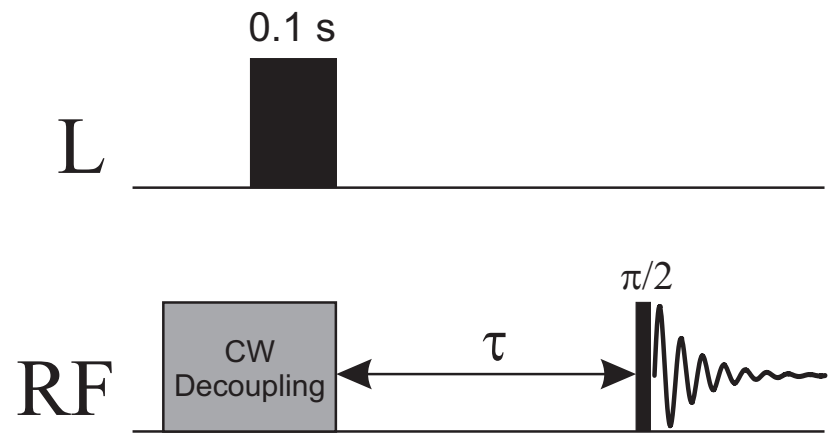

**Figure 2.26.** Pulse sequence used to estimate $^{19}$F-$^{1}$H cross-relaxation magnitude in the case of chemically pumped $^{19}$F magnetization.

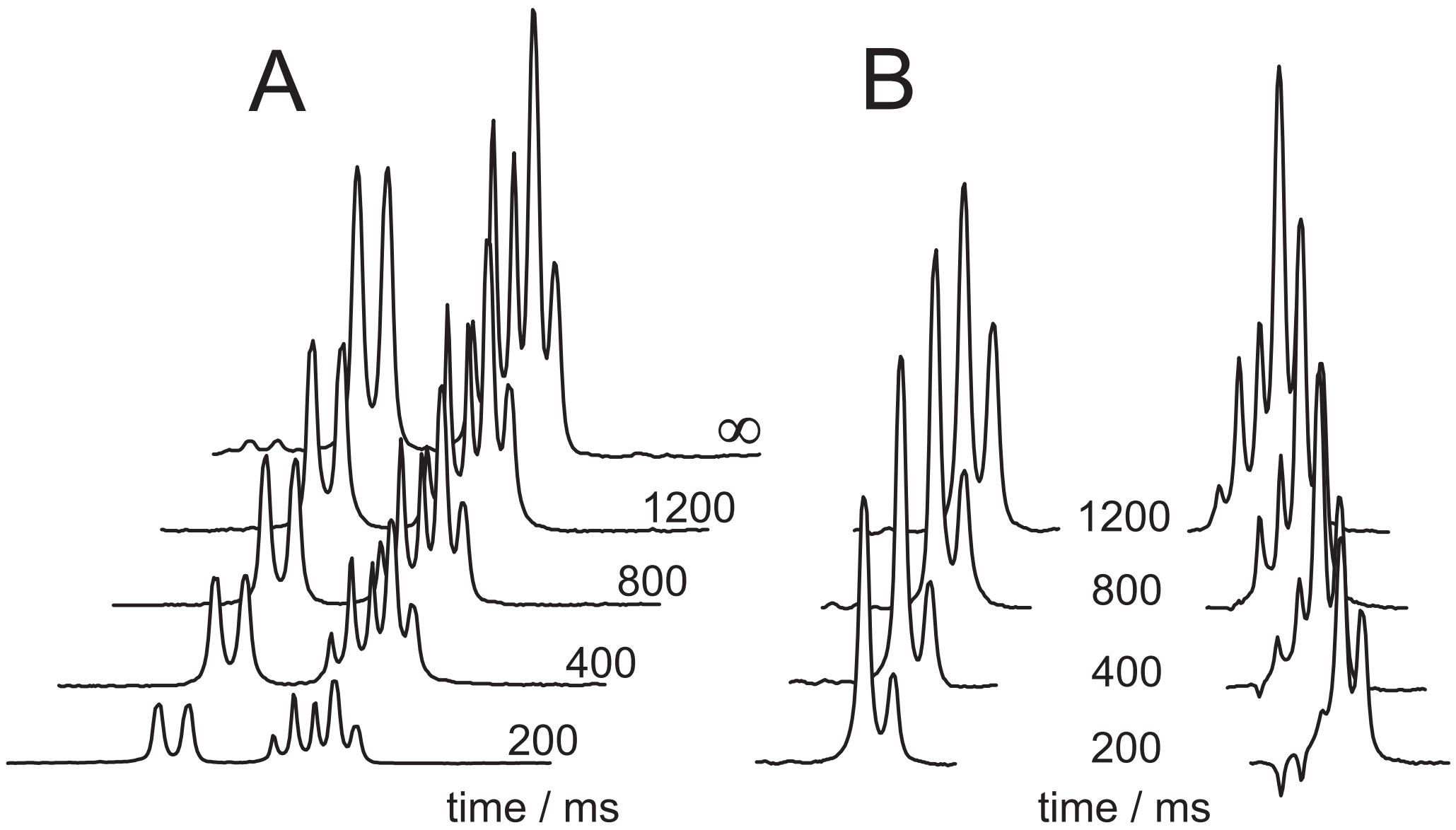

**Figure 2.27. (A)** 3-fluorotyrosine aromatic proton relaxation pattern recorded with sequence in Figure 2.26 with laser switched off (making it a simple saturation-recovery experiment). **(B)** Aromatic proton relaxation pattern for 3-fluorotyrosine recorded with sequence in Figure 2.26 with a 100 ms laser flash.

Figure 2.27 shows the spectra obtained from the sequence shown in Figure 2.26. When the laser irradiation of the sample is not performed (Figure 2.27A), the proton relaxation pattern looks like a typical saturation-recovery experiment with little, if any, cross-relaxation present. However, the saturation-recovery experiment performed in the presence of a nearby strongly polarized $^{19}$F nucleus leads to dramatic changes (Figure 2.27B). There are both strong nuclear Overhauser effect (H(5) proton magnetization briefly goes negative) and DD-CSA cross-correlated cross-relaxation effect (weakly coupled H(2) doublet components acquire unequal intensities) from fluorine to protons. The amplitude of the effects and the spectral resolution are sufficient for an accurate quantitative analysis, which is performed in Chapter 3.

If the laser irradiation period is extended into the mixing time in the sequence on Figure 2.26, the cross-relaxation and cross-correlation effects become even more dramatic (Figure 2.28). The behaviour of the H(2) magnetization, which does not receive direct pumping, becomes governed almost entirely by cross-correlation with the result that it becomes antiphase with respect to the $^{19}$F-$^{1}$H *J*-coupling and in that shape gets accumulated in the spin system (Figure 2.28A). Even after the laser is switched off (Figure 2.28B), the accumulation of this antiphase signal continues for some time, driven by the strong polarization that has accumulated on fluorine.

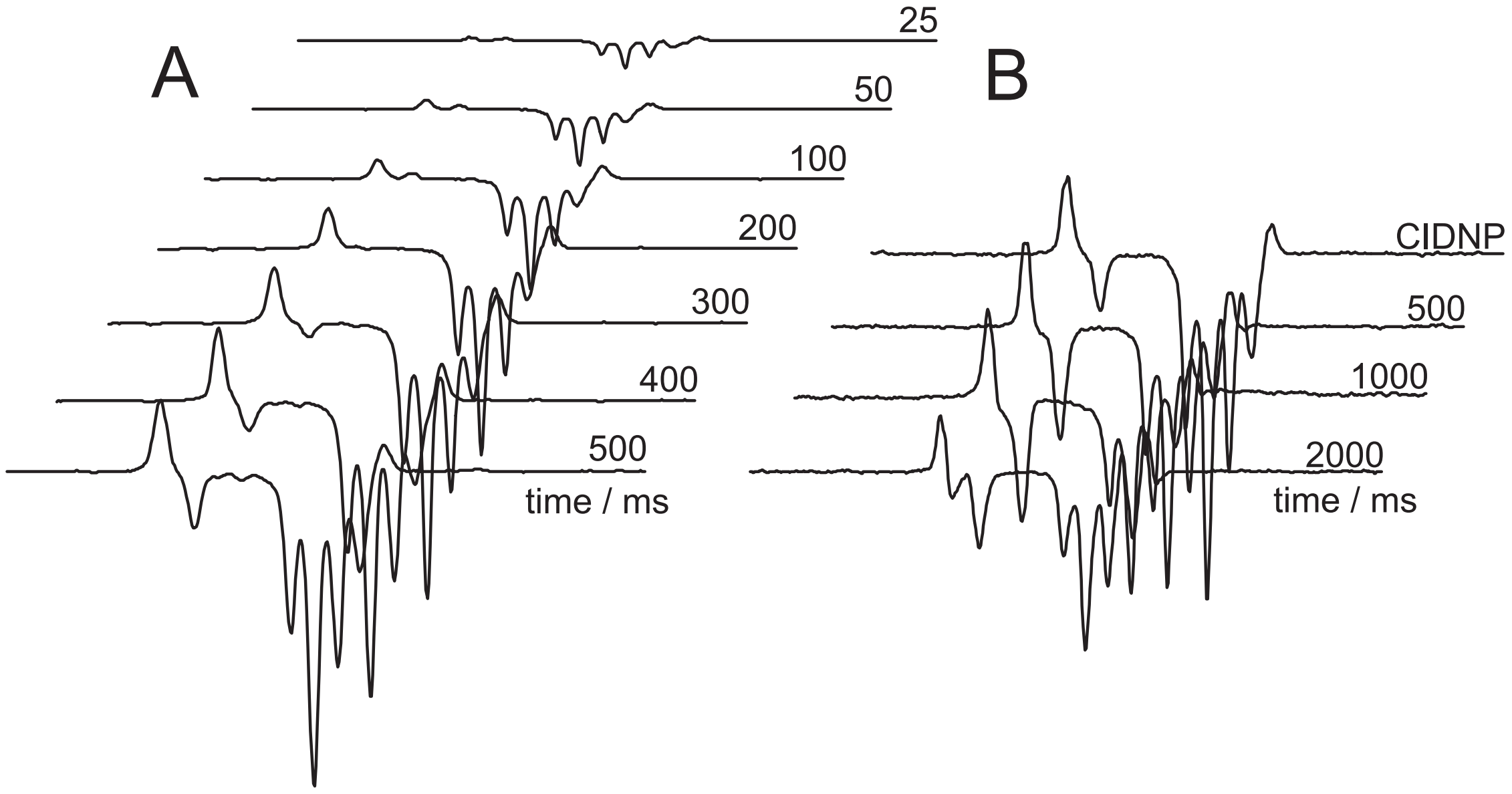

**Figure 2.28. (A)** CIDNP spectra recorded with the pulse sequence in Figure 2.25 with the laser irradiation occurring during the mixing time. **(B)** Continuing evolution of the magnetization generated in (A) after the laser has been switched off.

The amplitudes of the relaxation effects shown in figures 2.27 and 2.28 are remarkably large, with Overhauser effects and cross-correlations of the order of 50% of the equilibrium signal intensity. Such amplitudes of, for example, the $2H_zF_z$ longitudinal two-spin order that the H(2) signal demonstrates in Figure 2.28A, are impossible to generate if one starts from the thermal equilibrium spin polarization. The situation clearly merits further investigation (which is reported in Chapter 3), and 3-fluorotyrosine appears to be particularly convenient for that purpose.

## 2.6 Summary

The reconnaissance experiments described in this chapter have highlighted the following conditions as important in the choice of an informative photo-CIDNP system:

*Ability for a 4 mM donor, 0.2 mM FMN $D_2O$ solution to withstand at least 50 laser flashes at 488 nm, 100 ms, 5 W output power without substantial photodegradation. Pass: fluorophenols, fluorotyrosines. Fail: fluorotryptophans.*

*Weak temperature dependence of $^1H$ and $^{19}F$ chemical shifts. Pass: fluorotyrosine. Fail: fluorophenols, fluorotryptophans.*

*Simple spin system and well resolved signal multiplets. Pass: 4-fluorophenol, 3-fluorotyrosine.*

Based on these criteria, the best system for quantitative investigation appears to be 3-fluorotyrosine.

The predictive power of *ab initio* hyperfine tensor calculations appears to be impressive, with no experimentally detected deviations from the expected photo-CIDNP behaviour.

From the qualitative relaxation experiments it appears that the diamagnetic dipolar and DD-CSA relaxation do not inherit the complexity of the CIDNP generation processes and may be described by a simple extension to the Redfield relaxation theory, specifically, by including the constant "magnetization source" terms into the density matrix evolution equation for the duration of the laser flash.

# Chapter 3
# *Chemically pumped $^{19}$F-$^{1}$H cross-relaxation*

*Based on: I. Kuprov, P.J. Hore, J. Magn. Res. 168 (2004) 1-5.*

### 3.1 Introduction

Based on the reconnaissance experiments and calculations described in Chapter 2, this Chapter details a systematic investigation into the relaxation processes in a chemically pumped $^{19}$F-$^{1}$H spin system of 3-fluorotyrosine. In particular, it explores the $^{19}$F photo-CIDNP effect as a source of strong long-range $^{19}$F-$^{1}$H heteronuclear Overhauser effects (NOEs). The measurements described below characterise the $^{19}$F-$^{1}$H dipole-dipole (DD) cross-relaxation and DD-CSA (chemical shift anisotropy) cross-correlation effects associated with $^{19}$F photo-CIDNP in 3-fluorotyrosine.

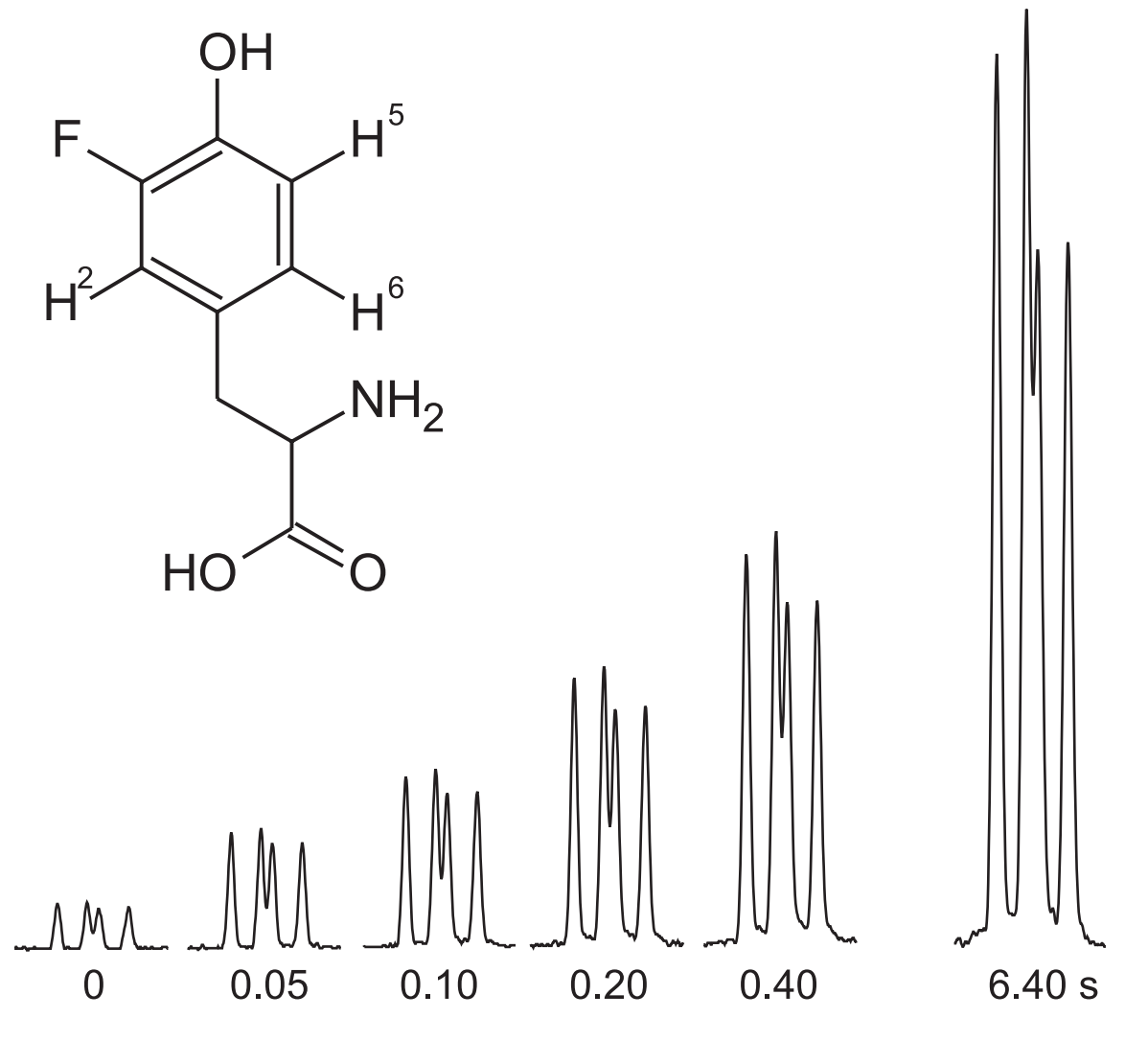

**Figure 3.1** $^{19}$F NMR spectra of 3-fluorotyrosine as a function of laser irradiation time. The spectrum labelled "0" is the conventional NMR spectrum. Irradiation was performed prior to the 90° pulse in a single-scan pulse-acquire experiment.

Beyond the field of CIDNP, the fluorinated tyrosine (Figure 3.1) has long been popular in biophysical applications of $^{19}$F NMR and several research groups in the world have expression systems set up to produce fluorotyrosine-labelled proteins [87, 90, 93, 106-115]. An investigation into the photo-CIDNP pumped $^{19}$F NMR experiments using 3-fluorotyrosine as a test compound benefits this established community and ensures that the more complicated biophysical systems are at hand, should the exploratory experiments appear promising.

From a practical point of view, 3-fluorotyrosine is convenient because it has a very weak dependence of the $^{19}$F chemical shift on temperature and is photochemically more robust than fluorophenols or fluorotryptophans (see Chapter 2 for details). At a magnetic induction of 14.1 T (600 MHz proton frequency) another valuable feature emerges: the proton on the left hand side of the aromatic ring (Figure 3.1) becomes weakly *J*-coupled to the protons on the right hand side. This eliminates a fair deal of complicated mathematics from the relaxation analysis of the spin system. Last, but not least, 3-fluorotyrosine is cheap.

### 3.2 Chemically pumped $^{19}$F-$^{1}$H cross-relaxation – theoretical setup

From the preliminary experiments described in Chapter 2 and prior work on $^{1}$H photo-CIDNP pumped cross-relaxation [116] it appears that for the purpose of nuclear spin relaxation analysis the production of CIDNP by continuous low-power illumination of an NMR sample in a strong magnetic field can be described by adding constant magnetisation-source terms to the equation of motion of the density operator of the spin system in the absence of the photochemical reaction (i.e. in the dark)

$$\frac{d\hat{\rho}}{dt} = \left(\frac{d\hat{\rho}}{dt}\right)_{\text{dark}} + \sum_i p_i \hat{I}_z^i \qquad (3.1)$$

where $\hat{I}_z^i$ is the *z*-magnetisation operator for the *i*-th nucleus and $p_i$ is the corresponding photo-CIDNP magnetisation pumping rate, which can be either positive or negative. The CIDNP source terms are non-stochastic and time-independent, so they appear unchanged in the magnetisation mode evolution equations obtained after treating the spin system in the Redfield relaxation matrix formalism [9]. CIDNP multiplet effects [41, 117], if present, can be introduced by adding similar source terms for the longitudinal multi-spin orders.

Although a rigorous analysis of the longitudinal cross-relaxation of the fluorine and the H(2) proton in 3-fluorotyrosine (see Figure 3.1 for numbering scheme) should strictly include all three protons in the aromatic ring, and possibly also the β-$CH_2$ and α-CH protons, it will be shown below that the experimental data are fully described by a simpler relaxation model that includes the interaction of the anisotropically shielded $^{19}$F nucleus with the adjacent H(2) as an AX spin system and neglects the contributions of the remote protons.

The longitudinal relaxation equations for a weakly coupled two-spin spin system when one spin has an anisotropic chemical shift tensor[7] have been obtained by Goldman [36]. After introduction of the CIDNP source term, these expressions become

$$\frac{d}{dt}\begin{bmatrix} 1 \\ H_z \\ F_z \\ 2H_zF_z \end{bmatrix} = -\begin{bmatrix} 0 & 0 & 0 & 0 \\ -p_H & \rho_{HH} & \sigma_{HF} & 0 \\ -p_F & \sigma_{HF} & \rho_{FF} & \delta_{F,HF} \\ 0 & 0 & \delta_{F,HF} & \rho_{HFHF} \end{bmatrix}\begin{bmatrix} 1 \\ \Delta H_z \\ \Delta F_z \\ 2H_zF_z \end{bmatrix} \tag{3.2}$$

where $\Delta H_z$ and $\Delta F_z$ are the deviations of the $^1$H and $^{19}$F *z*-magnetizations from equilibrium ($\Delta H_z = H_z - H_{z0}$, and similarly for $\Delta F_z$) and $2H_zF_z$ is the longitudinal $^1$H-$^{19}$F two-spin order. Although the self-relaxation parameters $\rho_{HH}$, $\rho_{FF}$, and $\rho_{HFHF}$ contain contributions from relaxation mechanisms other than the DD and CSA mechanisms, the dipolar cross-relaxation rate $\sigma_{HF}$ and the rate of accumulation of longitudinal two-spin order $\delta_{F,HF}$ arise in this system solely from dipolar interactions and DD-CSA cross-correlation. These parameters may be written as [36, 37]

$$\sigma_{HF} = \frac{1}{10}\left(\frac{\mu_0}{4\pi}\right)^2 \frac{\gamma_H^2\gamma_F^2\hbar^2\tau_c}{r_{HF}^6}\left(\frac{6}{1+(\omega_F+\omega_H)^2\tau_c^2} - \frac{1}{1+(\omega_F-\omega_H)^2\tau_c^2}\right)$$

$$\delta_{F,HF} = \frac{2}{5}\frac{\mu_0}{4\pi}\frac{\gamma_F^2\gamma_H\hbar B_0}{r_{HF}^3}\frac{\tau_c}{1+\omega_F^2\tau_c^2}\Delta\sigma_F^g \tag{3.3}$$

where $r_{HF}$ is the proton-fluorine distance, $\tau_c$ is the rotational correlation time, and $\Delta\sigma_F^g$ denotes

$$\Delta\sigma_F^g = \tfrac{1}{2}\sigma_{XX}(3\cos^2\theta_{X,HF} - 1) + \tfrac{1}{2}\sigma_{YY}(3\cos^2\theta_{Y,HF} - 1) + \tfrac{1}{2}\sigma_{ZZ}(3\cos^2\theta_{Z,HF} - 1) \tag{3.4}$$

in which $\sigma_{XX}$, $\sigma_{YY}$, $\sigma_{ZZ}$ are the principal elements of the anisotropic part of the chemical shift tensor and $\theta_{X,HF}$, $\theta_{Y,HF}$, $\theta_{Z,HF}$ are the angles between the corresponding principal axes and the H-F vector. Although it is often assumed that $^{19}$F shielding tensors in fluorine-substituted benzenes are axially symmetric, calculations reveal a significant rhombicity [35]. Equation (3.4), rather than the expression for an axial tensor [37], is therefore preferred.

In the extreme narrowing limit and taking $B_0$ = 14.1 T (600 MHz proton frequency), one obtains the following expressions for the cross-relaxation rates

[7]The anisotropy of the aromatic proton chemical shielding tensors is typically about 10 ppm, more than an order of magnitude smaller than the anisotropy of $^{19}$F shielding. Aromatic protons also exhibit much smaller photo-CIDNP enhancements. On these grounds, the contribution of $\delta_{H,HF}$ term to the evolution of the longitudinal two-spin order in Equation (3.3) has been neglected and the term was explicitly zeroed to reduce variational freedom.

$$
\begin{aligned}
\sigma_{HF}/\ \text{Hz} &= 2.52 \times 10^{11} \frac{\tau_c/\text{s}}{(r_{HF}\,/\,\text{Å})^6} \\
\delta_{F,HF}/\ \text{Hz} &= 1.01 \times 10^{9} \frac{\tau_c/\text{s}\ \ \Delta\sigma_F^g/\text{ppm}}{(r_{HF}\,/\,\text{Å})^3}
\end{aligned}
\tag{3.5}
$$

The integrated form of Equation (3.2) can be directly fitted to experimental data to obtain values of the parameters. Once both $\sigma_{HF}$ and $\delta_{F,HF}$ are known, simple arithmetic gives $\Delta\sigma_F^g$ and $\tau_c$. Care must be taken to ensure that the appropriate multipliers are introduced to account for the difference in the equilibrium polarisations of protons and fluorine if both are normalised to unity.

### 3.3 Experimental methods

$^{1}$H and $^{19}$F photo-CIDNP spectra were recorded on a 600 MHz (14.1 T) NMR spectrometer equipped with a 5 mm $^{19}$F-{$^{1}$H} probe. The light source was a Spectra-Physics Stabilite 2016-05 argon ion laser, operating in multi-line mode at 5 W output power, principally at 488 and 514 nm. A mechanical shutter controlled by the spectrometer was employed to produce light pulses of 0.05-6.40 s duration. The light was brought into the 5 mm sample tube from above via a 1 mm diameter optical fibre (Newport F-MBE), positioned inside a coaxial insert (Wilmad WGS 5BL) whose tip was 2 mm above the top of the NMR coil [75]. After shimming, this arrangement has no adverse effect on the NMR resolution or line shape and requires no modification to the NMR probe [118]. An air-tight sample re-injection system was employed to counteract depletion of the photosensitizer [118], such that there was negligible photoreduction for up to 12 s total irradiation time.

A $D_2O$ solution containing 4 mM 3-fluoro-DL-tyrosine (Lancaster) and 0.2 mM FMN (Sigma Aldrich) at pH 5.0 (uncorrected for deuterium isotope effect), purged with argon for 20 minutes, was used in all experiments. During each experiment the sample was irradiated for a prescribed time and, after a variable relaxation delay, subjected to a 90° pulse on either $^{1}$H or $^{19}$F followed by immediate acquisition of the free induction decay. No paramagnetic broadening was detected in any of the 3-fluorotyrosine spectra, consistent with the low steady state concentration of radicals produced during the irradiation periods and their rapid recombination when the light is extinguished.

After application of a shifted Gaussian window function, zero-filling and Fourier transformation, the spectra were integrated using mixed Lorentzian-Gaussian line fitting. The magnitudes of the $F_z$, $2H_z^{(2)}F_z$, $2H_z^{(5)}F_z$ and $4H_z^{(2)}H_z^{(5)}F_z$ magnetization modes at the end of the relaxation delay were calculated from the integrals of individual lines in the $^{19}$F multiplet. Equation (3.2) was solved using a 4$^{th}$-order adaptive grid

Runge-Kutta method with $10^{-10}$ relative error on each integration step. Minimization of the weighted least squares error functional was performed using the Nelder-Mead simplex method. A Monte-Carlo method was employed to estimate errors in the fitting parameters. Libraries supplied with the Matlab 6.0 software package were used for all these computations.

### 3.4 Computational methods

Computation of *in vacuo* equilibrium geometry and hyperfine coupling constants were performed using the GAMESS program [77, 119] at the DFT B3LYP 6-311G**/EPR-III level. The *in vacuo* $^{19}$F magnetic shielding tensor was estimated using Gaussian98 [76] at the GIAO HF 6-311++G(2d, 2p) level using the 3-fluorotyrosine geometry obtained from a separate B3LYP 6-311G** GAMESS calculation. The relatively inexpensive Hartree-Fock based computation of the shielding tensor was used because previous studies had revealed no obvious advantage of MPn, CI or other methods accounting for electron correlation for the $^{19}$F nuclei of the molecules in question [120]. The calculated hyperfine couplings for the neutral radicals derived from tyrosine and 3-fluorotyrosine by abstraction of hydrogen from the phenolic OH group are given in Table 3.1 together with the experimental values for the tyrosyl radical [121].

The computing power was provided by a home-made heterogeneous GAMESS cluster built to harvest computing power from the laboratory UltraSparcIII+/Solaris and (Pentium4, AthlonMP)/Linux workstations. Built-in socket-based parallelisation was used in the case of GAMESS-US and LAM/MPI based parallelisation in the case of PC GAMESS [119]. The interlink was provided by gigabit Ethernet. A built-in load levelling mechanism was used for all computations.

### 3.5 Results

Figures 3.2 to 3.5 illustrate the changes in the $^{19}$F and $^{1}$H NMR spectra and polarizations of 3-fluorotyrosine produced by laser irradiation. Under continuous irradiation the fluorine CIDNP magnetisation pumping rate is $p_F = +34\ s^{-1}$ (Figures 3.1 and 3.4A, left hand panel), i.e. the initial build-up rate of the $F_z$ mode corresponds to the accumulation of 34-fold $^{19}$F hyperpolarization per second. The CIDNP magnetisation pumping rate for the adjacent H(2) proton is $p_H = +1.3\ s^{-1}$, hence the initial slow rise (Figure 3.4A, right hand panel). After four seconds of irradiation, the $^{19}$F *z*-magnetization reaches a steady state, in which the CIDNP magnetisation pumping is balanced by relaxation, corresponding to 24 times the equilibrium magnetization (Figure 3.4A, left hand panel). Under continuous laser irradiation the $^{19}$F magnetisation stays at this level. Under identical conditions, the polarization of H(3) in tyrosine itself has a much smaller pumping rate, $p_H \approx -5\ s^{-1}$.

Increasing the laser output power from 5 to 25 W gives up to 40-fold $^{19}$F equilibrium hyperpolarization in the 3-fluorotyrosine-FMN system. Higher polarisation is difficult to obtain due to sample heating by the intense laser light.

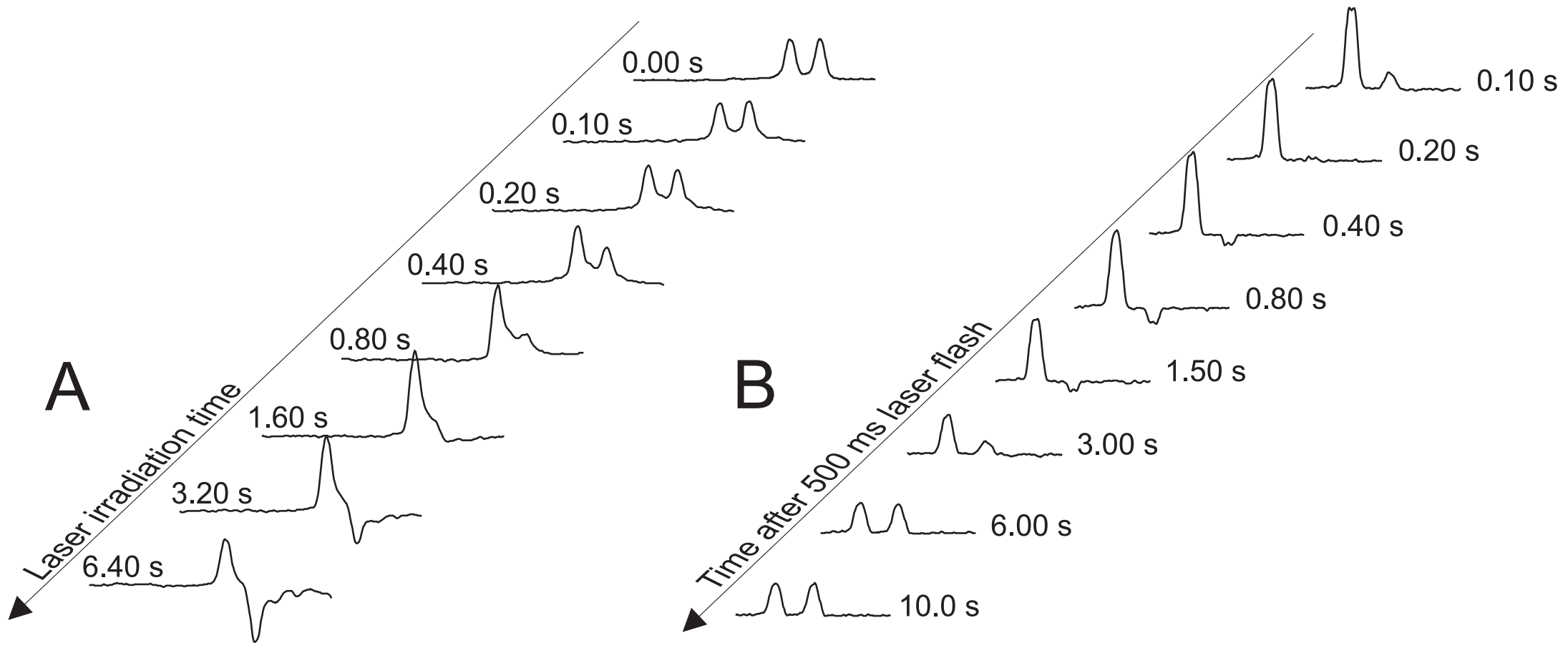

**Figure 3.2.** The H(2) resonance taken from 600 MHz $^{1}$H photo-CIDNP NMR spectra of 3-fluorotyrosine as a function of: (A) laser irradiation time and (B) the time after a 0.5 s laser flash. The gradual deterioration in the spectral resolution in set (A) is due to non-uniform sample heating by the laser.

As expected from Equation (3.2) and demonstrated by the unequal multiplet component intensities in Figures 3.1 and 3.2, longitudinal two-spin order, $2H_z^{(2)}F_z$, accumulates as a result of cross-correlation between the F-H(2) DD interaction and the fluorine CSA. In the case of continuous irradiation (Figure 3.4A, centre panel) the amplitude of $2H_z^{(2)}F_z$ after several seconds of irradiation attains a steady state value of $2H_z^{(2)}F_z / F_{z0} = -2.0$. Such a magnitude of $^{1}$H-$^{19}$F two-spin order is impossible to generate using traditional NMR methods. The maximum value of $2H_zF_z / F_{z0}$ obtained by Dorai and Kumar [122] for similar molecules by inversion of the $^{19}$F equilibrium magnetisation was about 0.15.

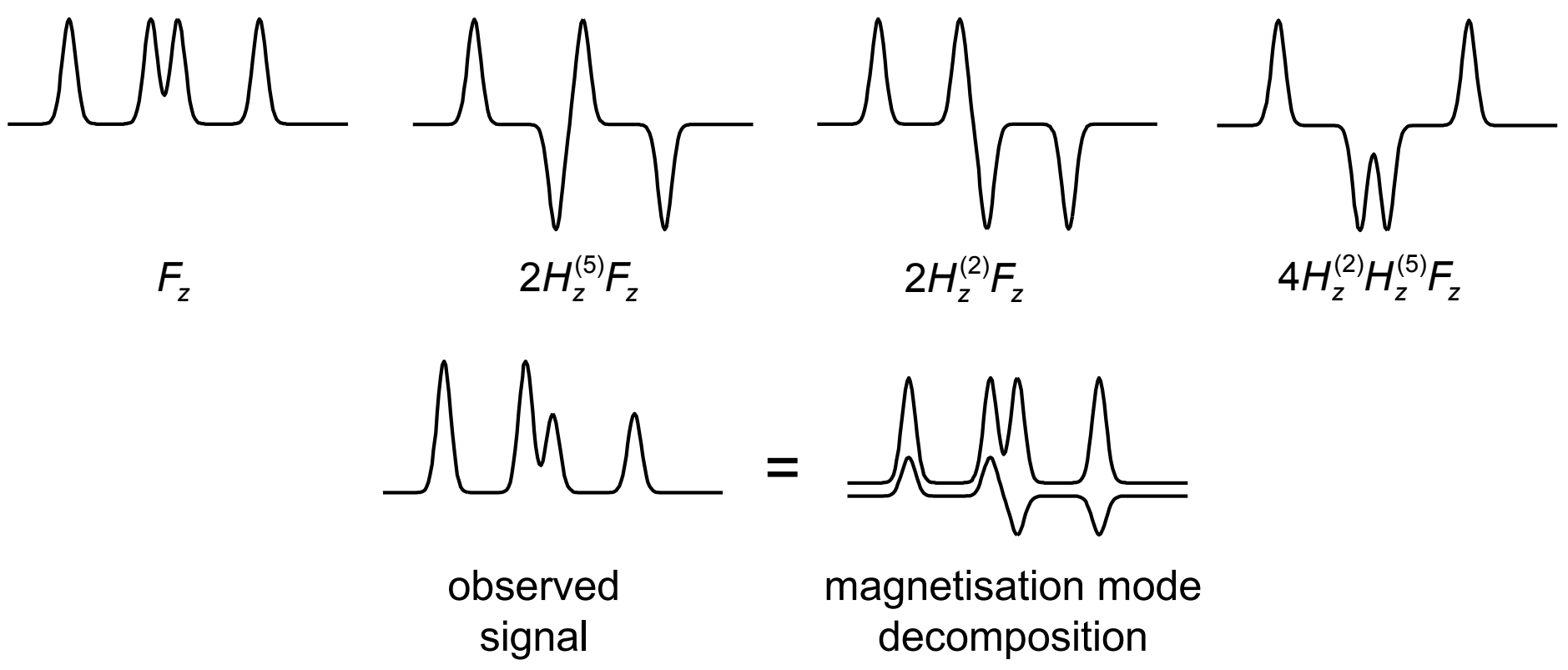

**Figure 3.3** An illustration of the $^{19}$F spectra corresponding to different longitudinal multi-spin orders in a $^{19}$F-$^{1}$H spin system.

The amplitudes of the longitudinal multi-spin magnetisation orders of fluorine with other protons, such as $2H_z^{(5)}F_z$ and $4H_z^{(2)}H_z^{(5)}F_z$ as determined from the relative integrals of the $^{19}$F multiplet components, are barely above the noise level of the integration (Figure 3.5), justifying the approximation (*vide supra*) that these modes do not contribute significantly to the longitudinal cross-relaxation within the F-H(2) spin system. As can be seen from Figure 3.4, the experimental data for 3-fluorotyrosine are very satisfactorily described using this approximation.

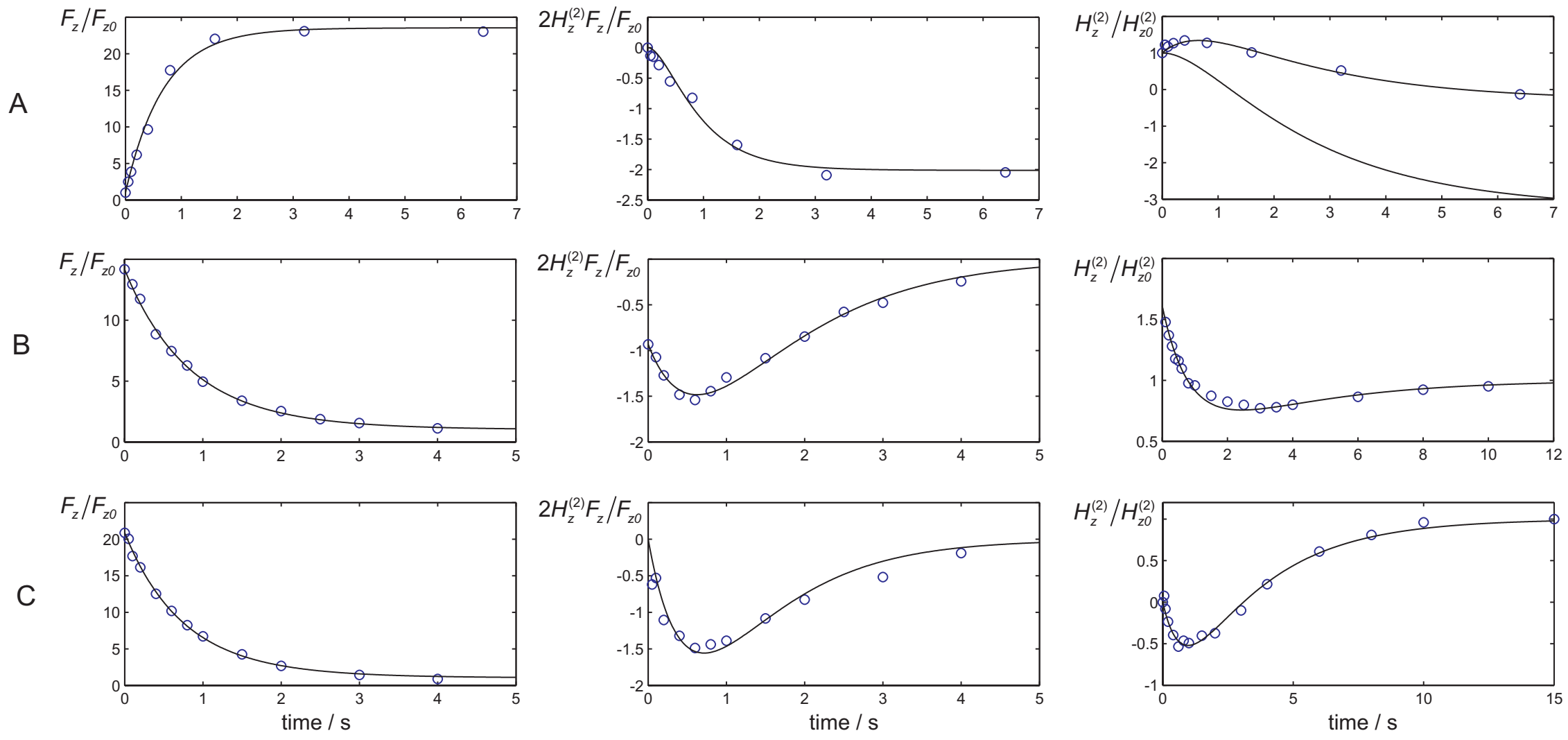

**Figure 3.4** The time dependence of three longitudinal magnetizations in the F-H(2) spin system in 3-fluorotyrosine. (A) With continuous laser irradiation. The lower curve in the right hand panel is the polarisation computed in the absence of $^{1}$H CIDNP. (B) As a function of mixing time after a 0.5 s laser flash. (C) As a function of mixing time after a 1.0 s laser flash with $^{1}$H decoupling during irradiation. The solid lines are the results of fitting Equation (3.2) to the data. The pulse sequences are shown on figure 6.

The time dependence of the H(2) *z*-magnetisation is a result of competition between direct CIDNP pumping and $^{19}$F-$^{1}$H dipolar cross-relaxation (Figure 3.4A, right hand panel). The CIDNP of H(2) in tyrosine and 3-fluorotyrosine is weakly absorptive, and causes the $H_z$ polarisation to rise initially to about 150% of the equilibrium polarisation. However, the negative NOE from the highly polarised $^{19}$F nucleus gradually overwhelms the $^{1}$H CIDNP: after about six seconds of irradiation the observed $^{1}$H polarisation is pushed below zero, corresponding to a NOE of around –100%, resulting from competition between $^{1}$H CIDNP pumping and the $^{19}F$-$^{1}H$ NOE. In complex molecules (e.g. a $^{19}F$-labelled protein) most protons have no CIDNP of their own. Solving Equation (3.2) with $p_H$ set to zero leads to an actual NOE magnitude of around –400% after six seconds of irradiation (Figure 3.4A, right hand panel).

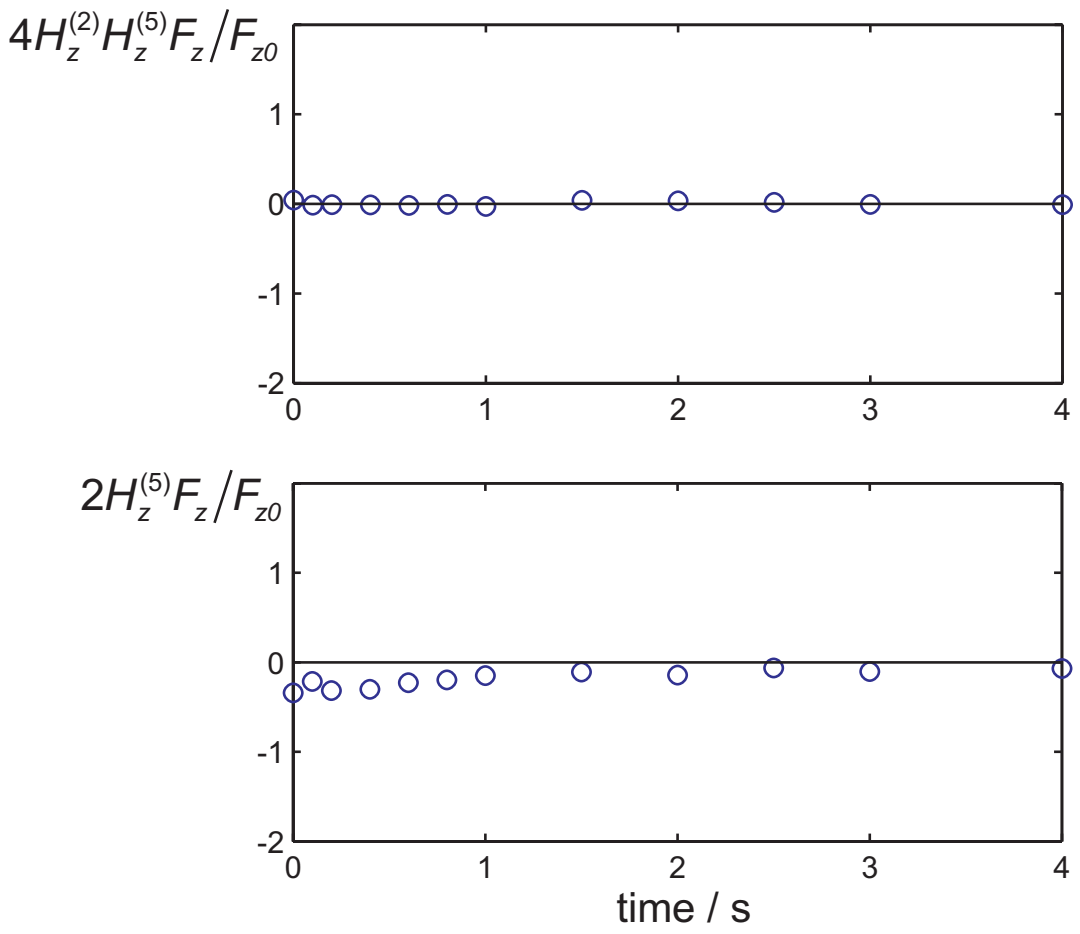

**Figure 3.5.** The amplitudes of two fluorine longitudinal multi-spin orders with protons other than H(2) as a function of time after a 0.5 s laser flash.

The $^{19}F$-$^{1}H$ spin system shows similar behaviour in dynamic CIDNP NOE experiments. The three panels in Figure 3.4B show the magnetization evolution after a 500 ms laser flash. Because the $^{19}F$ CIDNP enhancement after 500 ms irradiation is only about 15-fold and decays quickly, the cross-relaxation and cross-correlation effects are weaker, but still very pronounced. In particular, it can be seen that the $2H_z^{(2)}F_z/F_{z0}$ two-spin order first grows to an amplitude of −1.5 and then decays to zero. The longitudinal $^{1}H$ magnetisation $H_z^{(2)}$ first loses the CIDNP that accumulated during the irradiation, briefly dips about 25% below its equilibrium value as a result of the $^{19}F$-$^{1}H$ NOE, and gradually relaxes to equilibrium. Solving Equation (3.2) using $p_H = 0$ gives a NOE of −55%.

**Table 3.1.** Calculated isotropic hyperfine coupling constants for the 3-fluorotyrosyl radical compared to the computed and experimental [121] HFCs of the tyrosyl radical.

| | tyrosyl | | 3-fluorotyrosyl |
|---|---|---|---|
| Nucleus | Experimental HFC/mT[1] | Computed HFC/mT | Computed HFC/mT |
| H(2) | 0.15 | 0.26 | 0.28 |
| H(3) / F(3) | 0.65 | –0.68 | 1.57 |
| H(5) | 0.65 | –0.68 | –0.62 |
| H(6) | 0.15 | 0.26 | 0.20 |

[1] The signs of the experimental values were not determined.

As mentioned above, the $^{19}F$-$^{1}H$ NOE is considerably more pronounced when the proton has no counteracting CIDNP of its own. This can be demonstrated in the case of 3-fluorotyrosine by destroying the $^{1}H$ CIDNP by decoupling the protons during the laser flash. Figure 3.4C shows the evolution of the magnetisation modes after a 1.0 s laser flash, with $^{1}H$ decoupling. Once the decoupler is switched off (at $t = 0$), the $H_z^{(2)}$ magnetisation abruptly goes negative as a result of the $^{19}F$-$^{1}H$ NOE, reaching a value of $H_z^{(2)}/H_{z0} = -0.5$ which is above the maximum theoretical NOE for conventional NMR and exceeds the magnitudes of $^{19}F$-$^{1}H$ NOEs usually observed in such systems by a factor of five.

The DD and DD-CSA cross relaxation rates measured for the F-H(2) spin system are presented in Table 3.2. The values of the correlation time $\tau_c$ and the $^{19}F$ CSA were calculated using a F−H(2) distance of 2.60 Å, obtained by *ab initio* computation *in vacuo*. Within experimental error, the three sets of measurements give the same values for $\tau_c$ and $\Delta\sigma_F^g$. The values of $\Delta\sigma_F^g$ agree well with those reported for similar molecules by Dorai and Kumar [122].

**Table 3.2.** Experimental DD and DD-CSA cross relaxation rates, rotational correlation times and $^{19}F$ geometrically weighted shielding anisotropy for the F-H(2) spin system of 3-fluorotyrosine.

| Experiment | $\delta_{F,HF}$ / $s^{-1}$ | $\sigma_{HF}$ / $s^{-1}$ | $\tau_c$ / ps | $\Delta\sigma_F^g$ / ppm |
|---|---|---|---|---|
| Continuous irradiation | 0.27 ± 0.07 | 0.088 ± 0.003 | 108 ± 4 | 44 ± 11 |
| After 0.5 s irradiation | 0.24 ± 0.04 | 0.087 ± 0.012 | 106 ± 15 | 39 ± 8 |
| After 1.0 s irradiation with $^{1}H$ decoupling | 0.30 ± 0.05 | 0.094 ± 0.008 | 115 ± 10 | 45 ± 8 |

The orientation of the computed $^{19}F$ nuclear shielding tensor is schematically shown in Figure 3.6. The computation yields the absolute chemical shielding tensor with the following eigenvalues: $\sigma_{11}$ = 326 ± 10 ppm, $\sigma_{22}$ = 357 ± 10 ppm, $\sigma_{33}$ = 436 ± 10 ppm. The error estimates are based on the discrepancies between calculated and experimental shielding tensor components in a study by de Dios and Oldfield [35]. The most shielded component $\sigma_{33}$ was found to be perpendicular to the benzene ring plane, and the least shielded component $\sigma_{11}$ to be at a 25° angle to the direction of the F-H(2) vector. The resulting computed geometrically weighted anisotropy, $\Delta\sigma_F^g$, was 62 ± 30 ppm.

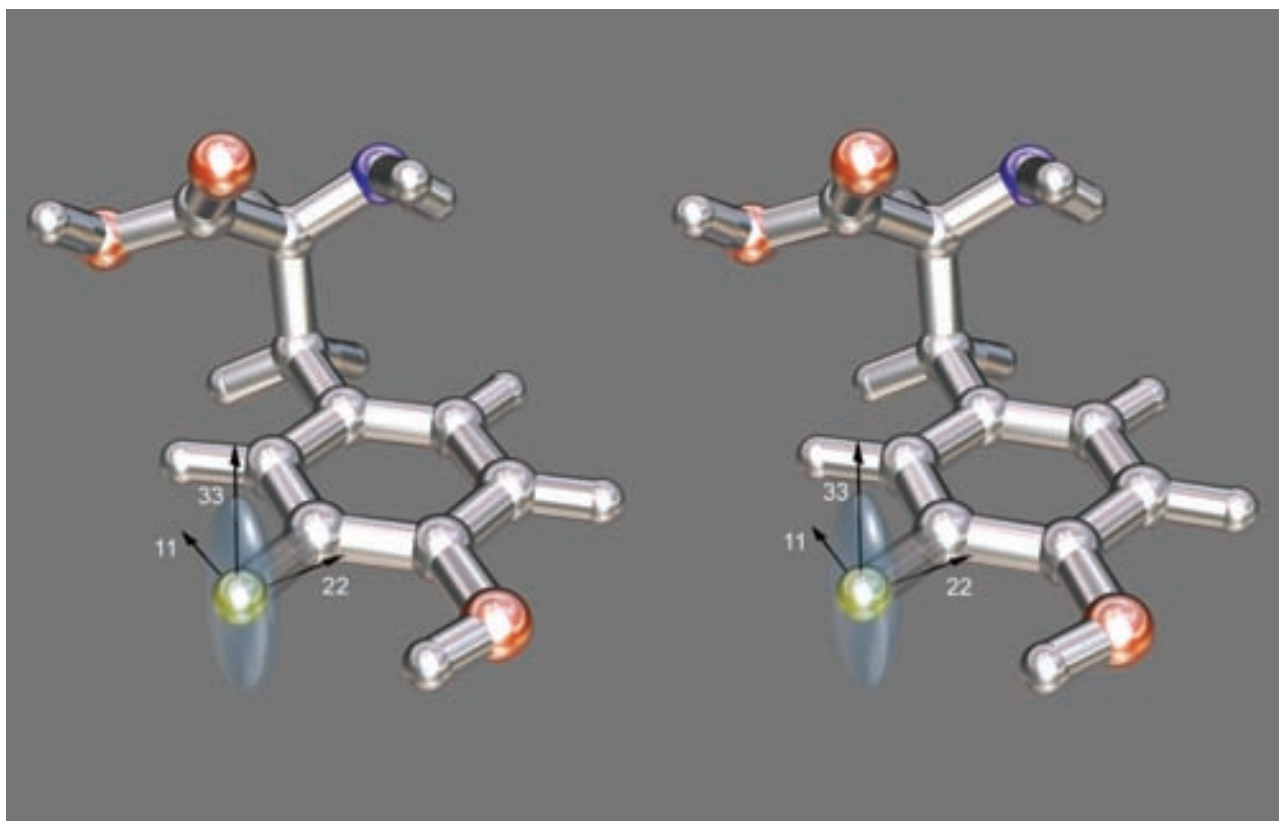

**Figure 3.6.** Stereo view of the orientation of the fluorine magnetic shielding tensor in 3-fluorotyrosine in vacuum as computed using GIAO HF 6-311++G(2d, 2p) theory on a DFT B3LYP 6-311G** optimized geometry.

### 3.6 Discussion

The relatively large $^{19}F$ CIDNP build-up rate reported above (+34 $s^{-1}$) compared to the corresponding proton in tyrosine (−5 $s^{-1}$) is at least partially due to the different magnetic properties of the intermediate radicals. The $^{19}F$ hyperfine coupling constant computed for the 3-fluorotyrosyl radical is +1.57 mT (Table 3.1) as opposed to −0.65 mT for the corresponding proton in the tyrosyl radical [121]. For non-viscous solutions and in a strong magnetic field, the geminate CIDNP effect of a particular nucleus is proportional to its hyperfine coupling constant in the intermediate radical [116]. Larger hyperfine interactions also lead to faster nuclear spin-lattice relaxation in the radical and hence less cancellation of recombination and escape polarization in the cyclic reaction scheme [47]. Although a full explanation of this effect will require experiments with microsecond time-resolution [123], its consequences may nonetheless be put to good use.

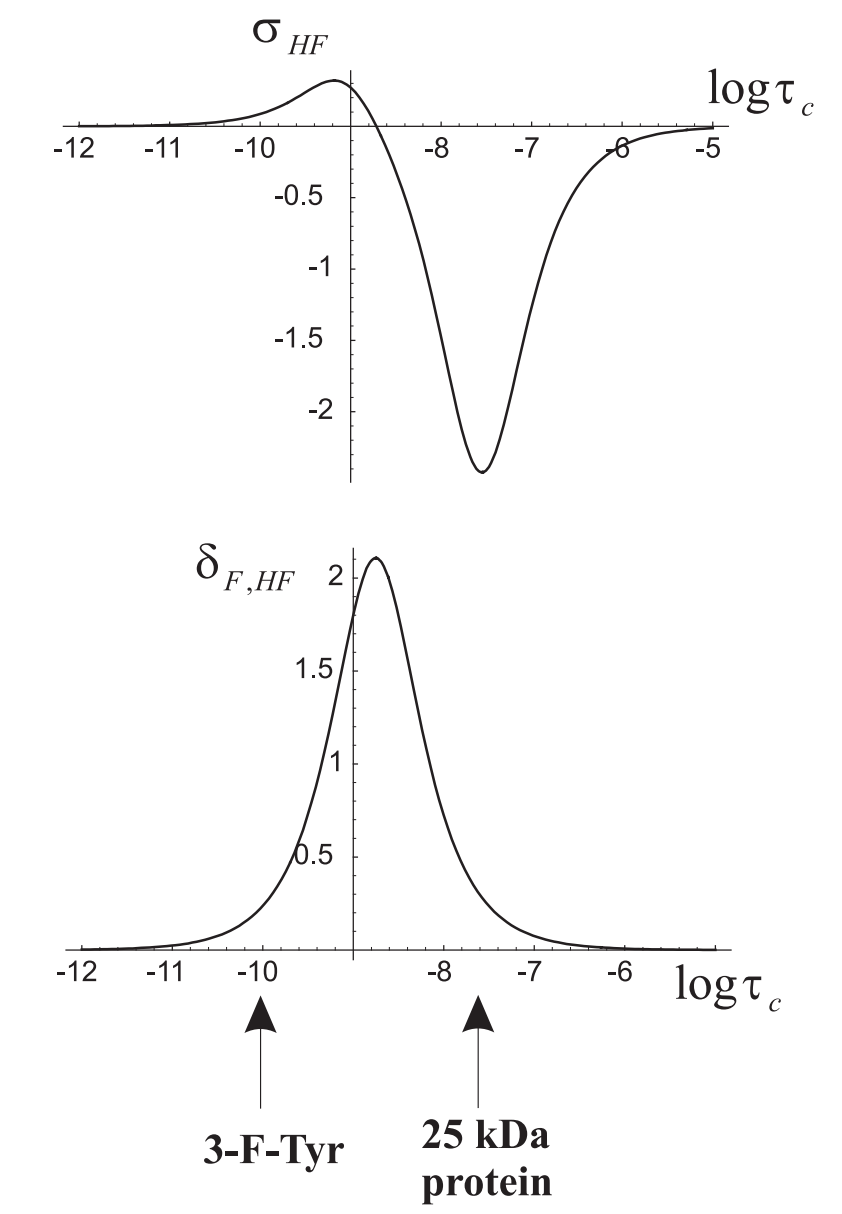

**Figure 3.7.** Dependence of the DD cross-relaxation rate and the rate of accumulation of longitudinal multi-spin order on the rotational correlation time. Calculated from Equations (3.3) using $r_{HF}$ = 2.60 Å and $\Delta\sigma_F^g$ = 40 ppm.

Figure 3.4A shows that 20-fold $^{19}F$ hyperpolarization can be sustained for at least seven seconds by continuous laser irradiation with no damage (or indeed any change at all) to the 3-fluorotyrosine sample. If the same holds true in a complex molecular system, such as a fluorine-labelled protein, this implies that long-distance dipolar energy transfer could occur, either directly or mediated by intervening protons.

The measured dipolar cross-relaxation rates and NOE magnitudes allow one to speculate about CIDNP-induced $^{19}F$-$^{1}H$ NOEs when 3-fluorotyrosine is incorporated into a larger molecule. Assuming that the destination atom has no CIDNP of its own (which would normally be the case) and taking the detectable NOE enhancement to be about 1%, one can make a rough estimate that an NOE effect that reaches a magnitude of 400% at 2.60 Å distance would be still detectable at distances of up to about $2.60(400)^{1/6}$ = 7 Å.

A consideration of the dependence of the polarisation transfer rates on the rotational correlation time (Figure 3.7) provides another argument in favour of chemically pumped $^{19}F$-$^{1}H$ NOEs. The isotropic correlation time for 3-fluorotyrosine is about $10^{-10}$ s, with the result that both $\sigma_{HF}$ and $\delta_{F,HF}$ are comparable and relatively small. For a (~25 kDa) protein with $\tau_c$ of about 20 ns, however, the DD cross-relaxation rate $\sigma_{HF}$ would be approximately 30 times larger, while the cross-correlation rate is barely changed. This does not necessarily imply a 30-fold increase the CIDNP NOE, because longitudinal self-relaxation will also be faster in the more slowly tumbling molecule, but it certainly means that the drain of fluorine magnetization into longitudinal multi-spin orders will be reduced.

In most one- and multi-dimensional experiments, using CIDNP as a source of strong nuclear polarization amounts simply to adding photosensitizer to the sample and inserting a laser flash at the start of the pulse sequence [48], accompanied, if necessary, by simultaneous decoupling to remove undesired direct CIDNP effects. For experiments requiring a large number of scans, photosensitizer depletion can be countered by employing a sample re-injection device [118].

The phenomenon explored herein is known in the proton-proton case as CINOE (Chemically Induced NOE) [124]. This abbreviation seems unsatisfactory for two reasons. First, NOE is not induced, only amplified by chemical pumping. And second, this abbreviation does not match the NMR abbreviation standards in that it has no (preferably awkward) second meaning. I therefore suggest that this Chemically Amplified Nuclear Overhauser Effect be referred to as CANOE.

# Chapter 4

# *Uniform illumination of optically dense NMR samples*

*Based on: I. Kuprov, P.J. Hore, J. Magn. Res. 171 (2004) 171-175.*

### 4.1 Introduction

A variety of photochemical NMR experiments require illumination of the sample inside the NMR probe, including but not limited to: photo-CIDNP (Chemically Induced Dynamic Nuclear Polarization) [42, 47, 125], metal ion release from photolabile cage compounds [72, 126], photochemical kinetics [57, 127, 128] and studies of photoactive proteins [129-132].

One of the biggest technical challenges faced in such experiments is to deliver light efficiently into the active region of the NMR sample. Ideally, the entire sample volume should be uniformly illuminated to maximise sensitivity and to avoid concentration and temperature gradients, which might distort the observed kinetics. Such problems are likely to be most severe for optically dense samples. It is highly desirable that uniform illumination is achieved without extensive probe modifications, which may compromise the NMR performance, and in a manner that allows facile transfer from one spectrometer to another. Several techniques have been devised in recent years, including illumination from above, below or the side, some of which are shown schematically in Figure 4.1.

Bringing light in from the side, through the radiofrequency (RF) coil, via a quartz light guide and a prism or a mirror (Figure 4.1A), has the disadvantage of non-uniform irradiation of the sensitive region even if the sample has low optical density. Such an arrangement is incompatible with certain RF coil designs and can conflict with the presence of field gradient coils. Irradiation from below using a flat-bottomed NMR tube (Figure 4.1B) is somewhat less demanding technically, but would still normally require the re-location of electronic components and variable temperature equipment inside the probe body. It also suffers from inhomogeneous illumination for optically dense samples, a problem that can be reduced by using a bespoke NMR sample tube with a tapered ("V-cone") interior [51] to extend illumination into the active region (Figure 4.1C) at the expense of both spectral resolution and sensitivity (the filling factor is reduced by ~20%). The attraction of these three approaches is that they are readily compatible with short, intense visible or ultraviolet pulses from YAG or excimer lasers as well as continuous wave (CW) light sources.

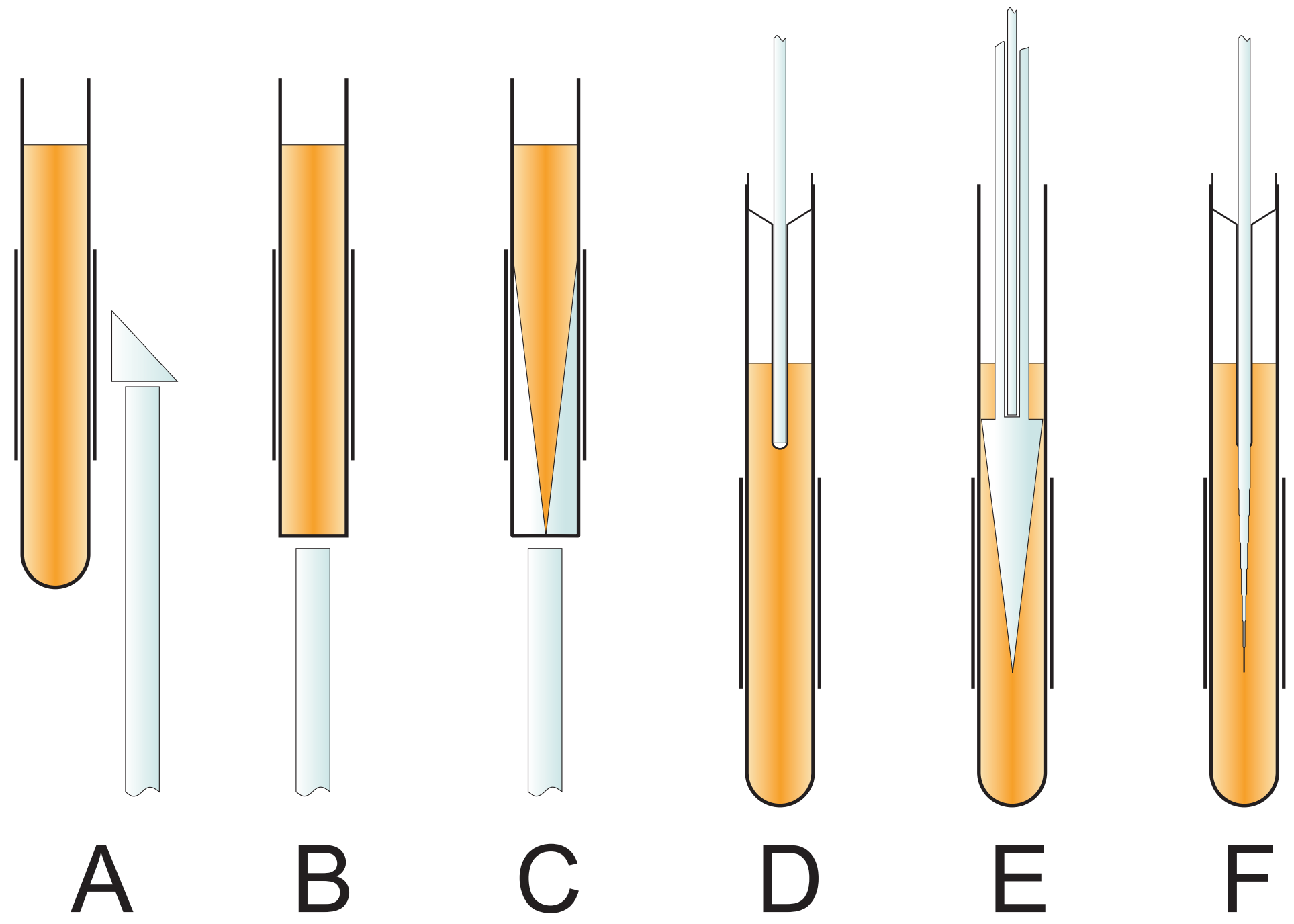

**Figure 4.1.** Schematic drawings of NMR sample illumination methods. **(A)** Illumination from the side through the receiver coil via a cylindrical quartz rod installed inside the probe body surmounted by a prism or mirror. **(B)** Illumination from below with a quartz light guide and a flat-bottomed NMR tube. **(C)** A variant of B using a "V-cone" NMR tube to permit more homogeneous irradiation of optically dense samples. **(D)** Illumination from above using an optical fibre held inside a coaxial glass insert. **(E)** A variant of D in which light is distributed by means of a "pencil tip" insert. **(F)** The arrangement demonstrated here with a stepwise tapered optical fibre.

To circumvent the requirement for probe modification, various methods have been suggested for illumination from above. Almost twenty years ago, Berliner and colleagues demonstrated the use of an optical fibre [75] held inside the NMR tube by means of a coaxial capillary insert (Figure 4.1D), an arrangement that has been used extensively in Oxford for photo-CIDNP studies of proteins [48, 52, 118, 133]. Because it introduces a magnetic susceptibility jump, the end of the insert must be positioned at least 2 mm above the NMR coil to avoid undue lineshape distortion. This approach is convenient, technically straightforward and particularly suitable for CW lasers. Pulsed lasers may be used only insofar as the optical power density is compatible with the fibre. Improved uniformity of illumination is afforded by combining the optical fibre with a “pencil tip” insert as demonstrated by Schwalbe *et al.* [72, 126] (Figure 4.1E). While retaining the simplicity of the Berliner method, this design, like the V-cone, sacrifices some sensitivity and resolution.

In this chapter a straightforward and inexpensive method is proposed for illumination from above in which the light is distributed along the axis of the NMR tube by means of a tapered optical fibre (Figure 4.1F). While illumination from above the coil (Figure 4.1D) gives rise to an exponential fall in light intensity from the top of the sensitive region to the bottom (optical path length ~20 mm), the tapered tip (with a path length of ~3 mm) gives almost uniform illumination. This approach leads to minimal degradation of spectral resolution, less than a 5% loss of filling factor and requires no probe modifications. The amount of light reaching the NMR sample can be conveniently monitored via the $^{19}$F photo-CIDNP enhancement of 4-fluorophenol sensitised by flavin mononucleotide (FMN). For light flashes between 1 ms and 100 ms, the photo-induced $^{19}$F magnetisation is a direct measure of the local light intensity [52] allowing the use of NMR imaging techniques to reveal the spatial distribution of light within the sample tube.

## 4.2 Materials and methods

The end of a Newport F-MBE optical fibre (1 mm outer diameter) was tapered using the following procedure. A 10 cm length of the plastic cladding was mechanically stripped from the end of the fibre, the final 2 cm of which was then immersed for 1 min in a mixture containing 30% HF, 20% $H_2SO_4$ and 50% $H_2O$ at 60 ºC to detach the fibre sheath from the core. The exposed core was rinsed with water and extruded stepwise (typically 1.5 mm every 30 min) into the same acid mixture from a plastic pipette tip, again at 60 ºC (Figure 4.2A). After about 5 hours the tip was removed, washed with water and dried. The resulting cone tapers in typically ~11 steps down to ~50 μm over a length of ~20 mm (Figure 4.2B). Smaller and more frequent adjustments of the fibre in the etching medium result in a smother cone. It should be noted that the hydrofluoric acid solution is extremely toxic and volatile when hot.

$^{19}$F photo-CIDNP spectra were recorded on a Varian Inova 600 MHz (14.1 T) NMR spectrometer equipped with a 5 mm $^{19}$F{$^1$H} *z*-gradient probe. The light source was a Spectra Physics BeamLok 2080 argon ion laser, operating in multi-line mode at 10 W output power, principally at 488 and 514 nm. A mechanical shutter (NM Laser Products LS200) controlled by the spectrometer was used to produce 100 ms light pulses. The light was focused into a 6 m length of optical fibre (Newport F-MBE), using a Newport M-5X objective lens, the other end of which was attached (via Newport

SMA connectors) to a 2 m section of the same fibre whose tapered tip was held inside a 5 mm NMR tube by a truncated Wilmad WGS 5BL coaxial insert (Figure 4.1F).

A 4.0 mM solution of 4-fluorophenol (Lancaster) in $D_2O$ containing 0.2-3.2 mM FMN (Sigma) at pH 5.0 (uncorrected for deuterium isotope effect) was used in all experiments. Samples were purged with argon for 20 min prior to use. Profiles of the CIDNP intensity along the tube were recorded by applying a constant *z*-gradient during detection of the free induction decay in a one-scan flash-pulse-acquire experiment.

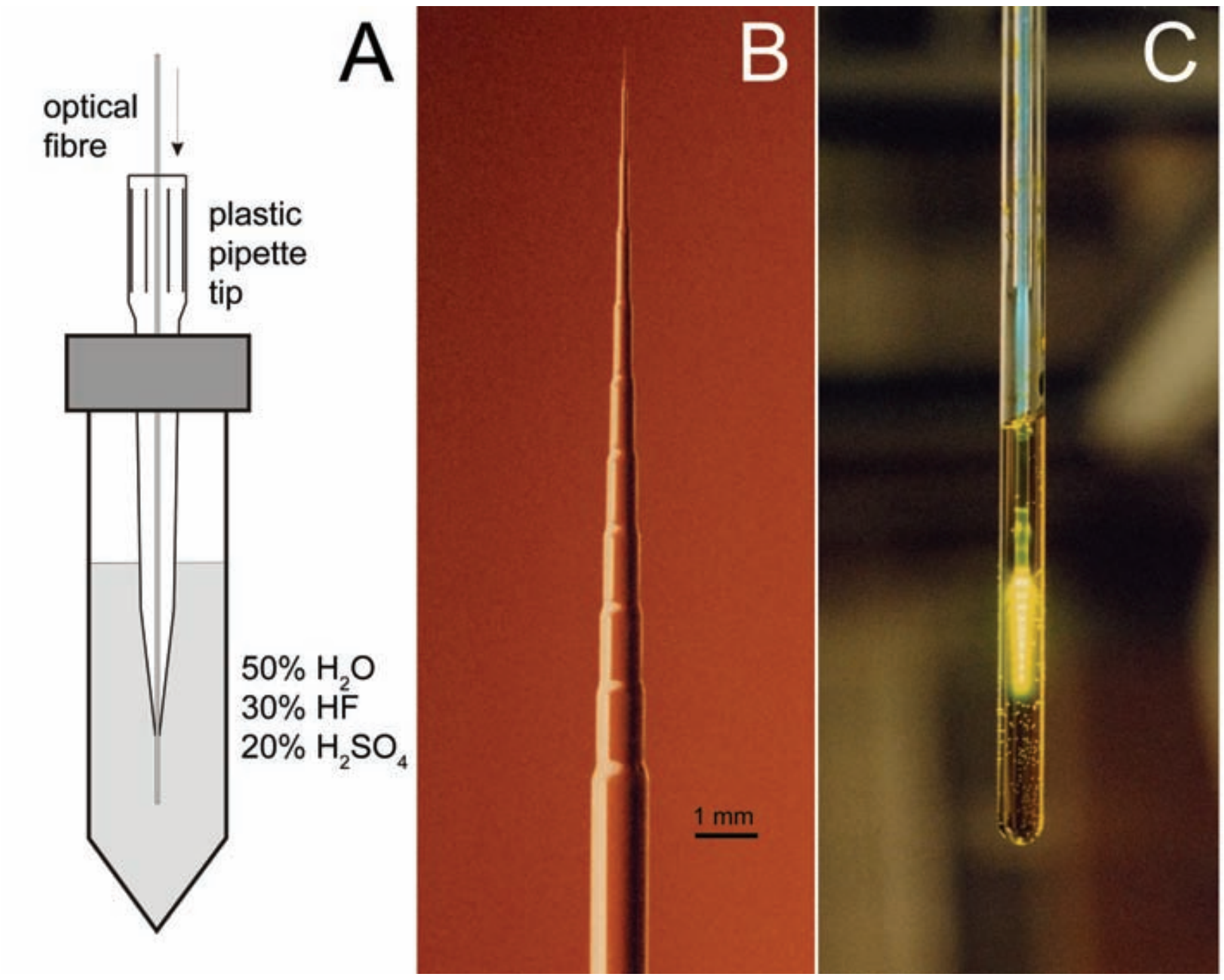

**Figure 4.2. (A)** The production of the tapered fibre tip; **(B)** the resulting tip; and **(C)** an assembled NMR sample irradiated principally at 488 and 514 nm. The yellow colour is FMN fluorescence excited by the 488 nm laser line. The terminal diameter of the fibre cone is less than 50 μm. In C, the optical fibre appears much thicker than it actually is because of the lensing effect of the solution.

### 4.3 Results and discussion

The shape of a fibre tip resulting from the treatment described above is shown in Figure 4.2B. This particular specimen contains 11 steps; in different experiments, cones with 7 to 15 steps were produced. The best results were obtained for 10-15 steps of 1-1.5 mm. In the absence of scratches the tips are quite robust, the most frequent damage being to the thinnest sections, caused by inaccurate insertion into an NMR tube. Such breakages are easily repaired using the etching procedure described above to produce a new tip by tapering the last few sections and at the same time creating a few new steps further up the fibre.

The light intensity distribution depends very strongly on the length of the fibre cone and the step size. The shorter cones, which have steeper shoulders, tend to emit light predominantly from the shoulders leading to a non-uniform illumination pattern. At the other extreme, very long cones with a large number of small steps and shoulders transmit light by total internal reflection to the very end of the tip, whence almost all of the light is emitted. The fairly uniform light emission shown in Figure 4.3 is a result of experimental optimisation of both the length of the cone and the size of the steps. Figure 4.3 also shows that the emission of light from an optimum cone occurs at an angle of ~45° to the fibre axis, predominantly from the vertical sections of the fibre between the shoulders, suggesting that the light paths are similar to those shown in Figure 4.3C. There is also some emission at the steeper shoulders.

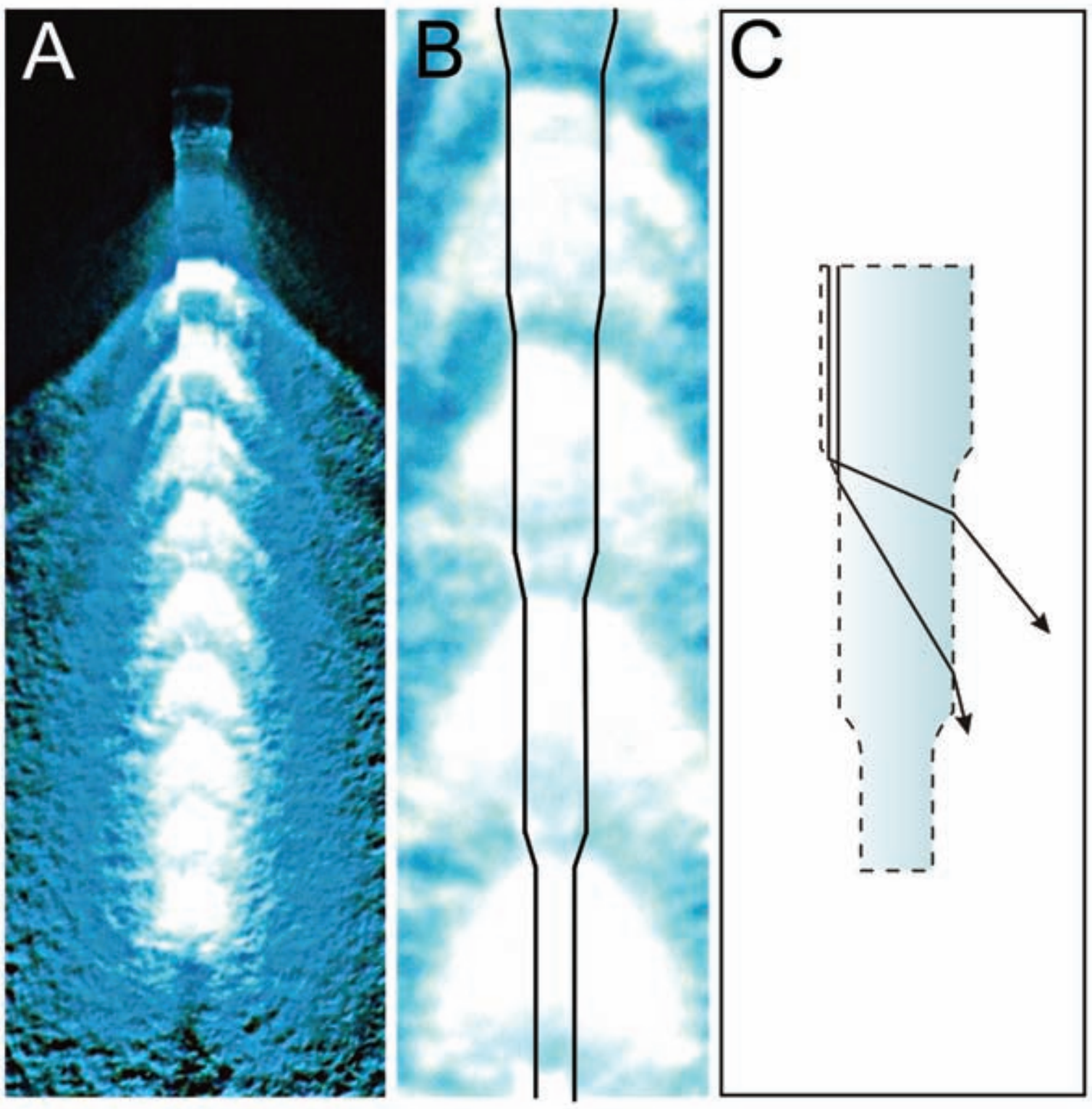

**Figure 4.3. (A)** Emission of argon laser light from the tapered fibre tip placed on a white surface. **(B)** The central part of A magnified with the fibre outlined for clarity. **(C)** Likely paths of light rays in the fibre cone.

Light intensity distributions along the axis of the sample, $L(z)$, measured using $^{19}F$ photo-CIDNP of 4-fluorophenol, are shown in Figure 4.4. Illumination from above using the arrangement in Figure 4.1D yields the expected Beer-Lambert dependence (Figure 4.4A):

$$L(z) = A + Be^{-Cz} \tag{4.1}$$

where $z$ is distance, $B$ and $C$ depend on the sample concentration and the tip position, while the constant contribution $A$ comes from the equilibrium nuclear magnetisation, from laser light reflections from the inner parts of the probe and from the 514 nm output of the $Ar^+$ laser which is absorbed by FMN to a far lesser extent than that at 488 nm. In the case of the tapered fibre tip the light intensity profile is much more even (Figure 4.4B). The light distribution image is somewhat narrower in the case of the tapered fibre, because the tip employed in this experiment is 2 mm shorter than the receiver coil. At high resolution, the images in Figure 4.4B have a low amplitude sinusoidal modulation superimposed on the top of the rectangular distributions, with the number of maxima matching the number of steps in the cone (not shown).

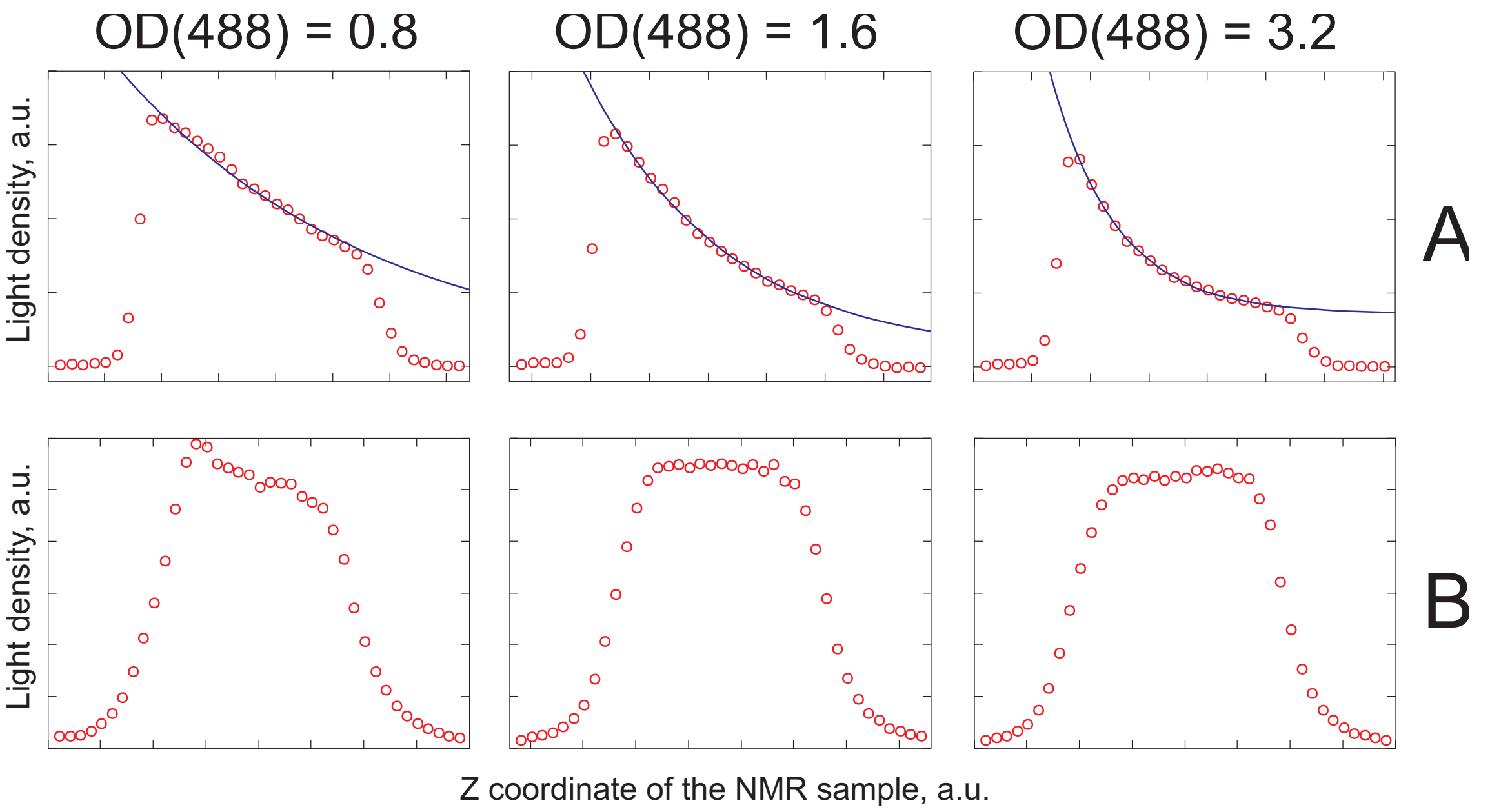

**Figure 4.4.** The light intensity profiles along the *z*-axis of the sample measured using $^{19}F$ photo-CIDNP of 4-fluorophenol (/FMN/ = 0.2, 0.4 and 0.8 mM) for **(A)** illumination from above using the arrangement shown in Figure 4.1D and **(B)** for a tapered fibre (Figure 4.1F).

As expected, the overall $^{19}F$ CIDNP signal intensity is attenuated at high optical densities (Figure 4.5). This is the result of both light absorbance and attenuation of the CIDNP intensity at high photosensitizer concentrations (e.g. due to triplet quenching by ground state FMN). For the most strongly coloured solutions, irradiation from above with a square cut fibre results in very little light reaching the sensitive region of the sample inside the receiver coil (Figure 4.5A), whereas the light escaping from the tapered tip always illuminates the active region (Figure 4.5B).

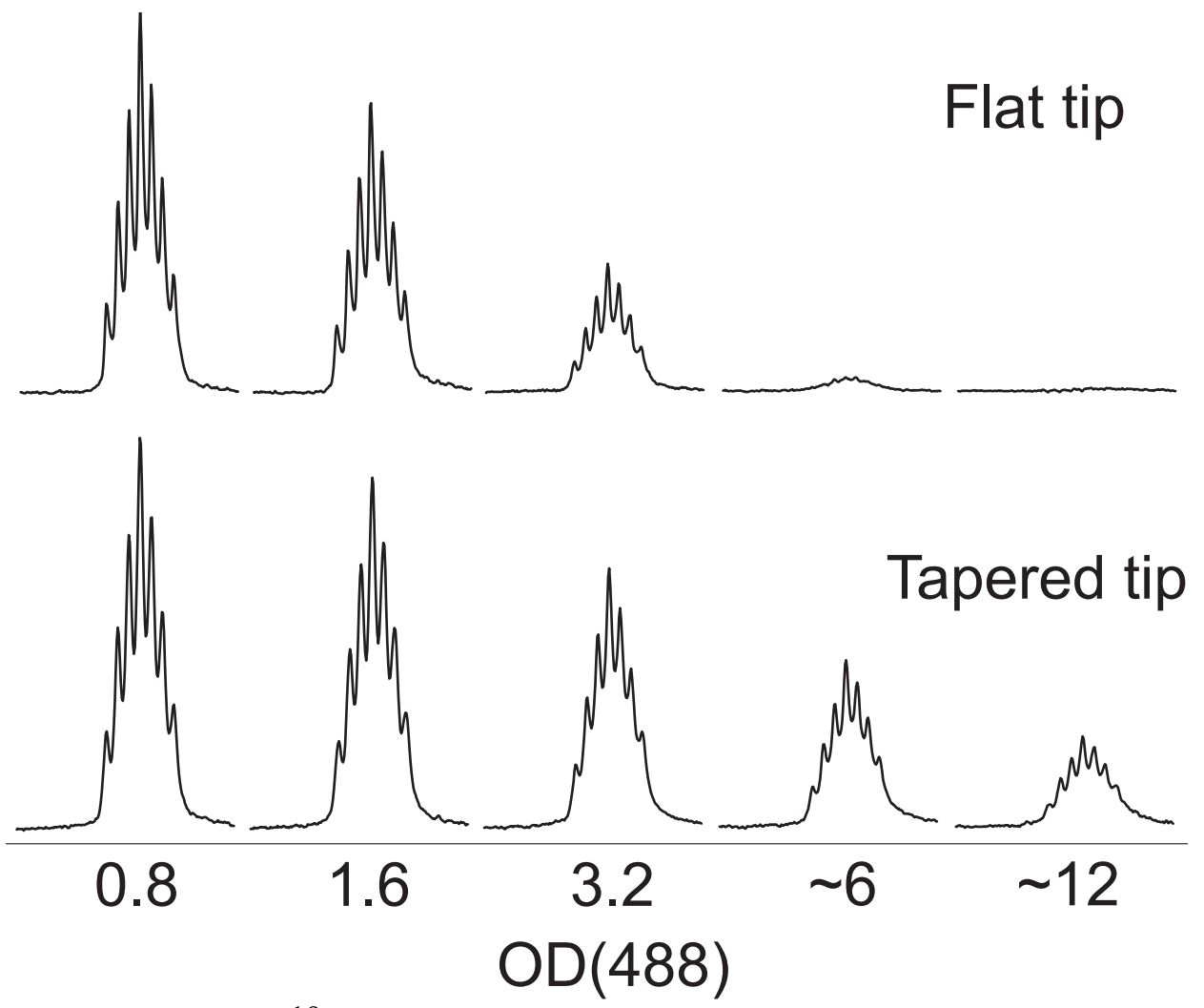

**Figure 4.5.** $^{19}F$ photo-CIDNP signal amplitude as a function of photosensitizer concentration ([FMN] = 0.2-3.2 mM) for **(A)** illumination from above using the arrangement shown in Figure 4.1D and **(B)** for a tapered fibre (Figure 4.1F).

Somewhat surprisingly, the presence of the fibre cone in the centre of the NMR sample does not lead to a significant deterioration in spectral resolution (Figure 4.6). After appropriate shimming, the width at half-height of the HDO in $D_2O$ increases from 0.8 to 1.2 Hz and there is some deviation of the lineshape from Lorentzian form. However, as this broadening should be the same for all lines in the spectrum, mild reference deconvolution [134] would, if necessary, restore the normal lineshape and width. For the linewidths commonly encountered in macromolecular NMR, the presence of the fibre should have a negligible effect on the resolution. The simplex-based shimming algorithms supplied with Varian NMR software usually yield a sufficiently homogeneous field in less than 10 minutes.

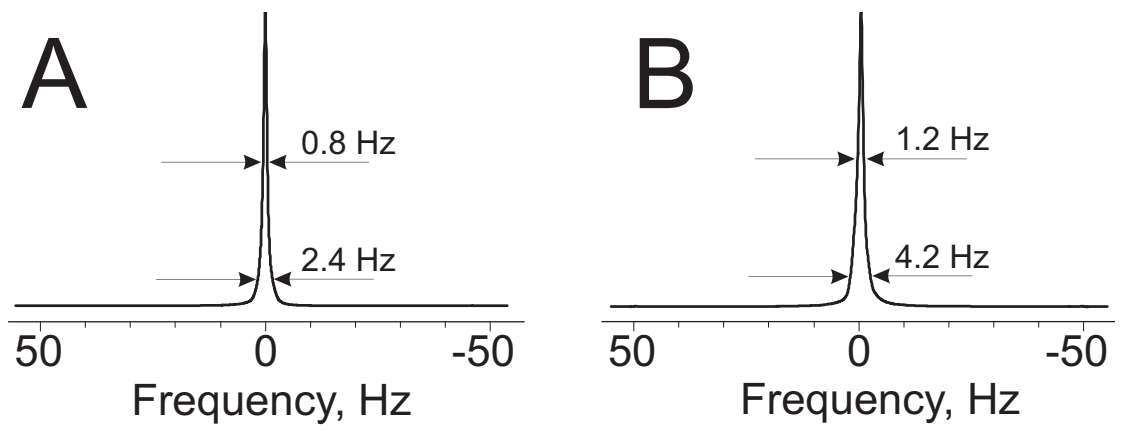

**Figure 4.6.** $^1$H NMR lineshapes of the residual HDO signal in a sample of $D_2O$, **(A)** without and **(B)** with the fibre present inside the NMR tube. The field was reshimmed after the insertion of the fibre.

The volume of the sample displaced by the cone is remarkably small. In Figure 4.2C, the fibre appears thicker than it actually is because the surrounding liquid acts as a magnifying lens. Once the sheath has been removed, the untreated core of the fibre has a diameter of about 0.9 mm: simple arithmetic shows that the volume of the cone is less than 5% of the total sample volume. This may be one of the reasons why shimming is relatively straightforward. Another is that the discontinuities in magnetic susceptibility produced by introducing the fibre are almost exclusively perpendicular to the axis of the tube [135].

An important feature of the cone is its large light emission area: around 25 mm$^2$ as opposed to 0.8 mm$^2$ in the case of the square cut tip. As a consequence, the sample heating arising from high-power illumination is much more homogeneous. For the square cut fibre, the water signal has a long tail (Figure 4.7A) indicating the presence of localised hot regions ($\Delta T = 6$ K for a 40 Hz shift). The thin layer between the end of the fibre and the top of the coil (see Figure 4.1D) is probably even hotter. Supplying the same amount of energy to the sample via the tapered fibre tip results in a much more uniform heating as evidenced by a more symmetrical lineshape, narrower at the base (Figure 4.7B).

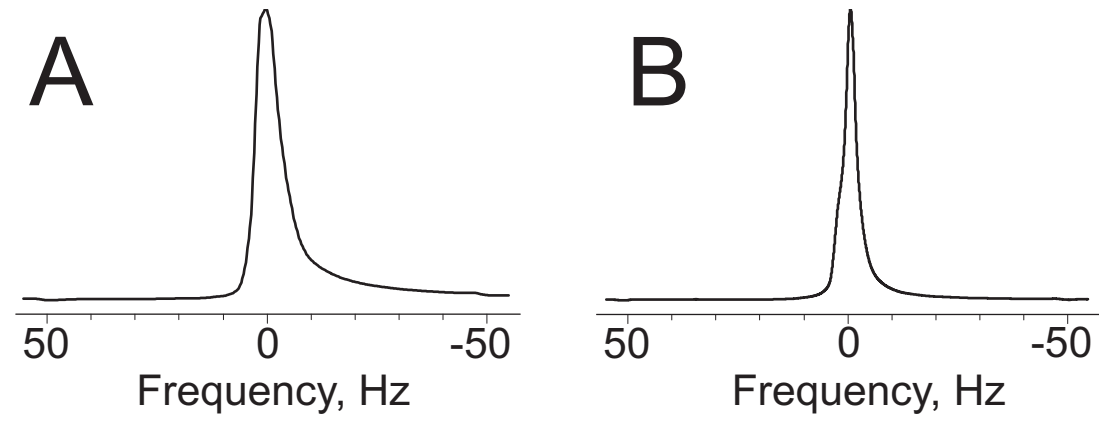

**Figure 4.7.** The lineshape of the HDO signal in a sample of $D_2O$ containing 0.2 mM FMN after one second of high-power (10 W) illumination for **(A)** illumination from above using the arrangement shown in Figure 1D and **(B)** for a tapered fibre (Figure 4.1F).

It should be noted that the steps on the fibre cone are essential. Our experience with a smooth cone (produced by slow continuous extrusion of the fibre into the etching medium) was that most light comes out of the very tip, (diameter ~50 μm). When such a fibre was immersed into an absorptive fluid, the solution at the tip boiled and the end of the fibre disintegrated. It thus seems that the additional light reflections provided by the shoulders in a stepped tip are essential for efficient distribution of light to the sample.

For low concentrations of the photosensitizer, additional uniformity of the light intensity distribution is likely to be obtained by using an NMR tube made of white-coloured glass, which would scatter the escaping light back into the active region. Whitening commercial NMR tubes turned out to be a challenging task, as the process should not introduce any proton-containing and/or paramagnetic compounds and should not significantly alter the tube diameter. Attempts to use either liquid hydrofluoric acid or HF gas have so far been unsuccessful. Sandblasting the tube was also unsuccessful. The option of using an NMR tube made of white material therefore remains to be explored.

## 4.4 Prolonged sample illumination: ways and consequences

Although the photochemical reaction that gives rise to the photo-CIDNP effect in the fluorophenol/FMN system is cyclic, it is not perfectly cyclic. After a prolonged illumination period the photosensitizer does get photochemically reduced or destroyed, a phenomenon referred to as *bleaching* [136]. There are several techniques that aim to minimize the extent of photosensitizer degradation and to prolong the sample lifetime. One may attempt to oxidize the sensitizer back into the reactive form by using an added reagent [48], in some cases just by letting the sample consume atmospheric oxygen [137]. Alternatively, one may attempt to thoroughly purify the reagents to remove the undesired side reactions, or to increase the photosensitizer concentration to minimize depletion effects.

Remarkably, it has recently been demonstrated that lumiflavin, which is a compound with the same chromophore as FMN, shows surprisingly little photo-degradation under continuous laser illumination. The photo-degradation rate of lumiflavin was found by Holzer et al. [136] to be at least an order of magnitude smaller than that of FMN. Lumiflavin, however, is far less soluble in water than FMN, and its utility is therefore limited. Also, in the experiments reported by Holzer et al. the solutions did not contain any electron donor, and therefore the results may be inapplicable to a photo-CIDNP system.

Whatever means of prolonging the sample lifetime is used, one is always faced with the question of illumination uniformity and the extent of photosensitizer degradation in different parts of the sample, as well as the overall rate of that degradation. All these parameters may be conveniently monitored using CIDNP-based imaging. It is instructive to consider the time dependence of the CIDNP image of a sample illuminated from above. Figure 4.8 shows the CIDNP polarization distribution in the Z-direction of the sample as a function of argon ion laser illumination time up to a total of 80 seconds. The photo-CIDNP images show the expected exponential intensity distribution originating from Beer-Lambert's law. However, at a low (0.2 mM) FMN concentration a peculiar phenomenon emerges after several hundred flashes. Due to the higher light intensity in the upper parts of the sample, the photosensitizer degradation there is faster. After eight hundred scans it leads to a non-uniform photosensitizer concentration distribution that compensates the irradiation non-uniformity and the CIDNP image becomes rectangular. At higher concentrations the required extent of sensitizer depletion becomes progressively more difficult to attain and we see only a weak equilibration of this kind for the 0.8 mM FMN sample.

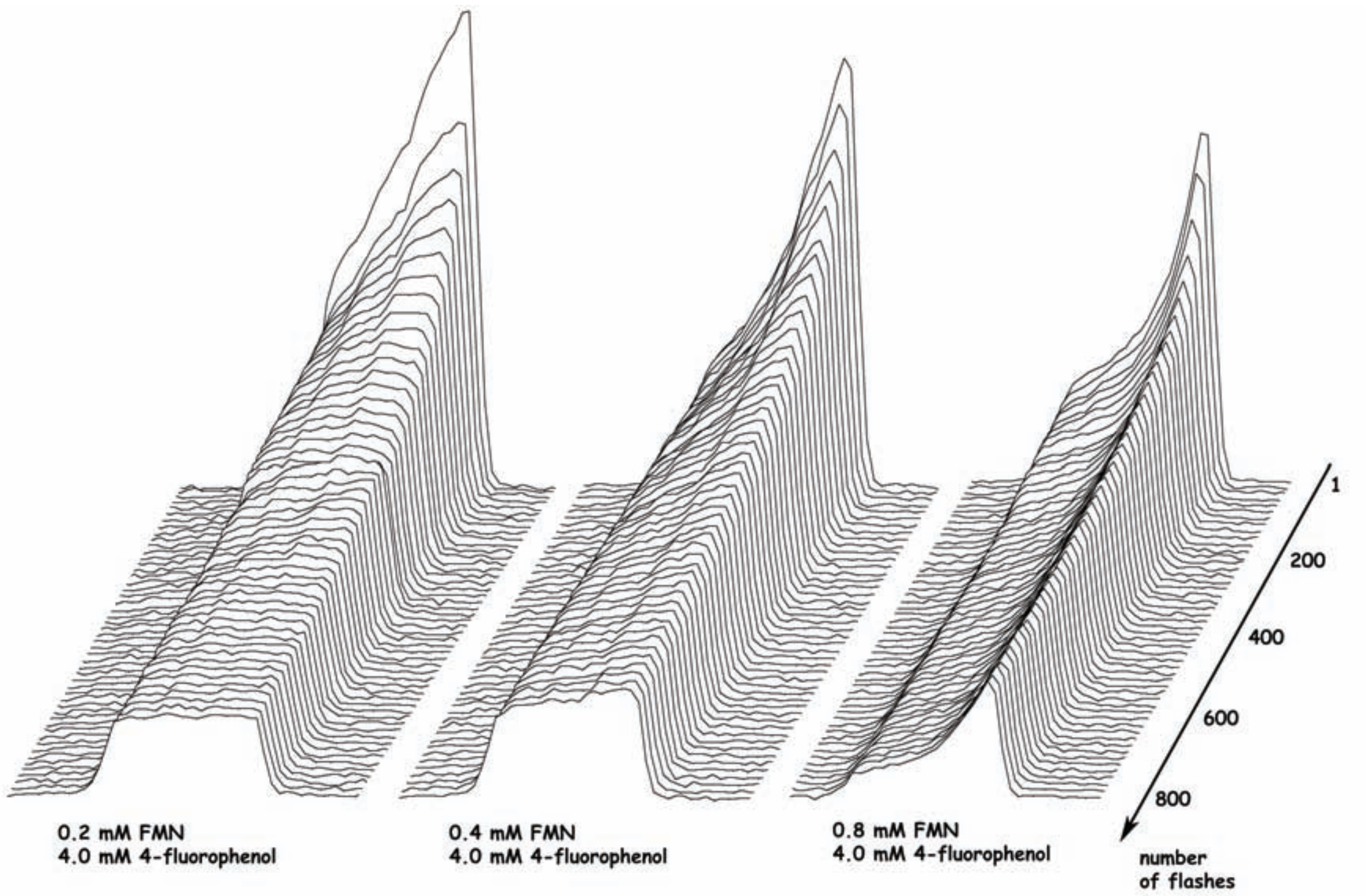

**Figure 4.8.** $^{19}$F Z-axis images of the 4-fluorophenol sample irradiated for 100 ms prior to the 90-degree pulse (square cut fibre tip). The way the image is arranged the light is brought from the right. The two visible ridges on each of the image sequences are due to the Z-gradient coil arrangement. The interval between scans was 20 seconds to allow the sample to cool down before the next flash.

Illumination from above, therefore, generates reagent concentration gradients after prolonged illumination periods. The consequences of this may vary from none whatsoever for first-order photo-induced kinetics measurements to severe data distortion in the case where second-order kinetics is monitored by recording the overall nuclear magnetization dynamics. Photo-induced radical reactions are almost never simple, it is therefore imperative that concentration gradients be avoided. This puts a restriction on the maximum optical density of the sample in a photo-CIDNP kinetics measurement, and demands that the initial radical concentration be considered a distributed, rather than constant, quantity in the analysis of CIDNP kinetics.

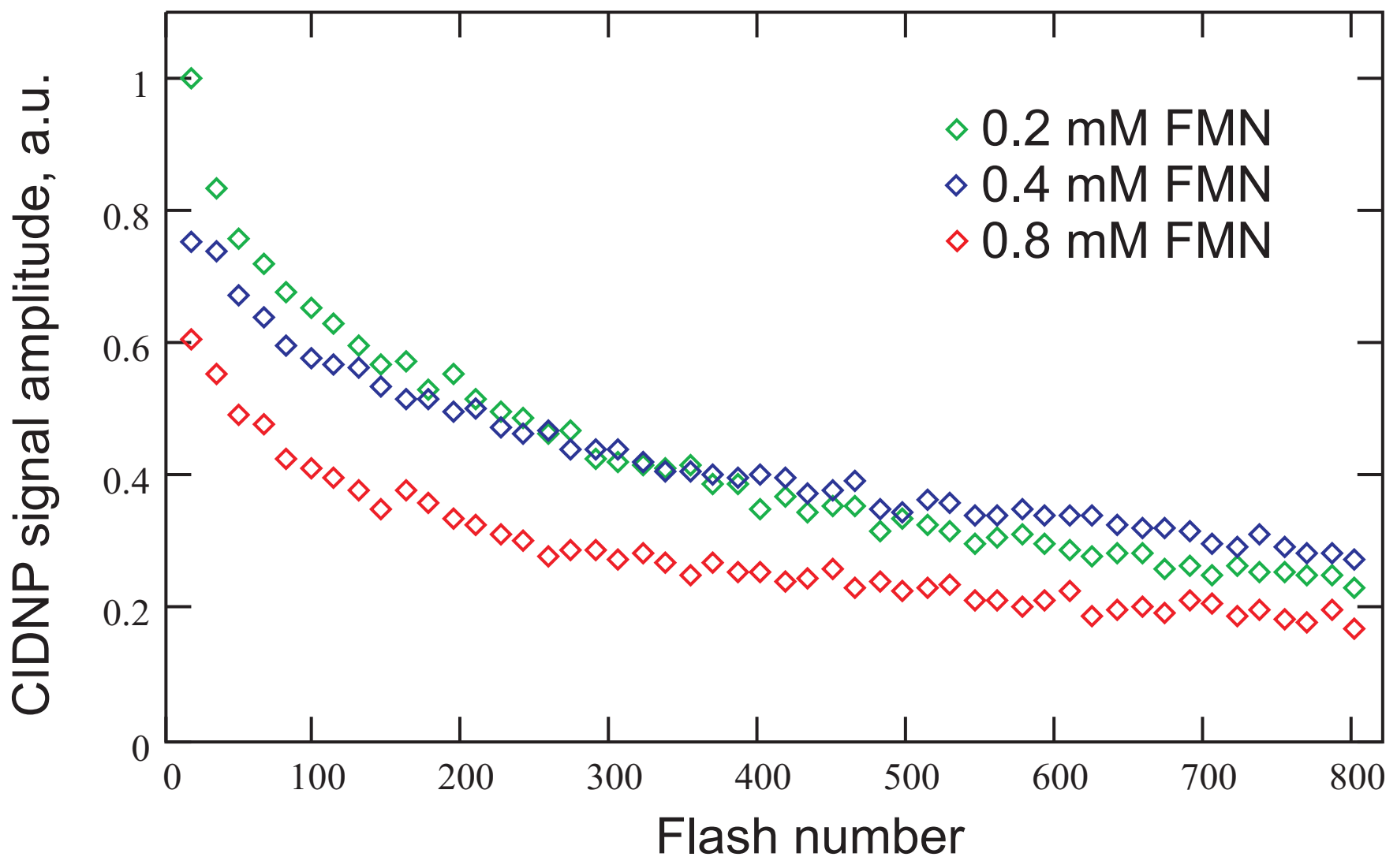

**Figure 4.9.** $^{19}F$ photo-CIDNP magnitude dependence on the flash number for the 4-fluorophenol sample (4 mM in $D_2O$, degassed) with three different photosensitizer concentrations in case of irradiation from above. The laser flashes are 100 ms duration, 5 W output power at 488 nm. The interval between scans was 20 seconds to allow the sample to cool down before the next flash.

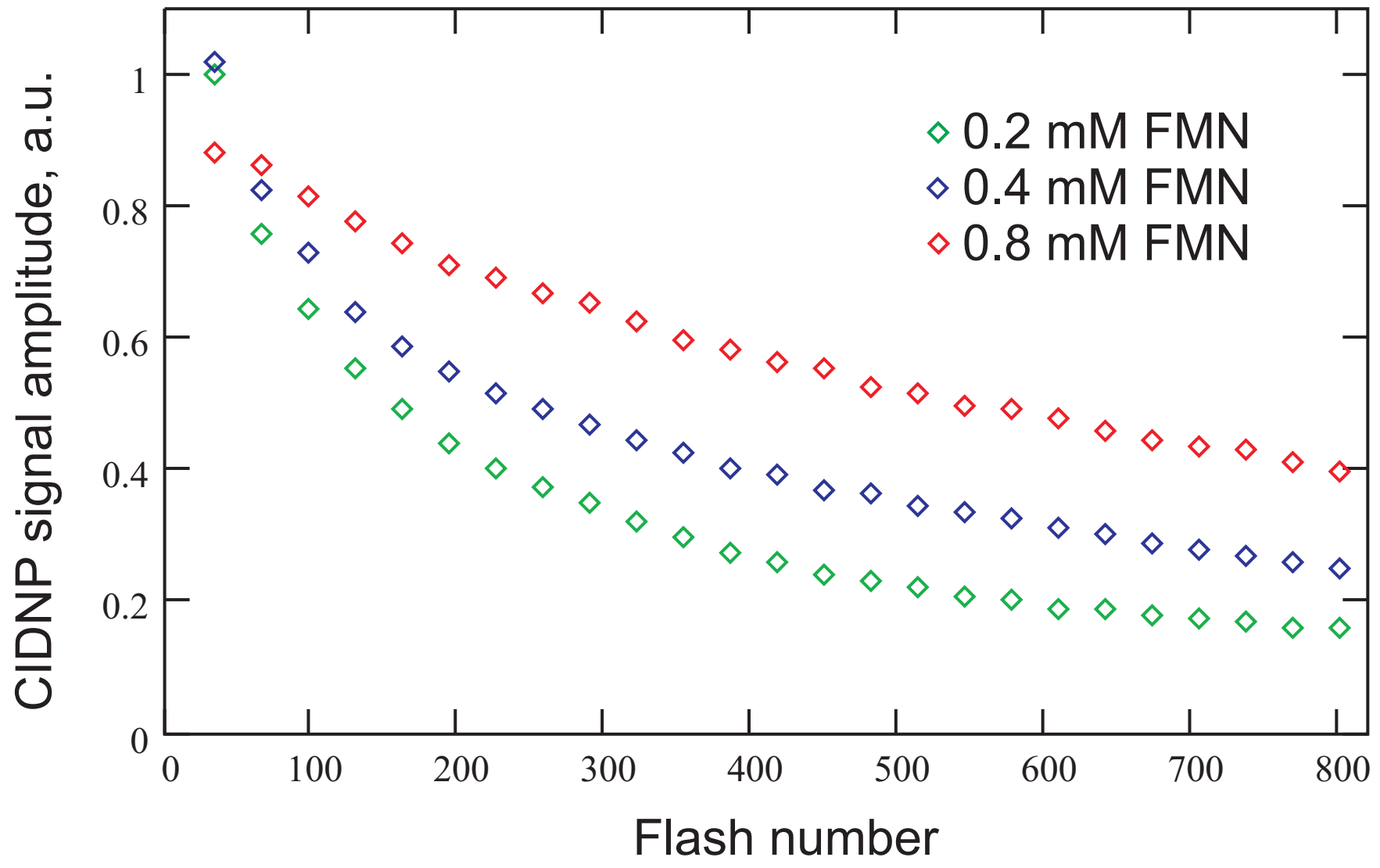

**Figure 4.10.** Photo-CIDNP magnitude dependence on the flash number for the 4-fluorophenol sample (4 mM in $D_2O$, degassed) with three different photosensitizer concentrations in case of sample irradiation using a tapered fibre tip. The laser flashes are 100 ms duration, 5 W output power at 488 nm. The interval between scans was 20 seconds to allow the sample to cool down before the next flash.

The time dependence of the integrated $^{19}F$ photo-CIDNP amplitude in a multi-flash photo-CIDNP experiment is shown on Figures 4.9 and 4.10. The intensity of the photo-CIDNP effect follows the expected pattern – a gradual decay of the signal as a result of photo-degradation of the sensitizer. After a total of 80 seconds of irradiation less than 50% of the initial amplitude of the photo-CIDNP effect remains. Remarkably, the most densely coloured 0.8 mM FMN sample has the lowest CIDNP amplitude when irradiated from above and by far the highest amplitude when irradiated with a tapered fibre. In the latter case, after 800 flashes the reduction in the CIDNP amplitude is only about 50%. Assuming a signal-to-noise ratio of 10-30 in a typical one-scan photo-CIDNP spectrum, employing a tapered fibre tip and 0.8 mM concentration of FMN photosensitizer should allow one to record a 2D $^{1}$H-$^{1}$H or $^{1}$H-$^{19}$F CIDNP-COSY or CIDNP-NOESY spectrum without resorting to sample re-injection techniques [48]. All 2D spectra will, however, have some additional line broadening caused by the attenuation of the CIDNP effect along the indirect dimension. This broadening may be reduced by performing an experiment with interleaved acquisition, when the indirect dimension increments precede phase cycling and signal accumulation increments.

### 4.5 Square-cut *vs.* tapered illuminating tip: performance benchmarks

Figures 4.9 and 4.10 have shown that, as well as providing a more uniform sample illumination, the tapered fibre tip increases the observed amplitude of the photo-CIDNP effect at high photosensitizer concentrations. The signal amplitude benchmark experiments described below show that for a wide variety of both photosensitizer and target compound concentrations, the detected CIDNP magnitude with the tapered tip is higher or equal to that achieved by illumination with a square-cut tip. The photo-CIDNP experiments described below are of two types:

**Initial rate regime** experiments, which employ light flashes much shorter than the longitudinal relaxation time of the nuclei in question. The detected photo-CIDNP amplitude in such experiments is directly proportional to the photo-CIDNP magnetization pumping rate and is to a first approximation independent of the longitudinal relaxation time.

**Asymptotic regime** experiments which employ light flashes much longer than the longitudinal relaxation time of the monitored nucleus and represent the equilibrium nuclear polarisation resulting from a competition between magnetization pumping and relaxation processes.

A 2D grid of photo-CIDNP experiments spanning the active compound concentration range of 1.0-6.0 mM and a photosensitizer concentration range of 0.2-1.4 mM (OD = 2-10 cm$^{-1}$) in the initial rate regime is shown on Figure 4.11. An 50-100% advantage of the tapered fibre tip is apparent for the entire concentration grid. The advantage becomes less apparent in the asymptotic regime (Figure 4.12), probably because after 2 seconds of intense irradiation the layer of solution immediately under the square cut fibre tip bleaches out and becomes transparent so that more light reaches the active region.

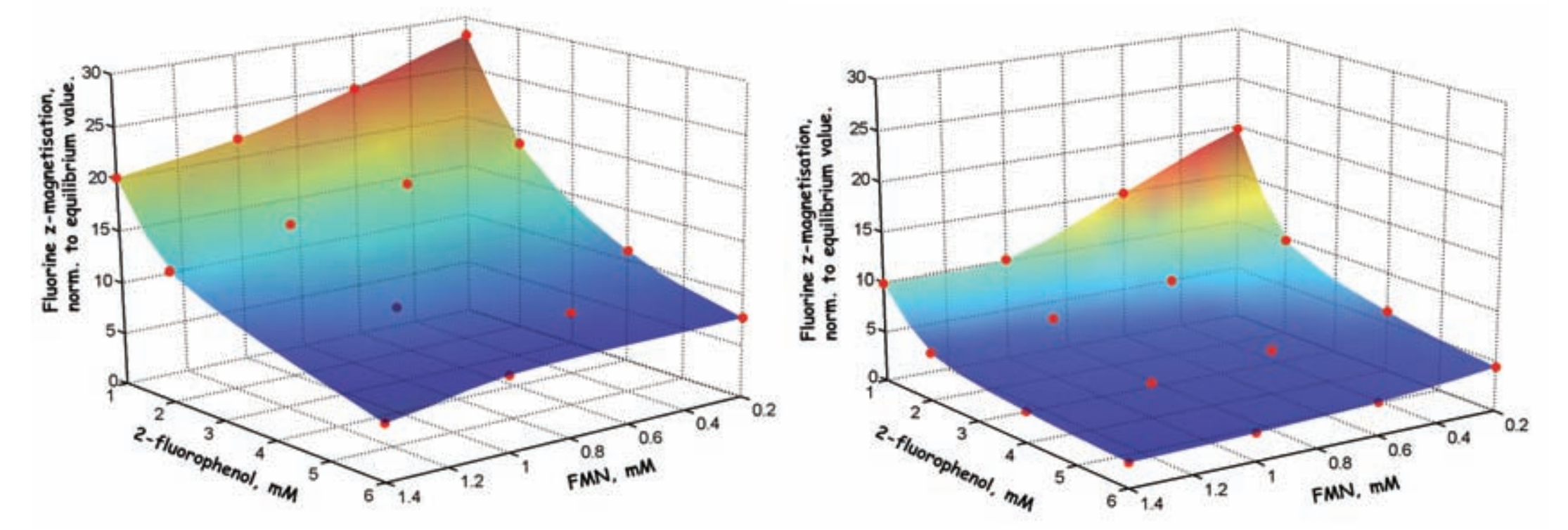

**Tapered tip** **Square tip**

**Figure 4.11.** Dependence of the $^{19}F$ photo-CIDNP enhancement factor on the concentrations of FMN photosensitizer and electron donor (2-fluorophenol) in the initial rate CIDNP accumulation regime.

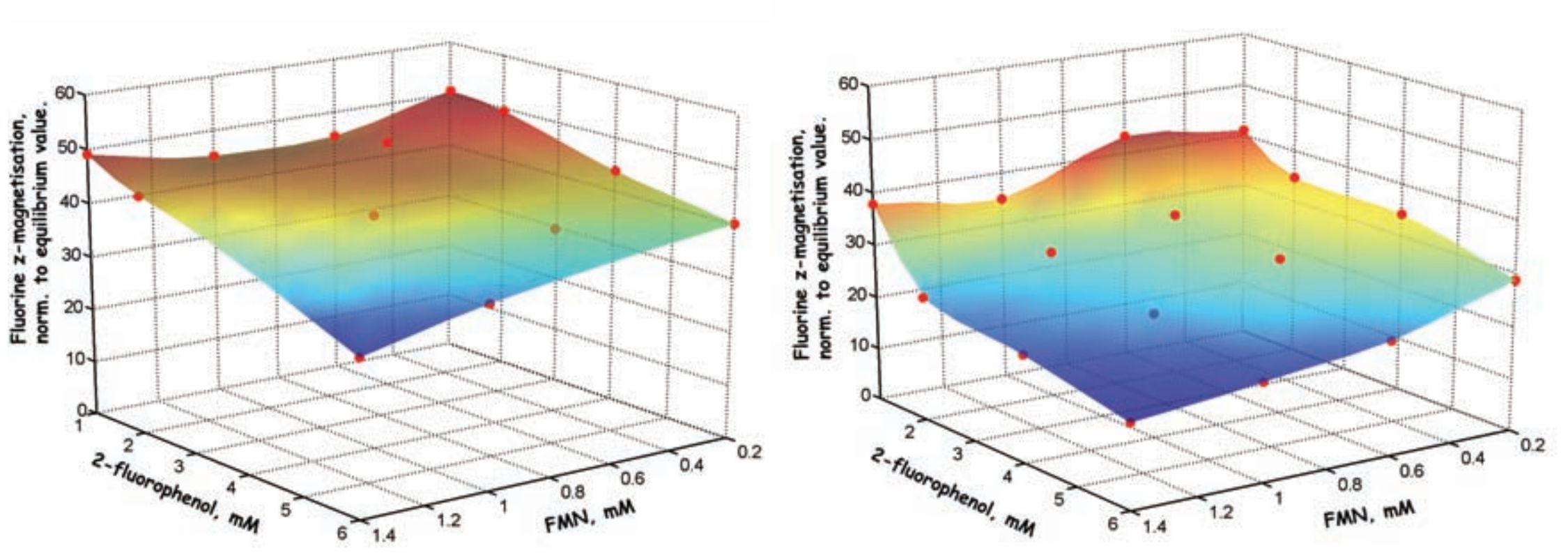

**Tapered tip** **Square tip**

**Figure 4.12.** Dependence of the $^{19}F$ photo-CIDNP enhancement factor on the concentrations of FMN photosensitizer and electron donor (2-fluorophenol) in the asymptotic CIDNP accumulation regime.

Another prominent feature of initial regime plots on Figure 4.11 is the fact that the relative photo-CIDNP enhancement has an inverse dependence on the concentration of the active compound. This agrees very well with the kinetic model of the CW-pumped photo-CIDNP effect, which suggests that even at much smaller concentrations of the active compound all the FMN triplets get intercepted. The detected CIDNP polarization in a CW irradiation experiment thus originates from a small sub-ensemble of a large pool of molecules and does not depend strongly on the size of this pool (see Figure 4.13 for the absolute detected magnitude data). Altering the concentration of the active compound thus has a weak effect on the magnitude of photo-CIDNP magnetization, but increases the normalization divisor. A $1/C$ or similar behaviour is therefore expected. A weak dependence of the absolute detected signal magnitude on the concentration of the active compound is a very good news, meaning that high concentrations of the investigated compounds are unnecessary. This is important in the case of sparingly soluble, expensive or aggregating compounds.

In the asymptotic regime (Figure 4.14) the enhancement becomes approximately proportional to the active compound concentration, and the normalized surface becomes flatter. The absolute detected intensity plot becomes proportionally ascending. This is again consistent with the kinetic model, because under a steady supply of the active photosensitizer molecules the number of polarized molecules in the steady state linearly depends on the concentration.

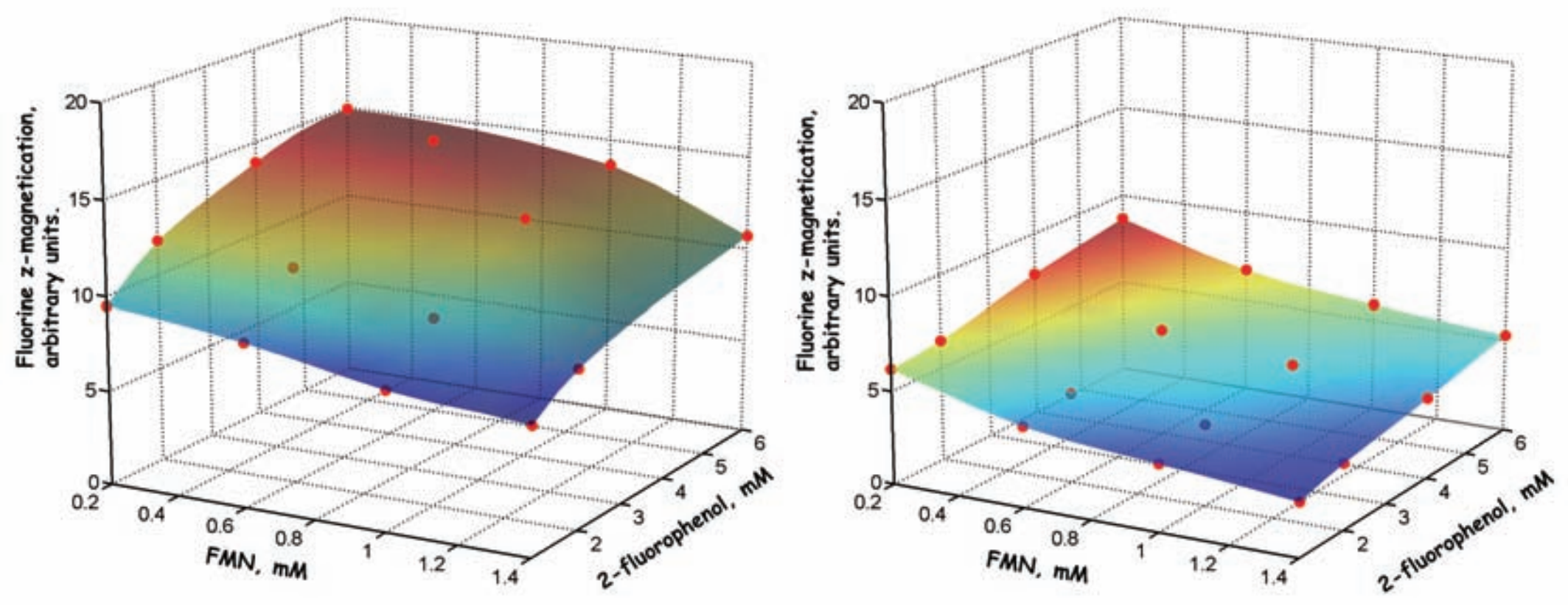

**Tapered tip** **Square tip**

**Figure 4.13.** The dependence of the absolute detected $^{19}$F NMR signal amplitude on the concentration of FMN photosensitizer and the electron donor (2-fluorophenol) in the initial rate CIDNP accumulation regime.

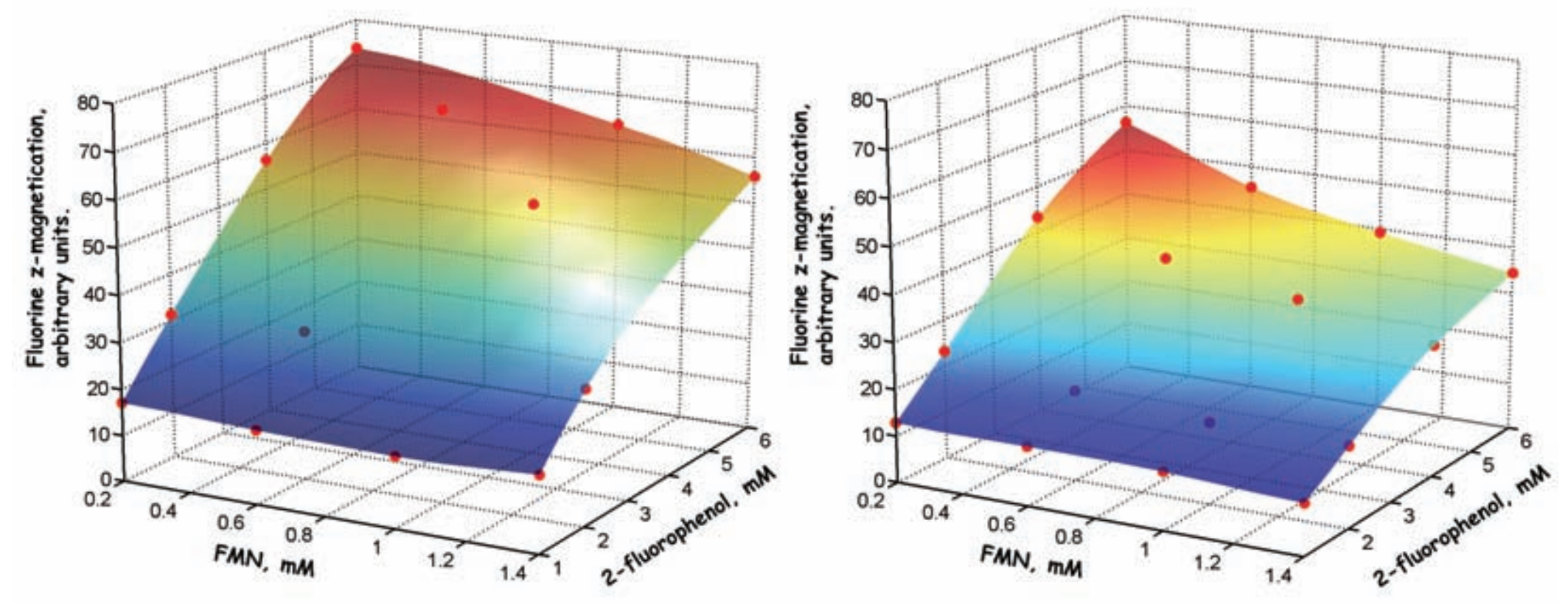

**Tapered tip** **Square tip**

**Figure 4.14.** The dependence of the absolute detected $^{19}$F NMR signal amplitude on the concentration of FMN photosensitizer and the electron donor (2-fluorophenol) in the asymptotic CIDNP accumulation regime.

## 4.6 Concluding remarks

The simplicity and efficiency of the tapered fibre illumination system is quite remarkable. In all cases the tapered fibre irradiation scheme is either 50-100% better (short flashes) or approximately as good (long flashes) as irradiation from above. Given the possibility of at least a seven-fold increase in the photosensitizer concentration compared to the traditional illumination from above and at least an order of magnitude weaker temperature gradients due to greater surface area, the tapered fibre illumination scheme may in some cases (2D CIDNP experiments, very densely coloured samples) be highly advantageous. An example of application of the tapered fibre illumination will be given in Chapter 7, which describes photo-CIDNP experiments on a strongly coloured green fluorescent protein. $^{19}$F CIDNP experiments on this protein turned out to be impossible to perform with the traditional illumination setup because the concentrations required to get any light into the active region were too low to permit signal detection. Utilizing a tapered fibre allowed an order of magnitude increase in concentration, making the photo-CIDNP signal easily detectable.

# Chapter 5

# *Fluorotyrosine-labelled Trp-cage protein*

### 5.1 Introduction

This Chapter describes the chemical synthesis of 3-fluoro-N-fluorenylmethoxycarbonyl-O-tert-butyl-L-tyrosine, the precursor for FMOC-protected solid phase synthesis of fluorotyrosinated peptides, and describes NMR and photo-CIDNP experiments on the fluorotyrosine-labelled Trp-cage peptide, which is the smallest fully folded amino acid sequence known to date [138].

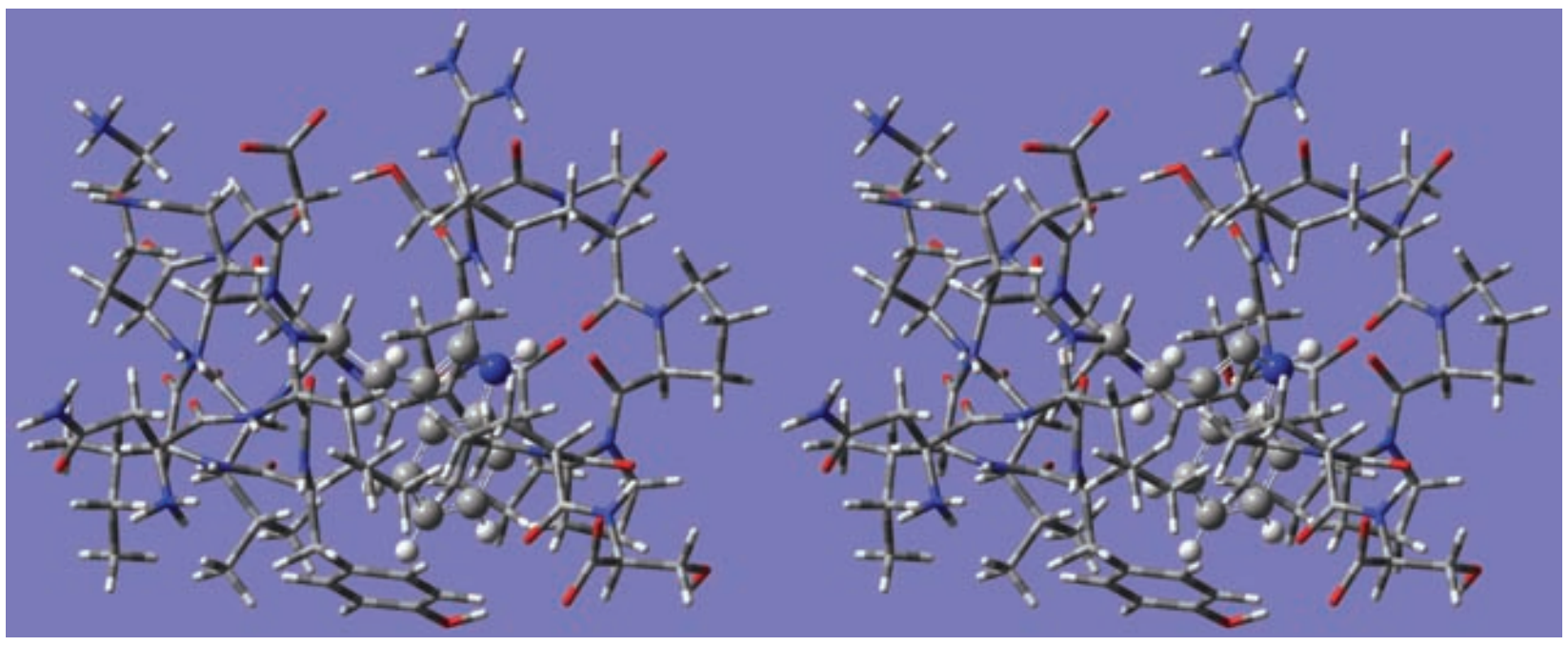

**Figure 5.1.** Stereo plot of folded Trp-cage peptide.

Also referred to as tc5b, Trp-cage is a 20-residue proline-rich amino acid sequence, derived from the C-terminal fragment of the 39-residue exendin-4 peptide and stabilized by further mutations [138]. The Trp-cage folds extremely fast (4.1 μs, [139]) and has a well-defined hydrophobic core consisting of tryptophan encaged by two prolines (Figure 5.1). From NMR and CD data it appears to be 85% folded in neutral aqueous solution at 20°C and shows a cooperative melting transition at 315 K [138]. Owing to its small size, Trp-cage is an ideal testing ground and calibration system for molecular dynamics and *ab initio* methods, with a number of works already published in this area [140-142]. Among the notable results is the fact that an AMBER force field folds the entire structure correctly starting from the fully elongated sequence, even in implicit water. However, the melting point predicted by the simulation appears to be unphysically high [143]. The Trp-cage has also recently been used to investigate the low solvent viscosity limit of the protein folding speed [139]. Small size and the presence of a single solvent-accessible tyrosine make Trp-cage a convenient testing ground for the $^{19}$F photo-CIDNP method outlined in the previous chapters. Specific questions that this small testing system is likely to answer are:

*Given the large hyperfine anisotropy of the fluorine nucleus in the intermediate radicals, is there any dependence of the observed $^{19}$F photo-CIDNP effect on the rotational correlation time, perhaps through relaxation contribution to either geminate or escaped radical dynamics?*

*Would the buried tryptophan residue exhibit any photo-CIDNP effect, either directly or through multi-staged electron transfer via the nearby tyrosine?*

*Despite the unfavourable rotational correlation time in the folded state of this peptide, would it be possible to detect chemically pumped nuclear Overhauser effect from fluorine to the surrounding protons?*

Tyrosine was chosen as a site for the fluorine label placement for a number of reasons. Tyrosine has a much greater photostability compared to tryptophan, as well as much simpler spin system (see Chapters 2 and 3 for details). The fluorotyrosine/FMN photo-CIDNP system has received a detailed theoretical and experimental treatment [52]. It is also much cheaper than the fluorinated tryptophans, especially in the racemic form. In the specific case of Trp-cage, fluorinating the tyrosine is also much less likely to destabilize the overall peptide fold, because the fluorine label is introduced into the peripheral tyrosine residue, rather than into the tryptophan core (Figure 5.1).

Fluorinated amino acids are toxic to bacteria [144], the simplest way to prepare a 20-residue fluorotyrosinated peptide is therefore via solid phase synthesis, meaning that an appropriately protected chemical precursor is needed for fluorotyrosine incorporation. In the most popular case of fluorenylmethoxycarbonyl (FMOC) protected peptide synthesis this precursor is 3-fluoro-N-FMOC-O-tBu-L-tyrosine, which is not commercially available and therefore had to be synthesized.

## 5.2 Solid phase peptide synthesis

Before we proceed to the description of the protected precursor synthesis, the so-called *FMOC chemistry*, i.e. the fluorenylmethoxycarbonyl amino protecting group based solid phase peptide synthesis will be briefly outlined here. The general scheme of solid phase peptide synthesis is given on Figure 5.2. The functionalised resin denoted with a blue ball is usually either polystyrene cross-linked with 1% divinylbenzene or polyacrylamide-based copolymer. The resins may be functionalised with a wide variety of linker groups (usually TFA-cleavable, such as Wang linker [145]), serving as initial attachment points for the growing amino acid chain. All chemical steps are performed by passing a solution containing the necessary chemical through this resin.

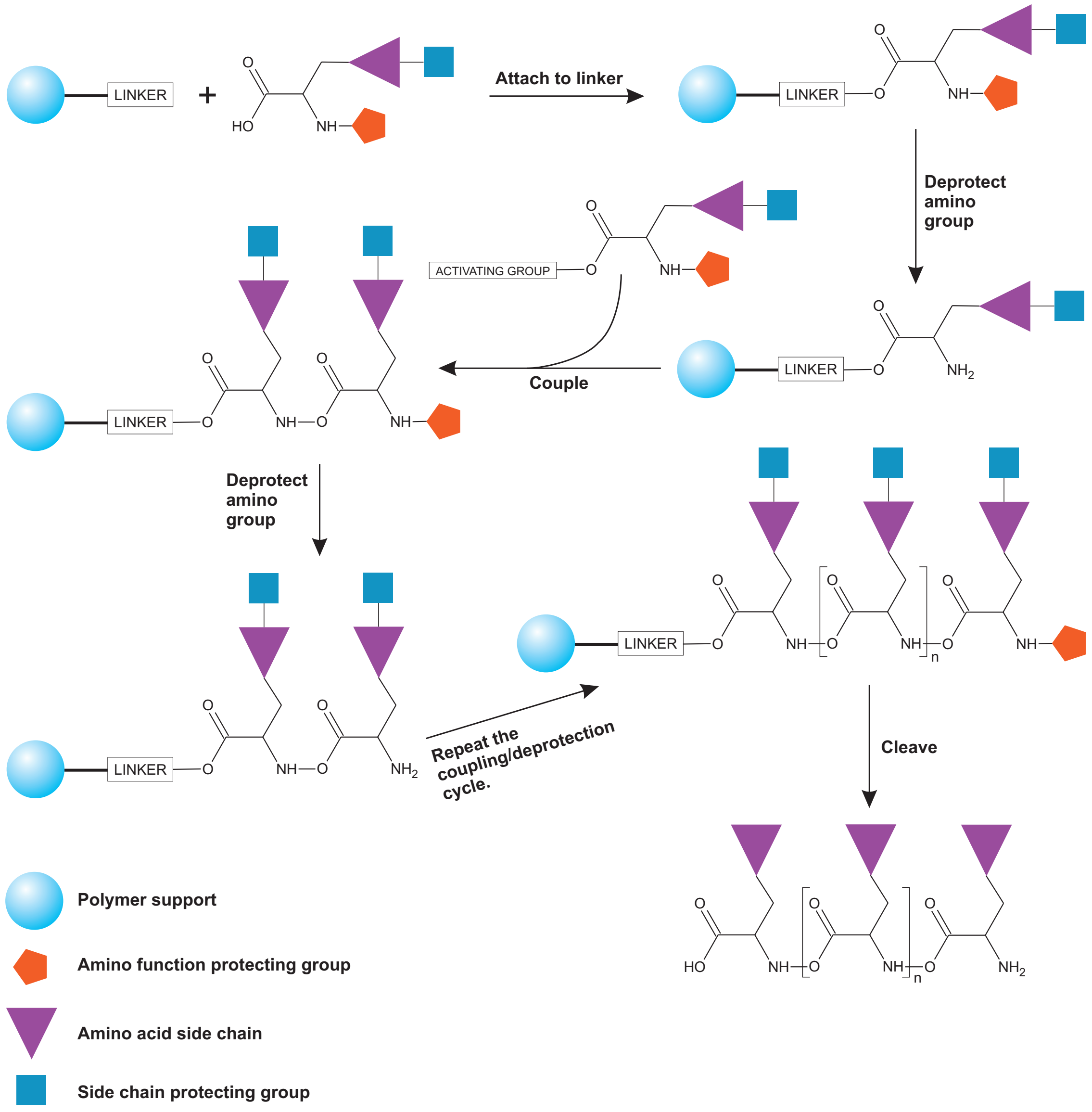

**Figure 5.2.** A general scheme of the solid phase peptide synthesis.

After the first amino acid is attached to the support, the following elongation step is repeated until the necessary peptide chain has been prepared: the amino group is deprotected, then coupled with the pre-activated carbonyl group of the next amino acid. After the last amino group has been deprotected, the peptide is cleaved from the resin and all the side chain protecting groups are removed.

A comprehensive list of resins, linkers and side chain protective groups is given, for example, in Novabiochem's catalogue [146], together with a review of most popular solid phase chemistries.

FMOC-based protection is an example of the carbamate protection of the amino group. Carbamates may be prepared by either treating the amine with an alkyl chloroformate under basic aqueous conditions, or by an alkyl alcohol activated by phosgene. By delocalizing the lone electron pair of nitrogen, the carbamate group deactivates nucleophilicity of the amino group, also becoming itself quite stable towards the nucleophiles. The O-alkyl group of the carbamate may be chosen to allow mild and specific deprotection by means that do not involve nucleophilic attack on the carbonyl moiety [147].

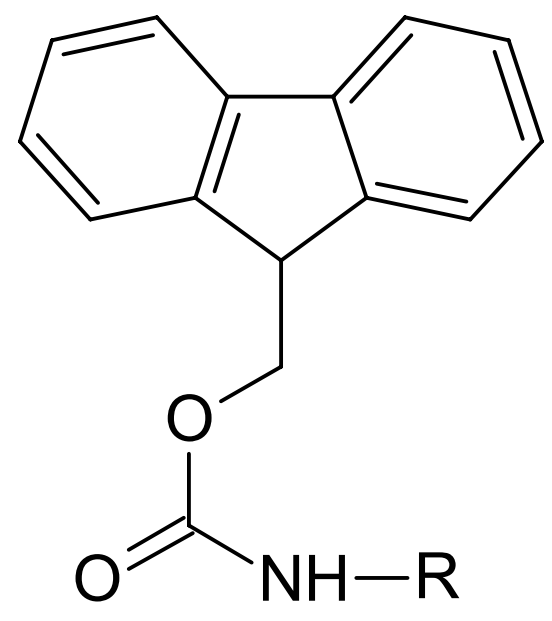

**Figure 5.3.** Fluorenylmethoxycarbonyl amino protecting group.

FMOC protection was developed by Carpino et al. in early 1970s [148-151]. These early works introduced the FMOC protecting group using fluorenylmethylchloroformate (FMOC-Cl), a toxic, corrosive and lachrymatory substance. A substantial improvement came with the introduction of FMOC N-succinimide (FMOC-NSu), a much less toxic and more stable compound [152]. FMOC-NSu is also superior to FMOC-Cl, in that it does not produce dipeptide byproducts for amino acids with less hindered side chains (Gly, Ala), and also causes less amino acid racemisation [153].

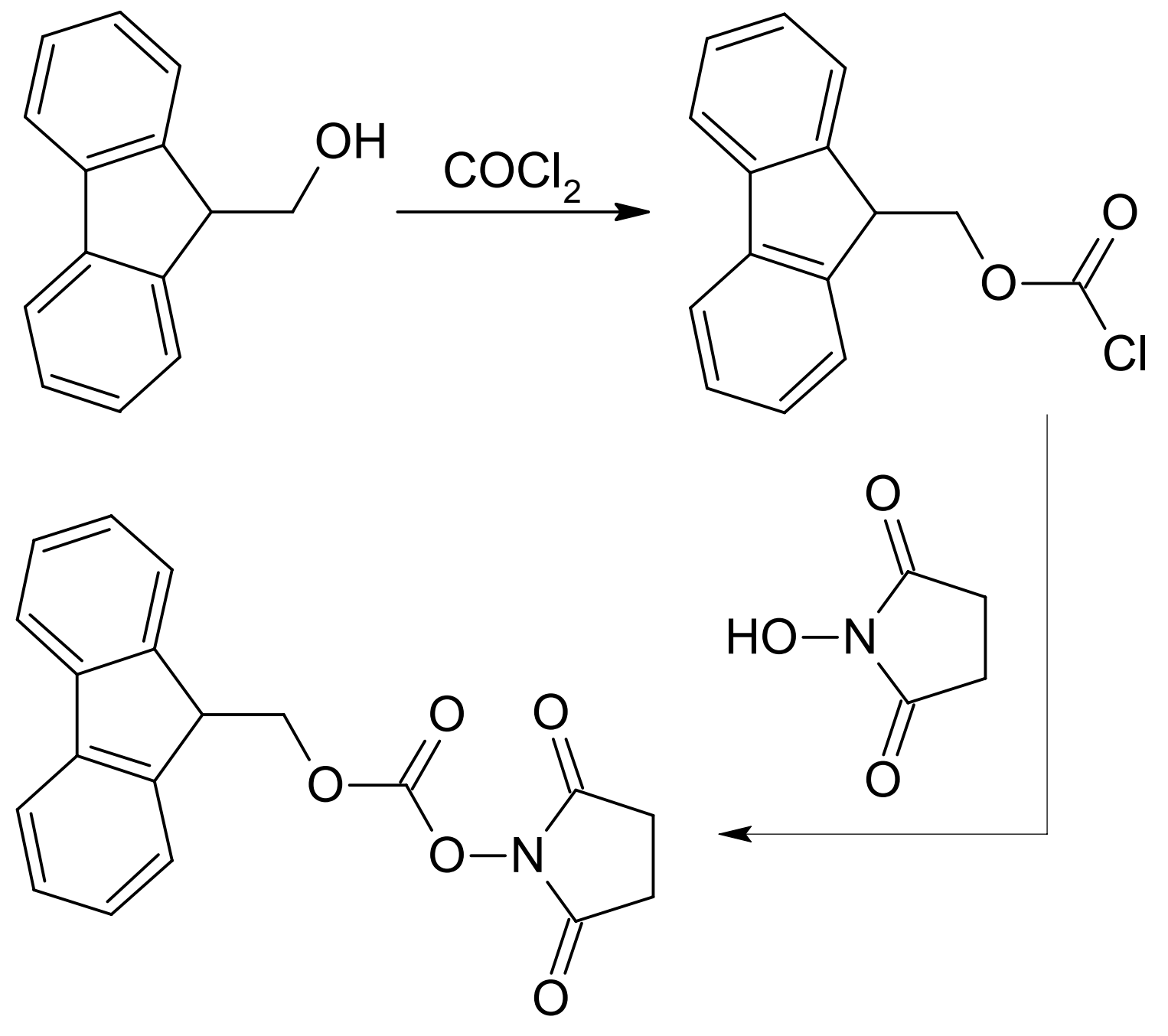

**Figure 5.4.** Preparation of FMOC-ONSu.

The original suggestion for the FMOC protecting group removal procedure was β-elimination with liquid ammonia at –33°C [148-151]. Some years later, however, it was found that a sufficiently selective cleavage may be performed at room temperature with aminoaliphatic compounds such as piperidine, morpholine and ethanolamine [149, 150], making FMOC group removal very convenient.

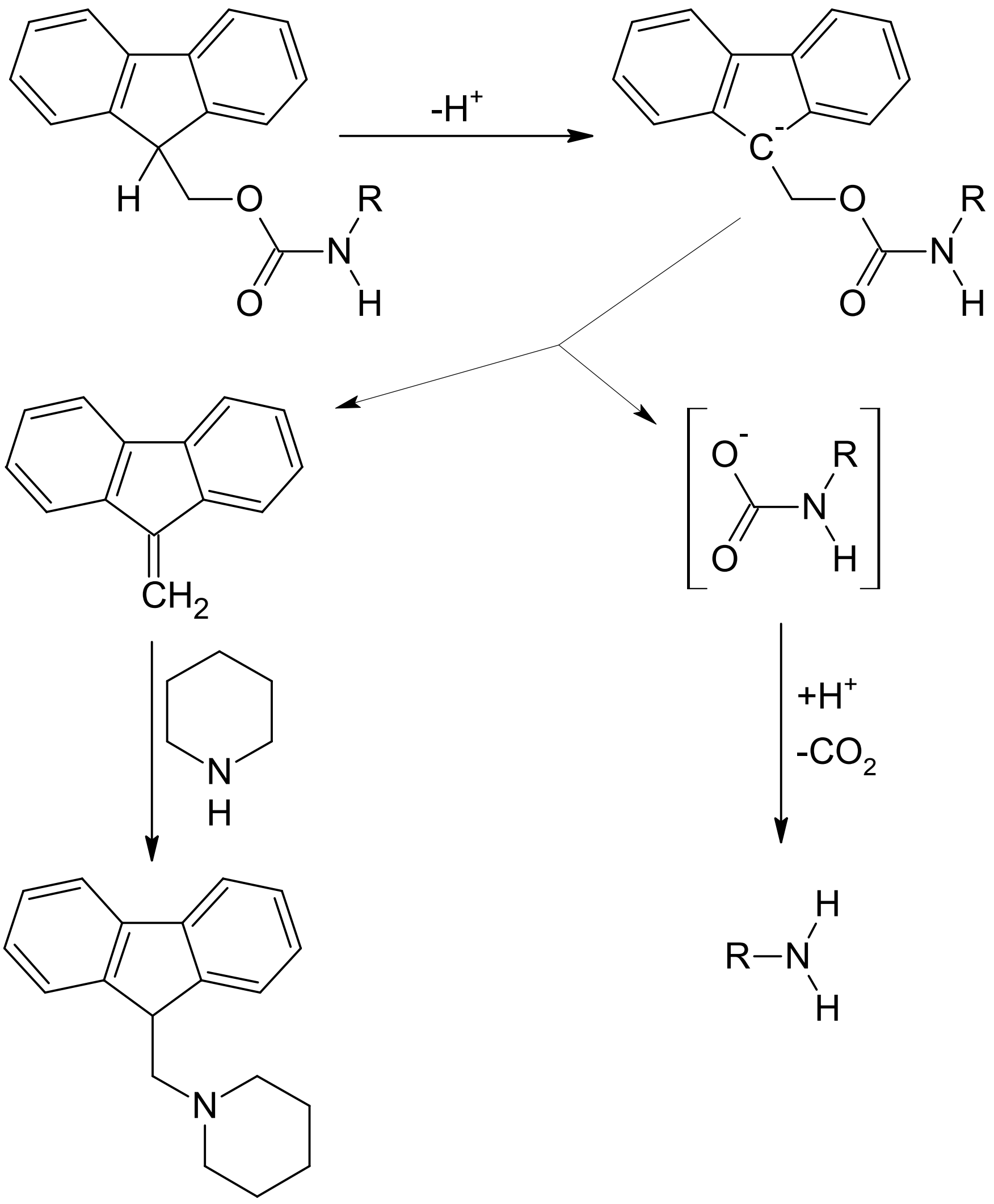

**Figure 5.5.** FMOC deprotection using piperidine in dimethylformamide.

OH groups of the side chains of serine, threonine and tyrosine amino acids are commonly protected by converting them into tert-butyl ethers. tBu provides excellent steric shielding of alcohols and can be easily removed in acidic conditions because of its propensity to form a tertiary cation. tBu can be conveniently placed onto an alcohol using isobutylene gas and either Lewis or protic acid. Deprotection takes place rapidly and irreversibly upon treatment with trifluoroacetic or hydrochloric acid, with a formation of isobutylene gas.

In FMOC-protected peptide synthesis it is not necessary to protect the carboxylic acid group of the amino acid. In the precursor synthesis described below, however, it was briefly protected to avoid formation of dipeptides on the FMOC- and tBu-protection stages. The protection method chosen was conversion of the carboxylic group into the methyl ester (a nucleophile/base labile protecting group), which can be easily cleaved by $S_N2$ type reactions. The most popular ways are base cleavage in an aqueous solution or treatment with $I^-$ or $CN^-$ in a polar aprotic solvent (pyridine, DMSO).

### 5.3 (N)-FMOC-(O)-tBu-3-fluoro-L-tyrosine synthesis

Stage 1.

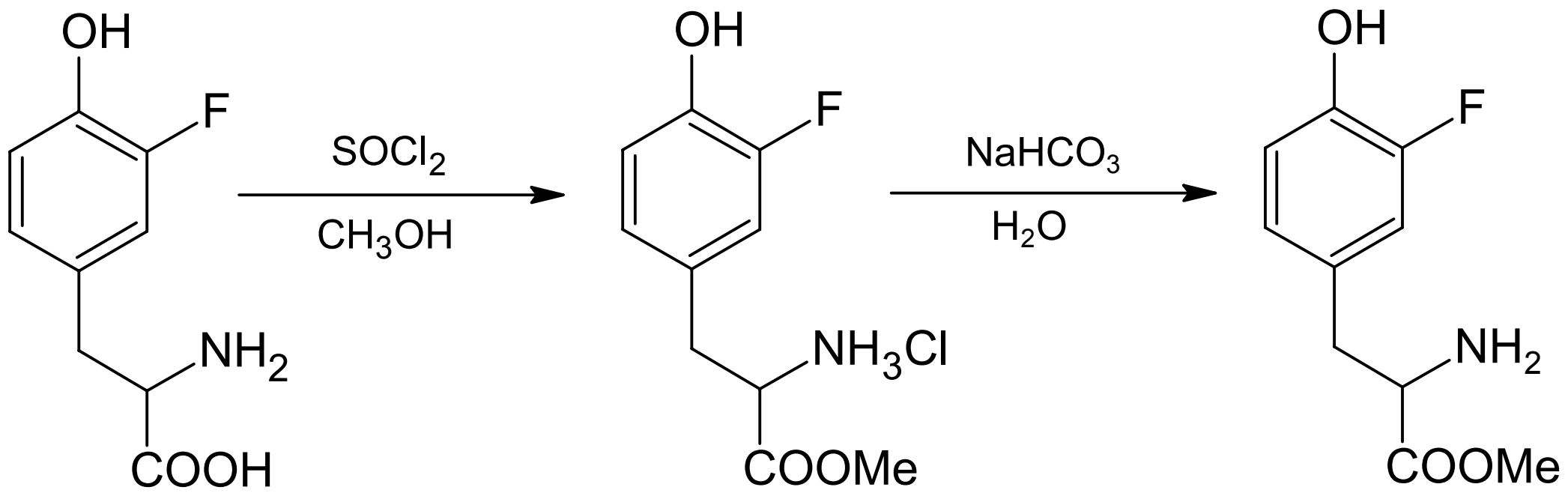

*1.5 mmol (0.299 g) of 3-fluoro-L-tyrosine was suspended in 10 ml of dry methanol. The suspension was cooled to –30 °C, and 2.0 mmol (0.15 ml) of $SOCl_2$ (density: 1.64 g $ml^{-1}$) was added dropwise with constant stirring. The clear yellowish solution was heated to 50°C and stirred for two hours. The resulting solution was poured into 20 ml of cold water and neutralized with aqueous $NaHCO_3$. The white precipitate was filtered off, washed with methanol and dried at 60°C and 40 mbar for 3 hours. Yield: ~100%.*

Stage 2.

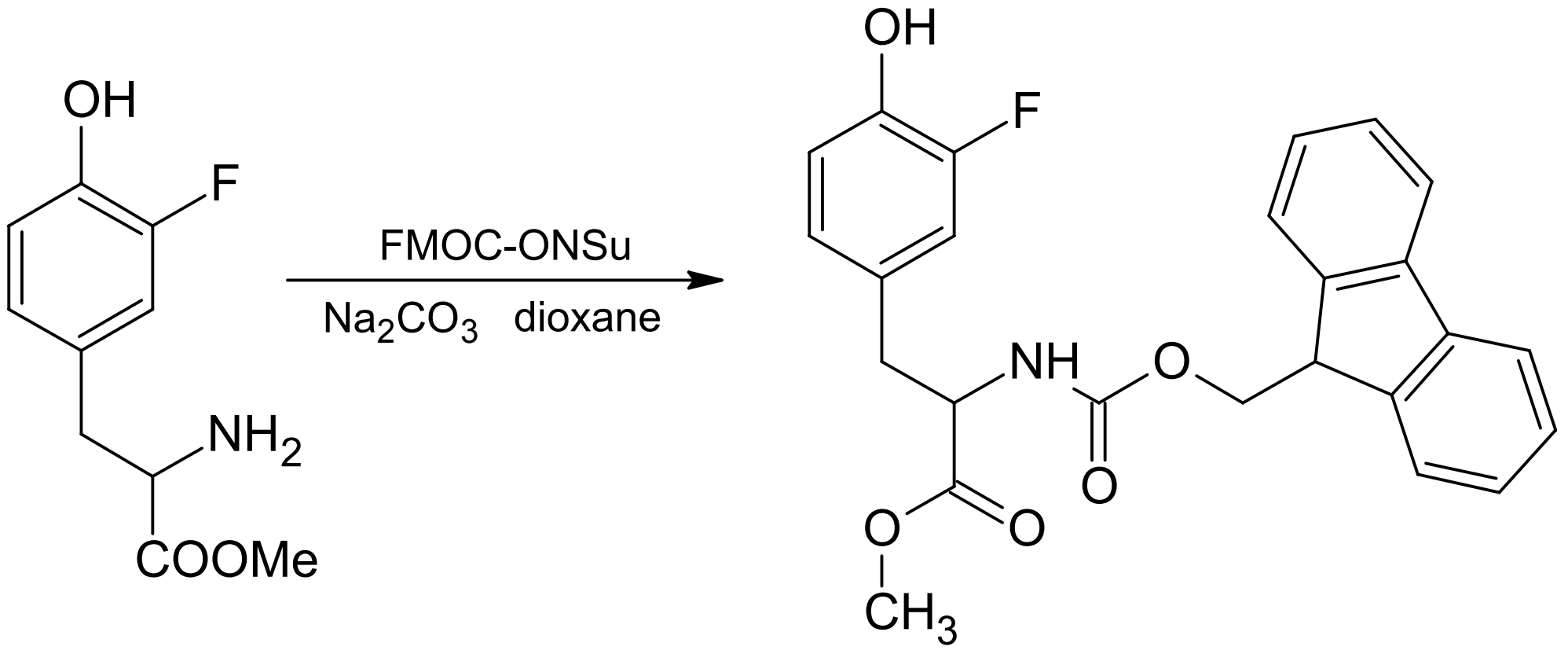

*2 mmol (0.426 g) of 3-fluoro-L-tyrosine methyl ester and 4.5 mmol (0.477 g) $Na_2CO_3$ were dissolved in 30 ml of water and cooled down to 0°C. 2.5 mmol (0.843 g) of FMOC-ONSu was added dropwise as a dioxane solution. The mixture was gradually warmed to room temperature and stirred for 2 hours. The resulting solution was poured* *into excess cold water and extracted with ether. The aqueous phase was acidified to pH 1 with 6 M HCl and triply extracted with ethyl acetate. The combined organic phases were washed with brine, dried over $Na_2SO_4$ and evaporated to an oil which crystallised from cyclohexane. The product was dried at 60°C and 40 mbar for 3 hours. Yield: 95%.*

Stage 3.

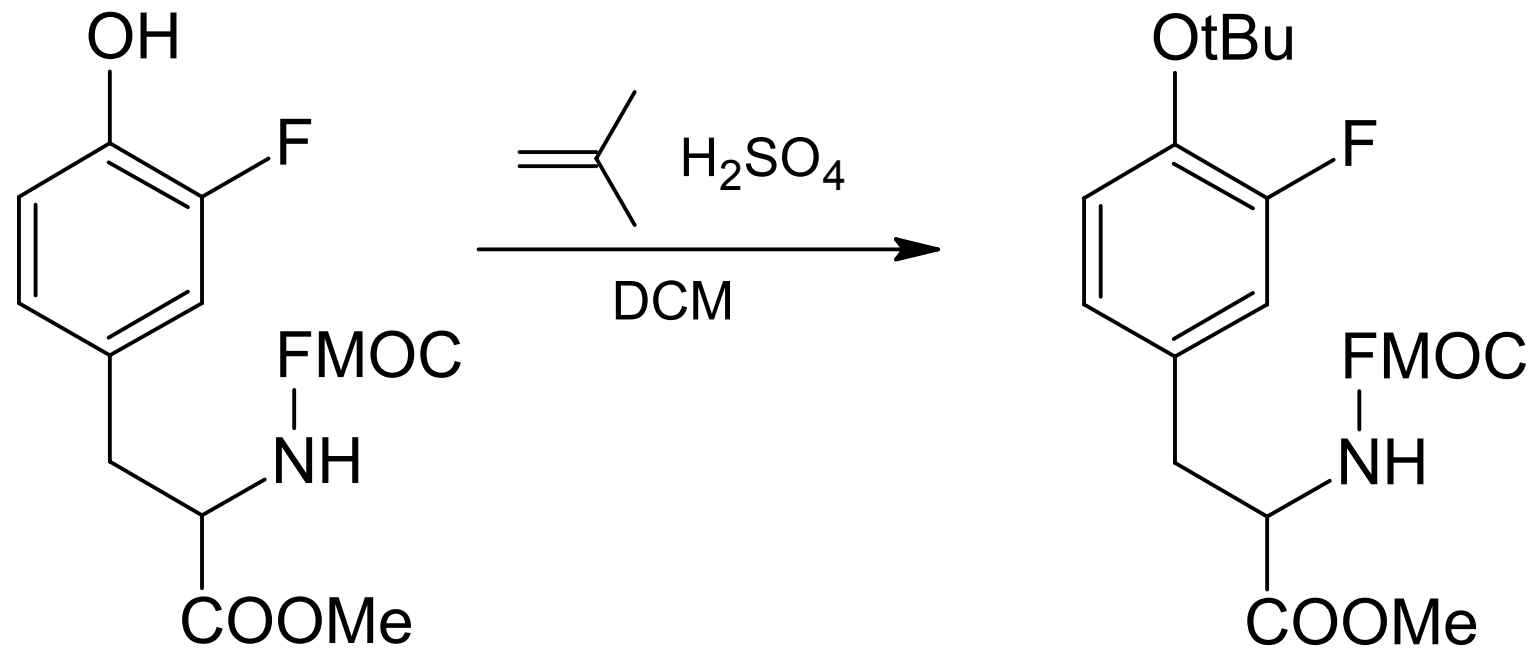

*2 mmol (0.871) g of (N)-FMOC-3-fluorotyrosine methyl ester were dissolved in 40 ml dichloromethane, 50μl of $H_2SO_4$ was added, and the solution was cooled down to –40°C. Approximately 1 ml of liquid isobutylene was condensed into the solution from the attached cylinder. The flask was then sealed, warmed up to room temperature and the mixture was vigorously stirred for 6 hours. The resulting mixture was washed with 3% $Na_2CO_3$, then brine and dried over $Na_2SO_4$. The organic phase was evaporated to an oil which crystallised from cyclohexane. Yield: 50%.*

Stage 4.

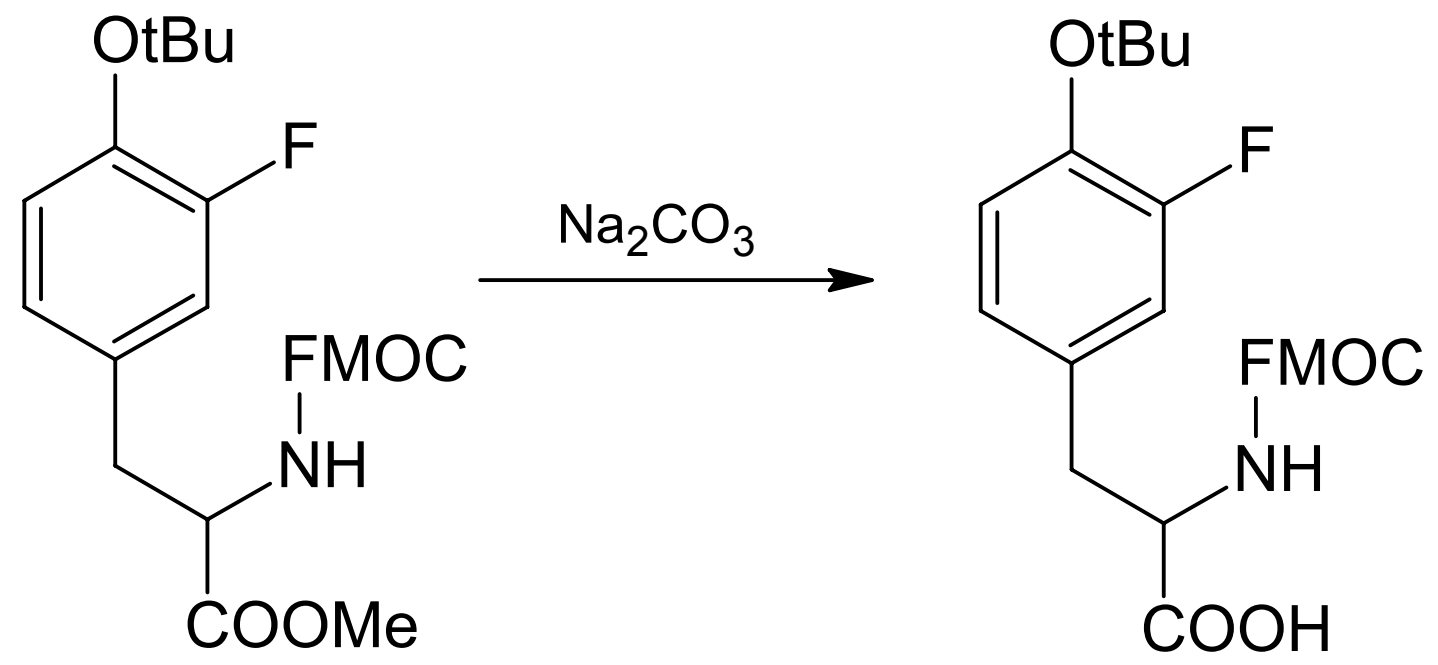

*The product of the previous stage (1 mmol, 0.556 g) was dissolved in the mixture of 100 ml 3% aqueous $Na_2CO_3$ and 67 ml acetonitrile and left to reflux at room temperature until the product spot has developed on the TLC plate (12 hours). The solution was then acidified to pH 3.5 and extracted with ethyl acetate. The organic phase was washed with brine, dried over $Na_2SO_4$ and evaporated to an oil which crystallized from cyclohexane. Yield: 90%.*

The product of Stage 4 (overall yield 42%, structure confirmed by $^{1}H$ NMR spectroscopy) was sent to Professor Niels Andersen's group at the University of Washington for use as a precursor for the solid phase synthesis of 3-fluorotyrosine-labelled Trp-cage protein. For various reasons, only 6 mg of labelled protein became available, allowing only a limited number of experiments to be performed.

## 5.4 Results and discussion

Prior to the photo-CIDNP experiments, a number of common NMR spectra were recorded to characterise the protein. The aromatic part of the proton spectrum shown in Figure 5.6 appears to be very neat, with both fluorotyrosine and tryptophan spin systems clearly resolved. A theoretical spectrum of the aromatic protons perfectly fits the experimental data with all the scalar couplings in the expected ranges.

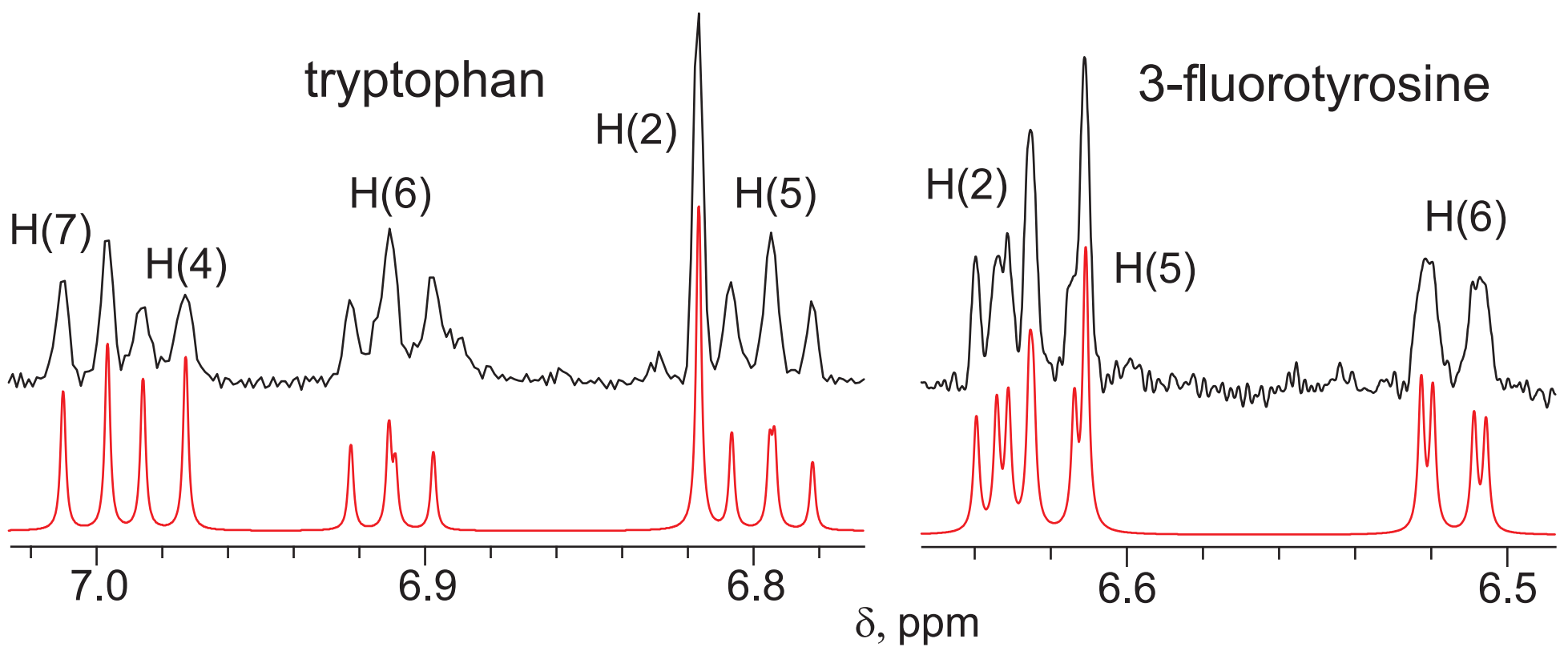

**Figure 5.6.** Theoretical fit of the 3-fluorotyrosine and tryptophan spin systems in the aromatic region of the Trp-cage protein $^{1}H$ NMR spectrum (1 mM protein in $D_2O$, 1 scan, Gaussian pseudoecho window function). The difference in H(5) chemical shift of tyrosine (c.f. Figure 2.25) is due to inductive deshielding by a closely positioned tryptophan aromatic system. In tryptophan: $J_{H(7)-H(6)}$=8.2 Hz, $J_{H(4)-H(5)}$=7.8 Hz, $J_{H(6)-H(5)}$=7.0 Hz. In fluorotyrosine: $J_{H(2)-H(6)}$=1.8 Hz, $J_{H(2)-F(2)}$=12.2 Hz, $J_{H(5)-F(2)}$=9.0 Hz, $J_{H(5)-H(6)}$=8.4 Hz. The assignment and the positive sign of all $J$-coupling constants were confirmed by a GIAO DFT B3LYP aug-cc-pVTZ-J *ab initio* calculation.

The $^{19}F$ spectrum however packs two surprises (Figure 5.7). First, there are two signals in the $^{19}F$ NMR spectrum, characterised by significantly different line widths. If we assume that the peptide is pure (which seems to follow from the proton spectra, as well as LC-MS data sent by Professor Andersen's group), the apparent signal doubling in the fluorine spectrum may have two possible explanations: fluorotyrosine ring-flip isomers (described in detail in Chapter 6) and proline isomers [154, 155][8]. Both these phenomena are well known; they generate multiple structural subensembles and would normally lead to several $^{19}$F signals with different relaxation behaviour. The other, and much more interesting, finding was that for the native state of the Trp-cage protein the $^{19}$F photo-CIDNP effect appears to have changed sign (Figure 5.7, right panel, c.f. Figure 3.1).

[8]What exactly is causing the observed $^{19}F$ signal doubling – tyrosine ring-flip isomers, proline cis-trans isomers, or both – is presently a disputed question. An $^{15}N$-labelled sample is required to record the $^{15}N$-$^{1}H$ HSQC spectrum and clear up the matter.

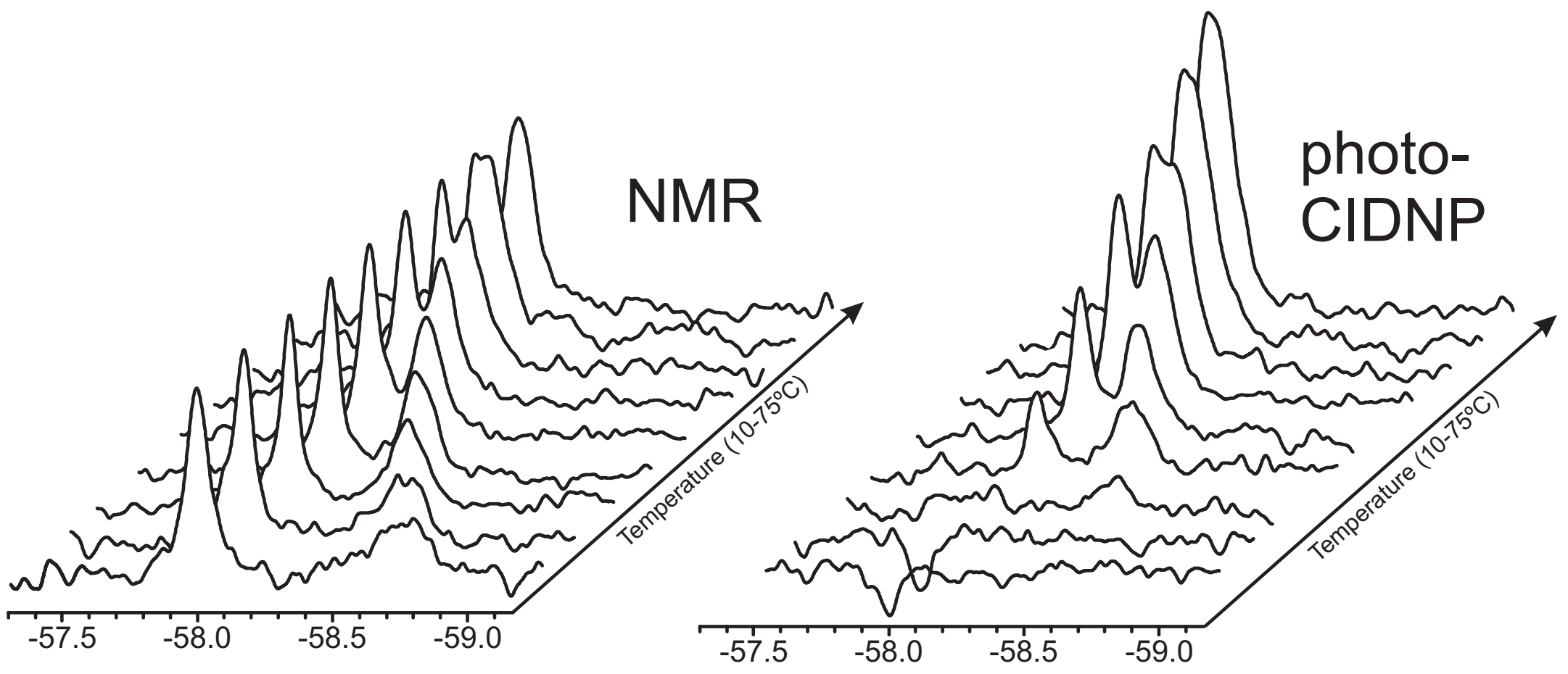

**Figure 5.7.** $^{19}$F NMR and photo-CIDNP spectra of the 3-fluorotyrosine-labelled Trp-cage protein (1 mM in $D_2O$, pH=5.0, 16 scans, CIDNP spectra are light – dark difference) as functions of temperature. The van't Hoff analysis of NMR intensities in the left panel results in ΔH°=7.6±1.6 kJ/mol, ΔS°=23±5 J/mol·K, assuming a two-state chemical equilibrium.

Upon temperature-induced (Figure 5.7, right panel) and urea-induced (Figure 5.8) denaturation, the sign of the $^{19}$F photo-CIDNP effect is slowly restored, with different rates for the two visible $^{19}$F signals. To show the reader quite how strange this phenomenon is, we will make a brief return to the theoretical description given in Sections 1.2 and 1.3.

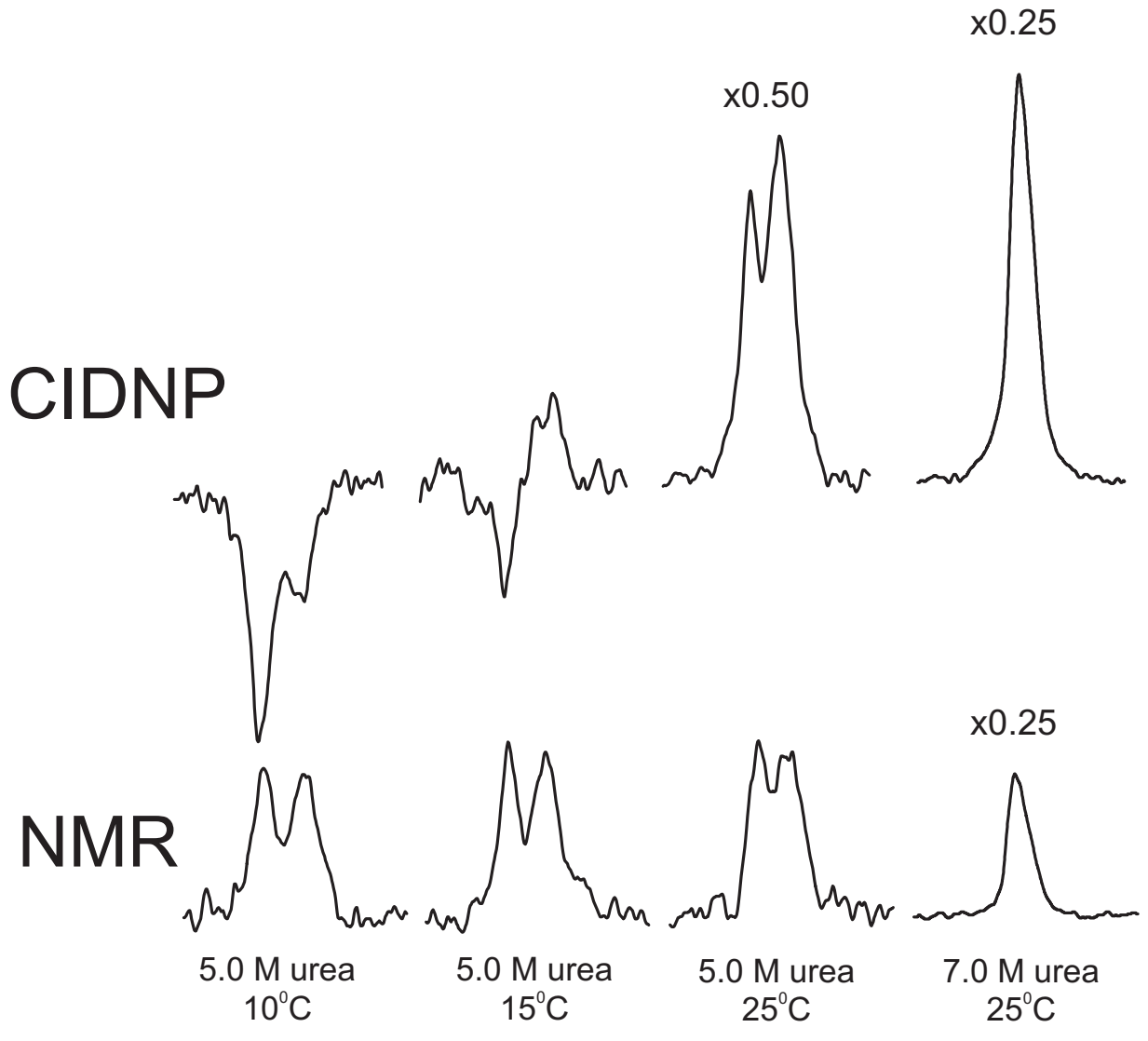

**Figure 5.8.** $^{19}$F NMR photo-CIDNP (right) spectra of 3-fluorotyrosine-labelled Trp-cage protein as functions of denaturant concentration and temperature.

For the fluorine nucleus in the fluorotyrosine/FMN system all four factors in the Kaptein's rules (Equation (1.49)) are positive, resulting in the prediction of absorptive enhancement, which is indeed observed for the individual 3-fluorotyrosine amino acid (Figure 3.1). To explain the pattern shown in Figures 5.7 and 5.8 without introducing any extra magnetization selection mechanisms, one of the factors in Equation (1.49) must bear a continuous dependence on temperature, possibly through another temperature-dependent parameter, such as the correlation time or the effective hydrodynamic radius.

Since the electron transfer reaction is initiated by a triplet flavin molecule and the recombination products are observed, the $\mu$ and $\varepsilon$ factors are very unlikely to change. Introducing sufficient electronic structure perturbation so as to alter the sign of the $^{19}$F hyperfine coupling or shift the *g*-factor requires either hundreds of kJ per mole of energy (hyperfine coupling) or introduction of stronger spin-orbit coupling (*g*-factor). Either of these things is extremely unlikely to be caused by simply heating the system up by 15°C (Figure 5.8). The conclusion therefore is that there is a different mechanism at work here. The chief suspect is spin relaxation, since one of the obvious parameters to change significantly upon thermal and chemical unfolding is the rotational correlation time. Indeed, recording the photo-CIDNP spectra of individual 3-fluorotyrosine in 30% and 70% glycerol as a function of temperature reveals the same type of dependence with the change of sign at high viscosities (Figure 5.9).

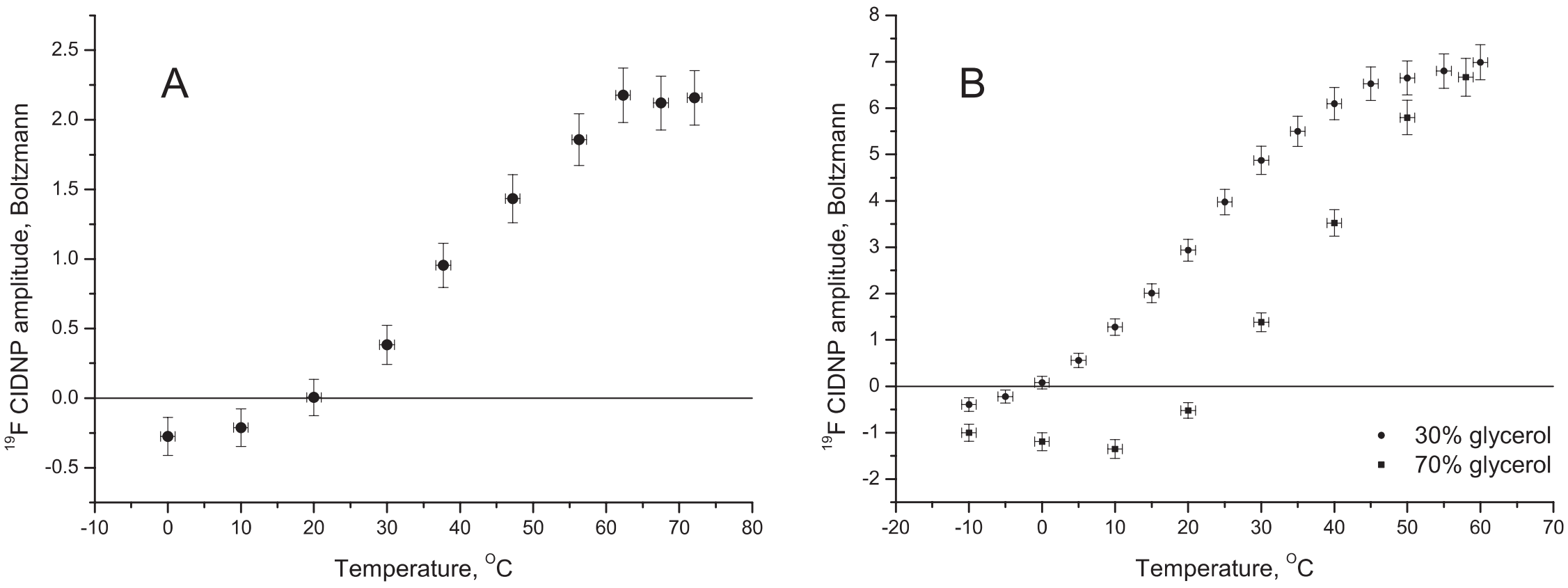

**Figure 5.9. (A)** Temperature dependence of the $^{19}$F CIDNP amplitude observed for the $D_2O$ solution (pH=5.0) of 3-fluorotyrosine labelled Trp-cage protein. **(B)** Temperature dependence of the $^{19}$F CIDNP amplitude observed for the $D_2O$/glycerol solution (pH=5.0) of 3-fluorotyrosine amino acid.

As such, paramagnetic relaxation effects in CIDNP, including that of fluorine-containing compounds, are nothing new [64, 65, 156-158]. Moreover, Overhauser-type processes are usually cited for most of the deviations from Kaptein's rules observed in photo-CIDNP experiments. The problem, however, is that at the field of 14.1 Tesla the Redfield theory calculations (described in Chapter 8) with cross-relaxation models suggested by Adrian et al. [64, 65, 156] and later by Tsentalovich et al. [157, 159, 160] all lead to the values of the paramagnetic cross-relaxation rates that are far too small to have any influence whatsoever on the nanosecond-scale geminate radical dynamics[9]. The fact that it is geminate $^{19}$F polarization which changes with the correlation time was established from direct observation of geminate polarization as a function of temperature in a water-glycerol solution (Figure 5.10) and from the absence of $^{19}$F escape cancellation demonstrated in Chapter 7 (Figure 7.6).

[9]This is a simple consequence of having a square of the electron Larmor frequency in the denominator of the spectral density $J(\omega \neq 0)$ and lack of $J(0)$ term in all of the longitudinal single- and multi-spin (cross-) relaxation rate expressions.

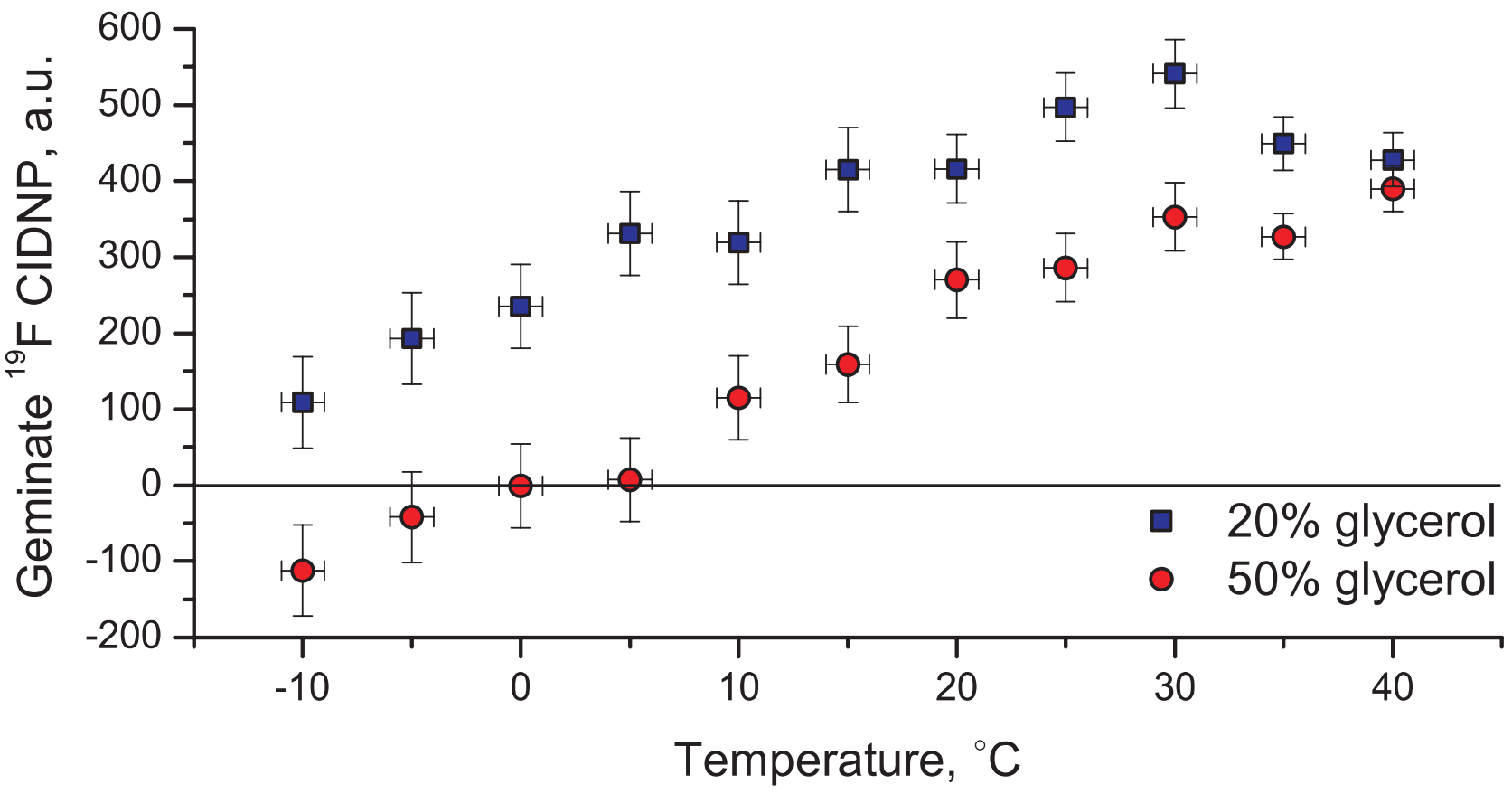

**Figure 5.10.** Temperature dependence of geminate $^{19}$F magnetization in water/glycerol solutions of 3-fluorotyrosine/FMN system, generated in a time-resolved photo-CIDNP experiment using the setup described in Chapter 7.

We must therefore conclude at this point that if it indeed is a relaxation effect that is responsible for this sign change, it must be more complicated than the longitudinal electron-nucleus cross-relaxation [157] or cross-correlation [159, 160] suggested in the earlier studies. Chapter 8 of this Thesis is devoted entirely to further exploration of this phenomenon, which turned out to be a complex transverse cross-correlated magnetization transfer process.

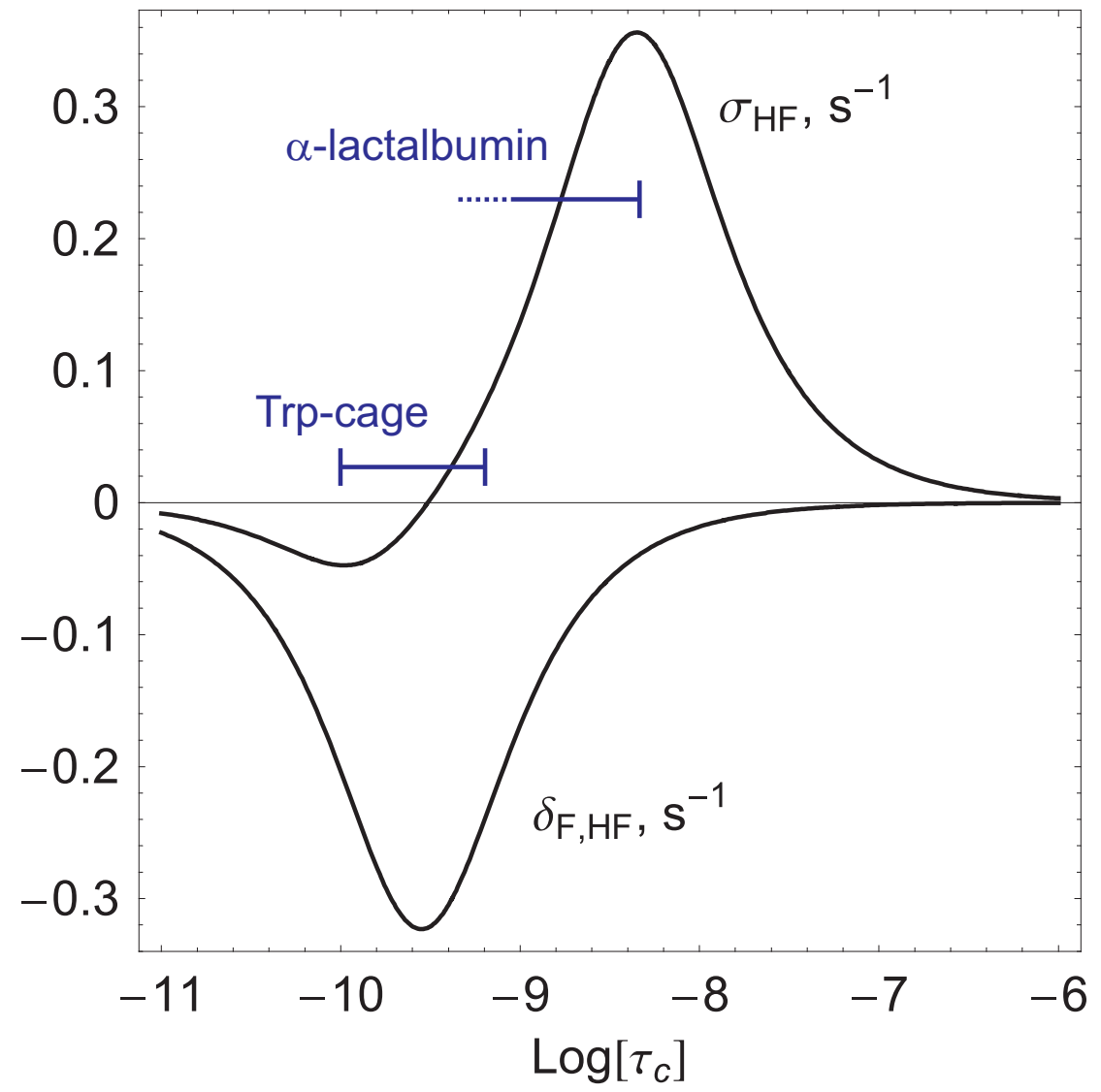

**Figure 5.11.** Correlation time dependence of $^{19}$F↔$^{1}$H dipolar cross-relaxation and DD-CSA cross-correlation rates (Redfield theory, isotropic tumbling approximation) and estimated correlation time intervals for Trp-cage and $\alpha$-lactalbumin proteins. **Trp-cage right bound**: Stokes-Einstein equation with $V_{mol}$=2800 Å$^3$ (Gaussian03 PCM calculation), $\eta(H_2O)=10^{-3}$ Pa·s, $T$=298 K, resulting $\tau_c$ = 0.7 ns. **Trp-cage left bound**: free aqueous 3-fluorotyrosine, $\tau_c$ = 0.10 ns [52]. **$\alpha$-lactalbumin right bound:** Stokes-Einstein equation with $V_{mol}$=18000 Å$^3$ (Gaussian03 PCM calculation), $\eta(H_2O)=10^{-3}$ Pa·s, $T$=298 K, resulting $\tau_c$=4.4 ns.

Leaving the sign change phenomenon until Chapter 8, we will mention in passing one other relaxation process which was found in Chapter 3 to be potentially useful in structural studies of fluorinated proteins – the diamagnetic $^{19}$F-$^{1}$H chemically pumped nuclear Overhauser effect (CANOE). When the $^{19}F$ photo-CIDNP enhancement is large, a strong and long-range $^{19}F \rightarrow ^{1}H$ NOE can potentially be observed for protons close to the $^{19}F$ nucleus. A theoretical estimates of the $^{19}F$-$^{1}H$ cross-relaxation and DD-CSA cross-correlation rates between F(3) and H(2) in the aromatic ring of 3-fluorotyrosine amino acid using Redfield theory in the isotropic tumbling approximation is shown in Figure 5.11. The cross-relaxation and cross-correlation rates with other protons will scale as $1/r^6$ and $1/r^3$ respectively.

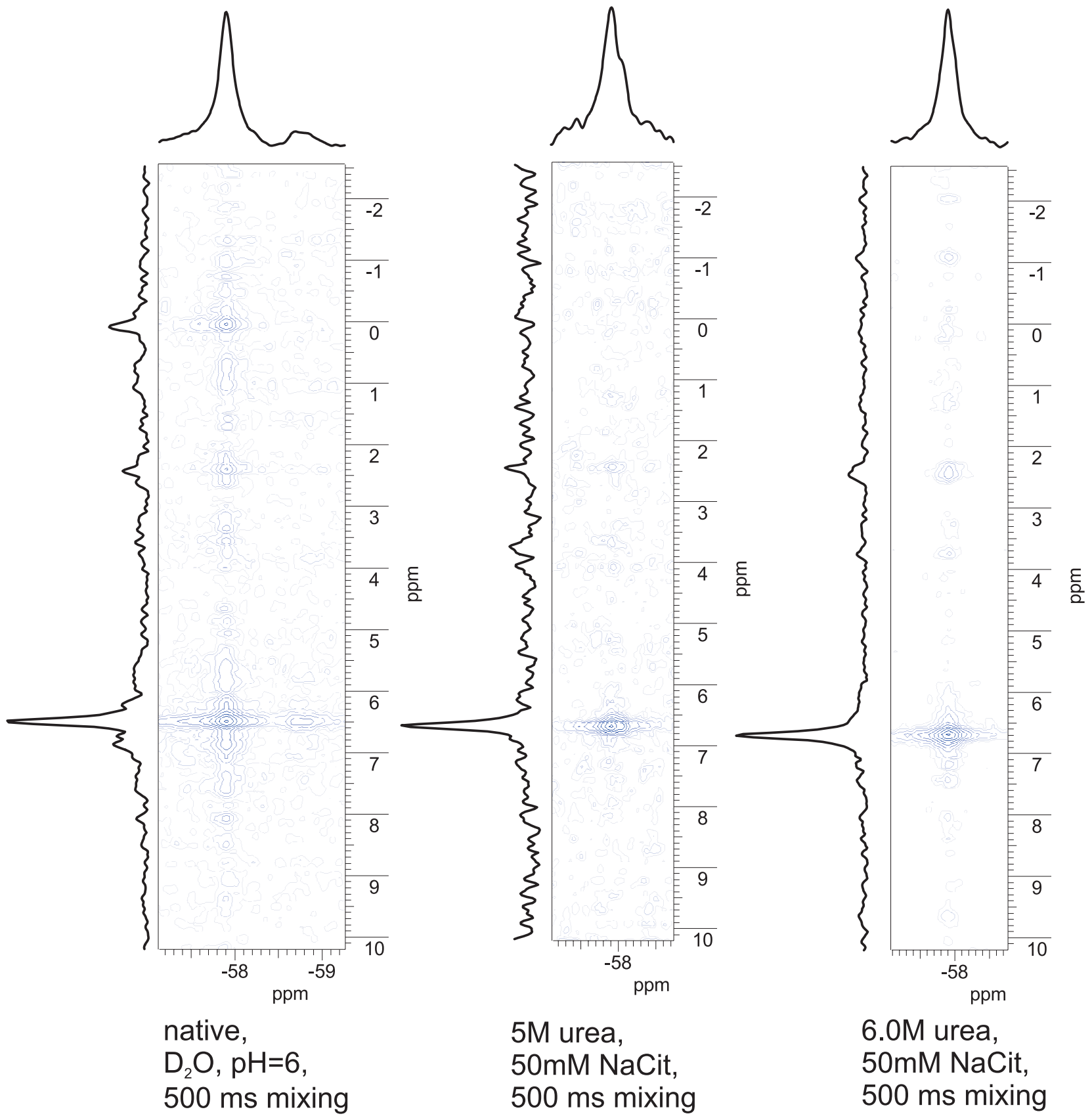

**Figure 5.12.** $^{1}H \rightarrow ^{19}F$ HOESY spectra of 1.0 mM $D_2O$ solution of fluorotyrosine-labelled Trp-cage protein in native, partially destabilized and strongly destabilized states. The 1D spectra are slices taken through points of maximum signal intensity. Each spectrum took over 24 hours to accumulate and a heavy (80 Hz Gaussian) window function was needed to extract anything out of the noise.

Figure 5.11 displays a rather pessimistic prediction for the Trp-cage protein and its partially unfolded states, with the cross-relaxation rate being very close to the zero crossing point of the spectral density function (Figure 5.11). The position of this zero crossing point is independent of the number of protons in the vicinity of the fluorine nucleus and only depends on the effective correlation time. To add to the problem, the expected range of correlation times for the Trp-cage is located at the maximum of the DD-CSA cross-correlation rate, meaning that $^{19}F$ magnetization will be strongly drained into unobservable multi-spin orders. The inverse cubic distance dependence of the DD-CSA cross-correlation rate means that protons in a very large (3-5 Å) radius will contribute substantially to this process, likely increasing the rate shown in Figure 5.11 by about an order of magnitude. The chances of observing a detectable $^{19}F \rightarrow {}^{1}H$ NOE in these circumstances are at best very slim, even if we assume that the actual hydrodynamic radii will be slightly larger than those used in Figure 5.11 due to molecular surface solvation. A heavier protein (e.g. α-lactalbumin), which shows both higher $^{9}F \rightarrow {}^{1}H$ NOE transfer rate and lower DD-CSA drain (Figure 5.11), is required to obtain practically useful data using this methodology. An alternative approach would be to use more viscous solvents and low temperatures, but this would take us away from the (experimentally preferable) near-physiological conditions.

To assess the feasibility of $^{19}F$-$^{1}H$ CANOE experiments on the fluorotyrosinated Trp-cage protein, several conventional HOESY [161, 162] spectra were recorded (Figure 5.12). The specific feature of the $^{19}F$ nucleus in proteins is extremely fast transverse relaxation, which short-circuits frequency labelling and coherence transfer stages in most 2D experiments involving $^{19}F$. The only way to record a protein HOESY spectrum with fluorine in one of the dimensions is to make sure that $^{19}F$ magnetization never goes into the transverse plane except at detection. A reverse ($^{1}H \rightarrow {}^{19}F$) HOESY sequence has this property and was therefore recorded.

As expected from the relaxation theory calculation (Figure 5.11), over 24 hours of pulsing and a very strong window function have only uncovered a trivial NOE from the $^{19}F$ nucleus to the nearest H(2) proton in fluorotyrosine (Figure 5.12). The native spectrum also has a very weak but detectable magnetization transfer from the H(5) of the nearby Trp6, methyl groups of Leu9, and possibly β-protons of Pro18 as well. Given that the native state has a near-zero $^{19}F$ photo-CIDNP effect (Figure 5.7) and the partially unfolded states lack detectable Overhauser effects even after very prolonged signal accumulation, it seems unreasonable to try $^{19}F$-$^{1}H$ CANOE experiments on the Trp-cage protein. The correlation time dependence of the $^{19}F$-$^{1}H$ cross-relaxation rate (Figure 5.11) suggests that a heavier fluorinated protein will likely present a much better system for a study using chemically pumped $^{19}F$-$^{1}H$ dipolar cross-relaxation.

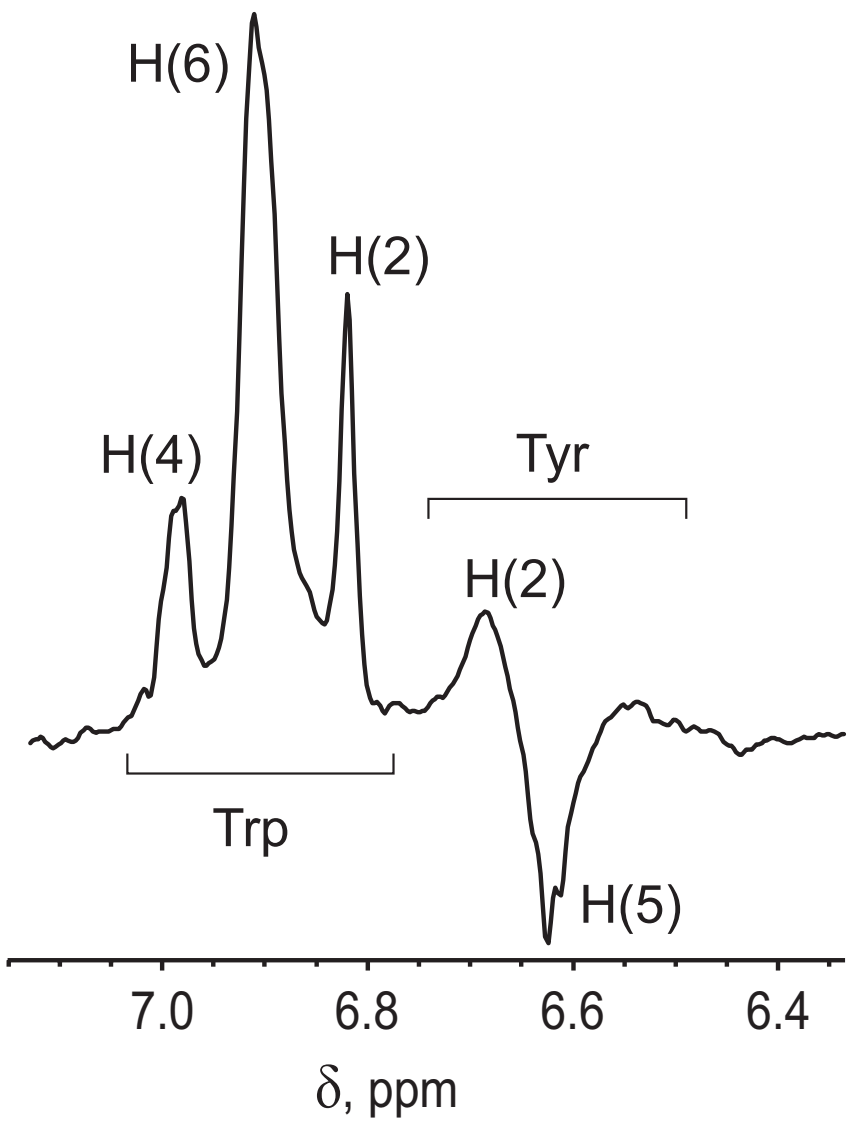

**Figure 5.13.** Aromatic region of the $^{1}H$ photo-CIDNP spectrum of the 3-fluoro-tyrosine-labelled Trp-cage protein.

The last thing to mention about the Trp-cage is the fact that even though the tryptophan aromatic ring is "encaged" there appears to be enough HOMO density available to the photosensitizer for the electron transfer to occur from the tryptophan residue as well as the more exposed tyrosine, resulting in the observed photo-CIDNP polarization on tryptophan as well as tyrosine in the Trp-cage (Figure 5.13). In apparent contradiction to the computed geometric solvent accessibilities, tryptophan shows higher photo-CIDNP polarization than tyrosine. This (and similar observations made in Chapter 6 for the green fluorescent protein) has prompted the author to revisit the controversial relation between the solvent accessibility and the photo-CIDNP effect [46, 163] and propose a modification (described in detail in Chapter 6), in which the CIDNP-geometry relation is based on amino acid HOMO accessibility to the photosensitizer, rather than the accessibility of any particular atoms or groups.

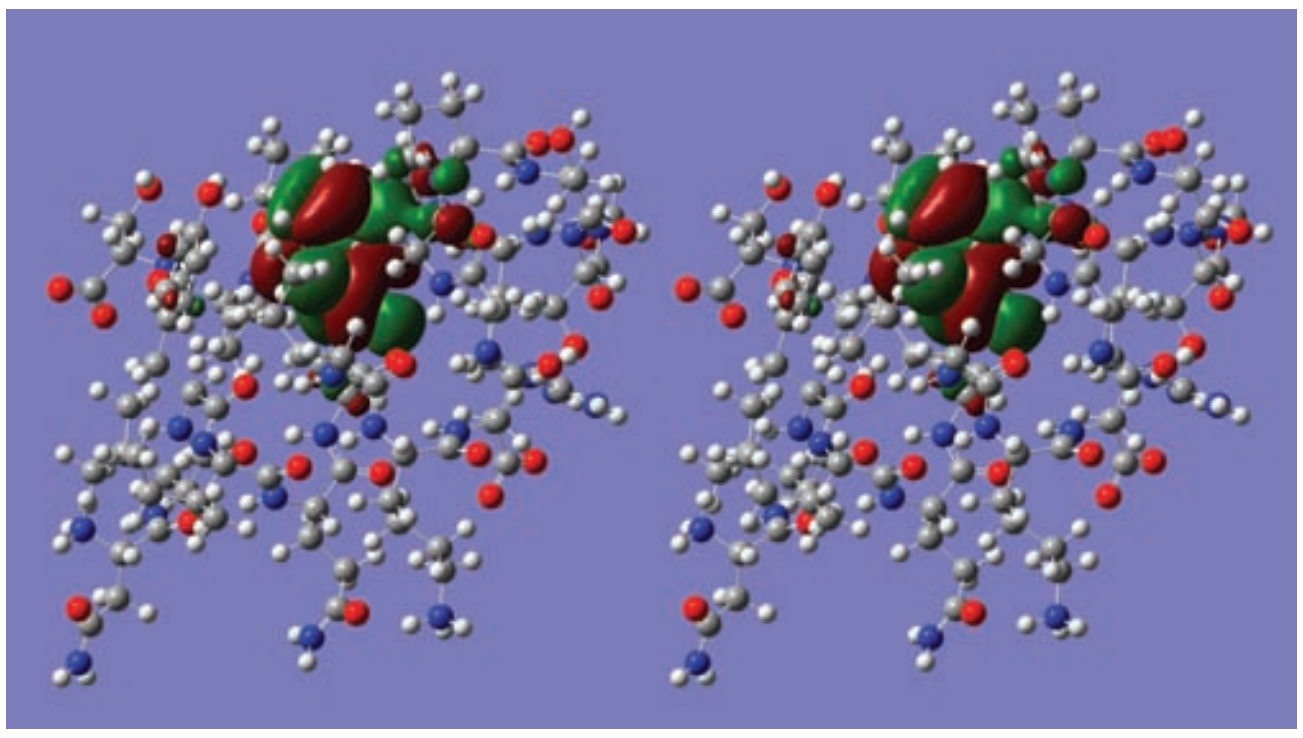

**Figure 5.14.** Stereo plot of the HOMO orbital of the Trp-cage protein (DFT B3LYP/6-31G using NMR structure geometry).

In the particular case of Trp-cage the observed photo-CIDNP polarization of tryptophan may have two (likely simultaneously active) causes. First, it is known that a significant sub-ensemble exists in the pool of Trp-cage molecules in which tryptophan solvent exposure is much greater than the NMR structure (Figure 5.1) suggests. The reason for this is that at room temperature Trp-cage is only ~85% folded [138]. The folding rate constant of $(4\ \mu s)^{-1}$ then implies an unfolding rate constant of about $(23\ \mu s)^{-1}$, meaning that the folding-unfolding equilibrium is very fast and allows tryptophan to participate in electron transfer. Second, even in the folded state the geometric accessibility of tryptophan HOMO may be sufficient for electron transfer to the photosensitizer to occur (Figure 5.14). A study of triplet photosensitizer quenching by tyrosine-tryptophan dipeptide performed by Morozova et al. [164, 165], as well as an earlier study on photosensitizer quenching competition effects [57] have found that, all other things being equal, tryptophan has a higher rate of electron transfer to the photosensitizer than tyrosine, meaning that a lesser degree of solvent exposure might lead to a comparable photo-CIDNP effect amplitude observed in Figure 5.13.

Overall, although the questions set out in the introduction to this Chapter have been answered, the study of the Trp-cage protein has generated further questions. Specifically, the origin of the fluorine signal doubling is unclear, as is the cause of the sign change in the $^{19}$F photo-CIDNP effect. While the latter question is dealt with in Chapter 8, the former is left for exploration by any interested party.

# Chapter 6
# $^{19}$F NMR and photo-CIDNP study of green fluorescent protein

*Based on: F. Khan, I. Kuprov, T.D. Craggs, S.E. Jackson, P.J. Hore, Biochemistry, submitted.*

### 6.1 Introduction

Green fluorescent protein (GFP), originally obtained from the jellyfish *Aequorea victoria*, is remarkable for its structural stability, nearly 100% fluorescence quantum yield and lack of cytotoxicity [166]. It is widely used as a reporter protein in biological and biochemical research, and is also interesting in its own right as a model system for photophysical [167] and protein folding studies [167-170].

The 238-residue amino acid sequence of GFP folds into an 11-stranded β-barrel (Figure 1) [171, 172] and undergoes an autocatalytic post-translational cyclization and oxidation around residues Ser65, Tyr66 and Gly67, forming an extended and rigidly immobilized conjugated π-system – the chromophore [167]. No cofactors are necessary for either formation or function of the chromophore, which is embedded in the interior of the protein [173]. In the native state, the chromophore is shielded from the bulk solvent and rigidly immobilized, the latter feature is thought to prevent vibrational relaxation and give GFP its re-markably high fluorescence quantum yield. On protein denaturation, the chromophore remains chemically intact but the fluorescence is lost. This green fluorescence is, therefore, a sensitive probe of the state of the protein.

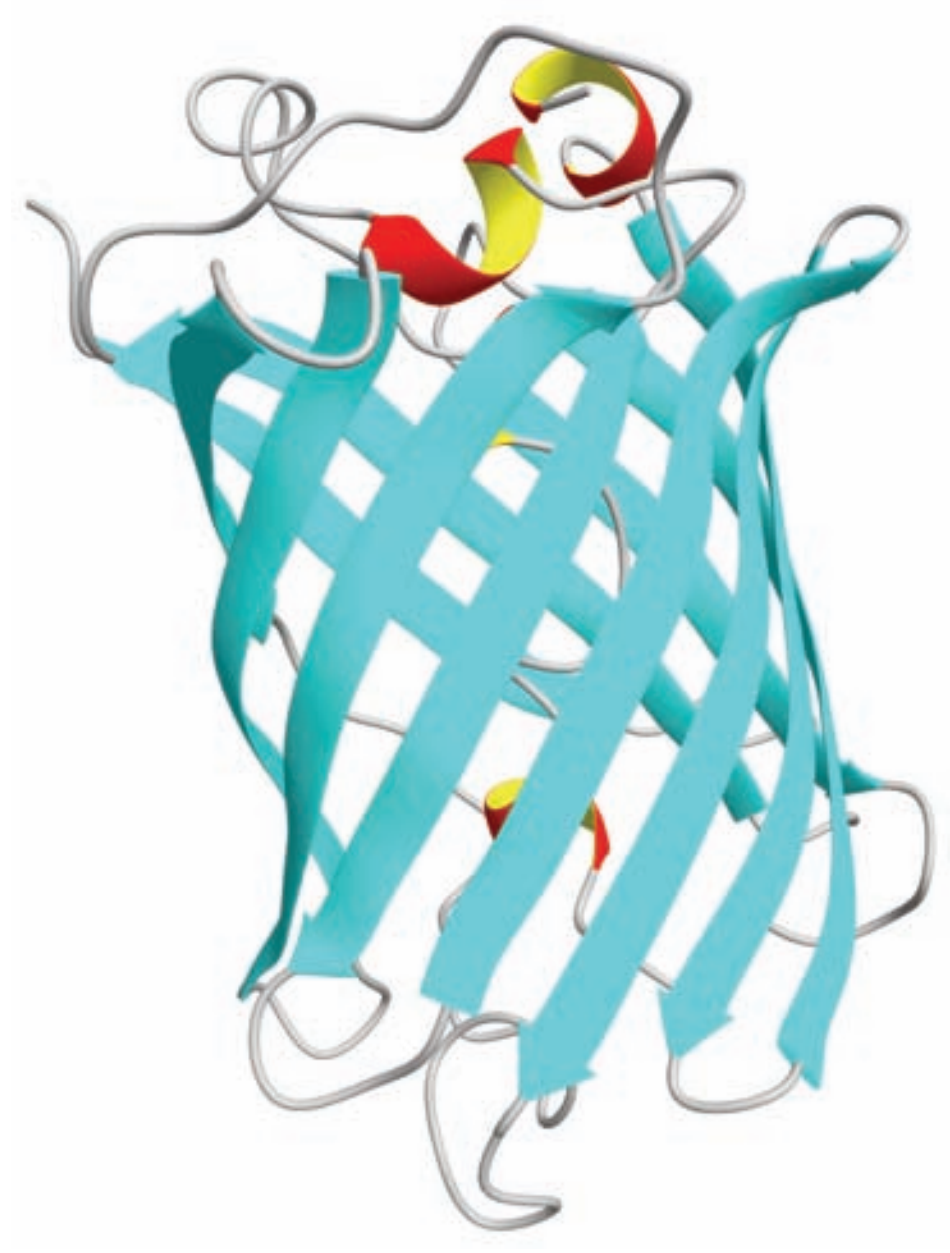

**Figure 6.1.** Ribbon diagram of the GFP structure.

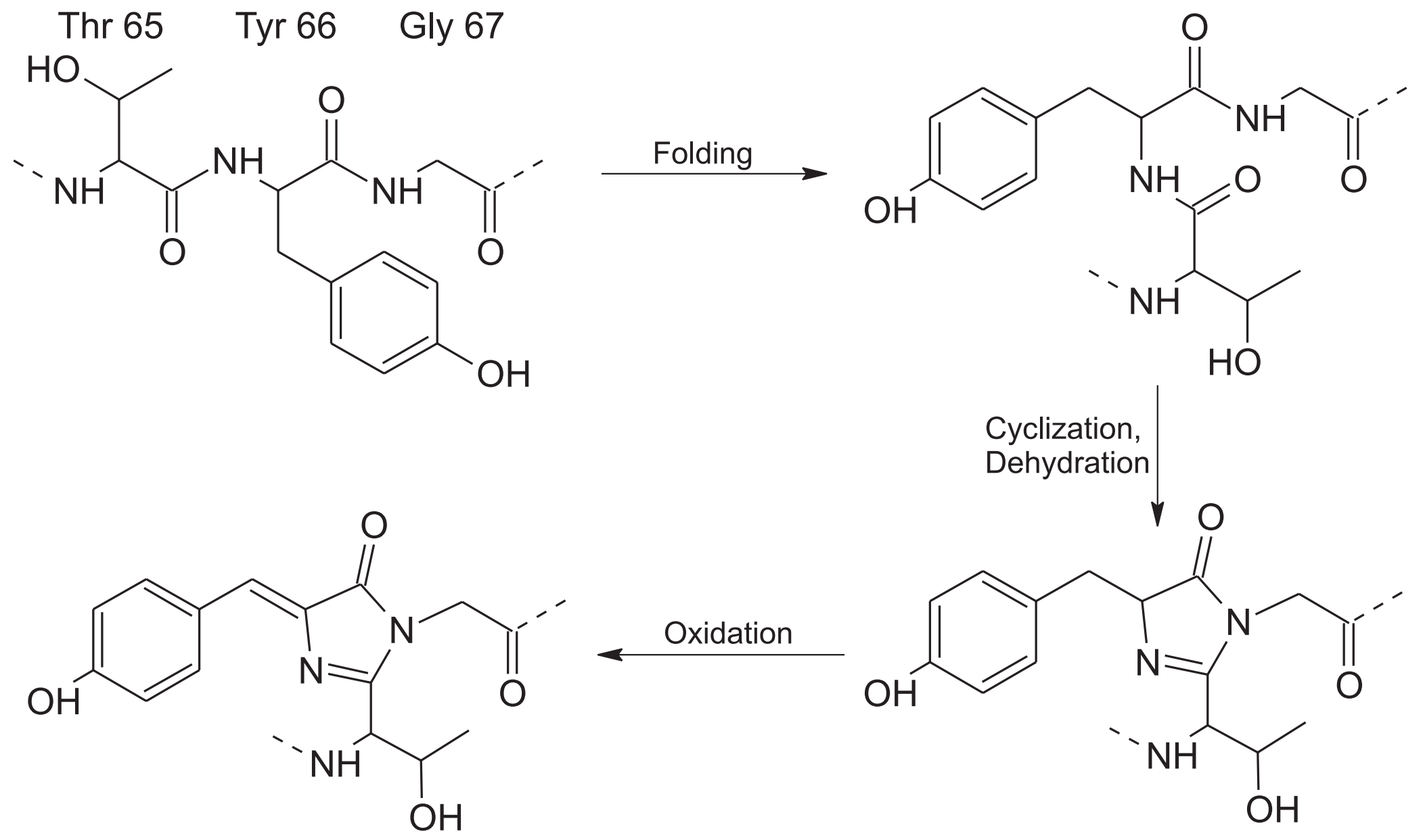

**Figure 6.2.** Chromophore formation in GFP. Adapted from Reid and Flynn [173].

GFP has been extensively engineered to alter and improve its properties and to facilitate its use as a reporter of gene expression, for protein localisation, as a biosensor and indicator of protein-protein interactions. To be fluorescent in these different biological assays, GFP must be completely folded.

Relatively little is known about folding of GFP, both *in vitro* and *in vivo*. The most detailed study to date is a work by Kuwajima *et al.*, who employed fluorescence and far-UV circular dichroism to probe phases of folding, unfolding and refolding reactions. Based on the UV and CD spectroscopic data, Kuwajima *et al.* proposed a folding model with several intermediate states [168]. The overall folding pattern appears to be quite complex and further studies using different techniques and probes are certainly necessary to obtain a complete understanding of GFP folding behaviour.

Nuclear magnetic resonance (NMR) spectroscopy has proven to be a powerful method of studying structured, partially structured and denatured states of proteins, including molten globules and kinetic intermediates [174]. In the cases where NMR can be used, it provides valuable residue-specific information on structure and dynamics and therefore complements optical techniques such as fluorescence and circular dichroism which only report the global properties. $^{1}$H, $^{13}$C and $^{15}$N are the most commonly used nuclei. Quite recently, however, much progress has been made on the study of proteins using $^{19}$F NMR [175-177].

A distinctive feature of $^{19}$F NMR spectroscopy is a strong chemical shielding anisotropy (CSA) of the fluorine nucleus, which is sometimes as high as 100 ppm. The nuclear relaxation rate due to the CSA is quadratic in magnetic induction, and, for a protein of the size of GFP at 10-20 Tesla, the transverse $^{19}$F relaxation is dominated by the CSA term and is very fast. This leads to broad spectral lines (60-400 Hz) and lowers the sensitivity. Fast transverse relaxation of $^{19}$F also short-circuits magnetization transfer stages in all but a few 2D NMR sequences [178].

Despite these seemingly formidable difficulties, $^{19}$F NMR of fluorine-labelled biological macromolecules is gaining popularity, the reason being that $^{19}$F nucleus has a large chemical shift dispersion and its chemical shielding is very sensitive to the changes in molecular environment [179]. For example, in a protein containing ten fluorinated tyrosines, one may generally expect that all ten will be resolved in the native state spectrum and that any structural transformation will be accompanied by significant changes in chemical shifts. Kinetic analysis of these changes can give direct information on the kinetics of the structural transformation in question [177]. In addition to the information obtained from chemical shifts, structural information can also be gathered from relaxation data. $^{19}$F relaxation behaviour is well understood and in principle allows extraction of local and global motion correlation times, order parameters and rotational diffusion tensors [178], which are useful indicators of the local mobility and stability of the macromolecule.

Another unusual property of the fluorine nucleus, namely the large $^{19}$F hyperfine coupling constant in fluorinated aromatic radicals, makes it particularly useful in NMR experiments with photochemical pumping of magnetization through the photo-CIDNP effect [52]. In high magnetic fields the amplitude of the photo-CIDNP effect is directly proportional to the hyperfine coupling constant in the intermediate radical. Fluorine-containing molecules are quite unique in this respect because the $^{19}$F hyperfine coupling constants in aromatic radicals are large, owing to both the high magnetogyric ratio of the $^{19}$F nucleus and the strong electro-negativity of the fluorine atom (see Chapter 2 for further details).

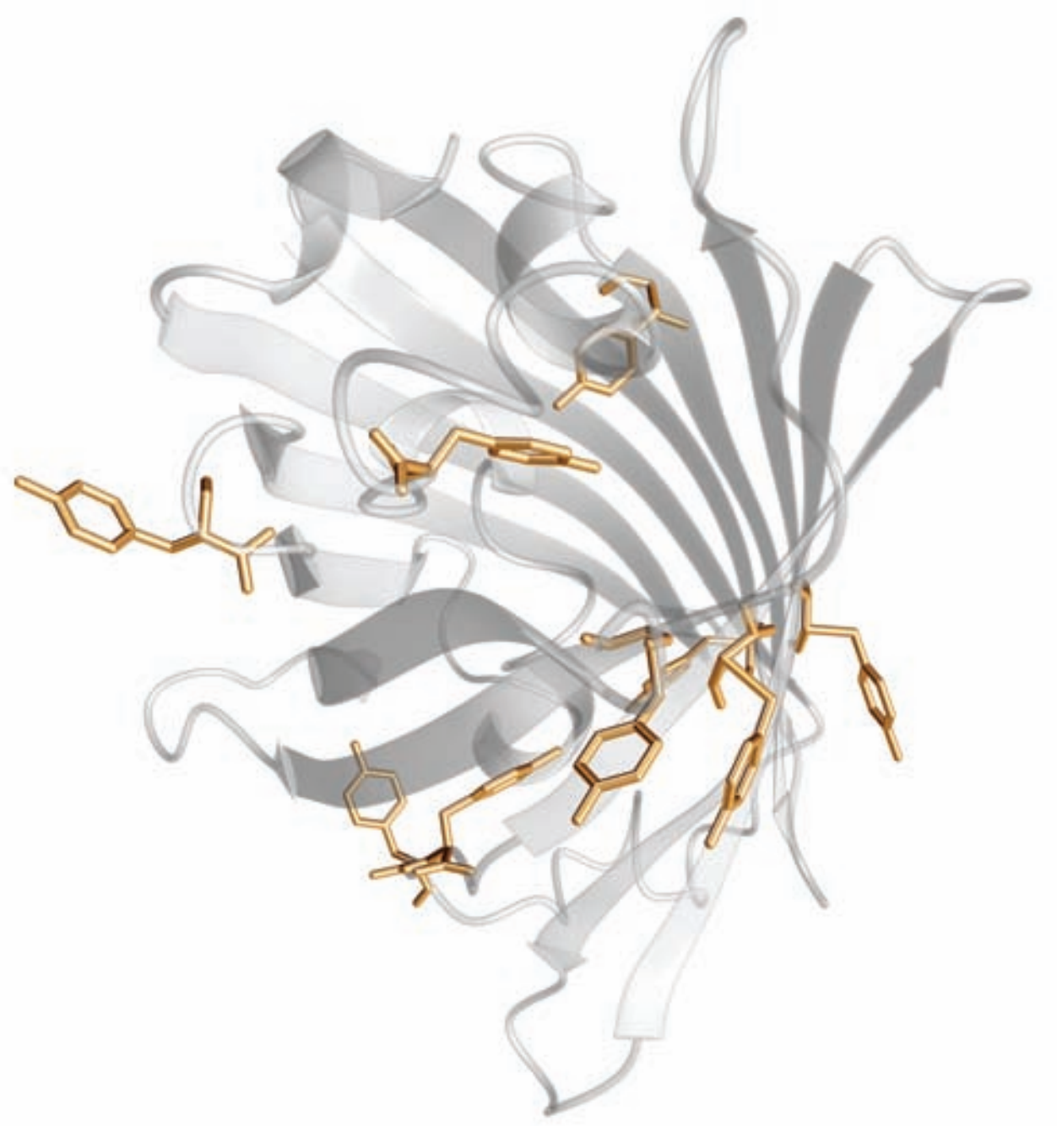

**Figure 6.3.** Nine tyrosines of GFP. The tenth tyrosine is a part of the chromophore (not shown). Out of nine proper tyrosines five are accessible to the solvent.

This Chapter describes the NMR and photo-CIDNP experiments on the uniformly 3-fluorotyrosine-labelled GFP. Signal assignment via site-directed mutagenesis and $^{19}$F photo-CIDNP techniques are described, as well as $^{19}$F CIDNP experiments performed to characterise the native and denatured states of GFP.

## 6.2 Materials and methods

The 3-fluorotyrosine-labelled GFP and all of the site-directed Tyr→Phe mutants were prepared by T.D. Craggs and F. Khan at the University of Cambridge using the QuikChange kit from Strategene. All mutations were confirmed by DNA sequencing of the truncated GFP gene. Chromatography and mass-spectroscopy data indicate > 99% purity and > 95 % fluorotyrosine incorporation. On sample storage, very small amounts of free 3-fluorotyrosine are accumulated, most likely due to slow decomposition. This admixture proved advantageous during the initial stages of the assignment as a chemical shift reference signal. However, 3-fluorotyrosine was found to disrupt the photo-CIDNP experiments as it preferentially reacted with the excited FMN photosensitizer. It was removed from the samples by dialysis prior to photo-CIDNP measurements.

377 MHz $^{19}$F NMR measurements were performed on 1.0 mM protein samples in 10% $D_2O$ with PBS buffer at 300 K using a Bruker DRX400 spectrometer equipped with a 5 mm QNP probe. $^{19}$F chemical shifts were referenced to external TFA. 564 MHz $^{19}$F NMR spectra were recorded in 10% $D_2O$ with PBS buffer at 300 K on a Varian Inova 600 (14.1 Tesla) NMR spectrometer equipped with a 5 mm $^{19}$F{$^1$H} *z*-gradient probe. In all cases, the spectra were recorded with sufficient sweep width to accommodate all signals without reflections. The acquisition time was set to longer than five times the reciprocal width of the narrowest signal. The hardware oversampling rate was set to the instrumental maximum. All FIDs were weighted by either an exponential or a Gaussian pseudoecho window function matched to the narrowest peak. For the line fitting analysis, the line positions were first determined by fitting a spectrum processed with a Gaussian pseudoecho window function, the resulting peak position list was then used for fitting the unfiltered spectrum to extract the line widths. The typical number of scans in the $^{19}$F spectra reported herein is 16384. $^{19}$F photo-CIDNP spectra were recorded in pure $D_2O$ at pH 6.0 (phosphate buffer). The light source was a Spectra Physics BeamLok 2080 argon ion laser, operating in single-line mode at 5 W output power at either 488 nm or 514 nm wavelength. A mechanical shutter (NM Laser Products LS200) controlled by the spectrometer was used to produce 100 ms light pulses. The light was focused into a 6 m length of optical fibre (Newport F-MBE), using a Newport M-5X objective lens. The other end of the fibre was attached (*via* Newport

SMA connectors) to a 2 m section of the same fibre whose stepwise-tapered tip (described in detail in Chapter 4 and published in ref. [180]) was held inside a 5 mm NMR tube by a truncated Wilmad WGS 5BL coaxial insert.

During each CIDNP experiment, the sample was irradiated for 100 ms and subjected to a 90° radiofrequency pulse on $^{19}$F followed by immediate acquisition of the free induction decay. After application of a shifted Gaussian window function, zero-filling and Fourier transformation, the spectra were analyzed using mixed Lorentzian-Gaussian line fitting.

### 6.3 Theoretical framework

The relaxation of $^{19}$F nuclei in GFP-sized molecules in solution occurs primarily due to rotational modulation of two anisotropic interactions: the dipole-dipole interaction with nearby protons (DD mechanism) and the Zeeman interaction with the applied magnetic field (chemical shift anisotropy, or CSA mechanism). A relaxation theory treatment described in Chapter 1 results in the following expressions for the longitudinal and transverse relaxation rates of the fluorine nucleus:

$$\frac{1}{T_1} = \frac{1}{4}\left(\frac{\gamma_{\mathrm{H}}\gamma_{\mathrm{F}}\hbar\mu_0}{4\pi r_{\mathrm{HF}}^3}\right)^2 \left(3J(\omega_{\mathrm{F}}) + 6J(\omega_{\mathrm{H}}+\omega_{\mathrm{F}}) + J(\omega_{\mathrm{H}}-\omega_{\mathrm{F}})\right) + \frac{\Delta\sigma^2 B_0^2 \gamma_{\mathrm{F}}^2}{3} J(\omega_{\mathrm{F}}) \tag{6.1}$$

$$\begin{aligned}\frac{1}{T_2} = {} & \frac{1}{8}\left(\frac{\gamma_{\mathrm{H}}\gamma_{\mathrm{F}}\hbar\mu_0}{4\pi r_{\mathrm{HF}}^3}\right)^2 \left(4J(0) + J(\omega_{\mathrm{H}}-\omega_{\mathrm{F}}) + 3J(\omega_{\mathrm{F}}) + 6J(\omega_{\mathrm{H}}) + 6J(\omega_{\mathrm{H}}+\omega_{\mathrm{F}})\right) + \\ & \frac{\Delta\sigma^2 B_0^2 \gamma_{\mathrm{F}}^2}{18}\left(4J(0) + 3J(\omega_{\mathrm{F}})\right)\end{aligned} \tag{6.2}$$

in which all the symbols have their usual meanings [20] and the expression for the spectral density, $J(\omega)$, depends on the motional model used. The isotropic tumbling approximation yields

$$J(\omega) = \frac{2}{5}\frac{\tau_{\mathrm{c}}}{1+\omega^2\tau_{\mathrm{c}}^2} \tag{6.3}$$

where $\tau_{\mathrm{c}}$ is the rotational diffusion correlation time, while the Lipari-Szabo restricted motion model [16, 26] has

$$J(\omega) = \frac{2}{5}\left(\frac{S^2\tau_{\mathrm{c}}}{1+\omega^2\tau_{\mathrm{c}}^2} + \frac{(1-S^2)\tau_{\mathrm{e}}}{1+\omega^2\tau_{\mathrm{e}}^2}\right) \tag{6.4}$$

in which $S^2$ is a generalized order parameter indicating the extent to which motion is restricted, $\tau_c$ is the global molecular rotational correlation time (assumed to be isotropic) and

$$\tau_e = \left(\frac{1}{\tau_c} + \frac{1}{\tau_i}\right)^{-1} \tag{6.5}$$

in which $\tau_i$ is the internal motional correlation time of the residue in question.

A direct estimate shows that for the fluorine nucleus of a 3-fluorotyrosine in GFP, the contribution to the relaxation rate from the CSA mechanism is always at least an order of magnitude greater than from the DD mechanism:

$$\frac{1}{8}\left(\frac{\gamma_H \gamma_F \hbar \mu_0}{4\pi r_{HF}^3}\right)^2 << \frac{\Delta\sigma^2 B_0^2 \gamma_F^2}{18} \tag{6.6}$$

The DD part of the relaxation rate may therefore be neglected, and the expression for the transverse relaxation rate takes the following form:

$$\frac{1}{T_2} = \frac{\Delta\sigma^2 B_0^2 \gamma_F^2}{18}\left(4J(0) + 3J(\omega_F)\right) \tag{6.7}$$

It has relatively recently become possible to estimate values of the chemical shift tensor anisotropy parameter, $\Delta\sigma$, from *ab initio* calculations [35]. Values of the anisotropy of the $^{19}F$ chemical shielding tensor were obtained using the Gaussian03 program [76] at three different levels of theory (Table 6.1). The most computationally expensive method, B3LYP/6-311++G(2d,2p), is a general-purpose technique for chemical shift calculation that has been shown to provide reasonably accurate results for most Period II elements; the CSGT DFT B3LYP/cc-pVDZ method is the one recommended by Tormena *et al.* [181] for accurate calculation of $^{13}C$ chemical shifts. While, in general, one would expect a better estimate from a higher level method, $^{19}F$ chemical shift calculations are known to disobey this rule [120], and the least computationally expensive GIAO HF / 6-311++G(2d,2p) method yields values of the $^{19}F$ chemical shifts that are superior to those obtained with DFT and MP2 [35]. We therefore chose to use the GIAO HF value for the relaxation analysis. DFT methods are known to overestimate $^{19}F$ shielding in fluorinated aromatics [120], something that we also see here.

Once $\Delta\sigma$ is known, measuring relaxation times at two or more different magnetic fields allows extraction of the order parameters and the correlation times. It is particularly convenient to use $T_2$ values for fluorinated proteins, both because of signal-to-noise and degassing difficulties associated with $T_1$ measurements, and because, for a signal broader than approximately 50 Hz, it is unnecessary to perform a spin-echo $T_2$ experiment, since accurate estimates of $T_2$ values may simply be obtained from line widths, assuming that the $B_0$ field has been shimmed properly.

**Table 6.1.** Computed values of $^{19}$F chemical shift anisotropy.

| Method (geometry / NMR) | $\left\lvert \sigma_{zz} - \frac{\sigma_{xx} + \sigma_{yy}}{2} \right\rvert$ / ppm |
|---|---|
| DFT B3LYP 6-311++G(2d,2p) in vacuum / GIAO DFT B3LYP 6-311++G(2d,2p) in vacuum | 135 |
| DFT B3LYP 6-311++G(2d,2p), zwitter-ion in PCM water / CSGT DFT B3LYP cc-pVDZ, zwitter-ion in PCM water | 119 |
| DFT B3LYP 6-311++G(2d,2p) in vacuum / GIAO HF 6-311++G(2d,2p) in vacuum | 95 |

## 6.4 Results and discussion

### *6.4.1 $^{19}$F signal assignment and NMR data*

The signal assignment was achieved using a combination of site-directed mutagenesis, photo-CIDNP methods and relaxation theory arguments. Wild-type GFP has ten tyrosine residues distributed throughout the structure (Figure 6.3). The $^{19}$F NMR spectrum of wild-type GFP shows a distinctive spread of $^{19}$F chemical shifts from −50 to −60 ppm (Figure 6.4), as expected for a folded protein. To assign the $^{19}$F NMR spectrum, each of the ten tyrosine residues was mutated (by Dr. F. Khan, Cambridge University), one at a time, to phenylalanine. If the mutation does not significantly perturb the structure, the spectrum of each of the tyrosine mutants should lack one or two (doubling may be caused by ring-flip isomers described below) signals compared to the spectrum of fully $^{19}$F-Tyr labelled wild-type protein.

For Y92F, Y106F, Y143F, Y182F, the spectra clearly lacked one or two signals (Figure 6.4) allowing these tyrosines to be assigned directly. In other cases, however, the mutation resulted in noticeable chemical shift changes (e.g. Y145F mutant, Figure 6.4), most likely a result of small structural perturbations caused by the loss of the polar hydroxyl group on mutation. This variation in the signal positions between mutants complicated the assignment of other tyrosine residues and, in some cases, lead to ambiguity. In these cases, the assignment was made on the basis of a line-fitting analysis of both NMR and photo-CIDNP data for each mutant and, in some cases also from the relaxation theory arguments, as described in detail below.

For Tyr66, which forms a part of the chromophore, the mutant protein (Y66F) could not be prepared in sufficient yield to acquire an NMR spectrum, most likely because the Tyr→Phe mutation in the chromophore results in the loss of the extensive hy-drogen bond network and destabilizes the protein. However, a relaxation theory argument may be used to assign this residue. After the post-translational cyclization, Tyr66 is incorporated into the chromophore, which is known to be a very rigidly immobilized structure. It should therefore be expected that Tyr66 would have the longest rotational correlation time among all tyrosines and, therefore the broadest signal. By far the broadest signal in the spectrum is located at around −55 ppm (Figure 6.4). However, in the Y74F mutant spectrum (Figure 3.4), that signal partially vanishes, indicating, along with its intensity and shape, that it is a superposition of Y66 and Y74 signals. The spectra labelled Y92F, Y151F, Y106F and Y39F in Figure 6.4 allow an unambiguous line-fitting analysis of this composite signal, resulting in the assignments and order parameter values reported in Table 6.2.

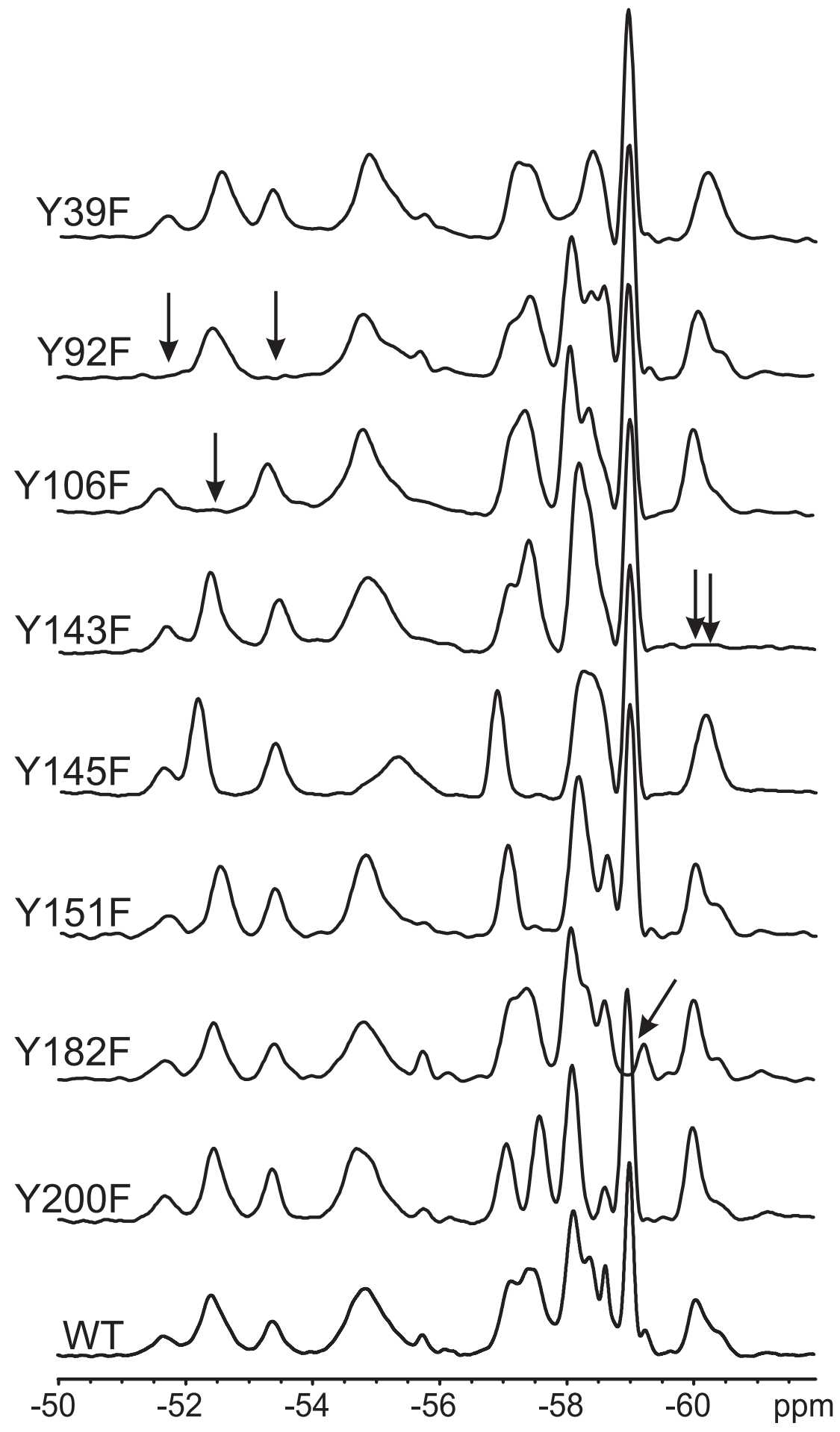

**Figure 6.4.** 564 MHz 19F spectra of wild-type and Tyr → Phe mutant GFPs used for signal assignment. For Y92F, Y106F, Y143F and Y182F the arrows indicate the loss of a signal corresponding to that residue. For Y92F and Y143F two signals are lost as these residues populate two distinct rotameric states (see section on ring-flip isomers).

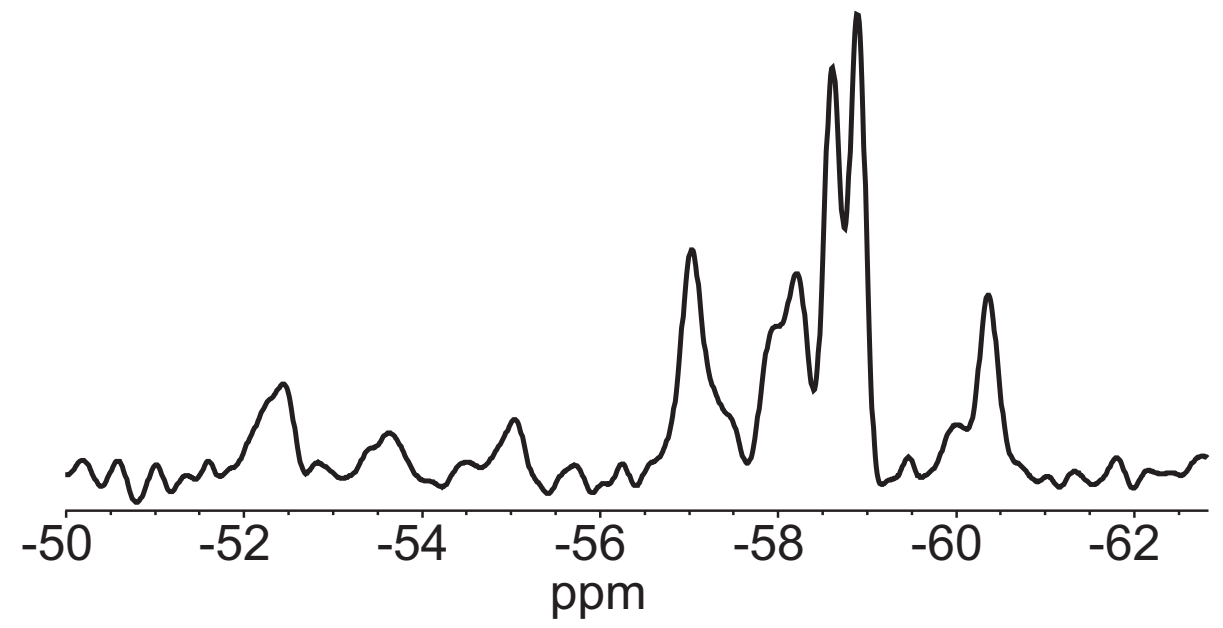

**Figure 6.5.** $^{19}$F NMR spectrum of the Y74F mutant, which was collected at a different temperature to the mutants shown in Figure 6.3 and consequently demonstrates a different ring-flip isomer distribution for Y92 and Y143, and also a lack of signal at around −55 ppm.

The chemical shift data presented in Table 6.2 represents the assignment that is consistent with over 95% of the available data. It is however conceivable that the chemical shift perturbations caused by mutation are substantial around 58 ppm, so the signal assignment in that region should be treated with caution.

### *6.4.2 Ring-flip isomers*

An interesting feature of 3-fluorotyrosinated proteins is the existence of fluorotyrosine ring-flip isomers. Two $^{19}$F NMR signals per fluorotyrosine residue can be expected, because the chemical environment around the two possible fluorine positions, interchanged by ring flips, will, in general, be different.

Two such signal pairs (Tyr92 and Tyr143) are observed in the wild-type GFP spectrum (Figure 6.4) with the rightmost pair of signals, belonging to Tyr143, displaying the characteristic behaviour: the relative intensity of the two signals is not constant, the signals sometimes coalesce into one apparent signal, and both signals vanish simultaneously when the residue is mutated (as in the Y143F spectrum, Figure 6.4). The same behaviour is also demonstrated by Tyr92; in this case, however, the signals of the two isomers are substantially further apart. All spectra also contain two to five low-intensity peaks of variable intensity which are likely to be weakly populated rotamers of the other fluorotyrosine residues.

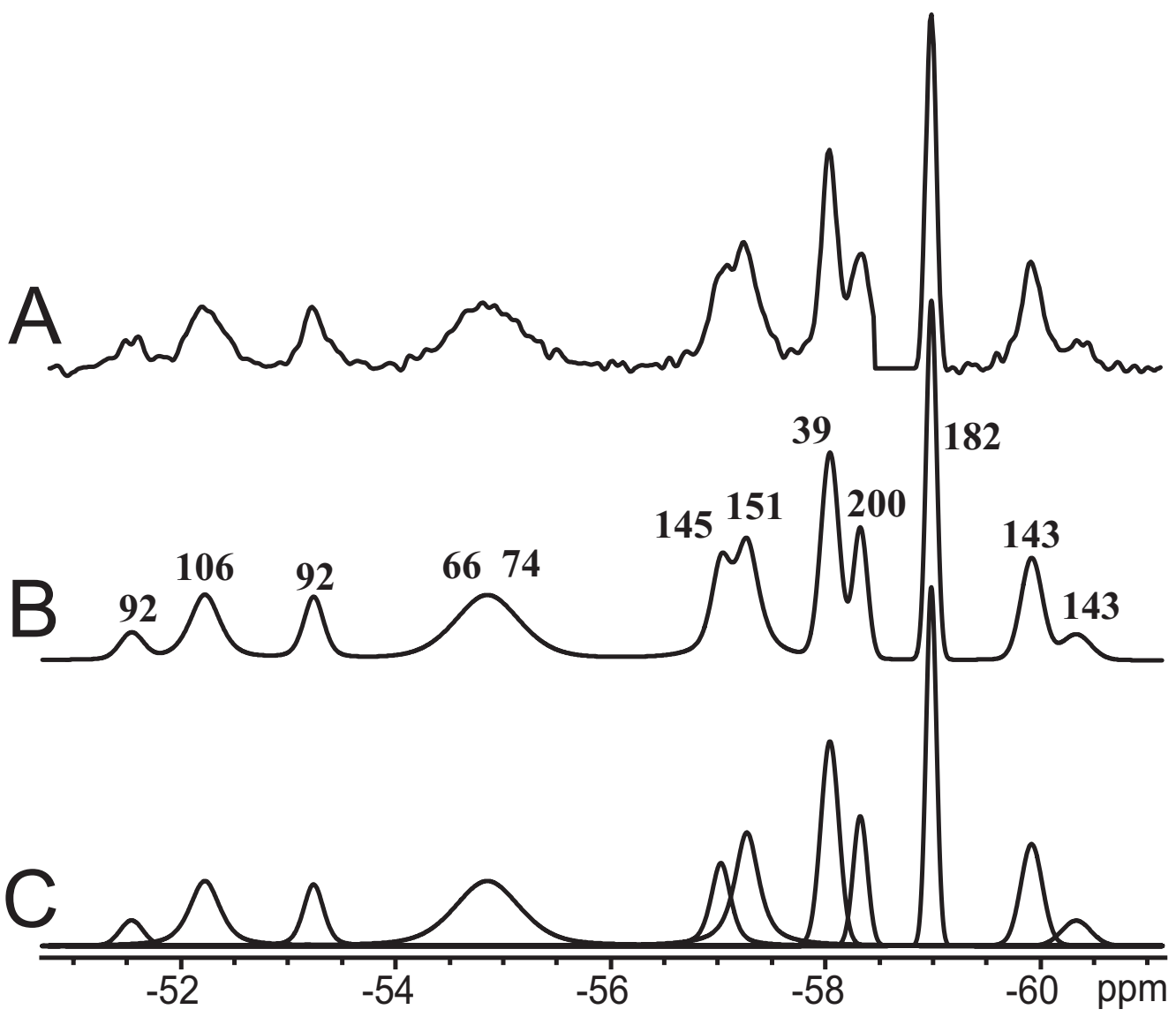

**Figure 6.6.** $^{19}$F assignment of GFP. **(A)** 564 MHz $^{19}$F spectrum of GFP with Lorentzian-to-Gaussian resolution enhancement. **(B)** Line fitting. **(C)** Individual lines in the fitting. The narrow signal from free 3-fluorotyrosine in spectrum (A) at −58.7 ppm becomes spurious under the action of the shifted Gaussian window function and has been zeroed to facilitate line fitting. Because of fast scan repetition rates the signal intensities are weighted by relaxation rates and are therefore not proportional to the number of spins.

The signals from Tyr106 and Tyr145, which are also buried (Table 6.2), do not appear to be split to such a degree[10]. This may be because the two isomers have identical chemical shifts, or because one isomer is highly favored over the other due to specific packing interactions in the native state. Alternatively, although the crystal structure indicates that these side-chains are buried and rigidly held, local breathing motions may allow the isomers to interconvert rapidly. In any case, the splitting of the $^{19}$F signal for Tyr92 and Tyr143 gives a sensitive probe of the local conformational flexibility of the protein, which can be utilized in future unfolding/folding experiments.

### *6.4.3 Relaxation analysis.*

It is known that the chromophore of GFP is a very rigidly immobilized structure [171, 172], which must therefore tumble with the overall molecular correlation time. Assuming $S^2$=1 for the chromophore in the isotropic overall tumbling approximation yields an overall molecular rotational correlation time of 17 ± 1 ns. The value obtained from the Stokes-Einstein equation is 16 ns; the experimental number is larger likely due to the elongated shape of the GFP barrel. Using $T_2$ values measured at two different field strengths, it is possible to solve Equation (6.7) for the Lipari-Szabo generalized order parameters ($S^2$) [16, 26]. The values of ($S^2$) for the ten fluorotyrosine residues of GFP are listed in Table 6.2. Tyr66 was assumed above to be immobile and the other residues show different extents of side-chain mobility, with the values of $S^2$ in good agreement with both CIDNP and the solvent accessibility data.

[10]There is, however, evidence of minor populations of other ring-flip isomers, as a number of unassigned low-intensity peaks are present in the spectrum.

**Table 6.2.** Assignment of the chemical shifts, the values of the generalized order parameters, the residue accessibilities and HOMO accessibility data for the ten tyrosines in GFP. The residues whose signals appear in the photo-CIDNP spectra are shown in bold.

| Residue | Chemical shift[a] /ppm | Line width, @ 14.1 T | Line width, @ 9.4 T | $S^2$ | SASA[b] % | SASA[c] % | HOMO acc-ty.[c,d] |
|---|---|---|---|---|---|---|---|
| Y92(I) | −51.65 ± 0.04 | 243 ± 17 | 101 ± 16 | 0.66 ± 0.08 | 0.6 | 0.0 | 0.0 |
| Y106 | −52.40 ± 0.08 | 259 ± 16 | 103 ± 9 | 0.69 ± 0.06 | 0.0 | 0.0 | 0.0 |
| Y92(II) | −53.36 ± 0.04 | 216 ± 8 | 84 ± 9 | 0.58 ± 0.03 | 0.6 | 0.0 | 0.0 |
| Y66/74 | −54.86 ± 0.10 | 381 ± 16 | 150 ± 14 | 0.99 ± 0.07 | 0.5/1.9 | 0.0 | 0.0 |
| Y145 | −57.06 ± 0.04 | 205 ± 9 | 82± 12 | 0.55 ± 0.05 | 6.4 | 2.6 | 0.0 |
| **Y151** | **−57.41 ± 0.06** | **194** ± 11 | **68** ± 12 | **0.51 ± 0.06** | **35.3** | **21.8** | **12.8** |
| **Y39** | **−58.11 ± 0.04** | **166** ± 13 | **68** ±2 | **0.43 ± 0.05** | **45.0** | **35.1** | **21.7** |
| **Y200** | **−58.36 ± 0.02** | **216** ± 17 | **125** ± 19 | **0.61 ± 0.10** | **21.2** | **12.8** | **9.9** |
| **Y182** | **−58.98 ± 0.02** | **86** ± 3 | **40** ± 1 | **0.24 ± 0.02** | **31.4** | **17.3** | **11.8** |
| Y143(I) | −60.08 ± 0.06 | 179 ± 4 | 109 ± 13 | 0.50 ± 0.04 | 22.8 | 10.6 | 1.0 |
| Y143(II) | −60.37 ± 0.06 | 158 ± 6 | 109 ± 13 | 0.45 ± 0.04 | 22.8 | 10.6 | 1.0 |

[a]The signal of residual free 3-fluorotyrosine was used as a chemical shift reference, at −58.60 ppm. [b]Solvent accessible surface area (SASA) 0.14 nm probe used. [c]0.30 nm probe used. [d]Atomic accessibility weighted by Mulliken atomic population for HOMO, the latter obtained from a DFT B3LYP 6-311G** *ab initio* calculation using GAMESS program [77].

The analysis above relies on the isotropic tumbling approximation of the GFP molecule. This is not strictly true, and an estimate using HYDRONMR program [182] and the X-ray structure of GFP results in the $D_{\|}/D_{\perp}$ ratio of 1.28, meaning that the order parameters will show additional dependence on the orientation of the chemical shielding tensor with respect to the diffusion tensor, which will in general be different for every residue. The resulting shift in the order parameter estimates may be as high as 20%. The experimental data collected in the present work is not sufficient to reliably estimate the rotational diffusion anisotropy and the angle between the diffusion tensor and the chemical shielding tensor. If and when such data becomes available, an analysis with the axial diffusion model would be very beneficial.

### *6.4.4 Photo-CIDNP experiments*

$^{19}$F photo-CIDNP spectra were recorded for native and denatured samples of GFP including denatured states in acid (pH 1.5, 2.5 and 2.9) and 6 M GdnHCl. These conditions were chosen as they are being used in kinetic refolding experiments [183]. Native GFP strongly absorbs at the usual photo-CIDNP excitation wavelength of 488 nm, so the illumination system was modified to use fibre optics specifically designed to deal with optically dense samples (see Chapter 4 for a detailed description). It was also

found that the UV lines of the argon ion laser degrade GFP, so the experiment was further adjusted to operate in a single-line rather than multi-line mode at a wavelength of 514 nm. The results for native and denatured GFP are shown in Figures 6.7 and 6.8. The folded GFP spectrum exhibits four photo-CIDNP peaks, corresponding to tyrosine residues 39, 151, 182, and 200. Three of the four mutants (Y92F, Y106F, Y143F) for which spectra were obtained also exhibit these same four peaks as expected. The spectrum of Y151F, however, has only three peaks, the peak corresponding to Tyr151 having been lost.

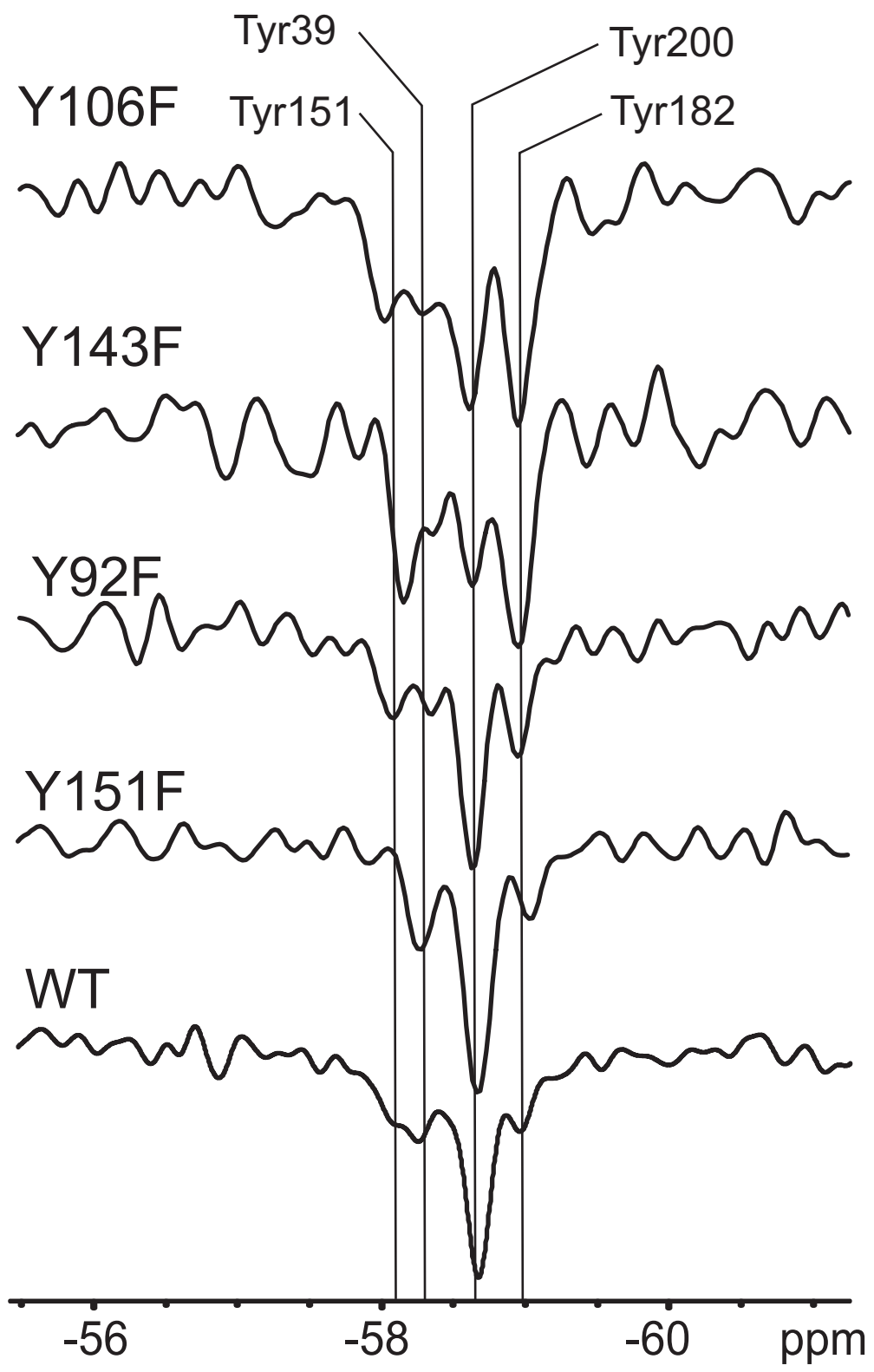

**Figure 6.7.** $^{19}$F photo-CIDNP spectra of wild-type and mutant GFP used for the signal assignment and the assessment of residue accessibilities.

In general, it is considerably more difficult to obtain detailed structural information on the denatured states of proteins than on native states. When the protein is unfolded, the chemically distinct $^{19}$F environments in the native state are lost, collapsing the NMR signals into a peak with the same chemical shift as free 3-fluoro-tyrosine (Figure 6.8). The small signal on the flank of this large peak belongs to the chromophore, which has a chemical shift slightly different from 3-fluorotyrosine due to the chemical modifications associated with chromophore formation. The photo-CIDNP spectra of denatured GFP at pH 1.5, 2.5 and 6 M GdnHCl are also very similar, and exhibit a positive enhancement. In contrast, the pH 2.9 denatured state exhibits both a positive and negative polarization. As the $^{19}$F CIDNP amplitude is related to the correlation time (see Chapter 8 for the theoretical details), it would appear that there is some heterogeneity in the size/mobility of the molecule at this pH. This could be due to two structures in equilibrium, one slightly more compact than the other. This is consistent with far-UV circular dichroism studies of the pH 2.9 denatured state which has considerably more secondary structure than the pH 1.5 or GdnHCl-unfolded state, and recent small-angle X-ray scattering data [183]. Residual structure, such as this, may be very important in the folding mechanism of GFP, effectively restricting conformational space at a very early stage during the folding process. It may also explain the results of folding measurements of GFP which suggest that folding is considerably faster from the pH 2.9 denatured state than from the pH 1.5 denatured state [183].

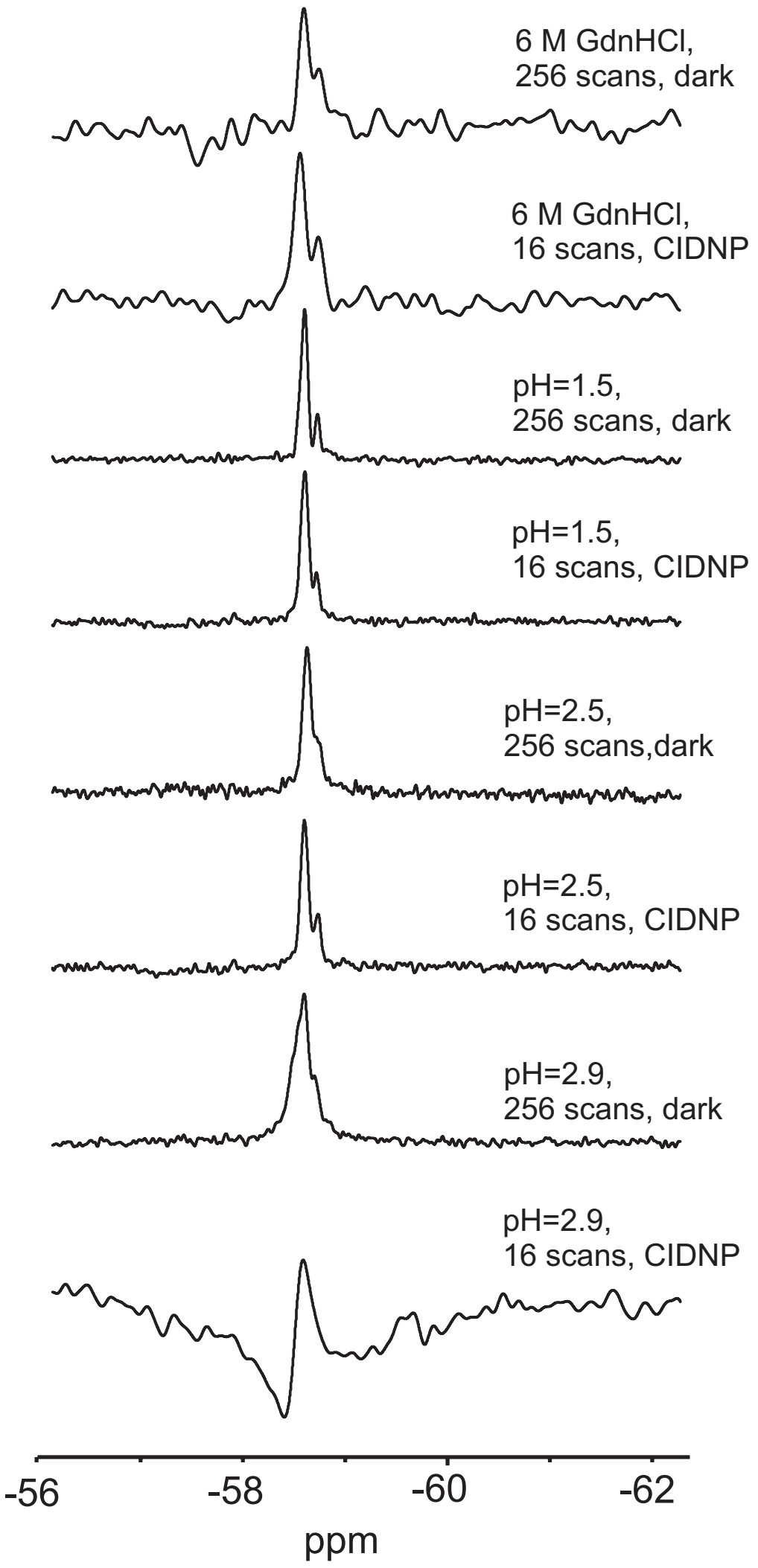

**Figure 6.8.** $^{19}$F NMR and photo-CIDNP spectra of denatured states of GFP.

### *6.4.5 Solvent accessibility data*

The CIDNP generation mechanism involves an electron transfer from an amino acid residue to the excited photosensitizer [184]. It has been recognized for a long time that electron transfer is only possible if the amino acid residue is exposed to the bulk solvent [60, 185]. However, attempts to correlate the solvent accessibility of a given residue with the observed photo-CIDNP effect are not always totally successful [47]. In the case of GFP, we also find that the amino acid accessibility does not completely predict either the existence of the CIDNP effect or its amplitude (Table 6.2). A more detailed analysis of the solvent accessibility data, in which the accessibility of the HOMO (from which the electron transfer actually occurs) is used rather than the solvent accessibility of the entire side chain, was therefore performed.

The accessibility data were calculated using the web-based GetArea program [186] and the crystal structure (PDB code 1b9c) [187], using either 1.4 or 3.0 Å as the radius of the solvent probe (to model accessibility by water and FMN, respectively). HOMO accessibilities were then computed as the weighted sum of atomic accessibilities, the weights corresponding to the Mulliken HOMO density on each atom.

The HOMO accessibility calculation produced a significant improvement in the prediction (Table 6.2). This finding is consistent with the fact that the electron transfer occurs from the HOMO of a solvent-exposed tyrosine to the excited photosensitizer. This can be illustrated by considering Tyr143 *versus* Tyr200. Tyr143 has the same overall solvent accessibility as Tyr200, but only the latter shows a photo-CIDNP effect. Visual inspection of the residues in the crystal structure [187] shows that the region of Tyr143 which protrudes into the solvent is an unreactive backbone fragment that bears no HOMO electron density, whereas it is the aromatic side chain that is exposed for Tyr200. In this case, the computed HOMO accessibilities are completely consistent with the observed photo-CIDNP effects.

# Chapter 7
# *Microsecond time-resolved photo-CIDNP instrument*

*Based on: I. Kuprov, M. Goez, P.A. Abbott, P.J. Hore,*
*Review of Scientific Instruments 86, 084103 (2005).*

This chapter describes the design and operation of microsecond time-resolved photo-CIDNP hardware, designed to go with a commercial 600 MHz NMR spectrometer. A frequency-tripled Nd:YAG laser is used as the light source, with a system of mirrors or prisms to route light to the NMR sample from above, removing the need for NMR probe modifications. The experiment has been designed for a shared NMR spectrometer and is straightforward and inexpensive to assemble and operate.

## 7.1 Introduction

The two main experimental techniques for photo-CIDNP NMR are the continuous-wave (CW) and time-resolved methods. CW photo-CIDNP is generally more popular, not least because it is relatively straightforward to implement using UV lamps [188] or CW lasers together with fibre optics [75, 189]. Prolonged irradiation of NMR samples can also lead to strong polarizations, 20- to 50-fold magnetization enhancements being quite common [52]. However, with illumination times of 50-5000 ms, a CW photo-CIDNP experiment detects a time-averaged or steady state value of the photochemically generated magnetization, which yields mechanistic information, but precludes kinetic studies.

The time-resolved photo-CIDNP technique, using a pulsed laser as the light source, is best suited for exploring events on a microsecond to millisecond timescale in photoinduced spin-selective reactions [57, 123, 127]. Time-resolved photo-CIDNP experiments generally require light pulses much shorter than a microsecond and pulse energies of at least 50 mJ. The resulting incident optical power densities are incompatible with optical fibres and rule out the use of illumination schemes that have been developed in the more mainstream CW CIDNP and optical NMR field [72, 189].

The pioneering work by Miller and Closs [190] a quarter of a century ago described a time-resolved photo-CIDNP installation based on an electromagnet 60 MHz FT NMR spectrometer, in which the laser- and RF-control electronics were custom-made and, of necessity, rather complex, but light routing was straightforward because the coil region of the probe was easily accessible from outside the magnet [191]. With the introduction of superconducting magnets the geometry around the sample volume has become much more constrained and difficult to access, such that sample irradiation can require, at the very least, "moving a few capacitors and drilling a few holes" in the probe [47], and sometimes also sacrificing one of the probe RF coils for synchronization purposes [192]. Hitherto, the method of choice for high-field time-resolved CIDNP studies has been to use a ≈5 mm diameter, cylindrical fused silica light guide passing off-axis through the body of the NMR probe, surmounted by a prism to bring the light into the NMR sample from the side through a window in the radiofrequency (RF) coil [45, 47, 193]. Another popular variant has the light guide on-axis, bringing light from below into the sample contained in a flat-bottomed NMR tube. In both cases, probe hardware modifications are inevitable [47]. Routing light from below has also become more difficult because the sophisticated and fragile temperature-control equipment needs to be relocated. Even if the latter can be achieved, the resulting axial asymmetry leads to field homogeneity distortions. Moreover, such probe modifications normally invalidate the manufacturer's warranty. With cryoprobes gaining in popularity, it is becoming obvious that "do-it-yourself" probe modification is no longer an attractive option, except for highly specialized applications.

The requirements for a modern time-resolved-CIDNP installation are: low cost, ease of assembly, use of readily available commercial components, and compatibility with multi-user high-field NMR spectrometers without the need for the modification of either probes or RF and synchronization circuitry. All of this has gradually become possible in recent years with the appearance of powerful and flexible pulse programming languages and commercial lasers with built-in TTL synchronization lines, thus eliminating the need for constructing the more complicated parts of the Miller and Closs design [190].

This Chapter details the design and operation of a time-resolved photo-CIDNP system based on a frequency-tripled Nd:YAG laser and a shared 600 MHz NMR spectrometer equipped with multi-user probes that could not be modified to permit sample irradiation. The utility of the approach is not restricted to photo-CIDNP measurements, other applications being microsecond time-resolved studies of photochemical reactions [57, 127, 128] and photo-induced transformations of biomolecules [129-132].

## 7.2 Time-resolved CIDNP hardware: general considerations

### *7.2.1 Light routing*

The main challenge in designing time-resolved CIDNP hardware is efficient and robust light routing. To preserve the magnetic field homogeneity, the laser cannot be brought closer to the superconducting magnet than about 1.5 m. The entrance to the magnet bore is 2 m or more above floor-level, and the sample, in the centre of the magnet, is more than a 1 m below the top of the bore. Modern NMR magnets are suspended on air bearings, which are normally sufficiently stable to keep the horizontal drift below 0.5 mm a day, the primary cause of drift being changes in mass due to cryogen evaporation and refill. With access only available from above, there are several possible means of bringing the high-intensity pulsed laser light into the active region of the NMR probe.

***Fused silica light pipes:*** an expensive and fragile solution that would require complete removal of the section of light pipe inside the magnet bore every time the sample was changed. Silica rods decollimate the light and are prone to gradual deterioration under intense UV light, forming cracks and “snow” along the beam path. Because of their rigidity, light pipes are an unattractive option since the vertical position of the magnet may change by as much as a few centimetres during a cryogen evaporation/refill cycle.

***Liquid light guides:*** in evaluating this option it was found that the decollimation and optical power loss (at 355 nm) in a 5 m length of a liquid light guide is unacceptable. The fused silica terminal windows of the light guide suffer the same kind of damage as solid silica light pipes. Furthermore, the iron alloy ferrules on most liquid light guides are incompatible with the superconducting magnet, and bespoke ferrules tend to be expensive. Liquid light guides are also insufficiently flexible and quite heavy, making it difficult to position a sample tube attached to a light guide, inside the magnet bore.

***Optical fibres and fibre bundles:*** attempts to send a 10 ns, 100 mJ UV light pulse through a single optical fibre of any diameter usually result in destruction of the fibre. Fibre bundles terminated by large diameter fused silica rods were also considered, but were ruled out for cost reasons.

***Through-air delivery using prisms and/or mirrors:*** was found to be the most satisfactory and cost-effective option. Three fused silica prisms or 355 nm dichroic mirrors and one long-focus fused silica lens are required to route the light to the sample and reduce the beam diameter from 8 mm at the laser aperture to 4 mm at the sample entrance (see detailed description below).

### *7.2.2 NMR sample considerations*

In a typical photo-CIDNP experiment it is often a requirement that the NMR sample is deoxygenated and isolated from the atmosphere [194]. Sample tubes can be kept air-tight by inserting a short, tightly fitting quartz rod, extending from the mouth of the tube to the coil region, which acts simultaneously as a light guide [188]. Alternatively, a smaller diameter $D_2O$-filled NMR tube placed inside the sample tube can fulfil the same role. With the latter, it is difficult to reproduce the optical power that reached the sample due to variable light scattering at the water meniscus. Fused silica rods, 20 cm long and 4 mm in diameter, with optically polished ends were therefore preferred. These are easily replaced if substantial photodamage occurs, and with cylindrically shaped rods, the bubbles that tend to form between the solution layer and the lower end of the rod can be expelled without difficulty.

### *7.2.3 Laser operation and synchronisation*

Pulsed gas (e.g. excimer) lasers — extensively used in the past for time-resolved CIDNP experiments [57, 190, 195] — are convenient in that they do not require an extended warm-up period to obtain a reproducible output, but are problematic due to their poor collimation and large beam cross-section. Doped YAG lasers, on the contrary, have excellent collimation, but require an extended warm-up period with the necessity to flash the lamp repetitively to achieve a stable optical output. Since efficient light routing is the primary concern, a frequency-tripled Nd:YAG laser was chosen as the light source in the hardware described below. The continuous lamp operation and the synchronization of the Q-switch and the RF pulse sequence turned out to be straightforward to implement in software.

## 7.3 Hardware arrangement and operation

The time-resolved photo-CIDNP setup described in this Chapter is shown schematically in Figures 7.1 (block diagram) and 7.2 (geometric layout). The Varian Inova 600 NMR spectrometer has five spare TTL lines that can be triggered from within the pulse sequence, allowing convenient control of external equipment. The arrangement described here uses three lines, two of which trigger the lamp and the Q-switch of the Nd:YAG laser; the third controls the optional safety interlock shutter. The required synchronization accuracy (~100 ns) is easily met by the built-in TTL lines of both spectrometer and laser.

At 355 nm, prisms have the advantage of being less expensive than high-energy dichroic mirrors, but each prism results in a ~10% optical power loss due to light scattering at its surfaces. A prism-based light routing system almost inevitably has some stray reflections, which may be put to good use by placing a pre-biased fast photodiode into the path of the stray beam and using the resulting signal to check the timing accuracy of the experiment. The pulse widths and timings specified in the pulse sequence were found to be reproduced by the Inova 600 hardware with an apparent accuracy of better than 10 ns.

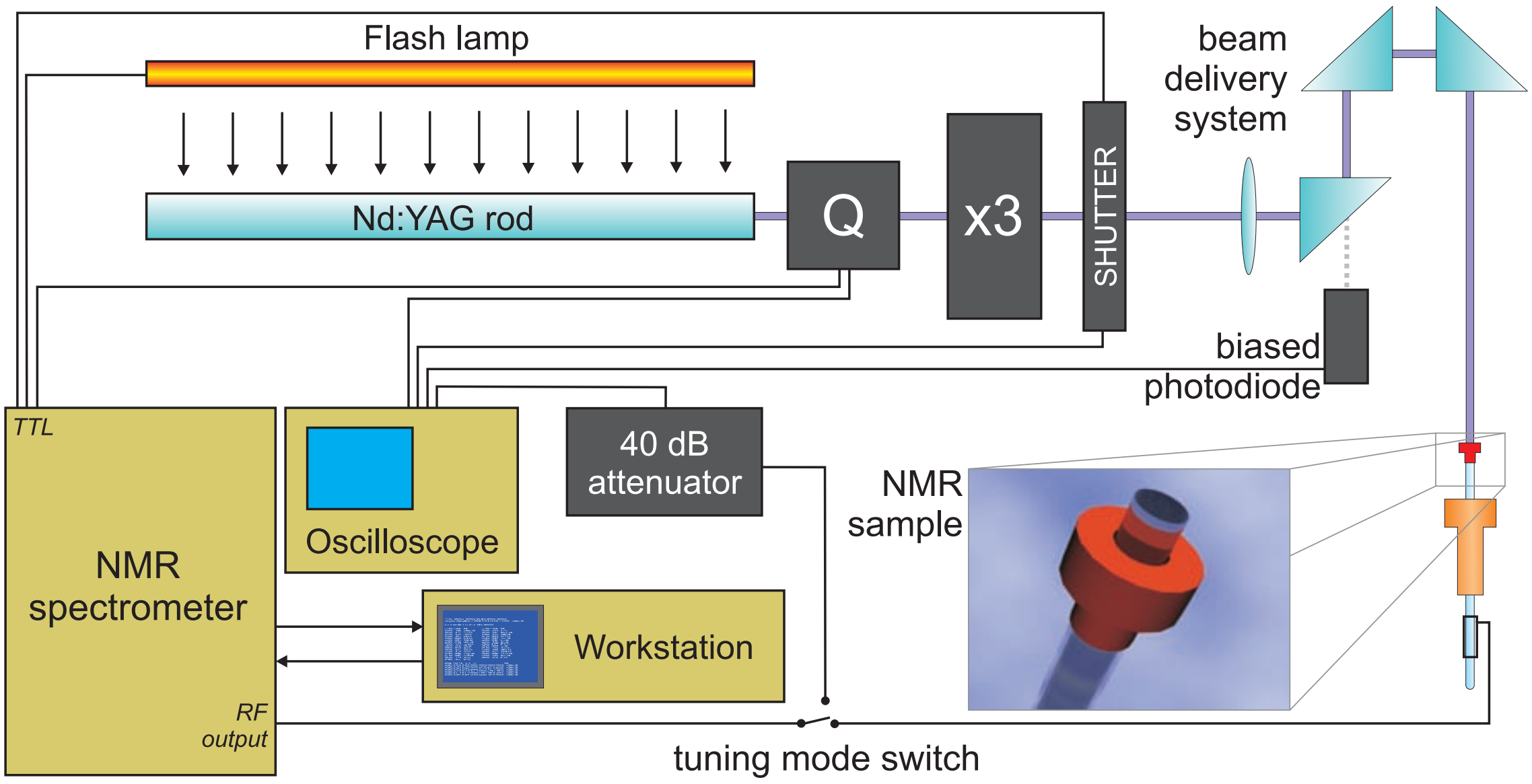

**Figure 7.1** Block diagram of the time-resolved photo-CIDNP experiment using a frequency-tripled (355 nm) Nd:YAG laser and sample illumination from above (to avoid the necessity for NMR probe modifications). The lamp, the Q-switch (Q) and the optional safety interlock shutter are controlled by the spectrometer via built-in TTL lines triggered from within the pulse sequence with event timing accuracy monitored on the oscilloscope. The beam-delivery system comprises frequency mixing crystals (x3), a long-focus silica lens and either three 355 nm dichroic mirrors or three fused silica prisms; the light is directed to the NMR sample from above through the magnet bore. A 4 mm diameter fused silica rod is used both to guide light to the active region of the sample and to isolate the sample from the atmosphere.

The beam delivery system may contain two or three deflectors (either prisms or mirrors). The two-prism configuration (without Prism B, Figure 7.2), requiring the beam to be sent at an oblique angle across the room, was ruled out on safety grounds. A three-prism configuration was therefore implemented. Prism A sends the beam vertically towards the ceiling, where Prism B directs it horizontally, well above eye-level, to a position above the magnet. Prism C, mounted exactly above the centre of the magnet bore, sends the beam downwards towards the sample. Ideally, all prisms need $X,Y,Z$ translation and $\theta,\phi$ rotation stages to facilitate alignment.

Alignment of the beam with respect to the sample may be achieved by putting an empty NMR tube plus spinner in the magnet, removing the probe and maximizing the signal from a photodiode placed coaxially beneath the magnet. A frosted glass plate in the beam path just above the photodiode ensures that it reports the overall light amplitude rather than accidental focussing spots generated by the bottom of the NMR tube. The minimum requirements for the degrees of freedom of the three prisms or mirrors are given in Figure 7.2. Once the arrangement has been set up, only Prism C needs regular (every few days) adjustment, a consequence of small magnet movements arising from cryogen evaporation. Seasonal changes in the geometry of the building may occasionally (once every few months) make it necessary to adjust Prisms A and B. It is convenient to monitor the beam position at the sample entrance using a paper collar placed around the quartz rod. Burn marks on the collar indicate the need for realignment.

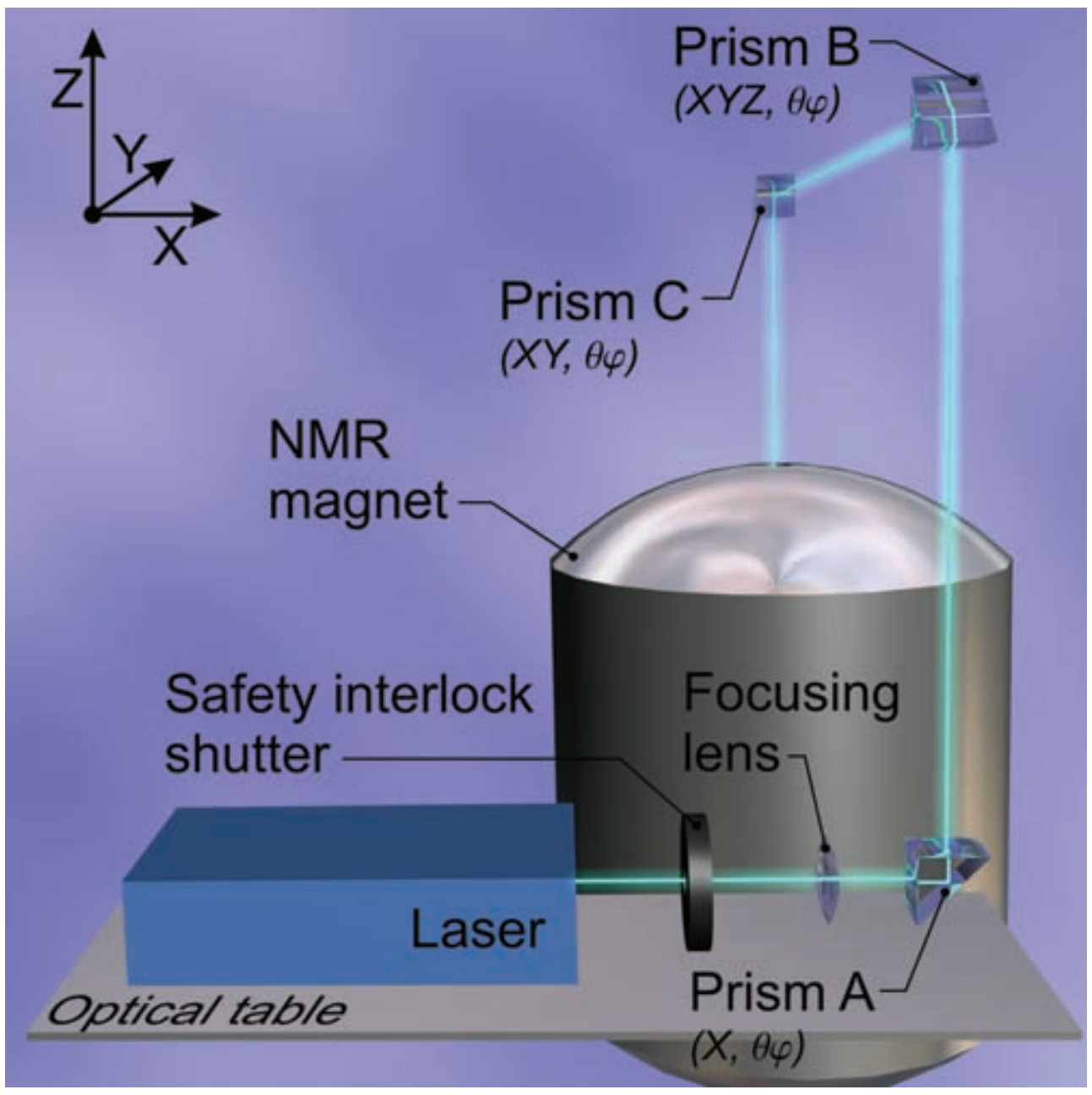

**Figure 7.2** Schematic drawing of the spatial hardware arrangement for the time-resolved photo-CIDNP experiment. Ideally, all three deflectors (prisms or mirrors) should have full rotation and translation stages to facilitate alignment (minimum requirements for the degrees of freedom are shown in the picture).

The lens (at least 5 m focal length was found to be required in practice) may be mounted anywhere along the beam path, the most convenient location being the optical table. Care must be taken to avoid focussing the beam to a point anywhere along its trajectory. If the beam diameter is reduced below 3 mm, snow-like lesions usually form within the fused silica optical components, resulting in a degraded performance. The beam must also pass far enough from the volume accessible to spectrometer users to comply with safety regulations, which have also imposed a need to black-out the windows of the spectrometer room with opaque material.

From the actual experience, acceptable magnetic field homogeneity can only be achieved if the lower end of the quartz rod is positioned at least 1 mm (up to 3 mm for very high resolution work) above the top of the receiver coil. This means that there is an absorptive layer of solution between the tip of the rod and the active region which attenuates the light, To maximize the sensitivity, the sample concentration would normally be adjusted for maximum light utilization, as outlined below.

Assuming that the laser light comes uniformly from above, the amount of laser light absorbed in the active NMR sample volume is:

$$I_{\mathrm{abs}} = I_0\left(1-10^{-\varepsilon cl}\right)10^{-\varepsilon cd} \tag{7.1}$$

where $c$ is the photosensitizer concentration, $l$ is the coil length, $d$ is the height of the liquid between the end of the quartz rod and the top of the coil, $I_0$ is the amount of laser light emerging from the quartz rod and $\varepsilon$ is the extinction coefficient of the photosensitizer. Finding the maximum of $I_{\mathrm{abs}}$ with respect to the photosensitizer concentration for given $l, d$ and $\varepsilon$ yields the optimum concentration, $c_{\mathrm{opt}}$ :

$$c_{\text{opt}} = \frac{1}{\varepsilon l} \log\left(1 + \frac{l}{d}\right) \tag{7.2}$$

Back-substitution of Equation (7.2) into Equation (7.1) gives the maximum fraction of the light that is absorbed productively:

$$f_{\max} = \frac{I_{\text{abs}}(c_{\text{opt}})}{I_0} = \frac{1}{1 + d/l}\left(1 + \frac{l}{d}\right)^{-d/l} \tag{7.3}$$

The quantity $f_{\max}$ has been is plotted in Figure 7.3 for a $l$ = 16 mm coil, and $d$ = 1 mm (solid line), $d$ = 2 mm (dashed line), $d$ = 3 mm (dotted line). This treatment allows one to choose the optimal photosensitizer concentration to ensure maximum utilization of the incident light.

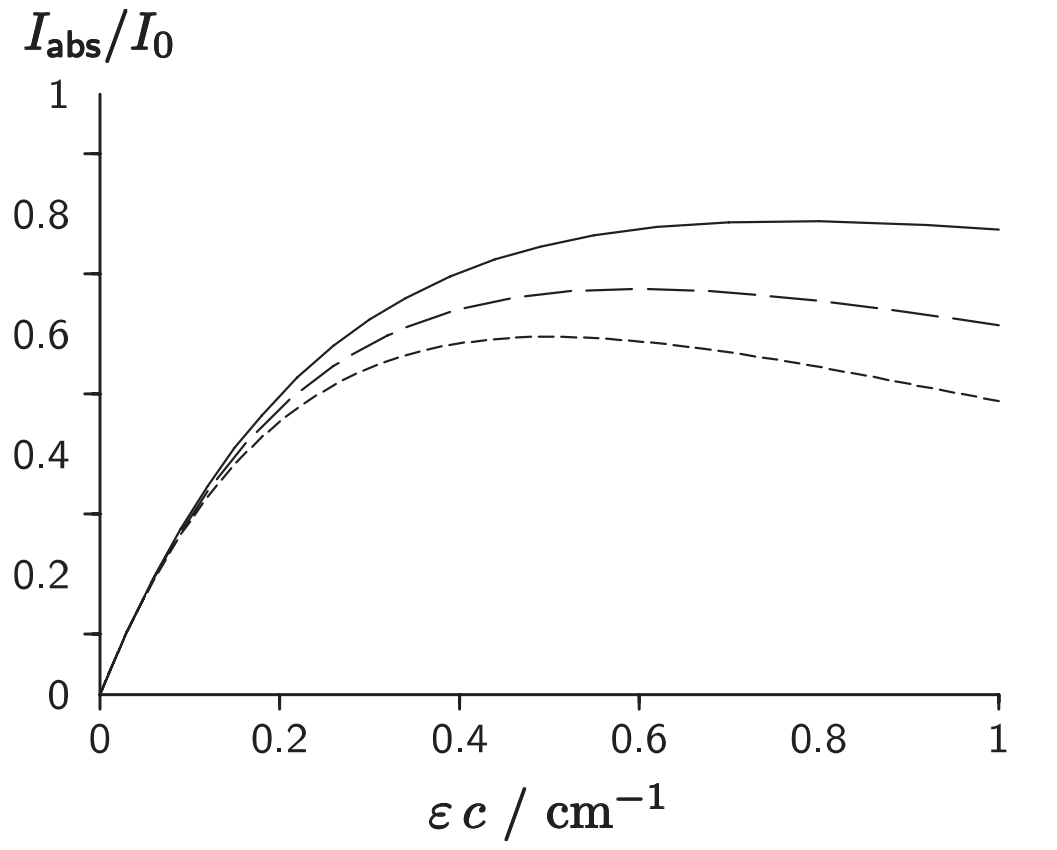

**Figure 7.3** Plot of Equation (7.3) for a 16 mm NMR coil and d = 1 mm (solid line), d = 2 mm (dashed line), d = 3 mm (dotted line).

### 7.4 Software description

A basic time-resolved CIDNP pulse sequence for a Nd:YAG laser is shown in Figure 7.4. The lamp and Q-switch are controlled using the spare TTL lines, triggered from within the pulse sequence. The requirement to keep the lamp running at a constant repetition rate results in the relaxation delay being replaced by a series of $n$ lamp cycles, with the delay $\tau_r$ being the reciprocal of the laser repetition frequency. The penultimate lamp cycle contains a background signal suppression scheme aligned to the end of the cycle. For this purpose, a series of 3-5 90° pulses followed by pulsed field gradients were used. The degree of suppression of unwanted NMR signals can be improved by using composite 90° pulses designed to minimize residual $z$-magnetization [12]. The suppression sequence is followed by a lamp flash and, after a population inversion period (labelled $\tau_{pi}$, typically 150-180 μs), the Q-switch is triggered. The light pulse (5-10 ns) follows after about 50 ns, the time taken for the Q-switch to respond to the TTL voltage ramp. After the light pulse, an incremented period $\tau$ is introduced, during which the photochemical reaction proceeds. The magnetization is then sampled by the RF pulse, which must be as short as possible (for the best time-resolution) and at the same time as close as possible to 90° (for optimum sensitivity). A reasonable compromise for a typical photo-CIDNP experiment is a 1.0 μs observation pulse at full available (or permissible, depending on the probe capacitor damage threshold) RF amplifier power. Given sufficient signal-to-noise ratio, the time resolution can be increased by deconvolution or iterative reconvolution [196].

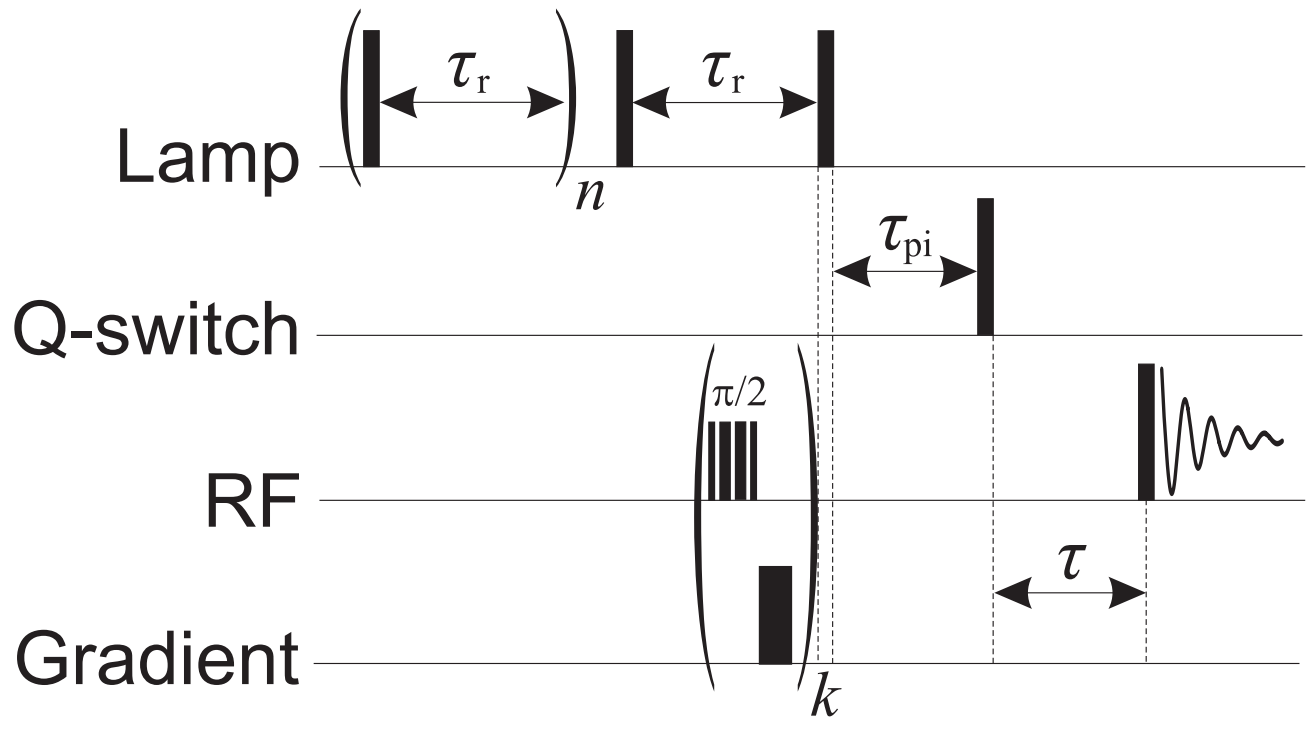

**Figure 7.4** Simple time-resolved photo-CIDNP pulse sequence written specifically to allow the Nd:YAG laser lamp to run continuously throughout the experiment. The radiofrequency pulses preceding the gradients are 45-90-90-45 composite pulses. For further details see text.

While Figure 7.4 represents all the essential parts of the pulse sequence for time-resolved photo-CIDNP, it is important to realise that commercial pulse sequence programming languages contain multiple "hidden" (but well-documented) delays in their pulse statements. These may be as large as several tens of microseconds and may be different for different amplifiers, probes and nuclei. This means that all the time-critical statements should be coded explicitly to avoid ambiguities. In particular, the amplifier unblanking delay which precedes the RF pulse must be constant to ensure that the pulse power is stable. The length of this period for modern amplifiers is about 10 μs, meaning that when the τ delay passes the 10 μs mark, the unblanking delay and the preceding Nd:YAG population inversion delay need to be appropriately reordered in the software.

### 7.5 Examples

Examples of the magnetokinetic data and time-resolved CIDNP spectra recorded with the pulse sequence of Figure 7.4 are shown in Figures 7.5-7.7. After the laser flash, the geminate radical pair spin-sorting process is finished in the first few hundred nanoseconds. The detected magnetization dynamics is therefore due to the magnetic and chemical transformations in those radicals that have escaped the primary cage recombination [41]. The particular case of a dye-sensitized pulsed photo-CIDNP experiment has been treated by Vollenweider and Fischer [43, 44] who suggested the following system of equations to describe the kinetics of the escaped radicals and the associated nuclear polarizations:

$$\begin{cases} R(t) = \dfrac{R_0}{1 + k_t R_0 t} \\ \dfrac{dP}{dt} = -k_t PR - k_t \beta R^2 - \dfrac{P}{T_1} - k_{ex} CP \\ \dfrac{dQ}{dt} = k_t PR + k_t \beta R^2 + k_{ex} CP \end{cases} \tag{7.4}$$

The first equation in (7.4) gives the concentration of the escaped radicals which are assumed to recombine with second order kinetics, with equal initial concentrations of donor cations and photosensitizer anions, $[A^+](0) = [F^-](0) = R_0$, and a second order termination rate constant $k_t$:

$$\begin{aligned} &F \xrightarrow{h\nu} {}^1F \longrightarrow {}^3F \\ &A + {}^3F \longrightarrow A^{\bullet+} + F^{\bullet-} \\ &A^{\bullet+} + F^{\bullet-} \xrightarrow{k_t} A + F \\ &A^{\bullet+} + A \xrightarrow{k_{ex}} A + A^{\bullet+} \end{aligned} \tag{7.5}$$

The second and third equations describe the magnetization of a given nucleus in the pool of radicals ($P$) and diamagnetic molecules ($Q$) with the first term describing recombination in singlet encounters and the second describing F-pair encounters that generate additional magnetization, specified by the factor $\beta = \gamma P^G / R_0$. The nuclear paramagnetic relaxation in the escaped radicals is assumed to be described by a single characteristic time $T_1$. The last terms of the second and third equations describe the transfer of magnetization between radicals and their diamagnetic parent molecules present at concentration $C$ by degenerate electron exchange with a rate constant, $k_{ex}$. The initial conditions for the magnetizations, $P(0) = -P^G$, $Q(0) = P^G$, reflect the spin-sorting nature of the radical pair mechanism, where the initial polarization ($P^G$) must be exactly opposite in geminate recombination products and escaped radicals [41].

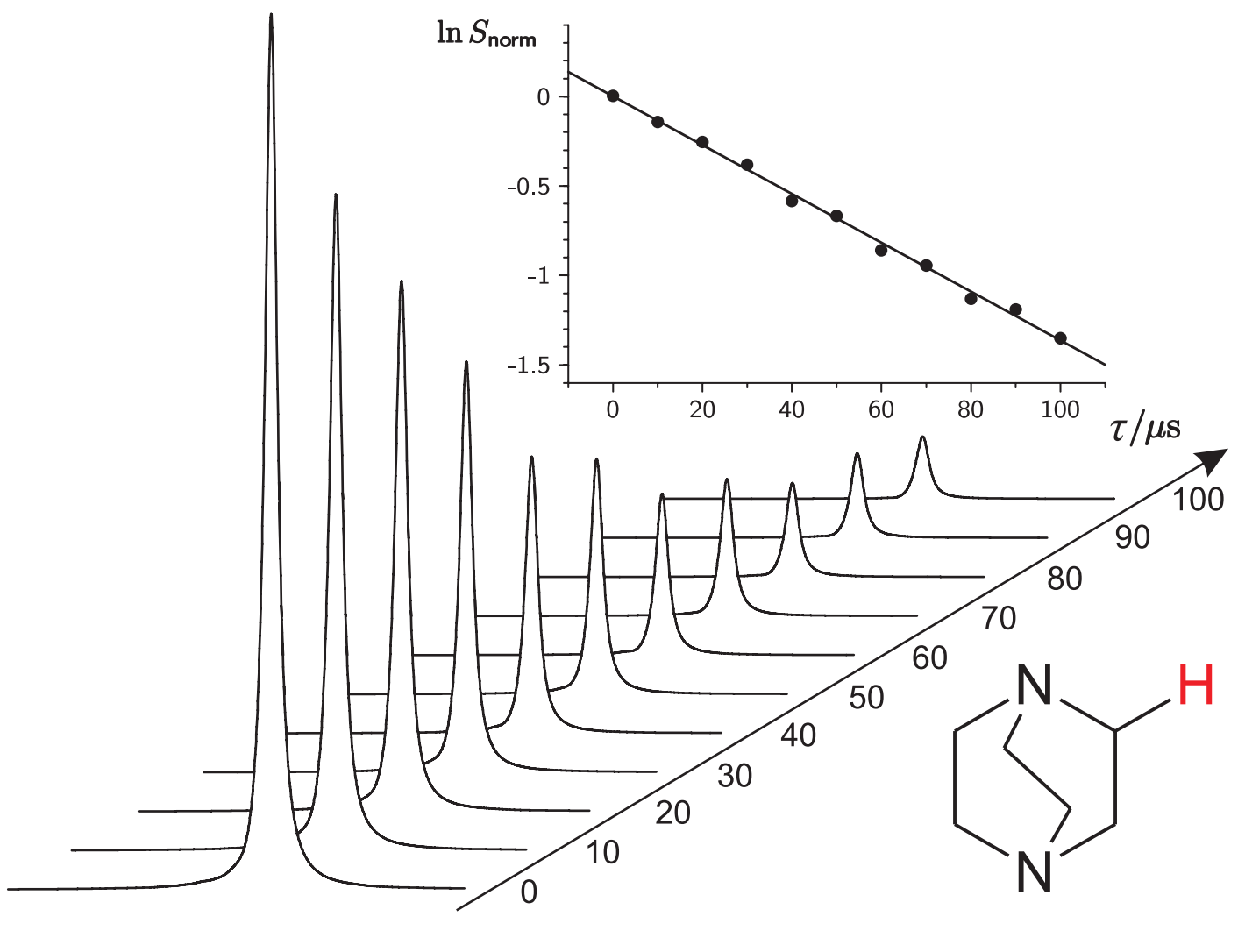

**Figure 7.5** $^1$H photo-CIDNP kinetics in the DABCO/FMN reaction. The signal belongs to the twelve equivalent protons of DABCO (experiment performed by Martin Goez).

Although the solution to Equations (7.4) does exist in a closed algebraic form, the fastest way to solve and fit them to the experimental data nevertheless appears to be numerical. The abundance of variable parameters in Equations (7.4) often necessitates global fitting of data sets in which, for example, the electron donor concentration is varied systematically [98]. Optionally, the initial guess may be first refined using a numerically efficient (4,4) Padé approximant around $t = 0$, which approximates the solution of Equations (7.4) to within about 3% within the operational interval. Starting the full minimization from this refined guess reduces the computation time by about an order of magnitude.

In the case of the photochemical reaction of the electron donor 1,4-diazabicyclo(2.2.2)octane (DABCO) with the photosensitizer flavin mononucleotide (FMN) (Figure 7.5) the dominant mechanism of magnetization transfer between the escaped radicals and the diamagnetic products is degenerate electron transfer [197], meaning that the $k_{ex}CP$ term in Equations (7.4) dominates. The result is pseudo-first order magnetization kinetics and hence an exponential decay in the nuclear polarization, $P$, and the nearly perfect cancellation of the geminate and escape polarization at long time delays (Figure 7.5, insert). In this particular case, the degenerate electron transfer rate constant was determined to be $8.0 \times 10^6$ $M^{-1}$ $s^{-1}$.

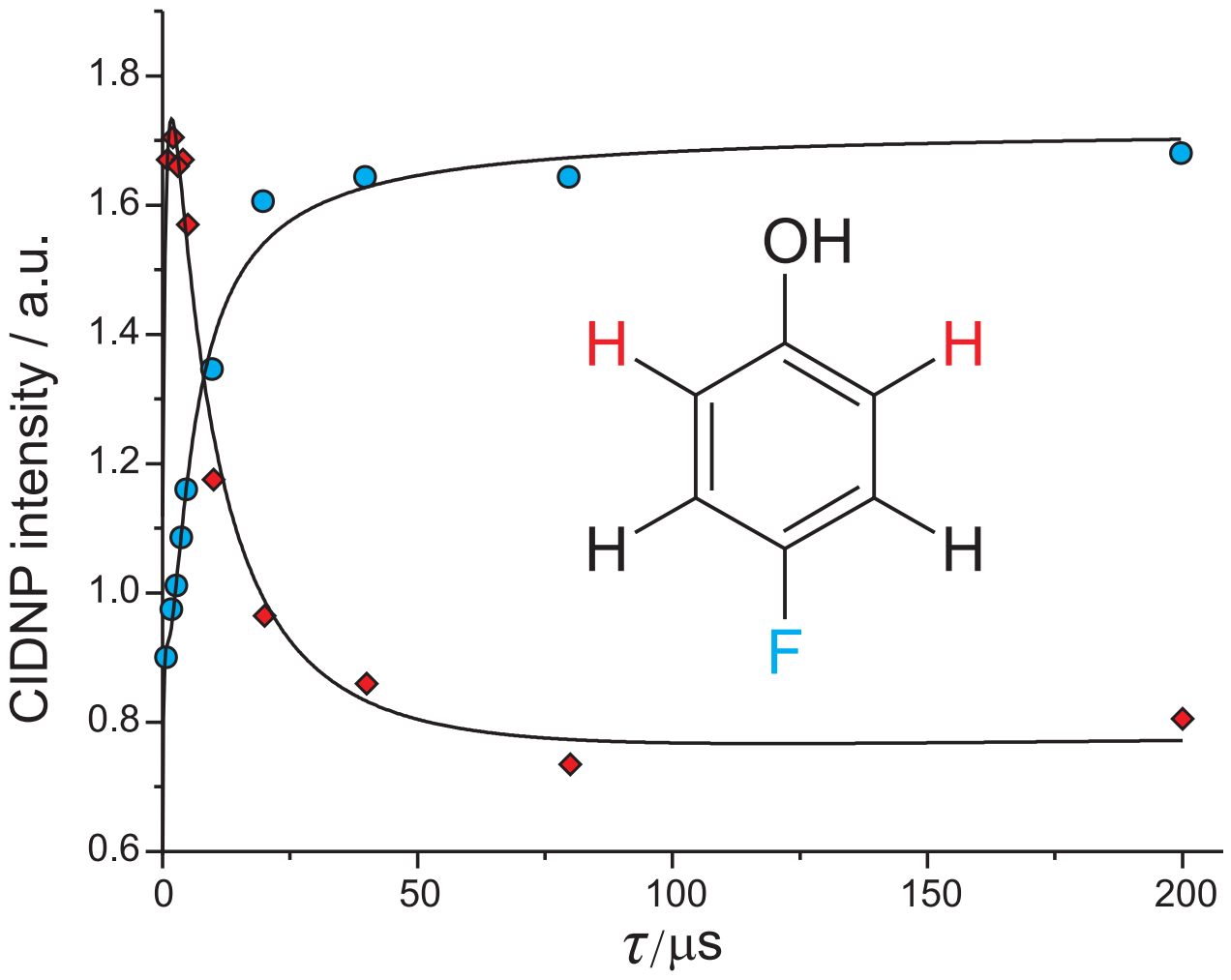

**Figure 7.6** $^1$H (diamonds) and $^{19}$F (circles) photo-CIDNP kinetics in the 4-fluorophenol/FMN reaction. The solid lines represent the theoretical best fit using the model suggested by Vollenweider and Fischer.

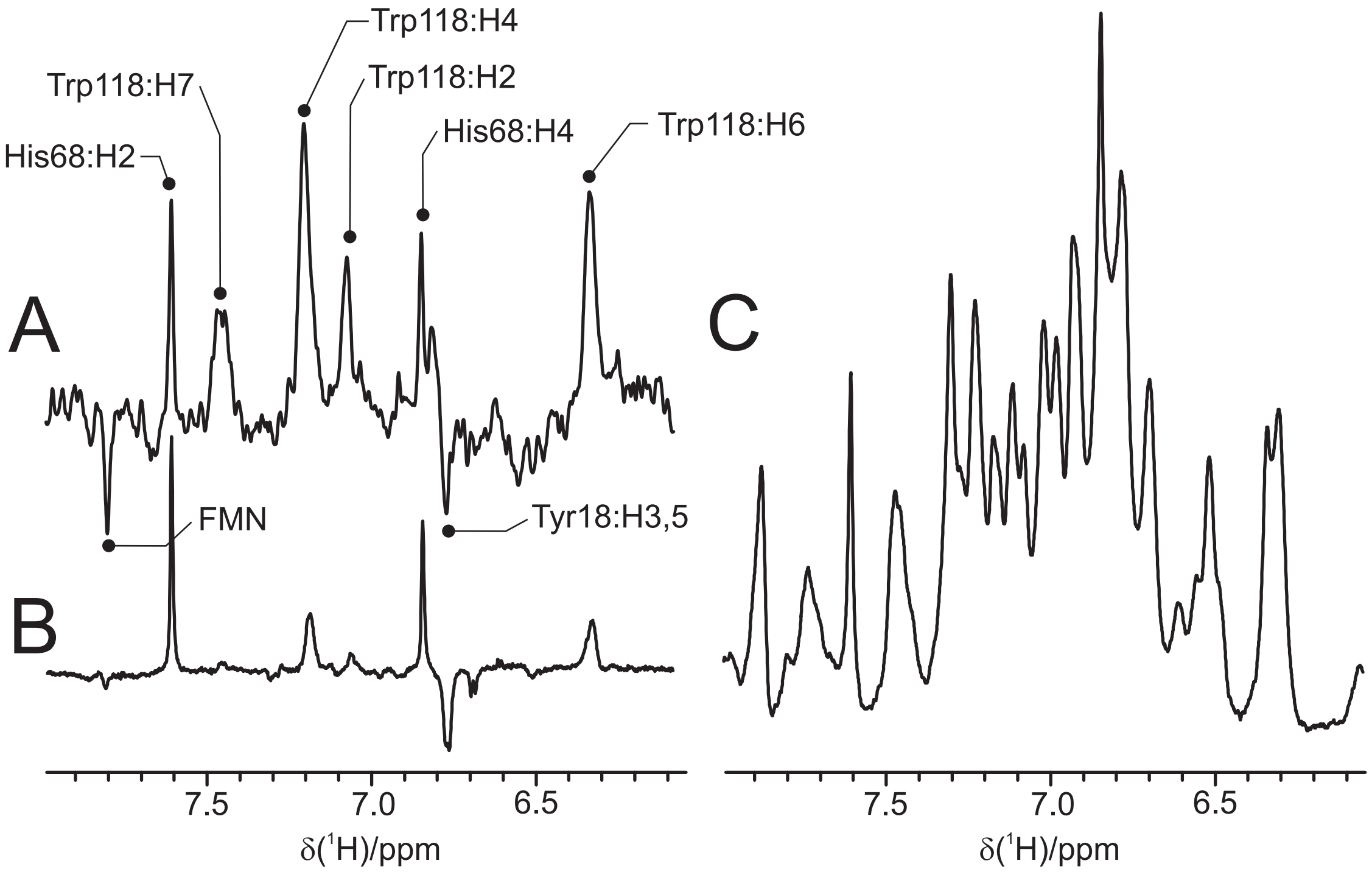

**Figure 7.7** The aromatic regions of $^1$H spectra of calcium-depleted bovine α-lactalbumin with FMN photosensitizer in $D_2O$ buffered by sodium cacodylate at pH 7.7 (experiment performed by Paul Abbott). **(A)** Time-resolved $^1$H photo-CIDNP spectrum (τ = 0) recorded as described in the text. **(B)** Continuous-wave $^1$H photo-CIDNP spectrum obtained using an $Ar^+$ laser as the light source (100 ms light flash at 488 nm with 15 W output power). **(C)** conventional $^1$H NMR spectrum. Spectra (A) and (B) show the difference between the illuminated and the "dark" spectrum with all other settings kept the same. The small peak located between His68:H4 and Tyr18:H3,5 in (A) likely belongs to a non-native conformation [198] and could not be assigned with certainty.

The magnetization dynamics for the 4-fluorophenol/FMN reaction (Figure 7.6) do not involve degenerate electron transfer due to rapid deprotonation of the initially formed phenol cation radical, so that the $k_{ex}CP$ term can be dropped from Equations (7.4). The initial sharp rise of both proton and fluorine signals is due to magnetization generation in F-pairs [41]. It is followed by either a plateau (fluorine) or partial bulk recombination cancellation (protons). The lack of magnetization cancellation in the case of the $^{19}$F nuclei is due to their extremely efficient spin-lattice relaxation in the phenoxyl radical caused by the large hyperfine coupling anisotropy [199]. Fitting Equations (7.4) to these data results in proton and fluorine $T_1$ values of 56 μs and 0.13 μs respectively. Such parameters would be difficult to measure by other means for such a short-lived radical [52].

Figure 7.7 shows the aromatic region of the 600 MHz $^1$H photo-CIDNP (A, B) and NMR (C) spectra obtained for calcium-depleted bovine α-lactalbumin (1.5 mM) with FMN photosensitizer (80 μM) in the presence of 50 mM sodium cacodylate buffer at pH 7.7 (in $D_2O$, uncorrected for deuterium isotope effect). In the case of the time-resolved spectrum (Figure 7.7A) the magnetization was sampled and acquired immediately after the light pulse to avoid secondary out-of-cage photochemical reactions [41] and so to ensure that only geminate CIDNP is present [123, 163]. This geminate polarization is directly related to the accessibility to the bulk solvent of the highest occupied molecular orbital of the aromatic amino acid residue and in principle allows a quantitative interpretation of CIDNP intensities in terms of side-chain solvent accessibilities. This would be difficult to do with steady-state CIDNP (Figure 7.7B) because of distortions caused by the secondary magnetization dynamics (an example of which is given in Figure 7.6) and diamagnetic cross-relaxation occurring on the timescale of the steady-state photo-CIDNP experiment. As one can see from comparing Figures 7.7A and 7.7B, the secondary processes introduce significant intensity changes and preclude simple quantitative analysis. The conclusion in this particular case is that the Trp118, His68 and Tyr18 residues are sufficiently exposed to the solvent for electron transfer (Tyr, Trp) or hydrogen atom transfer (His) to occur to the photosensitizer, with a rather low solvent exposure for Tyr18. A consideration of solvent accessibilities based on the steady-state photo-CIDNP spectrum (Figure 7.7B) would have overestimated the accessibility of Tyr18 and strongly underestimated that of Trp118. Another notable phenomenon which is absent from the CW CIDNP spectrum is the fact that the H(7) proton of the tryptophan aromatic ring (see Figure 2.2 for the numbering scheme) shows a substantial geminate CIDNP polarization, indicating that, contrary to PCM solvent ab initio calculations of tryptophan radicals [199] and in agreement with in situ protein EPR data [200], H(7) does have a hyperfine coupling of at least 0.3 mT in the intermediate tryptophan radicals of a protein/FMN reaction system.

# Chapter 8
## *Relaxation processes in the high-field $^{19}$F CIDNP*

### 8.1 Introduction

The unexpected correlation time dependence of the $^{19}$F photo-CIDNP phase observed in the protein experiments and high-viscosity solvent experiments on small molecules (Chapters 5 and 6) appears to have a complex and historically controversial explanation, mostly revolving around dipolar relaxation processes such as the Overhauser effect. A number of researchers in the past have noticed the relaxation contribution to the high-field CIDNP of Period II elements, manifested either at high viscosities, or for very long-lived radicals in non-viscous solvents [201] and attributed it to relaxation processes. Fluorine with its large hyperfine coupling has been cited as a classical example of Overhauser effect-driven CIDNP, as demonstrated by Adrian in 1971-1979 [64, 65, 156, 202, 203].

From the height of our current understanding of the spin dynamics underlying the photo-CIDNP effect, however, the phenomena observed by Adrian and others appear less clear than they seemed to be back in the 1970s. For example, Adrian's 1974 paper [64], analyzing the experimental work on 1-fluoro-4-[2-(4-fluoro-phenyl)ethyl]benzene (*p,p*-difluorobibenzyl) [66], reports detection of only $^{19}$F polarization, whereas, according to the EPR hyperfine coupling data, the proton polarization should also have been observed. Fluorine seemed to be special, and Adrian had attributed this to cross-relaxation caused by the large HFC anisotropy of the fluorine nucleus. My present-day calculation based on Equations (1.13) and (1.25), 60 MHz (proton frequency) magnetic field, *ab initio* $^{19}$F hyperfine coupling anisotropy and a more realistic $\tau_c = 10^{-10}$ s yielded the electron-$^{19}$F cross-relaxation rate of $4\times10^4$ $s^{-1}$. This is too slow to be manifested in either nanosecond-scale geminate or microsecond-scale escape dynamics, even more so if we note that the lower bound (electron relaxes through mechanisms other than dipolar as well) on the electron relaxation rate, computed with the same parameters, is $2\times10^5$ $s^{-1}$. The correct explanation, which Adrian does mention as an alternative, comes from the consideration of recombination and exchange cancellation described by Equations (1.50) and demonstrated in Figure 7.6, which is complete for the slowly relaxing protons (theoretical $T_1$ = 32 μs) and absent for the much faster relaxing fluorine (theoretical $T_1$ = 0.07 μs). Thus, while the dipolar interaction does of course dominate the self-relaxation behaviour of the experimental system described in [64] and [66], it is unlikely that cross-relaxation in the intermediate radicals occurs to any significant extent.

A further investigation of the matter by Adrian and co-workers [65] has proven, this time convincingly, that at *low magnetic field* ($B_0$ < 20 MHz) the Overhauser mechanism does contribute to the generation of photo-CIDNP polarization of the fluorine nuclei. At higher fields, the Overhauser effect can still be detected if the lifetime of the intermediate radicals is prolonged beyond several tens of microseconds, as it happens in the system reported by Roth *et al.* [69].

A number of later works have also found or suspected an Overhauser effect contribution to the CIDNP or CIDEP generation. Specifically, the 1992 paper by Borbat *et al.* argued that CIDEP swap from E/A to A/E pattern is likely caused by cross-relaxation with the nucleus at high viscosities [204]. Theoretical works describing the longitudinal dipolar cross-relaxation [157, 160] and longitudinal dipole-$\Delta g$ cross-correlation [159, 160] mechanism of CIDNP formation have been published by Tsentalovich and co-workers.

Some cold water had also been poured onto the Overhauser CIDNP idea, for example by Batchelor and Fischer [205], who reported that the earlier studies on radical photolysis CIDEP had incorrectly ascribed certain phenomena to cross-relaxation. After investigating unexpected CIDEP patterns, Valyaev *et al.* have ruled out the Overhauser mechanism on the grounds of it being too slow at the conditions of their experiments [158]. What exactly caused those patterns is still unclear.

In summary, it is well established that at fields lower than about 20 MHz (proton frequency) the electron does cross-relax with the nucleus during the lifetime of the radical pair, provided the HFC anisotropy is large enough. Why similar things appear to happen at higher fields as well, is a much less investigated question. A very simple calculation shows that all the longitudinal cross-relaxation rates in an electron-nucleus spin system contain the square of the electron Larmor frequency in their common denominator. At a 600 MHz magnetic field all these rates become too small and cannot exert any influence on the nano- to microsecond-scale dynamics of the photo-CIDNP effect generation. However, Nature seems to think otherwise, and the author of this work has had the questionable luck of encountering a number of these mysterious cases of strong-field relaxation effects (Chapters 5 and 6) and therefore embarked on a detailed theoretical investigation.

## 8.2 Relaxation analysis

### *8.2.1 Theoretical setup*

First of all, let us show that the existing models of Overhauser CIDNP all agree that it cannot take place at 600 MHz. We will start by performing the relaxation theory analysis of the three-spin system comprising two electrons with anisotropic *g*-tensors and a nucleus with an anisotropic hyperfine coupling to one of the electrons. Due care will be taken to account for the cross-correlation between the HFC tensor and one of the *g*-tensors. This model represents a radical pair in which one of the radicals is conformationally rigid and contains a magnetic nucleus. The *ab initio* calculation of hyperfine couplings in the 3-fluorotyrosyl radical (Chapter 2) shows that the one-nucleus approximation is sufficiently good, because the fluorine hyperfine coupling in the tyrosyl radical is by far the largest one and dominates the rest.

The algebraic core of the Bloch-Redfield-Wangsness relaxation theory, which is used in the treatment below, was described in detail in Chapter 1. The spin Hamiltonian of our system has the following form:

$$\hat{H} = \frac{\mu_B}{\hbar}\hat{\vec{I}}_{\mathrm{e1}}\cdot\mathbf{g}_{\mathrm{e1}}\cdot\vec{B}_0 + \frac{\mu_B}{\hbar}\hat{\vec{I}}_{\mathrm{e2}}\cdot\mathbf{g}_{\mathrm{e2}}\cdot\vec{B}_0 + \omega_{\mathrm{n}}\hat{\vec{I}}_{\mathrm{n}}\cdot\vec{B}_0 + \hat{\vec{I}}_{\mathrm{e1}}\cdot\mathbf{A}\cdot\hat{\vec{I}}_{\mathrm{n}} \tag{7.6}$$

in which the first and the second term describe the anisotropic Zeeman interactions of the two electrons with the external magnetic field. The third term accounts for the nuclear Zeeman interaction. The anisotropy of the nuclear shielding tensor is of the order of kHz and may be neglected. The last term describes the anisotropic electron-nucleus hyperfine interaction. After applying the transformation (1.22), the time dependence of the Hamiltonian (7.6) is condensed into two sets of Wigner coefficients:

$$\begin{aligned}
\hat{H}_0 &= \omega_{\mathrm{e1}}\hat{I}_{\mathrm{e1,z}} + \omega_{\mathrm{e2}}\hat{I}_{\mathrm{e2,z}} + A(\hat{\vec{I}}_{\mathrm{e1}}\cdot\hat{\vec{I}}_{\mathrm{n}}) \\
\hat{H}_1 &= \frac{Rh(\mathbf{g}_{\mathrm{e1}})}{2}\sum_{m'=-2}^{2}\hat{T}_{2,m'}\left(\hat{\vec{I}}_{\mathrm{e1}},\vec{B}_0\right)\mathfrak{D}^{(2)}_{m',-2}(t) + \frac{Rh(\mathbf{g}_{\mathrm{e1}})}{2}\sum_{m'=-2}^{2}\hat{T}_{2,m'}\left(\hat{\vec{I}}_{\mathrm{e1}},\vec{B}_0\right)\mathfrak{D}^{(2)}_{m',2}(t) + \\
&+\frac{Ax(\mathbf{g}_{\mathrm{e1}})}{\sqrt{6}}\sum_{m'=-2}^{2}\hat{T}_{2,m'}\left(\hat{\vec{I}}_{\mathrm{e1}},\vec{B}_0\right)\mathfrak{D}^{(2)}_{m',0}(t) + \frac{Rh(\mathbf{g}_{\mathrm{e2}})}{2}\sum_{m'=-2}^{2}\hat{T}_{2,m'}\left(\hat{\vec{I}}_{\mathrm{e2}},\vec{B}_0\right)\mathfrak{M}^{(2)}_{m',-2}(t) + \\
&+\frac{Rh(\mathbf{g}_{\mathrm{e2}})}{2}\sum_{m'=-2}^{2}\hat{T}_{2,m'}\left(\hat{\vec{I}}_{\mathrm{e2}},\vec{B}_0\right)\mathfrak{M}^{(2)}_{m',2}(t) + \frac{Ax(\mathbf{g}_{\mathrm{e2}})}{\sqrt{6}}\sum_{m'=-2}^{2}\hat{T}_{2,m'}\left(\hat{\vec{I}}_{\mathrm{e2}},\vec{B}_0\right)\mathfrak{M}^{(2)}_{m',0}(t) + \\
&+\frac{Rh(\mathbf{A})}{2}\sum_{m'=-2}^{2}\sum_{k=-2}^{2}\hat{T}_{2,k}\left(\hat{\vec{I}}_{\mathrm{e1}},\hat{\vec{I}}_{\mathrm{n}}\right)\mathfrak{N}^{(2)}_{k,m'}(\theta)\mathfrak{D}^{(2)}_{m',-2}(t) + \\
&+\frac{Rh(\mathbf{A})}{2}\sum_{m'=-2}^{2}\sum_{k=-2}^{2}\hat{T}_{2,k}\left(\hat{\vec{I}}_{\mathrm{e1}},\hat{\vec{I}}_{\mathrm{n}}\right)\mathfrak{N}^{(2)}_{k,m'}(\theta)\mathfrak{D}^{(2)}_{m',2}(t) + \\
&+\frac{Ax(\mathbf{A})}{\sqrt{6}}\sum_{m'=-2}^{2}\sum_{k=-2}^{2}\hat{T}_{2,k}\left(\hat{\vec{I}}_{\mathrm{e1}},\hat{\vec{I}}_{\mathrm{n}}\right)\mathfrak{N}^{(2)}_{k,m'}(\theta)\mathfrak{D}^{(2)}_{m',0}(t)
\end{aligned} \tag{7.7}$$

Despite its complicated appearance, this is in fact the form most suitable for symbolic processing, because the pairwise correlation functions for the Wigner coefficients are known. Let us consider the expression (7.7) term by term to reveal its uncanny simplicity.

The first three terms describe the modulation of the spin system energy caused by the overall molecular rotation of the first radical. The axiality and rhombicity of the *g*-tensor are defined similarly to (1.25):

$$\begin{aligned} Rh(\mathbf{g}_{\mathrm{e1}}) &= \frac{\mu_{\mathrm{B}} B_0}{\hbar}\left(g_{\mathrm{e1},xx} - g_{\mathrm{e1},yy}\right) \\ Ax(\mathbf{g}_{\mathrm{e1}}) &= \frac{\mu_{\mathrm{B}} B_0}{\hbar}\left(2g_{\mathrm{e1},zz} - \left(g_{\mathrm{e1},xx} + g_{\mathrm{e1},yy}\right)\right) \end{aligned} \tag{7.8}$$

and expression (1.23) is used to describe rotations of the spherical tensor operators. The molecular rotation takes place in the physical space, meaning that it is the energy coupling between the spin and the magnetic field that is modulated, while the basis tensors operating in the spin space remain intact. Because the two radicals are assumed to have moved sufficiently far apart to behave independently, the rotational functions $\mathfrak{M}_{m,k}^{(2)}(t)$ of the second radical, appearing in the further three terms in the Hamiltonian, are independent of $\mathfrak{D}_{m,k}^{(2)}(t)$, meaning that all cross-correlation functions between the two sets are zero. The last three terms in the Hamiltonian (7.7) describe the hyperfine interaction of the nucleus with the first electron and contain two consecutive rotations: one with a constant angle $\theta$ operating on the hyperfine tensor only and the other common with the overall molecular rotation. The first rotation describes the relative orientation of the hyperfine and *g*-tensor of the first radical. Because both tensors are assumed to be axial, only one angle is necessary to define their relative orientation.

The Hamiltonian (7.7) is completely equivalent to (7.6), except we now have explicit handles for the overall molecular rotations (*via* $\mathfrak{D}_{m,k}^{(2)}(t)$ and $\mathfrak{M}_{m,k}^{(2)}(t)$) and the angle between the hyperfine and the *g*-tensor axes for the first electron (*via* $\mathfrak{N}_{m,k}^{(2)}(\theta)$). The form (7.7) is automatically generated from (7.6) by the symbolic processing software [8], so in practical calculations the notational complexity never manifests itself.

Because the *g*-tensor of the first electron and the rotated HFC tensor share the same set of overall molecular rotation functions, the cross-correlation between the two anisotropies will be correctly accounted for. As we shall see below, it is this cross-correlation that is actually causing the unexpected CIDNP phase change observed in Chapters 5 and 6.

The other electron, which is assumed to have no significant hyperfine couplings to its nuclei, is independent and has its own rotational functions. Because it is not coupled to the other electron and its nucleus in any way, the relaxation theory problem is a direct product of the electron-nucleus problem and single electron problem, in which the latter relaxes through its own *g*-tensor anisotropy only. The relaxation superoperator of such system is a sum of relaxation superoperators of the two uncoupled and uncorrelated subsystems. It may be rigorously proven that for any multi-spin order involving both subsystems, the relaxation rate is just the sum of the relaxation rates of the constituent spin orders in the individual subsystems.

To sum up, we have an electron-nucleus two-spin system with cross-correlated HFC and *g*-tensor anisotropies, as well as an independent single-electron system that relaxes due to the anisotropy of its own *g*-tensor. Both these systems are fairly straightforward to treat if *Mathematica* is employed to deal with the notationally bulky Hamiltonian (7.7). Once implemented, this construction permits the calculation of any relaxation property of our spin system that falls within the applicability range of the Redfield theory.

### *8.2.2 Existing models*

The simplest model including dipolar cross-relaxation in an intermediate radical was proposed by Adrian and co-workers [65]. It only considers populations (neglecting coherences) and includes all the pathways shown in Figure 8.1.

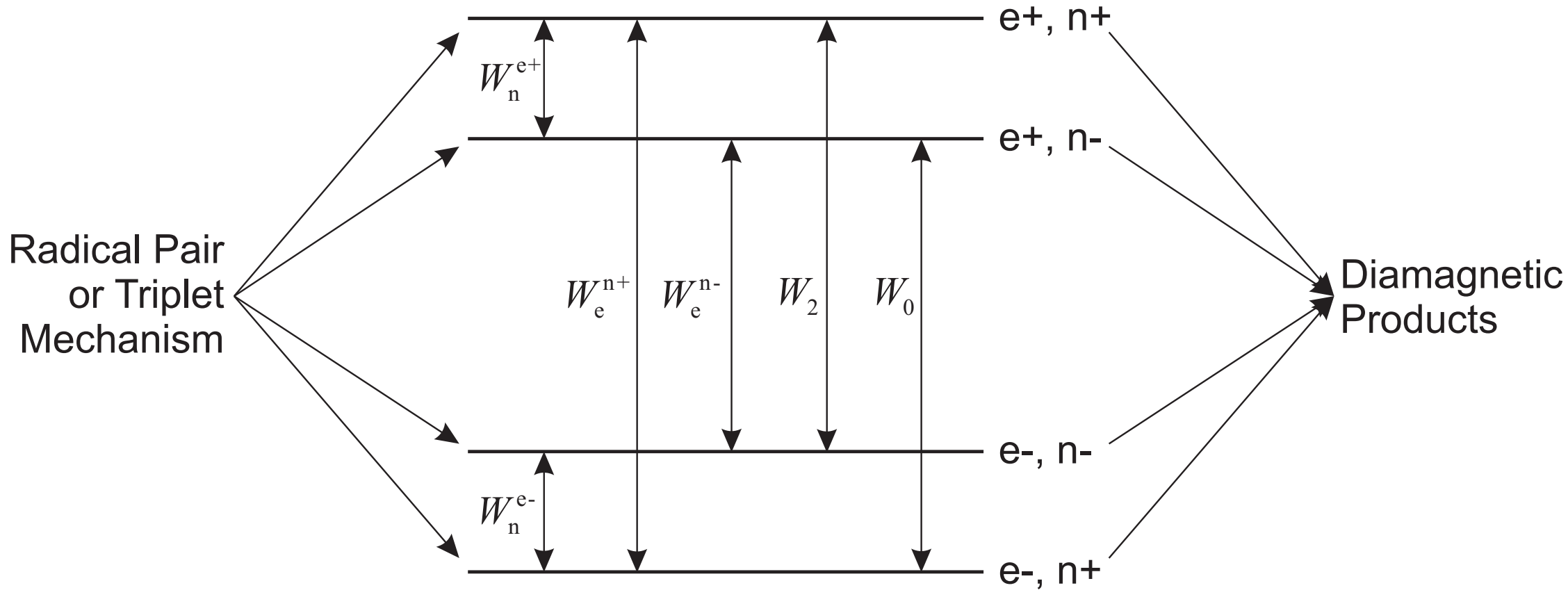

**Figure 8.1.** Energy levels and transitions in an electron-nuclear two-spin system. Adapted, with modifications, from reference [65].

It is easy to see that Adrian's model is the classical 4-level population dynamics model. It is presently used in many textbooks to discuss the origins of NOE and DNP effects. It includes single-quantum transitions, corresponding to a single spin flip, double-quantum (sometimes called flip-flip) and zero-quantum (flip-flop) transitions. The latter two are assumed to be caused by the rotational and translational modulation of the hyperfine coupling.

The relaxation theory treatment described above and in Chapter 1 results in the following values for the transition rates shown in Figure 8.1:

$$\begin{aligned}
W_{\mathrm{n}}^{\mathrm{e}+} &= W_{n}^{e-} = \frac{\Delta\mathbf{A}^2}{240}\frac{\tau_{\mathrm{c}}}{1+\tau_{\mathrm{c}}^2\omega_{\mathrm{n}}^2} \\
W_{\mathrm{e}}^{\mathrm{n}+} &= \frac{\Delta\mathbf{A}^2\tau_{\mathrm{c}}}{240}\left(\frac{1}{1+\tau_{\mathrm{c}}^2\omega_{\mathrm{e}}^2}\right)+\frac{\Delta\mathbf{g}^2\tau_{\mathrm{c}}}{60}\left(\frac{1}{1+\tau_{\mathrm{c}}^2\omega_{\mathrm{e}}^2}\right)+\frac{\Delta\mathbf{A}\Delta\mathbf{g}\tau_{\mathrm{c}}}{240}\left(\frac{1+3\cos(2\theta)}{1+\tau_{\mathrm{c}}^2\omega_{\mathrm{e}}^2}\right) \\
W_{\mathrm{e}}^{\mathrm{n}-} &= \frac{\Delta\mathbf{A}^2\tau_{\mathrm{c}}}{240}\left(\frac{1}{1+\tau_{\mathrm{c}}^2\omega_{\mathrm{e}}^2}\right)+\frac{\Delta\mathbf{g}^2\tau_{\mathrm{c}}}{60}\left(\frac{1}{1+\tau_{\mathrm{c}}^2\omega_{\mathrm{e}}^2}\right)-\frac{\Delta\mathbf{A}\Delta\mathbf{g}\tau_{c}}{240}\left(\frac{1+3\cos(2\theta)}{1+\tau_{\mathrm{c}}^2\omega_{\mathrm{e}}^2}\right) \\
W_2 &= \frac{\Delta\mathbf{A}^2\tau_{\mathrm{c}}}{60}\left(\frac{1}{1+\tau_{\mathrm{c}}^2(\omega_{\mathrm{e}}+\omega_{\mathrm{n}})^2}\right) \\
W_0 &= \frac{\Delta\mathbf{A}^2\tau_{\mathrm{c}}}{360}\left(\frac{1}{1+\tau_{\mathrm{c}}^2(\omega_{\mathrm{e}}-\omega_{\mathrm{n}})^2}\right)
\end{aligned} \tag{7.9}$$

in which the hyperfine tensor and *g*-tensor anisotropy parameters are now defined as

$$\begin{aligned}
&\Delta\mathbf{A} = 2A_{zz}-(A_{xx}+A_{yy}) &&\Delta\mathbf{g} = \left[2g_{zz}-(g_{xx}+g_{yy})\right]\frac{\mu_{\mathrm{B}}B_0}{\hbar} \\
&\Delta\mathbf{A}^2 = 4(\mathbf{A}:\mathbf{A}) &&\Delta\mathbf{g}^2 = 4(\mathbf{g}:\mathbf{g})\frac{\mu_{\mathrm{B}}B_0}{\hbar}
\end{aligned} \tag{7.10}$$

In the case of the 3-fluorotyrosyl radical both of these parameters may be computed *ab initio* (Chapter 2), the resulting values (angular frequency units) are $\Delta\mathbf{g} = 1.32\times10^{10}$, $\Delta\mathbf{A} = 2.75\times10^{9}$, with the $\theta = 90°$ angle between the principal axes of hyperfine and *g*-tensor. The computed hyperfine coupling anisotropy may be considered very accurate, as Chapter 2 demonstrates; the computed *g*-tensor anisotropy on the other hand should be treated as approximate with deviations from the true value possibly as large as ±30%.

The theoretical transition rates (7.9) for the case of the 3-fluorotyrosyl radical at a 14.1 Tesla magnetic field are plotted against the correlation time in Figure 8.2. It is obvious that within the entire range of experimentally available correlation times the electron-nucleus cross-relaxation rates $W_0$ and $W_2$ are too small to exert any influence whatsoever on the microsecond-scale radical spin dynamics – a consequence of having a square of the electron frequency in the spectral density denominators in Equations (7.9). Although the nuclear relaxation rate $W_{\mathrm{n}}$ appears to be sufficiently high, it does not lead to electron spin selection, since $W_n^{e+} = W_n^{e-}$. No high-field cross-relaxation is therefore occurring due to these transitions.

Thus, Adrian's relaxation model is unable to describe the correlation time dependence of $^{19}$F photo-CIDNP effect observed in Chapters 5 and 6. Specifically, at the magnetic field of 14.1 Tesla all the predicted dipolar cross-relaxation rates, even with dipole-$\Delta g$ cross-correlation accounted for, are too small to be active on the timescale of the radical pair evolution. If we decide to continue exploring the relaxation hypothesis, then it appears that restricting oneself to the populations and neglecting the evolution of multi-spin orders and coherences is not a good approximation.

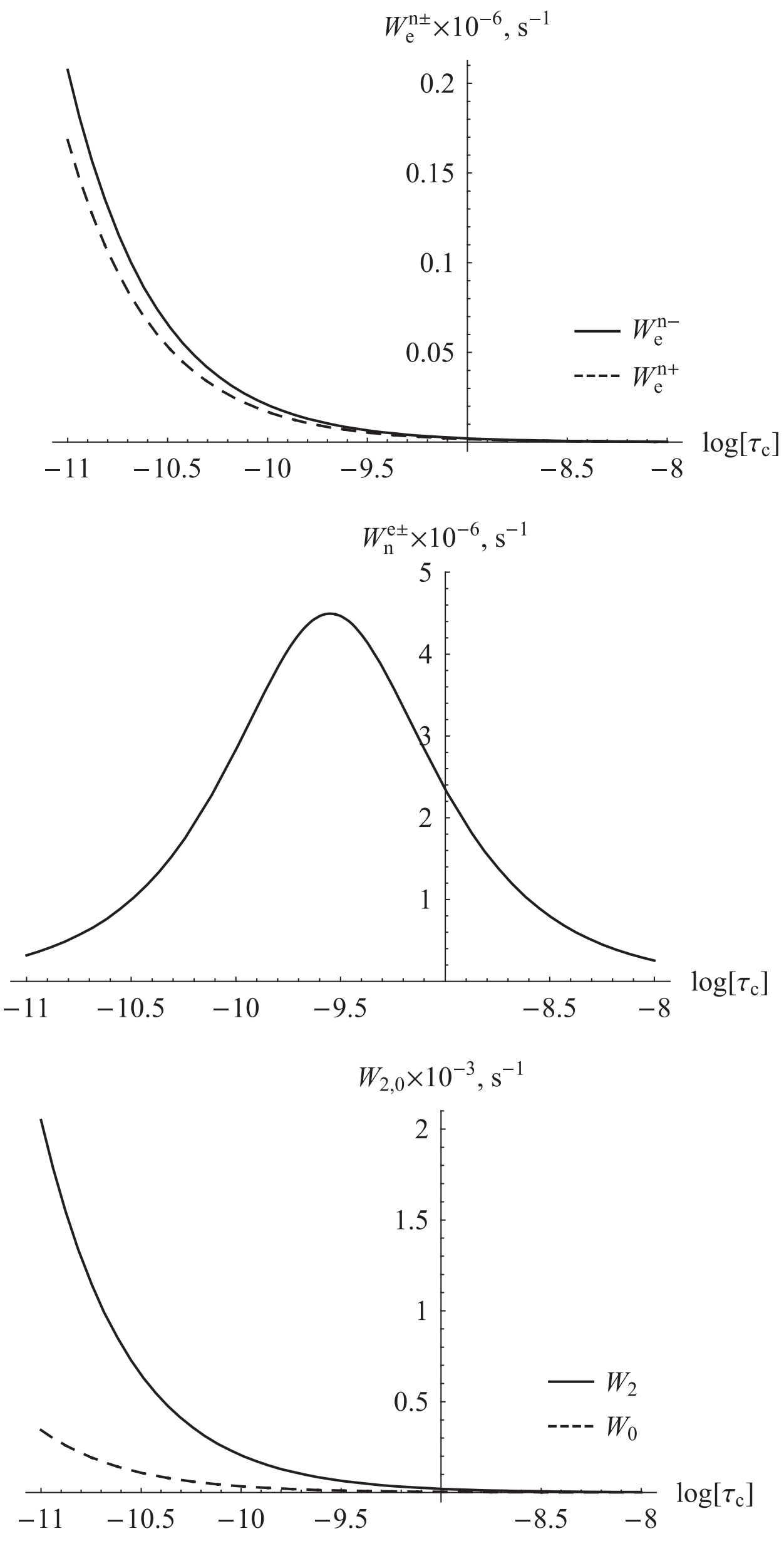

**Figure 8.2.** Correlation time dependence of the transition rates shown in Figure 8.1 (Equations (7.9)) for the 3-fluorotyrosyl radical at a magnetic field of 600 MHz (proton frequency), using *ab initio* values of hyperfine and *g*-tensor anisotropy.

A step forward from Adrian's description of Overhauser CIDNP has been taken by Tsentalovich and co-workers [157, 159, 160], who explicitly included the longitudinal multi-spin orders in the relaxation treatment, thereby expanding the number of spin states that are properly accounted for in the model. At 600 MHz however, their treatment suffers from the same critical fallacy as Adrian's – actual calculations lead to the values of the rates that are far too small to be of any influence on the microsecond time scale (Figure 8.3, all the other rates are linear combinations of those in Figure 8.2). Model fitting to the experimental data pulls out cross-relaxation rates that are incompatible with the theory [159]. All the theoretical rates still contain the square of the electron Zeeman frequency in the common denominator and therefore are small at the 14.1 Tesla magnetic field.

The algebraic complexity of the expressions reported by Tsentalovich *et al.* seems to have deterred further attempts to model the experimentally detected dependence of the high-field CIDNP on the correlation time, and no further studies have emerged that would deal with the case, even though the community is well aware of the fact that the high-field limit of the current theories is zero.

### 8.2.3 Proposed solution

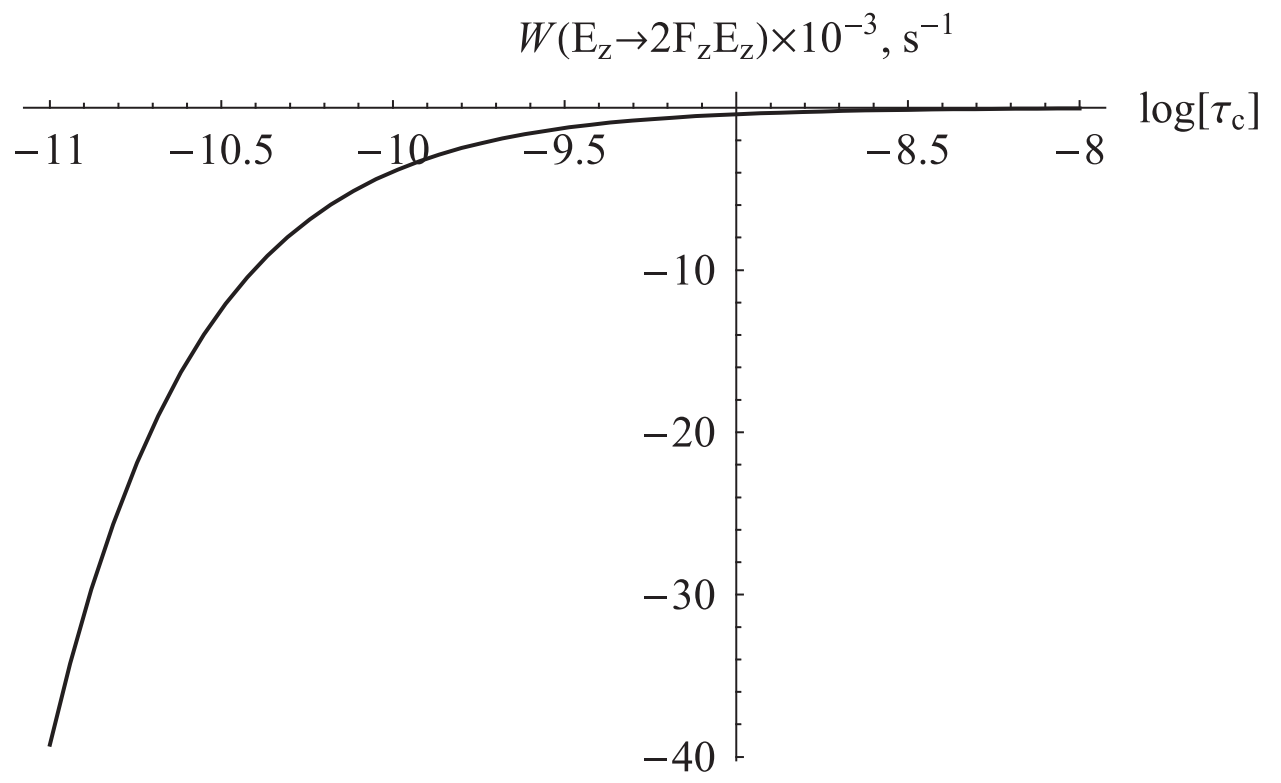

**Figure 8.3.** Correlation time dependence of the longitudinal dipole-Δ*g* cross-correlation rate for the 3-fluorotyrosine radical at a magnetic field of 600 MHz (proton frequency), using *ab initio* values of hyperfine and *g*-tensor anisotropy.

Seeing that prior research on the problem has only underlined its complexity, the author decided to take a brute force option, namely, to compute a *complete* (4096 elements) symbolic relaxation matrix for the three-spin system under consideration using symbolic processing software he recently developed[11], and perform the spin dynamics calculation in the complete operator space, including transverse magnetization and coherence operators. If the correlation time dependence of the high-field geminate $^{19}F$ photo-CIDNP is predicted, one can then try to find the relaxation pathway responsible.

If all the elements in a 64×64 relaxation superoperator were non-zero, the problem would have been formidable even to *Mathematica*. Fortunately only a few elements are non-zero and use can be made of the above mentioned fact that one radical relaxes independently from the other, even though they share a two-electron spin order. All the non-zero auto- and cross-relaxation rates in the spin system of the fluorine-containing radical are reported in Tables 8.1 and 8.2. The expressions for the relaxation of the second radical may be obtained by setting $\Delta\mathbf{A}=0$.

While the Tables 8.1 and 8.2 contain complete information about the relaxation of the two electrons and the nucleus (at least so far as the Redfield theory goes), they are uninformative to the naked eye. That is expected, since we are looking at the rather complicated evolution in a $16\otimes 4$-dimensional observable space. Fortunately, there are disjoint subspaces in the 16-dimensional space, four of them to be exact, and we can use the graphical representation for the relaxation and cross-relaxation rates, which makes the information comprehensible (Figure 8.4).

During the actual calculations care must be taken not to exceed the applicability range of the Redfield theory, which may be roughly outlined as $\tau_c \ll \max\left\{R_i^{-1}\right\}$, where $R_i$ are the computed relaxation rates. This criterion has been shown to have an ample safety margin and it is generally believed that Redfield theory remains quantitatively correct until $\tau_c \sim \max\left\{R_i^{-1}\right\}$, but no further than that [7, 9].

[11]The enclosed DVD contains the source code.

**Table 8.1** Non-zero self-relaxation rates of one- and two-spin orders in an electron-nucleus two-spin system with anisotropic and cross-correlated HFC and *g*-tensors[a].

| Spin order | Self-relaxation rate[b] |
|---|---|
| $F_z$ | $-\frac{\Delta\mathbf{A}^2}{360}\left(3J(\omega_F)+J(\omega_F-\omega_E)+6J(\omega_F+\omega_E)\right)$ |
| $F_+, F_-$ | $-\frac{\Delta\mathbf{A}^2}{720}\left(4J(0)+3J(\omega_F)+J(\omega_F-\omega_E)+6J(\omega_F+\omega_E)+6J(\omega_E)\right)$ |
| $E_z$ | $-\frac{\Delta\mathbf{A}^2}{360}\left(3J(\omega_E)+J(\omega_F-\omega_E)+6J(\omega_F+\omega_E)\right)-\frac{\Delta\mathbf{g}^2}{30}J(\omega_E)$ |
| $E_+, E_-$ | $-\frac{\Delta\mathbf{A}^2}{720}\left(4J(0)+6J(\omega_F)+J(\omega_F-\omega_E)+3J(\omega_E)+6J(\omega_F+\omega_E)\right)-$ $-\frac{\Delta\mathbf{g}^2}{180}\left(4J(0)+3J(\omega_E)\right)$ |
| $2F_zE_z$ | $-\frac{\Delta\mathbf{A}^2}{120}\left(J(\omega_F)+J(\omega_E)\right)-\frac{\Delta\mathbf{g}^2}{30}J(\omega_E)$ |
| $2F_zE_+$ $2F_zE_-$ | $-\frac{\Delta\mathbf{A}^2}{720}\left(4J(0)+J(\omega_F-\omega_E)+3J(\omega_E)+6J(\omega_F+\omega_E)\right)-$ $-\frac{\Delta\mathbf{g}^2}{180}\left(4J(0)+3J(\omega_E)\right)$ |
| $2F_+E_z$ $2F_-E_z$ | $-\frac{\Delta\mathbf{A}^2}{720}\left(4J(0)+3J(\omega_F)+J(\omega_F-\omega_E)+6J(\omega_F+\omega_E)\right)-\frac{\Delta\mathbf{g}^2}{30}J(\omega_E)$ |
| $2F_-E_-$ $2F_+E_+$ | $-\frac{\Delta\mathbf{A}^2}{240}\left(J(\omega_F)+J(\omega_E)+4J(\omega_F+\omega_E)\right)-\frac{\Delta\mathbf{g}^2}{180}\left(4J(0)+3J(\omega_E)\right)$ |
| $2F_-E_+$ $2F_+E_-$ | $-\frac{\Delta\mathbf{A}^2}{720}\left(3J(\omega_F)+3J(\omega_E)+2J(\omega_F-\omega_E)\right)-\frac{\Delta\mathbf{g}^2}{180}\left(4J(0)+3J(\omega_E)\right)$ |

[a]The electron magnetization operators are denoted *E*, the fluorine nucleus operators are denoted *F*.

[b]The Lorentzian spectral density function is defined as $J(\omega)=\tau_c/\left(1+\tau_c^2\omega^2\right)$. The HFC and *g*-tensor anisotropy parameters are defined as in (7.10).

Computing the geminate $^{19}F$ photo-CIDNP for the 3-fluorotyrosine/FMN system with this complete relaxation matrix based on known and computed spin system parameters using Noyes' re-encounter probability model [206, 207] results in the correlation time dependence shown in Figure 8.5. The computed geminate $^{19}F$ photo-CIDNP effect (Figure 8.5, blue curve) goes reassuringly negative as the correlation time is increased and its behaviour qualitatively matches the experimental curves shown in Figures 5.9 and 5.10. The direction of the correlation time dependence is determined by the sign of the $\Delta\mathbf{A}\Delta\mathbf{g}\left(1+3\cos(2\theta)\right)$ product, which is negative in the 3-fluorotyrosyl radical ($\Delta\mathbf{A}>0$, $\Delta\mathbf{g}>0$, $\theta=90°$).

**Table 8.2** Non-zero cross-relaxation rates between one- and two-spin orders in an electron-nucleus two-spin system with anisotropic and cross-correlated HFC and g-tensors[a].

| Spin orders | Cross-relaxation rate[b] |
|---|---|
| $F_z \leftrightarrow E_z$ | $\frac{\Delta\mathbf{A}^2}{360}\left(J(\omega_F-\omega_E)-6J(\omega_F+\omega_E)\right)$ |
| $F_z \leftrightarrow 2F_+E_-$<br>$F_z \leftrightarrow 2F_-E_+$ | $\frac{\Delta\mathbf{A}\Delta\mathbf{g}\tau_c\left(1+3\cos(2\theta)\right)\left(5+\tau_c^2\left(5\omega_F^2-6\omega_E\omega_F+3\omega_E^2\right)\right)}{1440\left(1+\tau_c^2\omega_F^2\right)\left(1+\tau_c^2(\omega_E-\omega_F)^2\right)}$ |
| $E_z \leftrightarrow 2F_+E_-$<br>$E_z \leftrightarrow 2F_-E_+$ | $\frac{\Delta\mathbf{A}\Delta\mathbf{g}\tau_c\left(1+3\cos(2\theta)\right)\left(1+\tau_c^2\left(3\omega_F^2-6\omega_E\omega_F+\omega_E^2\right)\right)}{1440\left(1+\tau_c^2\omega_E^2\right)\left(1+\tau_c^2(\omega_E-\omega_F)^2\right)}$ |
| $E_z \leftrightarrow 2F_zE_z$ | $-\frac{\Delta\mathbf{A}\Delta\mathbf{g}\tau_c\left(1+3\cos(2\theta)\right)}{120\left(1+\tau_c^2\omega_E^2\right)}$ |
| $2F_zE_z \leftrightarrow 2F_-E_+$<br>$2F_zE_z \leftrightarrow 2F_+E_-$ | $\frac{\Delta\mathbf{A}^2}{240}\left(J(\omega_F)+J(\omega_E)\right)$ |
| $2F_+E_- \leftrightarrow 2F_-E_+$ | $\frac{\Delta\mathbf{A}^2}{360}J(\omega_F-\omega_E)$ |
| $F_+ \leftrightarrow E_+$<br>$F_+ \leftrightarrow E_+$ | $-\frac{\Delta\mathbf{A}^2}{720}\left(2J(0)+3J(\omega_F)+2J(\omega_F-\omega_E)+3J(\omega_E)\right)$ |
| $F_+ \leftrightarrow 2F_zE_+$<br>$F_- \leftrightarrow 2F_zE_-$ | $-\frac{\Delta\mathbf{A}\Delta\mathbf{g}\tau_c\left(1+3\cos(2\theta)\right)\left(5+\tau_c^2\left(5\omega_F^2-6\omega_E\omega_F+3\omega_E^2\right)\right)}{1440\left(1+\tau_c^2\omega_F^2\right)\left(1+\tau_c^2(\omega_E-\omega_F)^2\right)}$ |
| $E_+ \leftrightarrow 2F_zE_+$<br>$E_- \leftrightarrow 2F_zE_-$ | $-\frac{\Delta\mathbf{A}\Delta\mathbf{g}\left(1+3\cos(2\theta)\right)}{720}\left(J(0)+6J(\omega_E)\right)$ |
| $E_+ \leftrightarrow 2F_+E_z$<br>$E_- \leftrightarrow 2F_-E_z$ | $-\frac{\Delta\mathbf{A}\Delta\mathbf{g}\tau_c\left(1+3\cos(2\theta)\right)\left(-1+2\tau_c^2\omega_F^2\right)}{1440\left(1+\tau_c^2\omega_F^2\right)}$ |
| $2F_zE_+ \leftrightarrow 2F_+E_z$<br>$2F_zE_- \leftrightarrow 2F_-E_z$ | $-\frac{\Delta\mathbf{A}^2\tau_c\left(2+\tau_c^2(\omega_E-\omega_F)^2\right)}{360\left(1+\tau_c^2(\omega_E-\omega_F)^2\right)}$ |

[a]The electron magnetization operators are denoted $E$, the fluorine nucleus operators are denoted $F$.

[b]The Lorentzian spectral density function is defined as $J(\omega)=\tau_c/\left(1+\tau_c^2\omega^2\right)$. The HFC and $g$-tensor anisotropy parameters are defined as in (7.10).

Figure 8.5 also shows the other two cases, namely when $\Delta\mathbf{A}\Delta\mathbf{g}\left(1+3\cos(2\theta)\right)$ is positive and zero. In the positive case there is a constructive interference between the RPM-generated and relaxation-generated geminate photo-CIDNP effect. When either the HFC or the g-tensor anisotropy is zero, the effect vanishes completely. This type of dependence is characteristic of a cross-correlated cross-relaxation process.

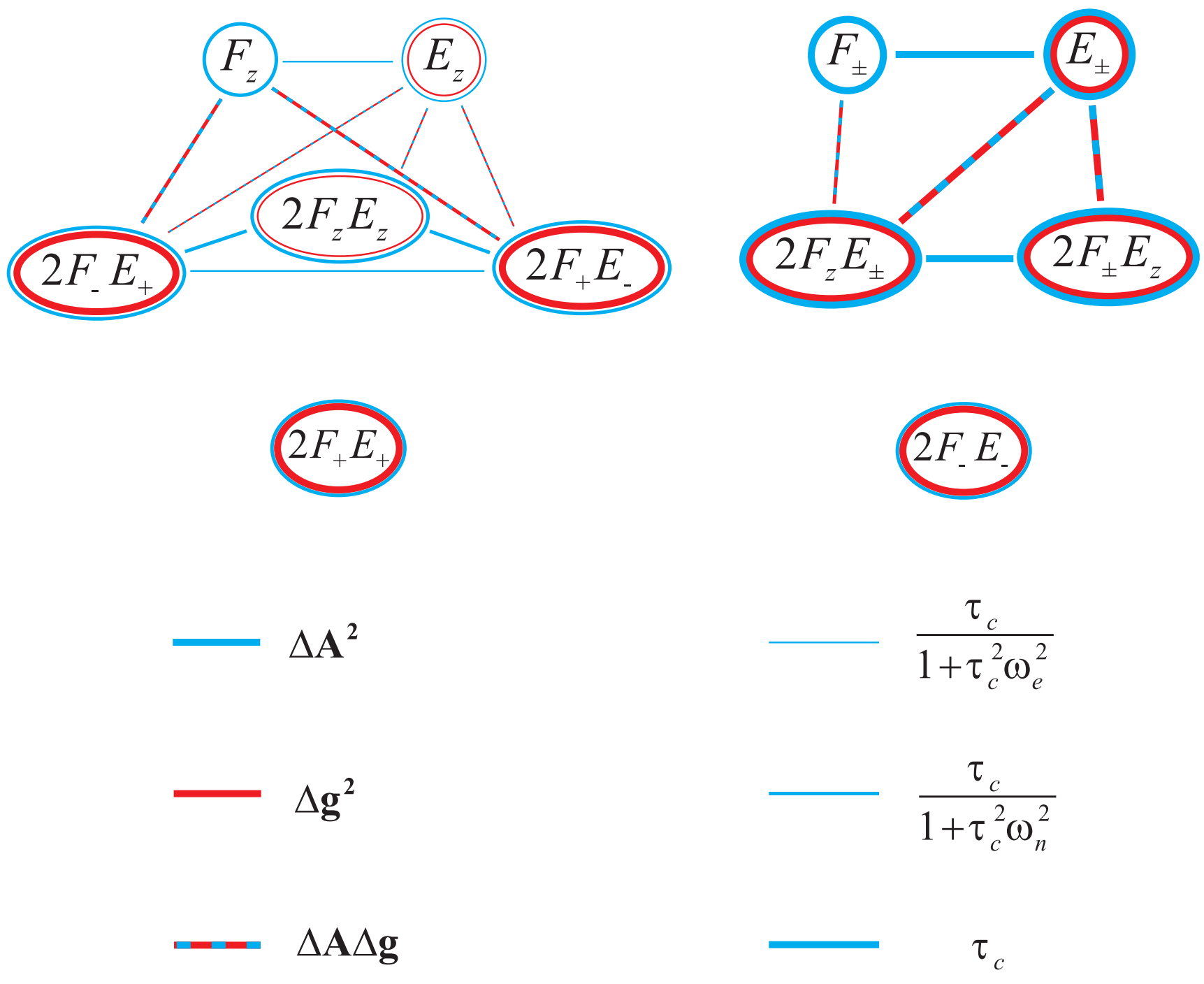

**Figure 8.4.** Relaxation map for the two-spin dipole-$\Delta g$ cross-correlated system. The circles around the spin orders encode self-relaxation rates and the lines encode cross-relaxation rates. The legend shows the relation between the colour and thickness of the line and the type and the leading (high-field limit) spectral density term of the corresponding rate.

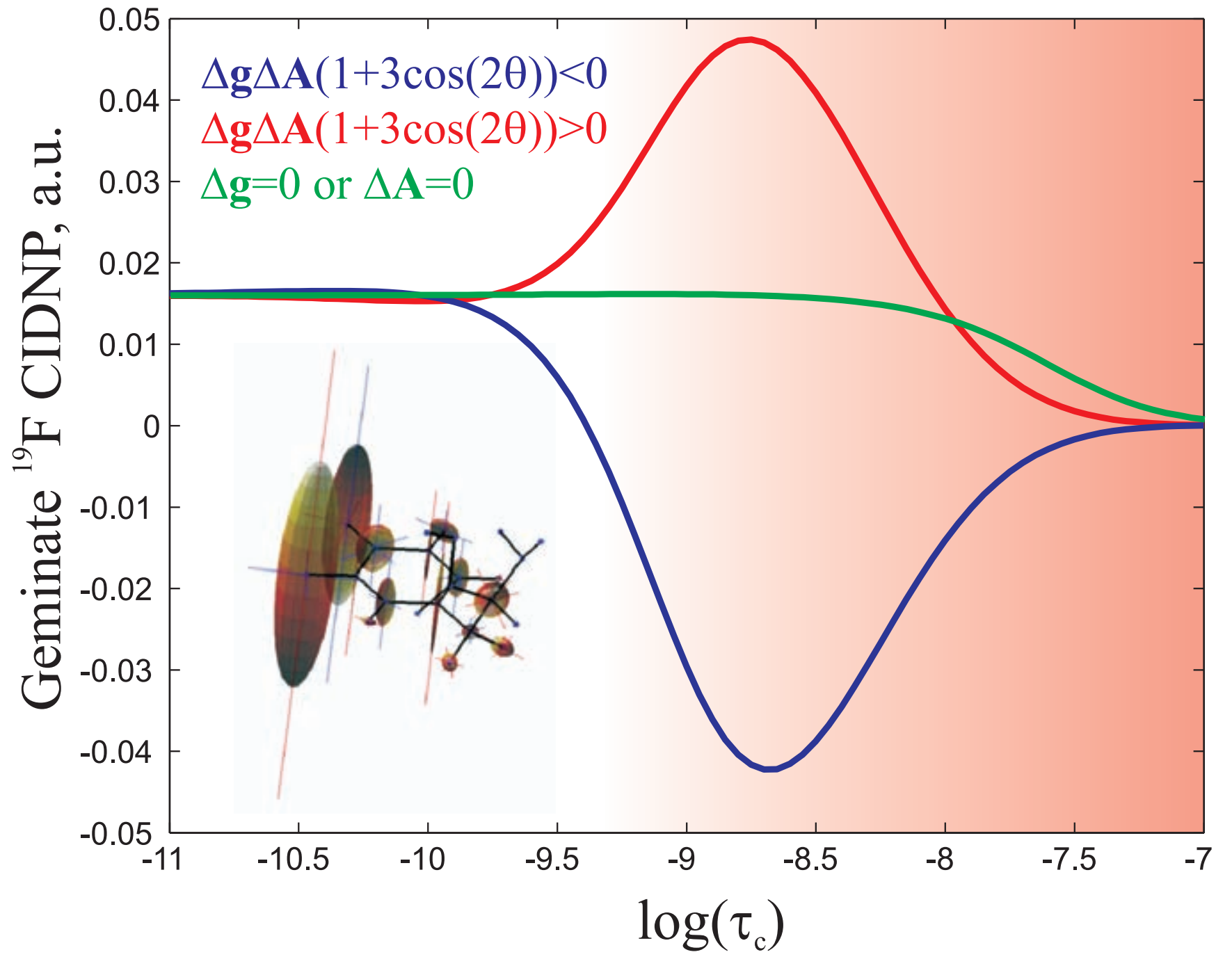

**Figure 8.5.** The correlation time dependence of the computed geminate $^{19}$F photo-CIDNP effect in the 3-fluorotyrosine/FMN system at 14.1 Tesla magnetic field. The calculation was performed with a complete relaxation matrix and Noyes' re-encounter probability model. The region marked in shades of red is outside the applicability interval of the Redfield theory.

Thus, quite encouragingly, it appears that, after all the relaxation pathways have been taken into account, the correlation time dependence of the high-field geminate $^{19}$F photo-CIDNP effect gets explained, completely *ab initio* and without any adjustable parameters. The important question now is which particular relaxation pathway is responsible for this phenomenon. A systematic search has been performed, when every element of the relaxation matrix was zeroed in turn and computation was repeated and the result inspected for significant differences to that in Figure 8.5. This search has come up with two relaxation pathways connecting eight pairs of spin orders that are wholly responsible for the observed correlation time dependence of geminate $^{19}$F CIDNP at 14.1 Tesla:

$$E_{1+}\begin{pmatrix} \mathbb{1}_2 \\ E_{2+} \\ E_{2z} \\ E_{2-} \end{pmatrix} \leftrightarrow F_z E_{1+}\begin{pmatrix} \mathbb{1}_2 \\ E_{2+} \\ E_{2z} \\ E_{2-} \end{pmatrix} \qquad E_{1-}\begin{pmatrix} \mathbb{1}_2 \\ E_{2+} \\ E_{2z} \\ E_{2-} \end{pmatrix} \leftrightarrow F_z E_{1-}\begin{pmatrix} \mathbb{1}_2 \\ E_{2+} \\ E_{2z} \\ E_{2-} \end{pmatrix} \tag{7.11}$$

These transitions are shown with a thick diagonal dashed line on the top right diagram in Figure 8.4. Physically they represent a relaxation transition from pure electron singlet or triplet state to the singlet or triplet that is conditional upon the nuclear spin state – remarkably similar to the usual CIDNP generation scheme, but stemming from relaxation. These are essentially transverse processes, which is why they were not picked up by the theories proposed by Adrian and Tsentalovich *et al.* When the radicals recombine, the electron component disappears, leaving the correlation time dependent nuclear magnetization behind.

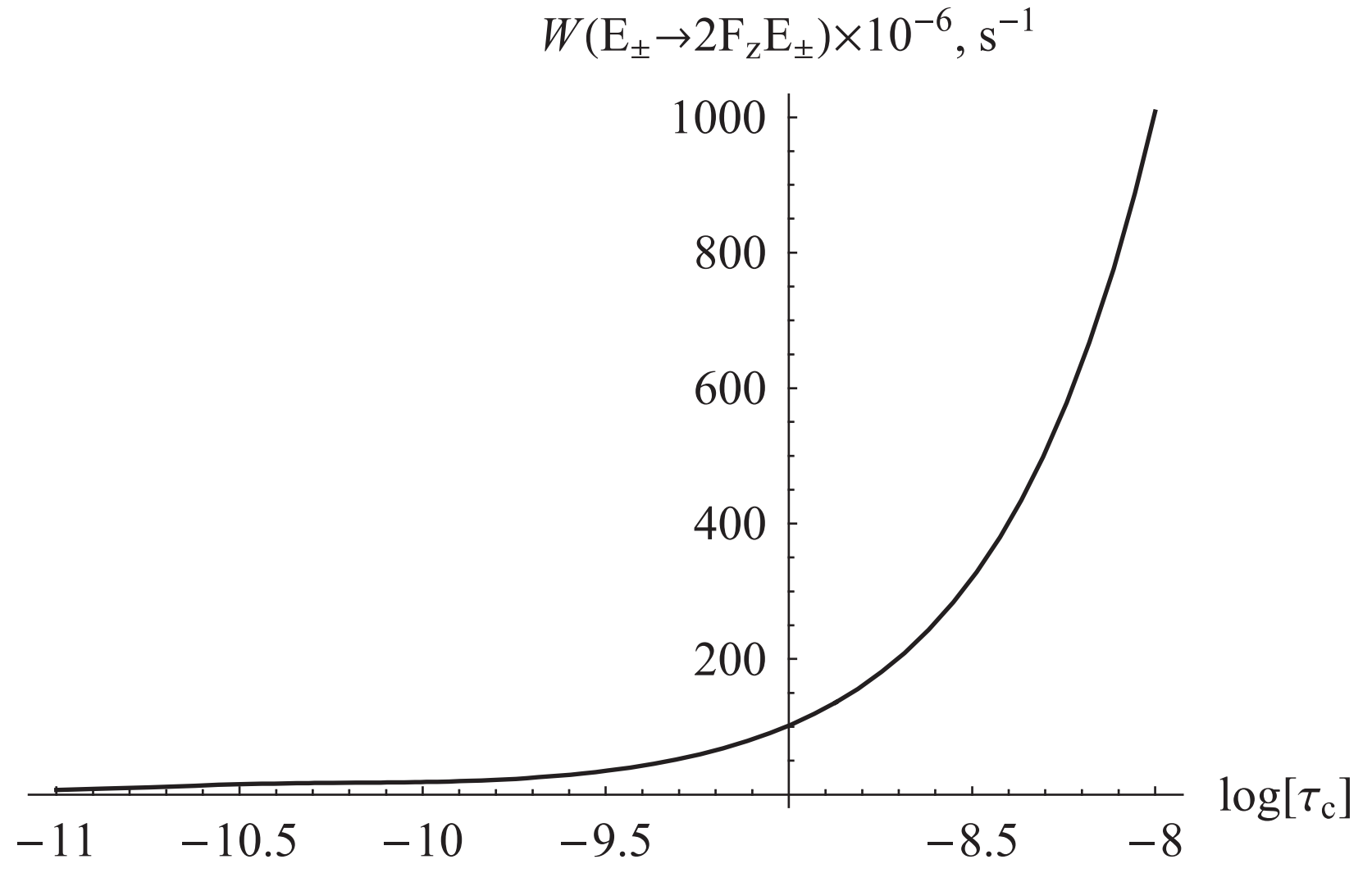

**Figure 8.6.** The theoretical correlation time dependence of the rates of relaxation transitions (7.11) in the 3-fluorotyrosine radical at 14.1 Tesla magnetic field using *ab initio* values of hyperfine and *g*-tensor anisotropy. The region to the left of $\log(\tau_c) = -8.5$ is outside the validity interval of the Redfield theory.

The correlation time dependence of the processes (7.11) is given in Table 8.2 and shown graphically for the 3-fluorotyrosyl radical in Figure 8.6 (*c.f.* Figures 8.2 and 8.3). As the correlation time increases, this cross-relaxation rate quickly becomes fast enough to be operational on the nanosecond timescale of geminate magnetization generation.

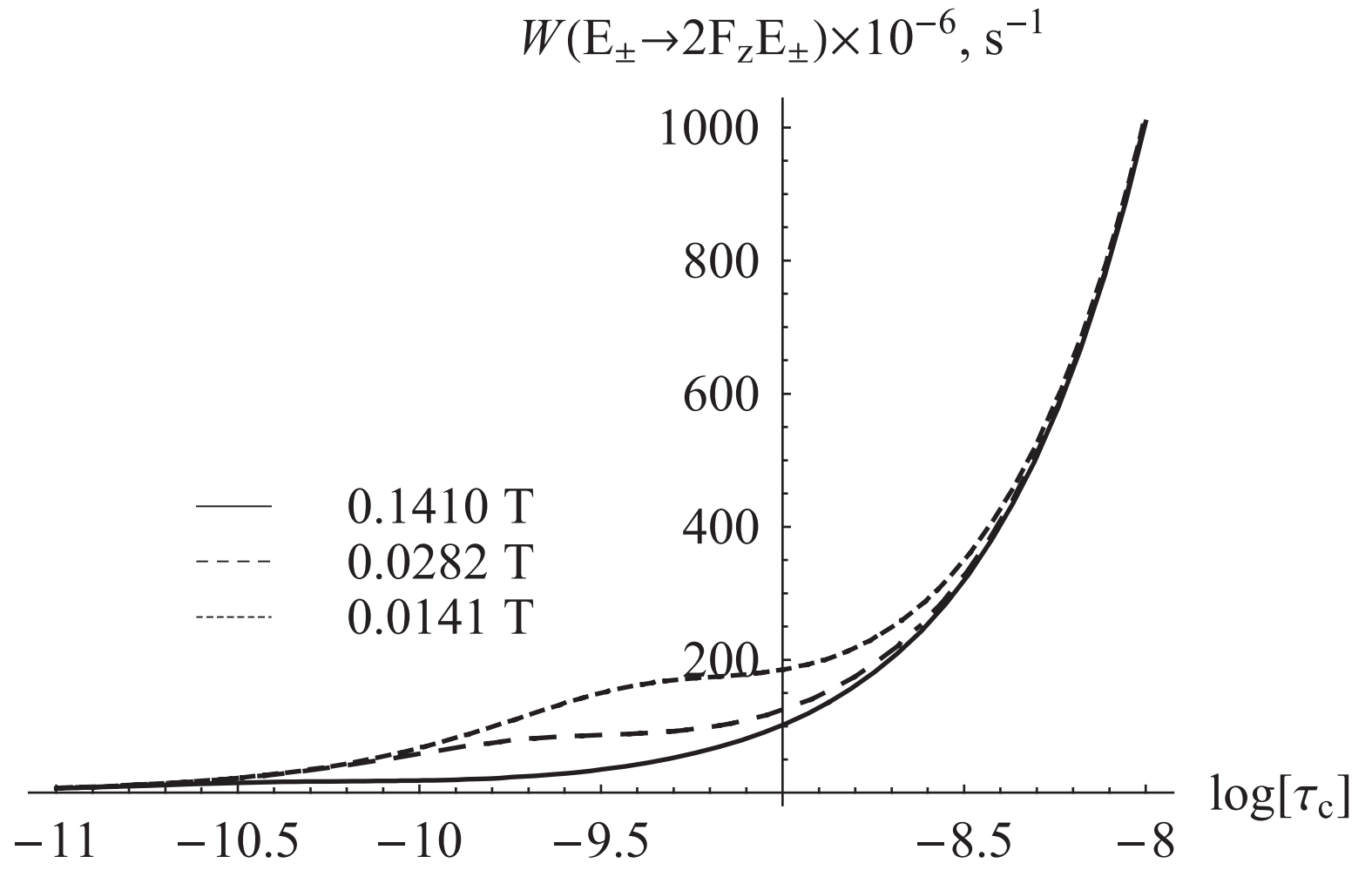

**Figure 8.7.** The theoretical correlation time dependence of the rates of relaxation transitions (7.11) in the 3-fluorotyrosine radical at three different magnetic fields using *ab initio* values of hyperfine and *g*-tensor anisotropy. The region to the left of $\log(\tau_c) = -8.5$ is outside the validity interval of the Redfield theory.

Because the leading term in the spectral power density multiplier in the rate expression does not depend on the magnetic field,

$$W\left(E_{\pm} \to 2F_Z E_{\pm}\right) = -\frac{\Delta\mathbf{A}\Delta\mathbf{g}\,\tau_c\left(1+3\cos(2\theta)\right)}{720}\left(1+\frac{6}{1+\tau_c^2\omega_E^2}\right) \quad (7.12)$$

the correlation time dependence displayed in Figure 8.6 remains completely unchanged all the way down to the fields of the order of 0.1 Tesla (Figure 8.7). At the fields lower than 0.1 Tesla the cross-relaxation rate (7.12) actually increases for short correlation times, suggesting that the same $E_{\pm} \to 2F_Z E_{\pm}$ relaxation-driven process may be operational (along with the now fast enough Adrian's and Tsentalovich' mechanisms) at the low fields as well.

### *8.2.4 Potential applications*

Beyond its conceptual value (which clearly merits further investigation, both theoretical and experimental), the correlation time dependence of the geminate $^{19}F$ photo-CIDNP effect may have practical uses, as it provides a direct measure of the correlation time in short-lived radical species. Another interesting question, in the context of some unexplained CIDEP patterns recently discussed by Borbat *et al.* [204, 208], is whether the complete relaxation matrix treatment predicts those CIDEP patterns as well and which relaxation pathways are responsible. More generally, because it is now possible to do brute force analytical relaxation theory on non-trivial spin systems, it seems worthwhile to launch a deeper general investigation into (hitherto rather sparsely researched) area of relaxation-driven radical spin dynamics. Because further $^{19}F$ photo-CIDNP studies of large proteins are being planned (for author's postdoc period with Professor Peter Hore), all of these questions will likely be investigated and answered. Stay tuned.

# References

1. McNaught, A.D., A. Wilkinson, and International Union of Pure and Applied Chemistry., *Compendium of chemical terminology: IUPAC recommendations*. 2nd ed. 1997, Oxford: Blackwell Science. vii, 450.
2. Martin, M.L., G.J. Martin, and J.-J. Delpuech, *Practical NMR spectroscopy*. 1980, London; Philadelphia: Heyden. xxxi, 460.
3. Simon, B. and M. Sattler, *De novo structure determination from residual dipolar couplings by NMR spectroscopy.* Angewandte Chemie, International Edition, 2002. **41**(3): p. 437-440.
4. Banci, L., I. Bertini, and C. Luchinat, *Nuclear and electron relaxation: the magnetic nucleus-unpaired electron coupling in solution*. 1991, Weinheim; Cambridge: VCH. xvi, 208.
5. Kumar, A. and P.K. Madhu, *Cross-correlations in multispin relaxation.* Concepts in Magnetic Resonance, 1996. **8**(2): p. 139-60.
6. Palmer, A.G., 3rd, *Probing molecular motion by NMR.* Current opinion in structural biology, 1997. **7**(5): p. 732-7.
7. Goldman, M., *Formal theory of spin-lattice relaxation.* Journal of Magnetic Resonance, 2001. **149**(2): p. 160-87.
8. *Mathematica, Version 5.1*. 2004, Wolfram Research, Inc.: Champaign, Illinois.
9. Redfield, A.G., *The Theory of Relaxation Processes*, in *Advances in Magnetic Resonance*, J.S. Waugh, Editor. 1965, Academic Press. p. 1-30.
10. Abragam, A., *The principles of nuclear magnetism*. 1961, Oxford: Clarendon Press. xvi, 599.
11. Slichter, C.P., *Principles of magnetic resonance*. 3rd ed. Springer series in solid-state sciences; 1. 1990, Berlin; London: Springer. xi, 655.
12. Ernst, R.R., G. Bodenhausen, and A. Wokaun, *Principles of nuclear magnetic resonance in one and two dimensions*. The International series of monographs on chemistry; 14. 1987, Oxford: Clarendon Press. xxiv, 610.
13. Brink, D.M. and G.R. Satchler, *Angular momentum*. 3rd ed ed. Oxford science publications. 1993, Oxford: Clarendon Press. xii, 170.
14. Palmer III, A.G., *Relaxation and dynamic processes*, in *Protein NMR spectroscopy, principles and practice*. 1996, Academic Press.
15. Sillescu, H. and D. Kivelson, *Theory of spin-lattice relaxation in classical liquids.* Journal of Chemical Physics, 1968. **48**(8): p. 3493-505.
16. Lipari, G. and A. Szabo, *Model-free approach to the interpretation of nuclear magnetic resonance relaxation in macromolecules. 1. Theory and range of validity.* Journal of the American Chemical Society, 1982. **104**(17): p. 4546-59.
17. Lipari, G. and A. Szabo, *Nuclear magnetic resonance relaxation in nucleic acid fragments: models for internal motion.* Biochemistry, 1981. **20**(21): p. 6250-6.
18. Lipari, G., A. Szabo, and R.M. Levy, *Protein dynamics and NMR relaxation: comparison of simulations with experiment.* Nature, 1982. **300**(5888): p. 197-8.
19. Overhauser, A.W., *Polarization of nuclei in metals.* Physical Review, 1953. **92**: p. 411-15.
20. Cavanagh, J., *Protein NMR spectroscopy: principles and practice*. 1996, San Diego; London: Academic Press. xxiii, 587.
21. Hu, K.-N., et al., *Dynamic Nuclear Polarization with Biradicals.* Journal of the American Chemical Society, 2004. **126**(35): p. 10844-10845.
22. Atkins, P.W. and R.S. Friedman, *Molecular quantum mechanics*. 4th ed. 2005, Oxford: Oxford University Press. xiv, 573.

23. Arfken, G.B. and H.-J. Weber, *Mathematical methods for physicists*. 5th ed. 2001, San Diego, Calif.; London: Academic Press. xiv, 1112.
24. Claridge, T.D.W., *High-resolution NMR techniques in organic chemistry*. Tetrahedron organic chemistry series; v. 19. 1999, Amsterdam; Oxford: Pergamon. xiv, 382.
25. Derome, A.E., *Modern NMR techniques for chemistry research*. Organic chemistry series (Pergamon Press). 1987, Oxford: Pergamon. xvii, 280.
26. Lipari, G. and A. Szabo, *Model-free approach to the interpretation of nuclear magnetic resonance relaxation in macromolecules. 2. Analysis of experimental data.* J. Am. Chem. Soc., 1982. **104**: p. 4559-4570.
27. Kruk, D. and J. Kowalewski, *Nuclear spin relaxation in solution of paramagnetic complexes with large transient zero-field splitting.* Molecular Physics, 2003. **101**(18): p. 2861-2874.
28. Kruk, D., T. Nilsson, and J. Kowalewski, *Nuclear spin relaxation in paramagnetic systems with zero-field splitting and arbitrary electron spin.* Physical Chemistry Chemical Physics, 2001. **3**(22): p. 4907-4917.
29. Bertini, I., et al., *Nuclear spin relaxation in paramagnetic complexes of S=1: Electron spin relaxation effects.* Journal of Chemical Physics, 1999. **111**(13): p. 5795-5807.
30. Atherton, N.M. and N.M. Atherton, *Principles of electron spin resonance*. Ellis Horwood PTR Prentice Hall physical chemistry series. 1993, New York: Ellis Horwood; PTR Prentice Hall. ix, 585.
31. Ramsey, N.F., *Magnetic shielding of nuclei in molecules.* Physical Review, 1950. **78**: p. 699-703.
32. Ramsey, N.F., *Chemical effects in nuclear magnetic resonance and in diamagnetic susceptibility.* Physical Review, 1952. **86**: p. 243-6.
33. Ramsey, N.F., *Dependence of magnetic shielding of nuclei upon molecular orientation.* Physical Review, 1951. **83**: p. 540-1.
34. Adamson, J.G., et al., *Simple and convenient synthesis of tert-butyl ethers of Fmoc-serine, Fmoc-threonine, and Fmoc-tyrosine.* Journal of Organic Chemistry, 1991. **56**(10): p. 3447-9.
35. de Dios, A.C. and E. Oldfield, *Evaluating 19F Chemical Shielding in Fluorobenzenes: Implications for Chemical Shifts in Proteins.* Journal of the American Chemical Society, 1994. **116**(16): p. 7453-4.
36. Goldman, M., *Interference effects in the relaxation of a pair of unlike spin-1/2 nuclei.* Journal of Magnetic Resonance (1969-1992), 1984. **60**(3): p. 437-52.
37. Kumar, A., R. Christy Rani Grace, and P.K. Madhu, *Cross-correlations in NMR.* Progress in Nuclear Magnetic Resonance Spectroscopy, 2000. **37**(3): p. 191-319.
38. Griesinger, C., *Methods for the measurement of angle restraints from scalar, dipolar couplings and from cross-correlated relaxation: application to biomacromolecues.* Methods and Principles in Medicinal Chemistry, 2003. **16**(BioNMR in Drug Research): p. 147-178.
39. Carlomagno, T., W. Bermel, and C. Griesinger, *Measuring the c1 torsion angle in protein by CH-CH cross-correlated relaxation: A new resolution-optimised experiment.* Journal of Biomolecular NMR, 2003. **27**(2): p. 151-157.
40. Schwalbe, H., et al., *Cross-correlated relaxation for measurement of angles between tensorial interactions.* Methods in Enzymology, 2001. **338**(Nuclear Magnetic Resonance of Biological Macromolecules, Part A): p. 35-81.
41. Muus, L.T., et al., *Chemically induced magnetic polarization: proceedings of the NATO Advanced Study Institute held at Sogesta, Urbino, Italy, April 17-30,*

*1977*. NATO advanced study institutes series. Series C, Mathematical and physical sciences; v. 34. 1977, Dordrecht, Holland; Boston: D. Reidel. xii, 407.
42. Goez, M., *Photochemically induced dynamic nuclear polarization.* Adv. Photochem., 1997. **23**: p. 63-163.
43. Vollenweider, J.K. and H. Fischer, *Absolute chemically induced nuclear polarizations and yields from geminate radical-pair reactions. A test of high-field radical-pair theories.* Chem. Phys., 1988. **124**(3): p. 333-45.
44. Vollenweider, J.K., et al., *Time-resolved CIDNP in laser flash photolysis of aliphatic ketones. A quantitative analysis.* Chem. Phys., 1985. **97**(2-3): p. 217-34.
45. Kaptein, R., *Photo-CIDNP studies of proteins.* Biol. Magn. Reson., 1982. **4**: p. 145-91.
46. Kaptein, R., K. Dijkstra, and K. Nicolay, *Laser photo-CIDNP as a surface probe for proteins in solution.* Nature, 1978. **274**(5668): p. 293-4.
47. Hore, P.J. and R.W. Broadhurst, *Photo-CIDNP of biopolymers.* Progr. NMR Spectrosc., 1993. **25**(4): p. 345-402.
48. Lyon, C.E., et al., *Two-dimensional $^{15}N$-$^{1}H$ photo-CIDNP as a surface probe of native and partially structured proteins.* Journal of the American Chemical Society, 1999. **121**(27): p. 6505-6506.
49. Lyon, C.E., et al., *Probing the Exposure of Tyrosine and Tryptophan Residues in Partially Folded Proteins and Folding Intermediates by CIDNP Pulse-Labeling.* Journal of the American Chemical Society, 2002. **124**(44): p. 13018-13024.
50. Mok, K.H., et al., *Rapid sample-mixing technique for transient NMR and photo-CIDNP spectroscopy: Applications to real-time protein folding.* Journal of the American Chemical Society, 2003. **125**(41): p. 12484-12492.
51. Sykes, B.D., *Biosynthesis and NMR characterisation of fluoroamino acid containing proteins.*, in *Magnetic Resonance in Biology*, J.S. Cohen, Editor. 1980, John Wiley & Sons. p. 171-196.
52. Kuprov, I. and P.J. Hore, *Chemically amplified $^{19}F$-$^{1}H$ nuclear Overhauser effects.* J. Magn. Reson., 2004. **168**(1): p. 1-7.
53. Ivanov, K.L., et al., *Investigation of the magnetic field dependence of CIDNP in multinuclear radical pairs. 1. Photoreaction of histidine and comparison of model calculation with experimental data.* Molecular Physics, 2002. **100**(8): p. 1197-1208.
54. Ivanov, K.L., et al., *Investigation of the magnetic field dependence of CIDNP in multi-nuclear radical pairs.Part II. Photoreaction of tyrosine and comparison of model calculation with experimental data.* Physical Chemistry Chemical Physics, 2003. **5**(16): p. 3470-3480.
55. van den Berg, P.A.W., et al., *Fluorescence correlation spectroscopy of flavins and flavoenzymes: photochemical and photophysical aspects.* Spectrochimica Acta, Part A: Molecular and Biomolecular Spectroscopy, 2001. **57A**(11): p. 2135-2144.
56. Martin, C.B., et al., *The Reaction of Triplet Flavin with Indole. A Study of the Cascade of Reactive Intermediates Using Density Functional Theory and Time Resolved Infrared Spectroscopy.* Journal of the American Chemical Society, 2002. **124**(24): p. 7226-7234.
57. Morozova, O.B., et al., *Time resolved CIDNP study of electron transfer reactions in proteins and model compounds.* Mol. Phys., 2002. **100**(8): p. 1187-1195.

58. Tsentalovich, Y.P., O.A. Snytnikova, and R.Z. Sagdeev, *Properties of excited states of aqueous tryptophan.* Journal of Photochemistry and Photobiology, A: Chemistry, 2004. **162**(2-3): p. 371-379.
59. Cintolesi, F., et al., *Anisotropic recombination of an immobilized photoinduced radical pair in a 50-mT magnetic field: a model avian photomagnetoreceptor.* Chemical Physics, 2003. **294**(3): p. 385-399.
60. Kaptein, R., *Simple rules for chemically induced dynamic nuclear polarization.* Journal of the Chemical Society D: Chemical Communications, 1971(14): p. 732-3.
61. Prudnikov, A.P., ë.A. Brychkov, and O.I. Marichev, *Integrals and series*. 2002, London: Taylor & Francis.
62. Cemazar, M., PhD thesis, University of Oxford, 2002.
63. Lyon, C.E., PhD thesis, University of Oxford, 1999.
64. Adrian, F.J., *Possible Overhauser mechanism for fluorine-19 nuclear spin polarization in the reaction of fluorobenzyl halides with sodium naphthalene.* Chemical Physics Letters, 1974. **26**(3): p. 437-9.
65. Adrian, F.J., H.M. Vyas, and J.K.S. Wan, *Magnetic field and concentration dependence of CIDNP in some quinone photolyses: Further evidence for an Overhauser mechanism.* Journal of Chemical Physics, 1976. **65**(4): p. 1454-61.
66. Rakshys, J.W., Jr., *Use of fluorine-19 CIDNP [chemically-induced dynamic nuclear spin polarization] in reaction mechanism studies. Evidence for the intermediacy of benzyl anion in the reaction of benzyl halides with sodium naphthalene.* Tetrahedron Letters, 1971(49): p. 4745-8.
67. Vyas, H.M. and J.K.S. Wan, *Chemically induced dynamic nuclear polarization in the photolysis of tetrafluoro-1,4-benzoquinone with plane polarized light. Evidence of phototriplet mechanism.* Chemical Physics Letters, 1975. **34**(3): p. 470-2.
68. Vyas, H.M. and J.K.S. Wan, *Electron spin resonance and kinetic study of the photolysis of p-fluoranil (tetrafluoro-p-benzoquinone) in dioxane.* International Journal of Chemical Kinetics, 1974. **6**(1): p. 125-32.
69. Thomas, M.J., et al., *Competing triplet and radical pair fluorine-19 polarizations in the electron transfer quenching of triplet a,a,a-trifluoroacetophenone.* Journal of the American Chemical Society, 1977. **99**(11): p. 3842-5.
70. Kohn, W., *Nobel lecture: electronic structure of matter-wave functions and density functionals.* Reviews of Modern Physics, 1999. **71**(5): p. 1253-1266.
71. Pople, J.A., *Nobel lecture: Quantum chemical models.* Reviews of Modern Physics, 1999. **71**(5): p. 1267-1274.
72. Wirmer, J., T. Kuhn, and H. Schwalbe, *Millisecond time resolved photo-CIDNP NMR reveals a non-native folding intermediate on the ion-induced refolding pathway of bovine a-lactalbumin.* Angew. Chem., 2001. **40**(22): p. 4248-4251.
73. Sykes, B.D., H.I. Weingarten, and M.J. Schlesinger, *Fluorotyrosine alkaline phosphatase from Escherichia coli. Preparation, properties, and fluorine-19 nuclear magnetic resonance spectrum.* Proceedings of the National Academy of Sciences of the United States of America, 1974. **71**(2): p. 469-73.
74. Winder, S.L., R.W. Broadhurst, and P.J. Hore, *Photo-CIDNP of amino acids and proteins: effects of competition for flavin triplets.* Spectrochimica Acta, Part A: Molecular and Biomolecular Spectroscopy, 1995. **51A**(10): p. 1753-61.

75. Scheffler, J.E., C.E. Cottrell, and L.J. Berliner, *An inexpensive, versatile sample illuminator for photo-CIDNP on any NMR spectrometer.* J. Magn. Reson., 1985. **63**(1): p. 199-201.
76. M. J. Frisch, G.W.T., H. B. Schlegel, G. E. Scuseria, et al., *Gaussian 03, Revision C.02*. 2004.
77. Schmidt, M.W., et al., *General atomic and molecular electronic structure system.* Journal of Computational Chemistry, 1993. **14**(11): p. 1347-63.
78. *DALTON, a molecular electronic structure program.* http://www.kjemi.uio.no/software/dalton/dalton.html. 2005.
79. Szabo, A. and N.S. Ostlund, *Modern quantum chemistry: introduction to advanced electronic structure theory*. 1996, Mineola, N.Y., London: Dover. xiv, 466.
80. Helgaker, T., M. Jaszunski, and K. Ruud, *Ab initio methods for the calculation of NMR shielding and indirect spin-spin coupling constants.* Chemical Reviews (Washington, D. C.), 1999. **99**(1): p. 293-352.
81. Barone, V., in *Recent Advances in Density Functional Methods*, D.P. Chong, Editor. 1995, World Scientific: Singapore. p. 287.
82. Rega, N., M. Cossi, and V. Barone, *Development and validation of reliable quantum mechanical approaches for the study of free radicals in solution.* Journal of Chemical Physics, 1996. **105**(24): p. 11060-11067.
83. Improta, R. and V. Barone, *Interplay of Electronic, Environmental, and Vibrational Effects in Determining the Hyperfine Coupling Constants of Organic Free Radicals.* Chemical Reviews (Washington, DC, United States), 2004. **104**(3): p. 1231-1253.
84. Holton, D.M. and D. Murphy, *Determination of acid dissociation constants of some phenol radical cations. Part 2.* Journal of the Chemical Society, Faraday Transactions 2: Molecular and Chemical Physics, 1979. **75**(12): p. 1637-42.
85. Lloyd, R.V. and D.E. Wood, *Free radicals in an adamantane matrix. VIII. EPR and INDO[intermediate neglect of differential overlap] study of the benzyl, anilino, and phenoxy radicals and their fluorinated derivatives.* Journal of the American Chemical Society, 1974. **96**(3): p. 659-65.
86. Lide, D.R., R.C. Weast, and Chemical Rubber Company., *CRC Handbook of chemistry and physics: a ready reference book of chemical and physical data*. 83rd ed. 2002, Boca Raton, Fla.: CRC Press. 1 v. (various pagings).
87. Dupureur, C.M. and L.M. Hallman, *Effects of divalent metal ions on the activity and conformation of native and 3-fluorotyrosine-PvuII endonucleases.* European Journal of Biochemistry, 1999. **261**(1): p. 261-268.
88. Wacks, D.B. and H.K. Schachman, *Fluorine-19 nuclear magnetic resonance studies of communication between catalytic and regulatory subunits in aspartate transcarbamoylase.* Journal of Biological Chemistry, 1985. **260**(21): p. 11659-62.
89. Wacks, D.B. and H.K. Schachman, *Fluorine-19 nuclear magnetic resonance studies of fluorotyrosine-labeled aspartate transcarbamoylase. Properties of the enzyme and its catalytic and regulatory subunits.* Journal of Biological Chemistry, 1985. **260**(21): p. 11651-8.
90. Monasterio, O., et al., *Tubulin-tyrosine ligase catalyzes covalent binding of 3-fluoro-tyrosine to tubulin: kinetic and [19F]NMR studies.* FEBS Letters, 1995. **374**(2): p. 165-8.

91. Sun, Z.Y., et al., *A fluorine-19 NMR study of the membrane-binding region of D-lactate dehydrogenase of Escherichia coli.* Protein Science, 1993. **2**(11): p. 1938-47.
92. Hinds, M.G., R.W. King, and J. Feeney, *Fluorine-19 NMR evidence for interactions between the cAMP binding sites on the cAMP receptor protein from E. coli.* FEBS Letters, 1991. **283**(1): p. 127-30.
93. Sixl, F., et al., *Fluorine-19 NMR studies of ligand binding to 5-fluorotryptophan- and 3-fluorotyrosine-containing cyclic AMP receptor protein from Escherichia coli.* Biochemical Journal, 1990. **266**(2): p. 545-52.
94. Dettman, H.D., J.H. Weiner, and B.D. Sykes, *19F nuclear magnetic resonance studies of the coat protein of bacteriophage M13 in synthetic phospholipid vesicles and deoxycholate micelles.* Biophysical journal, 1982. **37**(1): p. 243-51.
95. Hagen, D.S., J.H. Weiner, and B.D. Sykes, *Fluorotyrosine M13 coat protein: fluorine-19 nuclear magnetic resonance study of the motional properties of an integral membrane protein in phospholipid vesicles.* Biochemistry, 1978. **17**(18): p. 3860-6.
96. Hull, W.E. and B.D. Sykes, *Fluorine-19 nuclear magnetic resonance study of fluorotyrosine alkaline phosphatase: the influence of zinc on protein structure and a conformational change induced by phosphate binding.* Biochemistry, 1976. **15**(7): p. 1535-46.
97. Hull, W.E. and B.D. Sykes, *Fluorotyrosine alkaline phosphatase. 19F nuclear magnetic resonance relaxation times and molecular motion of the individual fluorotyrosines.* Biochemistry, 1974. **13**(17): p. 3431-7.
98. Tsentalovich, Y.P., et al., *Kinetics and mechanism of the photochemical reaction of 2,2'-dipyridyl with tryptophan in water: Time-resolved CIDNP and laser flash photolysis study.* J. Phys. Chem. A, 1999. **103**(27): p. 5362-5368.
99. Polenov, E.A. and B.I. Shapiro, *Angular dependence of a b-fluorine hyperfine interaction.* Zhurnal Strukturnoi Khimii, 1972. **13**(2): p. 329-32.
100. Polenov, E.A., B.I. Shapiro, and L.M. Yagupol'skii, *Anion radicals of nitrobenzene with substituents containing fluoroalkyl groups.* Zhurnal Strukturnoi Khimii, 1971. **12**(1): p. 163-7.
101. Polenov, E.A., et al., *Anion radicals of para-substituted nitrobenzenes with $CFH_2$, $CHF_2$, and $CF_3$ groups and angular-function parameters of the b-fluoro isotropic hyperfine interaction.* Zhurnal Fizicheskoi Khimii, 1993. **67**(1): p. 65-9.
102. Beregovaya, I.V., L.N. Shchegoleva, and V.E. Platonov, *Electronic and geometric structure of fluorinated benzyl radicals.* Izvestiya Akademii Nauk SSSR, Seriya Khimicheskaya, 1990(5): p. 1069-74.
103. Kestner, N.R. and J.E. Combariza, *Basis set superposition errors: theory and practice.* Reviews in Computational Chemistry, 1999. **13**: p. 99-132.
104. Van Duijneveldt, F.B., *Basis set superposition error.* Molecular Interactions, 1997: p. 81-104.
105. Beguin, C.G. and R. Dupeyre, *Nuclear spin relaxation in benzyl fluoride. I. Deuteron relaxation for internal rotational barrier determination and proton and fluorine-19 intra- and intermolecular relaxation in pure benzyl fluoride.* Journal of Magnetic Resonance (1969-1992), 1981. **44**(2): p. 294-313.
106. Bae, J.H., et al., *Crystallographic evidence for isomeric chromophores in 3-fluorotyrosyl-green fluorescent protein.* ChemBioChem, 2004. **5**(5): p. 720-722.

107. Xiao, G., et al., *Conformational changes in the crystal structure of rat glutathione transferase M1-1 with global substitution of 3-fluorotyrosine for tyrosine.* Journal of Molecular Biology, 1998. **281**(2): p. 323-339.
108. Xiao, G., et al., *Crystal Structure of Tetradeca-(3-Fluorotyrosyl)-Glutathione Transferase.* Journal of the American Chemical Society, 1997. **119**(39): p. 9325-9326.
109. Lui, S.M. and J.A. Cowan, *Studies of the Electronic and Dynamic Properties of High-Potential Iron Proteins by Substitution with Non-Natural Amino Acids. 3-Fluorotyrosine-Modified Chromatium vinosum High-Potential Iron Protein.* Journal of the American Chemical Society, 1994. **116**(10): p. 4483-4.
110. Kehry, M.R., M.L. Wilson, and F.W. Dahlquist, *A simple quantitative method for the determination of 3-fluorotyrosine substitution in proteins.* Analytical Biochemistry, 1983. **131**(1): p. 236-41.
111. Kimber, B.J., et al., *Fluorine-19 nuclear magnetic resonance studies of ligand binding to 3-fluorotyrosine- and 6-fluorotryptophan-containing dihydrofolate reductase from Lactobacillus casei.* Biochemistry, 1977. **16**(15): p. 3492-500.
112. Lu, P., et al., *lac Repressor: 3-Fluorotyrosine substitution for nuclear magnetic resonance studies.* Proceedings of the National Academy of Sciences of the United States of America, 1976. **73**(10): p. 3471-5.
113. Hull, W.E. and B.D. Sykes, *Fluorotyrosine alkaline phosphatase. Internal mobility of individual tyrosines and the role of chemical shift anisotropy as a fluorine-19 nuclear spin relaxation mechanism in proteins.* Journal of Molecular Biology, 1975. **98**(1): p. 121-53.
114. Hull, W.E. and B.D. Sykes, *Fluorotyrosine alkaline phosphatase. Fluorine-19 nuclear magnetic resonance relaxation times and molecular motion of the individual fluorotyrosines.* Biochemistry, 1974. **13**(17): p. 3431-7.
115. Munier, R.L. and G. Sarrazin, *Total substitution of 3-fluorotyrosine for tyrosine in the proteins of Escherichia coli.* Compt. Rend., 1963. **256**: p. 3376-8.
116. Hore, P.J., et al., *Cross-relaxation effects in the photo-CIDNP spectra of amino acids and proteins.* Journal of Magnetic Resonance (1969-1992), 1982. **49**(1): p. 122-50.
117. Goez, M., *An introduction to chemically induced dynamic nuclear polarization.* Concepts Magn. Reson., 1995. **7**(1): p. 69-86.
118. Maeda, K., et al., *Improved photo-CIDNP methods for studying protein structure and folding.* Journal of Biomolecular NMR, 2000. **16**(3): p. 235-244.
119. Granovsky, A., *PC GAMESS.* 2005.
120. Sanders, L.K. and E. Oldfield, *Theoretical Investigation of 19F NMR Chemical Shielding Tensors in Fluorobenzenes.* Journal of Physical Chemistry A, 2001. **105**(34): p. 8098-8104.
121. Tomkiewicz, M., R.D. McAlpine, and M. Cocivera, *Photooxidation and decarboxylation of tyrosine studied by EPR and CIDNP [chemically-induced dynamic nuclear polarization] techniques.* Canadian Journal of Chemistry, 1972. **50**(23): p. 3849-56.
122. Dorai, K. and A. Kumar, *Fluorine chemical shift tensors in substituted fluorobenzenes using cross correlations in NMR relaxation.* Chemical Physics Letters, 2001. **335**(3,4): p. 176-182.
123. Goez, M., *Pulse techniques for CIDNP.* Concepts Magn. Reson., 1995. **7**(4): p. 263-79.

124. Bargon, J. and G.P. Gardini, *Transfer of CIDNP via proton exchange and nuclear Overhauser effect.* Journal of the American Chemical Society, 1979. **101**(26): p. 7732.

125. Matysik, J., et al., *Photochemically induced nuclear spin polarization in reaction centers of photosystem II observed by $^{13}$C-solid-state NMR reveals a strongly asymmetric electronic structure of the $P_{680}^{\cdot +}$ primary donor chlorophyll.* Proceedings of the National Academy of Sciences of the United States of America, 2000. **97**(18): p. 9865-9870.

126. Kuhn, T. and H. Schwalbe, *Monitoring the Kinetics of Ion-Dependent Protein Folding by Time-Resolved NMR Spectroscopy at Atomic Resolution.* J. Am. Chem. Soc., 2000. **122**(26): p. 6169-6174.

127. Hore, P.J., et al., *Flash photolysis NMR. CIDNP time dependence in cyclic photochemical reactions.* Chem. Phys. Lett., 1981. **83**(2): p. 376-83.

128. Morozova, O.B., et al., *Consecutive biradicals during the photolysis of 2,12-dihydroxy-2,12-dimethylcyclododecanone: low- and high-field chemically induced dynamic nuclear polarizations (CIDNP) study.* J. Phys. Chem. A, 1998. **102**(20): p. 3492-3497.

129. Rubinstenn, G., et al., *Structural and dynamic changes of photoactive yellow protein during its photocycle in solution.* Nature Struct. Biol., 1998. **5**(7): p. 568-570.

130. Rubinstenn, G., et al., *NMR Experiments for the Study of Photointermediates: Application to the Photoactive Yellow Protein.* J. Magn. Reson., 1999. **137**(2): p. 443-447.

131. Harper, S.M., et al., *Conformational Changes in a Photosensory LOV Domain Monitored by Time-Resolved NMR Spectroscopy.* J. Am. Chem. Soc., 2004. **126**(11): p. 3390-3391.

132. Harper, S.M., L.C. Neil, and K.H. Gardner, *Structural basis of a phototropin light switch.* Science, 2003. **301**(5639): p. 1541-1544.

133. Cemazar, M., et al., *Oxidative folding intermediates with nonnative disulfide bridges between adjacent cysteine residues.* Proceedings of the National Academy of Sciences of the United States of America, 2003. **100**(10): p. 5754-5759.

134. Morris, G.A., *Reference Deconvolution*, in *Encyclopedia of Nuclear Magnetic Resonance*, D.M. Grant and R.K. Harris, Editors. 2002. p. 125-131.

135. Salomir, R., B.D. de Senneville, and C.T.W. Moonen, *A fast calculation method for magnetic field inhomogeneity due to arbitrary distribution of bulk susceptibility.* Concepts in Magnetic Resonance B, 2002. **19**: p. 26-34.

136. Holzer, W., et al., *Photoinduced degradation of some flavins in aqueous solution.* Chemical Physics, 2004. **308**(1-2): p. 69-78.

137. Silva, E., et al., *Riboflavin-sensitized photoprocesses of tryptophan.* Journal of Photochemistry and Photobiology, B: Biology, 1994. **23**(1): p. 43-8.

138. Neidigh, J.W., R.M. Fesinmeyer, and N.H. Andersen, *Designing a 20-residue protein.* Nature Structural Biology, 2002. **9**(6): p. 425-430.

139. Qiu, L. and S.J. Hagen, *A Limiting Speed for Protein Folding at Low Solvent Viscosity.* Journal of the American Chemical Society, 2004. **126**(11): p. 3398-3399.

140. Chowdhury, S., et al., *Ab initio Folding Simulation of the Trp-cage Mini-protein Approaches NMR Resolution.* Journal of Molecular Biology, 2003. **327**(3): p. 711-717.

141. Nikiforovich, G.V., et al., *Possible locally driven folding pathways of TC5b, a 20-residue protein.* Proteins: Structure, Function, and Genetics, 2003. **52**(2): p. 292-302.
142. Simmerling, C., B. Strockbine, and A.E. Roitberg, *All-Atom Structure Prediction and Folding Simulations of a Stable Protein.* Journal of the American Chemical Society, 2002. **124**(38): p. 11258-11259.
143. Pitera, J.W. and W. Swope, *Understanding folding and design: Replica-exchange simulations of \"Trp-cage\" miniproteins.* Proceedings of the National Academy of Sciences of the United States of America, 2003. **100**(13): p. 7587-7592.
144. Ghosh, R., R. Bachofen, and H. Hauser, *Incorporation of fluorine-19-substituted aromatic amino acids into membrane proteins from chromatophores of Rhodospirillum rubrum G9+.* FEBS Letters, 1985. **188**(1): p. 107-11.
145. Yan, L.Z. and J.P. Mayer, *Use of Trichloroacetimidate Linker in Solid-Phase Peptide Synthesis.* Journal of Organic Chemistry, 2003. **68**(3): p. 1161-1162.
146. Novabiochem, *Their catalogue.* 2002.
147. Robertson, J., *Protecting group chemistry*. Oxford chemistry primers; 95. 2000, Oxford: Oxford University Press. 96.
148. Carpino, L.A. and G.Y. Han, *9-Fluorenylmethoxycarbonyl function, a new base-sensitive amino-protecting group.* Journal of the American Chemical Society, 1970. **92**(19): p. 5748-9.
149. Carpino, L.A. and G.Y. Han, *(9-Fluorenylmethoxy)carbonyl-protected amino acids and peptides*, in *Ger. Offen.* 1972, (Research Corp.). DE. p. 29 pp.
150. Carpino, L.A. and G.Y. Han, *9-Fluorenylmethoxycarbonyl amino-protecting group.* Journal of Organic Chemistry, 1972. **37**(22): p. 3404-9.
151. Carpino, L.A. and G.Y. Han, *9-Fluorenylmethoxycarbonyl compounds*, in *U.S.* 1975, (Research Corp., USA). US. p. 8 pp. Division of U.S. 3,835,175.
152. Ten Kortenaar, P.B.W., et al., *Rapid and efficient method for the preparation of Fmoc-amino acids starting from 9-fluorenylmethanol.* International Journal of Peptide & Protein Research, 1986. **27**(4): p. 398-400.
153. Wellings, D.A. and E. Atherton, *Standard Fmoc protocols.* Methods in Enzymology, 1997. **289**(Solid-Phase Peptide Synthesis): p. 44-67.
154. Andreotti, A.H., *Native State Proline Isomerization: An Intrinsic Molecular Switch.* Biochemistry, 2003. **42**(32): p. 9515-9524.
155. Wedemeyer, W.J., E. Welker, and H.A. Scheraga, *Proline Cis-Trans Isomerization and Protein Folding.* Biochemistry, 2002. **41**(50): p. 14637-14644.
156. Adrian, F.J., *Triplet Overhauser mechanism of CIDNP.* NATO Advanced Study Institutes Series, Series C: Mathematical and Physical Sciences, 1977. **C34**(Chem. Induced Magn. Polariz.): p. 369-81.
157. Morozova, O.B., et al., *Cross-relaxation mechanism for the formation of nuclear polarization: a quantitative time-resolved CIDNP study.* Chemical Physics Letters, 1995. **246**(4,5): p. 499-505.
158. Valyaev, V.I., et al., *Cross relaxation in free radicals with chemically induced electron and nuclear polarization.* Molecular Physics, 1988. **63**(5): p. 891-900.
159. Tsentalovich, Y.P., et al., *A cross-correlation mechanism for the formation of spin polarization.* Journal of Physical Chemistry, 1993. **97**(35): p. 8900-8.
160. Tsentalovich, Y.P., et al., *Cross-relaxation and cross-correlation mechanisms leading to spin polarization.* Zeitschrift fuer Physikalische Chemie (Muenchen, Germany), 1993. **182**(1-2): p. 119-29.

161. Alam, T.M., D.M. Pedrotty, and T.J. Boyle, *Modified, pulse field gradient-enhanced inverse-detected HOESY pulse sequence for reduction of t1 spectral artifacts.* Magnetic Resonance in Chemistry, 2002. **40**(5): p. 361-365.

162. Yu, C. and G.C. Levy, *Two-dimensional heteronuclear NOE (HOESY) experiments: investigation of dipolar interactions between heteronuclei and nearby protons.* Journal of the American Chemical Society, 1984. **106**(22): p. 6533-7.

163. Morozova, O.B., et al., *Time-resolved CIDNP study of native-state bovine and human a-lactalbumins.* J. Phys. Chem. B, 2004. **108**(39): p. 15355-15363.

164. Morozova, O.B., A.V. Yurkovskaya, and R.Z. Sagdeev, *Reversibility of Electron Transfer in Tryptophan-Tyrosine Peptide in Acidic Aqueous Solution Studied by Time-Resolved CIDNP.* Journal of Physical Chemistry B, 2005. **109**(8): p. 3668-3675.

165. Morozova, O.B., et al., *Intramolecular Electron Transfer in Tryptophan-Tyrosine Peptide in Photoinduced Reaction in Aqueous Solution.* Journal of Physical Chemistry B, 2003. **107**(4): p. 1088-1096.

166. Tsien, R.Y., *The green fluorescent protein.* Ann. Rev. Biochem., 1998. **67**: p. 509-544.

167. Zimmer, M., *Green Fluorescent Protein (GFP): Applications, Structure, and Related Photophysical Behavior.* Chem. Rev., 2002. **102**: p. 759-781.

168. Enoki, S., et al., *Acid denaturation and refolding of green fluorescent protein.* Biochemistry, 2004. **43**: p. 14238-48.

169. Fukuda, H., M. Arai, and K. Kuwajima, *Folding of green fluorescent protein and the cycle3 mutant.* Biochemistry, 2000. **39**: p. 12025-12032.

170. Stepanenko, O.V., et al., *Comparative studies on the structure and stability of fluorescent proteins EGFP, zFP506, mRFP1, "dimer2", and DsRed1.* Biochemistry, 2004. **43**: p. 14913-14923.

171. Ormo, M., et al., *Crystal structure of the Aequorea victoria green fluorescent protein.* Science, 1996. **273**: p. 1392-1395.

172. Yang, F., L. Moss, and G. Phillips, *The molecular structure of green fluorescent protein.* Nat. Biotech., 1996. **14**: p. 1246-1251.

173. Reid, B.G. and G.C. Flynn, *Chromophore formation in green fluorescent protein.* Biochemistry, 1997. **36**: p. 6786-6791.

174. Redfield, C., *NMR studies of partially folded molten-globule states.* Methods in Molecular Biology (Totowa, NJ, United States), 2004. **278**(Protein NMR Techniques): p. 233-254.

175. Li, H. and C. Frieden, *NMR Studies of 4-19F-Phenylalanine-Labeled Intestinal Fatty Acid Binding Protein: Evidence for Conformational Heterogeneity in the Native State.* Biochemistry, 2005. **44**(7): p. 2369-2377.

176. Shu, Q. and C. Frieden, *Relation of Enzyme Activity to Local/Global Stability of Murine Adenosine Deaminase: 19F NMR Studies.* Journal of Molecular Biology, 2004. **345**(3): p. 599-610.

177. Bann, J.G. and C. Frieden, *Folding and domain-domain interactions of the chaperone PapD measured by 19F NMR.* Biochemistry, 2004. **43**(43): p. 13775-13786.

178. Peng, J.W., *Cross-Correlated 19F Relaxation Measurements for the Study of Fluorinated Ligand-Receptor Interactions.* Journal of Magnetic Resonance, 2001. **153**(1): p. 32-47.

179. Gakh, Y.G., A.A. Gakh, and A.M. Gronenborn, *Fluorine as an NMR probe for structural studies of chemical and biological systems.* Magnetic Resonance in Chemistry, 2000. **38**(7): p. 551-558.
180. Kuprov, I. and P.J. Hore, *Uniform illumination of optically dense NMR samples.* J. Magn. Res., 2004. **171**: p. 171-175.
181. Tormena, C.F. and G.V.J. da Silva, *Chemical shifts calculations on aromatic systems: a comparison of models and basis sets.* Chemical Physics Letters, 2004. **398**(4-6): p. 466-470.
182. Garcia de la Torre, J., M.L. Huertas, and B. Carrasco, *HYDRONMR: Prediction of NMR Relaxation of Globular Proteins from Atomic-Level Structures and Hydrodynamic Calculations.* Journal of Magnetic Resonance, 2000. **147**(1): p. 138-146.
183. Craggs, T.D. and S.E. Jackson, *Unpublished data.*
184. Tsentalovich, Y.P., et al., *Mechanisms of reactions of flavin mononucleotide triplet with aromatic amino acids.* Spectrochimica Acta, Part A: Molecular and Biomolecular Spectroscopy, 2002. **58A**(9): p. 2043-2050.
185. Berliner, L.J. and R. Kaptein, *Nuclear magnetic resonance characterization of aromatic residues of alpha-lactalbumins. Laser photo chemically induced dynamic nuclear polarization nuclear magnetic resonance studies of surface exposure.* Biochemistry, 1981. **20**(4): p. 799-807.
186. Fraczkiewicz, R. and W. Braun, *Exact and efficient analytical calculation of the accessible surface areas and their gradients for macromolecules.* J. Comp. Chem., 1998. **19**: p. 319-333.
187. Battistutta, R., A. Negro, and G. Zanotti, *Crystal structure and refolding properties of the mutant F99S/M153T/V163A of the Green Fluorescent Protein.* Proteins: Struct. Func. & Gen., 2000. **41**: p. 429-437.
188. Atkins, P.W., et al., *Electron spin resonance flash photolysis and chemically induced polarization study of the photolysis of benzaldehyde in solution.* J. Chem. Soc. (Faraday Trans. 2: Mol. Chem. Phys), 1973. **69**(10): p. 1542-57.
189. Kuprov, I. and P.J. Hore, *Uniform illumination of optically dense NMR samples.* J. Magn. Reson., 2004. **171**(1): p. 171-175.
190. Miller, R.J. and G.L. Closs, *Application of Fourier transform-NMR spectroscopy to submicrosecond time-resolved detection in laser flash photolysis experiments.* Rev. Sci. Instrum., 1981. **52**(12): p. 1876-85.
191. Blank, B., A. Henne, and H. Fischer, *CIDNP [chemically-induced dynamic nuclear polarization]. 17. Photochemical primary reactions of a-branched ketones and aldehydes in solution.* Helv. Chim. Acta, 1974. **57**(3): p. 920-36.
192. Yurkovskaya, A.V., Personal communication.
193. McDonald, G.G., *Photometer for aligning a photo CIDNP optical system.* J. Magn. Reson., 1983. **53**: p. 115.
194. Salikhov, K.M., Y.N. Molin, and A.L. Buchachenko, *Spin polarization and magnetic effects in radical reactions*. Studies in physical and theoretical chemistry; 22. 1984, Amsterdam; New York; Budapest: Elsevier. 419.
195. Yurkovskaya, A.V., et al., *The influence of scavenging on CIDNP field dependences in biradicals during the photolysis of large-ring cycloalkanones.* Chem. Phys., 1995. **197**(2): p. 157-66.
196. Goez, M., *Evaluation of flash CIDNP experiments by iterative reconvolution.* Chem. Phys. Lett., 1990. **165**(1): p. 11-14.

197. Roth, H.D. and A.A. Lamola, *Chemically induced dynamic nuclear polarization and exchange broadening in an electron-transfer reaction.* J. Am. Chem. Soc., 1974. **96**(20): p. 6270-5.
198. Schwalbe, H., Personal Communication.
199. Kuprov, I., *Chemically induced dynamic nuclear polarisation of 19F nuclei.* DPhil thesis, University of Oxford, 2005.
200. Lendzian, F., et al., *Electronic Structure of Neutral Tryptophan Radicals in Ribonucleotide Reductase Studied by EPR and ENDOR Spectroscopy.* J. Am. Chem. Soc., 1996. **118**(34): p. 8111-8120.
201. Maeda, K., *Personal communication.*
202. Adrian, F.J., *Contribution of S0-T1 intersystem crossing in radical pairs to chemically induced nuclear and electron spin polarizations.* Chemical Physics Letters, 1971. **10**(1): p. 70-4.
203. Adrian, F.J. and L. Monchick, *Theory of chemically induced magnetic polarization. Effects of S-T+-1 mixing in strong magnetic fields.* Journal of Chemical Physics, 1979. **71**(6): p. 2600-10.
204. Borbat, P.P., A.D. Milov, and Y.N. Molin, *Electron spin echo study of CIDEP in photolysis of di-tert-butyl ketone at low temperatures.* Pure and Applied Chemistry, 1992. **64**(6): p. 883-92.
205. Batchelor, S.N. and H. Fischer, *Time-Resolved CIDNP Investigation of the Cross-Relaxation Mechanism of 1H Nuclear Polarization.* Journal of Physical Chemistry, 1996. **100**(2): p. 556-64.
206. Noyes, R.M., *Models relating molecular reactivity and diffusion in liquids.* Journal of the American Chemical Society, 1956. **78**: p. 5486-90.
207. Noyes, R.M., *Kinetics of competitive processes when reactive fragments are produced in pairs.* Journal of the American Chemical Society, 1955. **77**: p. 2042-5.
208. Borbat, P.P., A.D. Milov, and Y.N. Molin, *Electron spin echo study of cross relaxation in free radicals with CIDEP.* Chemical Physics Letters, 1989. **164**(4): p. 330-4.

# Abbreviations

| | |
|---|---|
| A/E | absorption in low field, emission in high field |
| AMBER | assisted model building and energy refinement |
| B3LYP | Becke 3-term correlation, Lee-Yang-Parr exchange |
| BRW | Bloch-Redfield-Wangsness |
| CANOE | chemically amplified nuclear Overhauser effect |
| CD | circular dichroism |
| CGS | centimetre-gram-second |
| CIDEP | chemically induced dynamic electron polarization |
| CIDNP | chemically induced dynamic nuclear polarization |
| COSY | correlation spectroscopy |
| CPU | central processing unit |
| CSA | chemical shift anisotropy |
| CSGT | continuous set of gauge transformations |
| CW | continuous wave |
| DABCO | 1,4-diazabicyclo(2.2.2)octane |
| DD | dipole-dipole |
| DDI | distributed data interface |
| DDR | double data rate |
| DFT | density functional theory |
| DMSO | dimethyl sulfoxide |
| DNA | deoxyribonucleic acid |
| DNP | dynamic nuclear polarization |
| DVD | digital video disk |
| E/A | emission in low field, absorption in high field |
| EPR | electron paramagnetic resonance |
| FID | free induction decay |
| FMN | flavin mononucleotide |
| FMOC | fluorenylmethoxycarbonyl |
| GAMESS | general atomic and molecular electronic structure system |
| GB | gigabyte |
| GdnHCl | guanidinium chloride |
| GFP | green fluorescent protein |
| GIAO | gauge including atomic orbital |
| HF | Hartree-Fock |
| HFC | hyperfine coupling constant |
| HOESY | heteronuclear Overhauser effect spectroscopy |
| HOMO | highest occupied molecular orbital |
| IUPAC | international union for pure and applied chemistry |
| LAM | local area multicomputer |
| LC-MS | liquid chromatography with mass-spectroscopic detection |
| MPI | message passing interface |
| Nd:YAG | neodymium-doped yttrium/aluminium garnet |
| NMR | nuclear magnetic resonance |
| NOE | nuclear Overhauser effect |
| NOESY | nuclear Overhauser effect spectroscopy |
| ODE | ordinary differential equation |
| PBS | phosphate buffered saline |
| PCM | polarizable continuum model |
| PDB | protein data bank |
| PM3 | parameterized model 3 |
| QNP | quadruple-nucleus probe |
| R- | restricted |
| RAM | random access memory |
| RF | radiofrequency |
| RMS | root mean square |
| SASA | solvent-accessible surface area |
| SCF | self-consistent field |
| SHMEM | shared memory |
| SI | Système International |
| SO(3) | special orthogonal group of degree 3 |
| NSu | N-succinimide |
| SU(2) | special unitary group of degree 2 |
| tBu | tert-butyl |
| TCP/IP | transmission control protocol / internet protocol |
| TFA | trifluoroacetic acid |
| TLC | thin layer chromatography |
| TR | time-resolved |
| TTL | transistor-transistor logic |
| U- | unrestricted |
| UV | ultraviolet |
| ZFS | zero-field splitting |